\documentclass[acmsmall, screen=true]{acmart}

\usepackage{xspace}

\usepackage{url}

\usepackage{booktabs} %

\usepackage{listings,color}

\usepackage{subcaption}
\usepackage{dcolumn}
\usepackage{graphicx} 
\usepackage{amsmath}
\usepackage{amssymb}
\usepackage[final]{changes}
\def\BibTeX{{\rm B\kern-.05em{\sc i\kern-.025em b}\kern-.08emT\kern-.1667em\lower.7ex\hbox{E}\kern-.125emX}}

\newcommand{\Teamonly}{\textcolor{black}{Single-group}\xspace}
\newcommand{\teamonly}{\textcolor{black}{single-group}\xspace}
\newcommand{\Intergroup}{\textcolor{black}{Intergroup}\xspace}
\newcommand{\intergroup}{\textcolor{black}{intergroup}\xspace}

\newcommand{\teamonlyusers}{single-group members\xspace}
\newcommand{\Intergroupusers}{Intergroup members\xspace}
\newcommand{\intergroupusers}{intergroup members\xspace}
\newcommand{\teamonlyuser}{single-group member\xspace}
\newcommand{\intergroupuser}{intergroup member\xspace}
\newcommand{\communityname}[1]{{\sf #1}\xspace}

\newcommand{\variablename}[1]{\text{\it\small #1}}
\newcommand{\para}[1]{\noindent{\bf #1}\xspace}
\newcommand{\secref}[1]{Section~\ref{#1}\xspace}
\newcommand{\figref}[1]{Figure~\ref{#1}\xspace}
\newcommand{\tableref}[1]{Table~\ref{#1}\xspace}

\newcommand{\fightin}{Fightin-Words\xspace}

\newcolumntype{L}{D{.}{.}{2,5}}

\graphicspath{{graphics/}}
\setcopyright{acmcopyright}
\acmJournal{PACMHCI}
\acmYear{2019} 
\acmVolume{3} 
\acmNumber{CSCW} 
\acmArticle{193} 
\acmMonth{11} 
\acmPrice{15.00}
\acmDOI{10.1145/3359295}

\begin{document}

\title[Characterizing Language Differences between Intergroup and Single-group Members]
{Intergroup Contact in the Wild: Characterizing Language Differences 
between Intergroup and Single-group Members in NBA-related Discussion Forums}

\author{Jason Shuo Zhang}
\email{jasonzhang@colorado.edu}
\affiliation{%
  \institution{University of Colorado Boulder}
  \city{Boulder}
  \state{CO}
  \postcode{80309}
  \country{USA}
}

\author{Chenhao Tan}
\email{chenhao@chenhaot.com}
\affiliation{%
  \institution{University of Colorado Boulder}
  \city{Boulder}
  \state{CO}
  \postcode{80309}
  \country{USA}
}

\author{Qin Lv}
\email{qin.lv@colorado.edu}
\affiliation{%
  \institution{University of Colorado Boulder}
  \city{Boulder}
  \state{CO}
  \postcode{80309}
  \country{USA}
}

\thanks{We thank anonymous reviewers and members of the NLP+CSS research group at CU Boulder for their insightful comments and discussions; 
Maria Deslis for illustrating \figref{fig:illustration};
Scott Fredrick Holman from the CU Boulder Writing Center for his feedback 
and support during the writing process;
Jason Baumgartner for sharing the dataset that enabled this research.
This work is supported in part by the US National Science
Foundation (NSF) through grant CNS 1528138.}
\received{April 2019}
\received[revised]{July 2019}
\received[accepted]{September 2019}

\begin{abstract}
Intergroup contact has long been considered as an effective strategy to reduce prejudice between groups. However, recent studies suggest that exposure to opposing groups in online platforms can exacerbate \deleted{political} polarization.
To further understand the behavior of individuals who actively engage in intergroup contact in practice, we provide a large-scale observational study of intragroup behavioral differences between members with and without intergroup contact.
We leverage the existing structure of NBA-related discussion forums on Reddit to study 
the context of 
professional sports.
We identify fans of each NBA team as members of a group and trace whether they have intergroup contact. 
Our results show that members with intergroup contact use more negative and abusive language  in their affiliated group than those without such contact, after controlling for activity levels.
We further quantify different levels of intergroup contact and show that there may exist nonlinear mechanisms regarding how intergroup contact relates to intragroup behavior. 
Our findings provide complementary evidence to experimental studies in 
a novel context,
and also shed light on possible reasons for the different outcomes in prior studies.
\end{abstract}



\maketitle

\section{Introduction}

Driven by the growing concerns of tribalism and polarization 
in world politics~\cite{jamieson2008echo,chua2018political,sunstein2009going}, 
it is increasingly important to understand intergroup contact as a straightforward yet 
potentially powerful strategy to reduce prejudice between groups. 
Intergroup contact refers to interactions between members of different groups, and groups
can be defined using a variety of factors, including political ideology, 
place of origin, and race. 
A key hypothesis 
is that members with intergroup contact 
(henceforth ``\intergroupusers'') behave differently, e.g., by showing sympathy towards
other groups and voicing different opinions in their affiliated group~\citep{pettigrew2006meta,pettigrew2008does,dovidio2017reducing,dovidio2003intergroup,pettigrew1997generalized,pettigrew2011recent,Bail201804840,nyhan2010corrections,bail2014terrified,lee2018does,gillani2018me}. 
However, prior studies have observed different effects of intergroup contact.
For instance, self-reported surveys show that intergroup contact relates 
to reduced prejudice towards immigrants in European countries
~\cite{pettigrew1997generalized},
while a recent experimental study finds that exposure to opposing groups on Twitter can exacerbate 
political 
polarization~\cite{Bail201804840}.
Although self-reported surveys and experimental studies have been 
the main methods
for studying the effect of intergroup 
contact \cite{nyhan2010corrections,bail2014terrified,lee2018does,camargo2010interracial,boisjoly2006empathy,green2009tolerance,page2008little},
we believe that observational study allows researchers to characterize intergroup contact in the wild and provide valuable complementary evidence \added{in diverse contexts}.
Indeed, with the emergence of online groups, 
it has become possible to observe intergroup contact and individual behavior 
at a massive scale for substantial periods.
Our goal in this study is to investigate how individuals 
that choose to engage in intergroup contact behave differently from others without intergroup contact in their original affiliated group in online platforms.
We leverage the existing structure of NBA-related discussion forums on Reddit
to identify the group affiliations of users and intergroup contact \added{in the context of professional sports, a novel domain different from politics}. 
We choose online fan groups of professional sports teams as a testbed
for the following reasons. First, professional sports play a 
significant role in modern life~\cite{wenner1989media,cashmore2005making,guttmann2004ritual}.
People in the United States spent 
more than 31 billion hours watching sports games in 2015~\cite{nielsenreport}, 
and the attendance of the 2017-2018 National Basketball Association (NBA) season 
reached 22 million~\cite{nbaattend2018}.
Second, professional sports teams are unambiguously competitive in nature. 
Similar to other common contexts for studies on intergroup contact (e.g., political ideology),
fans of sports teams can treat fans of opposing teams 
as enemies and sometimes even engage 
in violence~\cite{roadburg1980factors,frosdick2013football}. 
Moreover, sports fans tend to think that the media and supporters from opposing teams
are likely to have unfair opinions against their favored teams, 
just like people with different \deleted{political} ideologies~\cite{fansbiased}.

\begin{figure}
    \centering
    \includegraphics[width=0.75\textwidth]{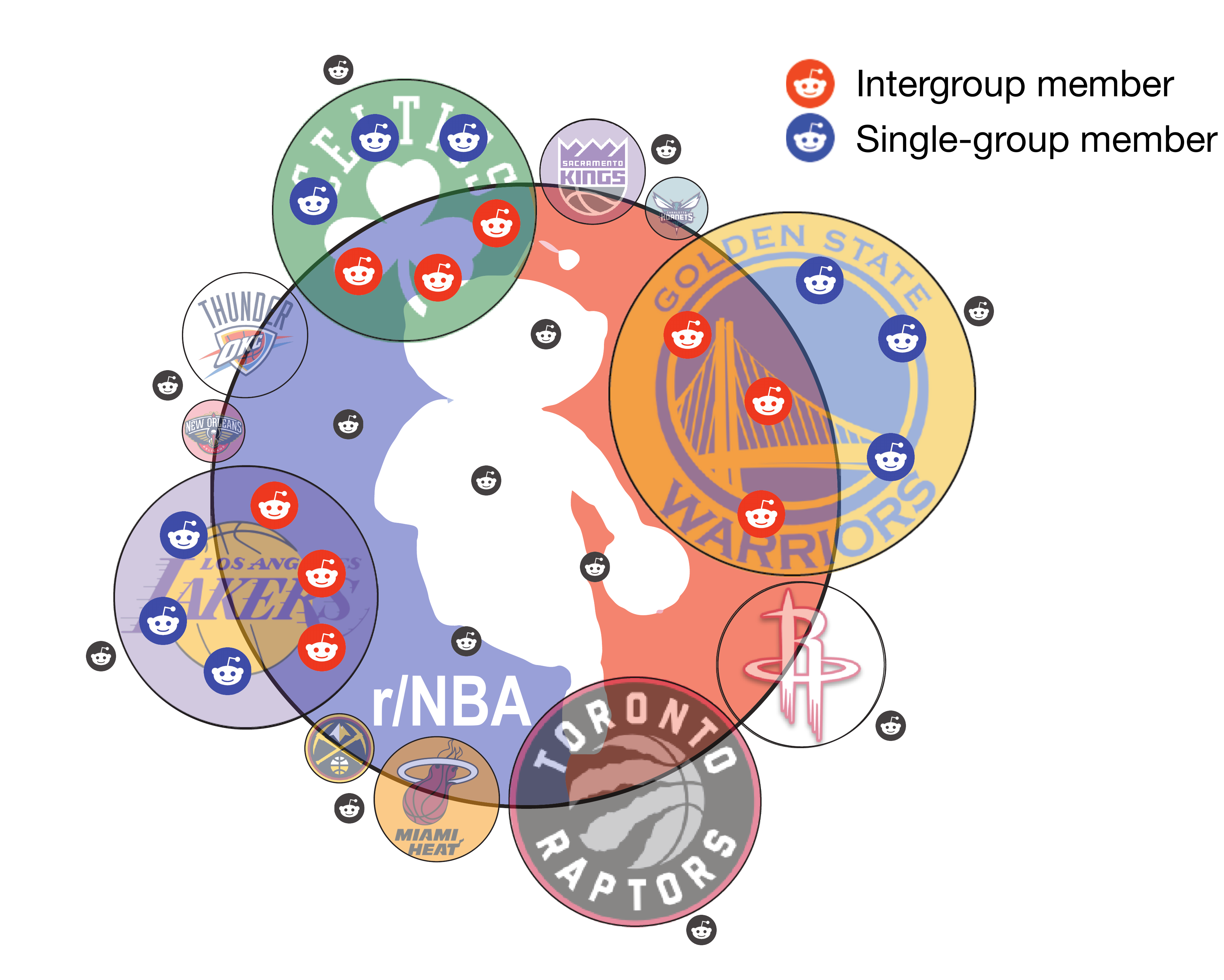}
    \caption{
    Illustration of NBA-related discussion forums (also known as subreddits) on Reddit.
    We identify group affiliation (i.e., whether a person is a fan of an NBA team) and intergroup contact based on the existing structure of NBA-related subreddits on Reddit.
    Each team has its team subreddit.
    Here, we present 11 of the 30 NBA teams (with corresponding team logos) to cover subreddits of different sizes.
    The central \communityname{/r/NBA} logo represents \communityname{/r/NBA}, where intergroup contact happens.
    The radius of each logo is proportional to the number of subscribers in the corresponding subreddit.
    Users in each team logo represent fans of a team based on their indicated support in NBA-related subreddits.
    Red icons refer to \intergroupusers who have engaged in intergroup contact and are thus also in the \communityname{/r/NBA} logo, 
    while blue icons refer to \teamonly users without such behavior.
    We only put red and blue icons in the three largest team subreddits due to space limitations, but every team subreddit has these two categories
    of users (see \figref{fig:overallnumofusers} for the number of \intergroup
    and \teamonlyusers of each team). 
    Note that not all users who participated in these discussion forums qualify as a fan of an NBA team (grey icons).
    }
    \label{fig:illustration}
\end{figure}

\figref{fig:illustration} illustrates our framework. 
There are 30 teams in the NBA, and every team has its
discussion forum (henceforth {\em team subreddit}) on Reddit, a place where fans of the corresponding team congregate and discuss
news, games, and any other topics that are relevant to the team. 
The low-access barrier on the Internet also enables users to communicate easily with fans from opposing teams.
In fact, \communityname{/r/NBA} is dedicated to interactions 
between fans of all NBA teams for any discussion related to the NBA. 
Contrary to each team's ``echo chamber,'' which is dominated by fans that support the same team,
\communityname{/r/NBA} represents an open and diverse environment where intergroup contact occurs. 
We can thus identify  \intergroupusers and \teamonlyusers based on whether they have
any activity in the intergroup setting (\communityname{/r/NBA}).
As posting comments is a major activity in online platforms such as Reddit,
analyzing the language used in online platforms provides an opportunity 
for capturing 
individuals' attitudes and emotions
~\cite{de2014characterizing,park2015automatic,schwartz2013personality,quercia2012personality}.
In particular, there has been growing concerns about hate speech 
and negative language in online communities \citep{hateoffensive,cheng+dnm+leskovec:2015,cheng2017anyone,seering_shaping_2017,chandrasekharan2017you}.
We thus focus on characterizing 
the differences in language usage between 
intergroup and \teamonlyusers in {\em their affiliated team subreddit} (the intragroup setting), 
e.g., whether \intergroupusers 
swear more than 
\teamonlyusers in their affiliated team subreddit.
As a result, we would be able to capture behavioral differences reflected in language use between intergroup and \teamonlyusers in the intragroup setting.
Note that in this work, we do not claim that intergroup contact causes such differences due to 
endogenous factors that may lead to individuals choosing to engage in intergroup contact 
in the wild (i.e., individuals who choose to engage in intergroup contact in practice 
may be inherently different from those who do not).

\smallskip
\para{Organization and highlights.}
We start by summarizing related work to put our work in context (\secref{sec:related}).
We then introduce our dataset and provide an overview of the framework 
for identifying group affiliations and intergroup contact in \secref{sec:setup}.
With \intergroupusers and \teamonlyusers identified, 
We investigate two research questions in the rest of the paper (methods in \secref{sec:methods} and results in \secref{sec:results}):

{\bf RQ1:} \textit{How do members with intergroup contact 
differ from those without such contact in intragroup language usage in NBA fan groups?}

{\bf RQ2:} \textit{How do different levels of intergroup contact relate to intragroup language usage?}

For \textbf{RQ1}, we first apply matching techniques to make sure 
the \intergroup and \teamonlyusers are comparable. 
We then analyze the behavioral differences between \intergroup and
\teamonlyusers by examining language usage of their comments
in their affiliated team subreddit.
We demonstrate intriguing contrasts between them: 
\intergroupusers tend to use more negative and swear words, and generate
more hate speech comments compared to \teamonlyusers in their affiliated 
team subreddit.

For \textbf{RQ2}, we are able to quantify different levels of intergroup contact 
for each \intergroupuser based on the frequency of intergroup contact.
Interestingly, we find varying mechanisms of how different levels of intergroup contact relate to intragroup behavioral differences.
For instance, although intergroup contact mostly monotonically relates to differences in language usage, the trends are not necessarily linear.
Such varying mechanisms provide complementary evidence to the seemingly conflicting results on intergroup contact in recent studies.

To explore the potential reasons behind the clear behavioral differences
in language usage between \intergroup and \teamonlyusers, 
we further compare the language usage of \intergroupusers
between the intragroup setting (affiliated team subreddit) and 
the intergroup setting (\communityname{/r/NBA}) in \secref{sec:twofaces}. 
This setup naturally controls for the subject because we compare the same
person across two different environments. 
We find that \intergroupusers are even more negative and more likely to swear
in the intergroup setting. 
Such negative intergroup contact may partly explain the observed differences
in intragroup language usage. 

Our work highlights the fact that individuals selectively choose to have intergroup contact in the wild, 
and in turn 
interact with people without intergroup contact in their original group.
We further demonstrate a variety of ways in which intergroup contact
levels can moderate intragroup behavior. 
These observations may reconcile recent conflicting results
with respect to intergroup contact.
Our findings 
indicate that observational studies can provide important
complementary evidence to experimental studies on this topic
because interventions can hardly result in deep and regular contacts.
We offer discussions in \secref{sec:discussion} and conclude our work in \secref{sec:conclusion}.

\section{Related Work}
\label{sec:related}

In this section, we first discuss 
studies that use language as a lens to understand human behavior, especially recent studies on the use of negative language in the context of antisocial behavior.
Next, we explain the growing concerns of tribalism, echo-chambers, and polarization, and highlight our specific context, sports, as a testbed for understanding these issues.
We 
then discuss the role of intergroup contact in affecting individual opinions towards opposing groups, 
including recent work on its backfire effect in online platforms.
\added{Finally, we point out opportunities in online sports discussion forums for 
understanding human behavior.}

\subsection{Language as a Lens of Human Behavior}
The proliferation of textual content online has inspired a vast body of literature to understand the language in online communication and its relationship with individual attributes.
Prior research in CSCW and related communities has investigated 
how language can reflect properties of individuals \citep{toma2010reading,naaman2010really,de2014characterizing,park2015automatic,schwartz2013personality,quercia2012personality,tan2015all}.
For instance, \citet{toma2010reading} 
show that linguistic emotions correlate with deception in online dating profiles;
\citet{de2014characterizing} uses linguistics style features to show that mothers with post-partum depression are more likely to use first-person singular pronouns and swear words;
\citet{naaman2010really} conduct a quantitative analysis of message content from over 350 Twitter users 
to characterize the type of messages posted on the platform and broadly classify users as
self-broadcasters and informers.
In general, users' demographic information and personality can also be predicted based on 
linguistic features extracted from textual social media data \cite{minamikawa2011blog,minamikawa2011personality,markovikj2013mining,golbeck2011predicting}.

Recently, 
negative language use
has been examined in the context of antisocial behavior in 
online communities \citep{cheng2017anyone,blackburn2014stfu,seering_shaping_2017,chandrasekharan2017you,cheng+dnm+leskovec:2015,hateoffensive,cheng2014community}. 
Conceptually related is a prior study on the effects of community feedback on user behavior, which reveals that
negative feedback can lead to future antisocial behavior \cite{cheng2014community}. 
\citet{cheng2017anyone} further design an experiment 
that shows negative mood expressed from textual content
increases the likelihood of trolling in online platforms.
A supervised learning model proposed by \citet{blackburn2014stfu} indicates that negative sentiments
are useful for predicting toxic players in online games.
In this work, we compare the differences in language usage
between intergroup and \teamonlyusers, with a focus on expressions of emotions.

\subsection{Tribalism, Echo Chambers, and Polarization}
A battery of studies in %
social sciences has shown 
that human behavior is shaped by our need to belong to a group
and by our proclivity to hate rival groups 
\citep{pettigrew2006meta,tajfel1982social,cikara2011usversusthem}.
Such behavior has been documented in a wide variety of contexts.
In the political context, for instance, recent studies find that 
``liberal group'' and ``conservative group'' on social media not only rarely talk to each other, 
but also use different hashtags and links to various websites within their tweets
~\citep{mason2018uncivil,mapping2014smith,gruzd2014investigating,grevet2014managing,pearce2014climate}.
Another commonly studied context is brand communities in the marketing literature \citep{cova2006brand,muniz2001brand,hickman2007dark,beal2001no,guerra2013measure}.
For example, \citet{hickman2007dark} show that in-groups
are strongly motivated to develop negative views of out-groups and engage in ``trash talking'' about out-groups. 
In-group members will also 
gain pleasure at the misfortune of rival brands and their users.

In the sports context, the team sports literature focus on the negative
consequences of rivalry, such as negative explicit and implicit attitudes
towards the opposing team \citep{cikara2011usversusthem,lehr2019outgroup}, 
\textit{schadenfreude}~\citep{havard2014glory,cikara2013their},
and even riot \citep{guilianotti2013football}.
These negative perceptions may even transfer to the sponsors of the rival team:
\citet{dalakas2005balance}
explore the negative sponsorship effects and find that sponsors of disliked NASCAR
drivers are viewed less positively than sponsors of liked drivers. 
Similarly, 
\citet{olson2018rival} finds that brands faced a steep decline 
in sales among Manchester City fans when they announced the sponsorship of the soccer club 
Manchester United, a fierce rival of 
Manchester City.
By examining the intergroup emotions of fans of the Boston Red Sox and 
New York Yankees, \citet{lehr2019outgroup} 
show that pleasure from 
a powerful rival's losses can outstrip that from gains of the supported team.
Given the competitive nature of professional sports and the importance of emotions in fan behavior, we believe that professional sports provide exciting opportunities 
for understanding polarization.

\subsection{Intergroup Contact}
Intergroup contact has long been considered as an effective strategy to reduce
prejudice between groups \citep{dovidio2017reducing}.
For instance, a seminal work by \citet{pettigrew1997generalized} shows 
that intergroup contact relates to reduced prejudice towards immigrants
based on self-reported surveys in France, Great Britain, the Netherlands, and West Germany.
\citet{wright1997extended} find 
correlational evidence that people who knew that an in-group member had an out-group friend 
had less negative intergroup attitudes. 
They also experimentally demonstrate
that providing this information induces more positive attitudes.
\citet{abbott2014makes} examine young people's assertive bystander intentions
in an intergroup (immigrant) name-calling situation and find that greater intergroup contact
is related to higher levels of empathy, higher levels of cultural openness, and reduced
intergroup bias.
From the perspective of language usage online, 
a field experiment designed by \citet{white2015emotion} demonstrates that
Muslim and Christian high-school students who have structured 
Internet intergroup interactions tend to use more affective and positive emotion words, 
and less anger and sadness words.
\citet{kim2018intergroup} test online contact with two distinct out-groups, 
undocumented immigrants and gay people. 
They find that direct online contact
improves attitudes towards both out-groups through positive and negative emotions,
whereas extended online contact reduces negative emotions and improve attitudes towards 
undocumented immigrants. 

However, recent studies on the ``backfire'' effect suggest that 
exposure to opposing groups in online platforms can exacerbate 
political polarization~\citep{Bail201804840,nyhan2010corrections,bail2014terrified,lee2018does}.
\footnote{\citet{wood2016elusive} show that backfire in \citet{nyhan2010corrections} 
is stubbornly difficult to reproduce, which further demonstrates the varying results in recent studies.}
For instance, \citet{Bail201804840} introduce intergroup contact by following a 
Twitter bot that aggregates tweets of opinion leaders from the opposing political ideology 
and find that Republicans who follow a liberal Twitter bot become substantially more conservative.
\citet{lee2018does} use panel data collected in South Korea to investigate the effects of social media usage on changing the political view.
They highlight the role of social media in activating political participation 
and pushing users toward ideological poles. 

A possible way to reconcile such differences in prior literature
is to review the mechanisms that contribute to the positive effects of intergroup contact:
(1) enhancing knowledge about other groups, 
(2) reducing anxiety when facing opposing groups, and 
(3) increasing empathy and perspective-taking~\citep{pettigrew2006meta,pettigrew2008does,dovidio2017reducing,dovidio2003intergroup,pettigrew1997generalized,pettigrew2011recent,wright1997extended}.
Depending on the motivations to engage in intergroup contact and the actual activities during the contact, intergroup contact in online platforms may not necessarily achieve these goals.
We aim to conduct a large-scale observational study to understand the differences between intergroup and \teamonlyusers in their original affiliated group, and also provide some insights on the nature of intergroup contact in \communityname{/r/NBA}.

It is useful to point out that there is 
little work on intergroup contact in the CSCW community.
In the meanwhile, several recent studies provide a characterization of intergroup conflict.
\citet{kumar2018community} examine cases of intergroup conflict across 36,000 communities on Reddit
where users of one community are mobilized by negative sentiment to comment in another community 
and show that less than 1\% communities start 74\% conflicts.
At the community level, by constructing a conflict network between subreddits, 
\citet{datta2019extracting} find that larger subreddits are more likely to be involved in conflicts with a large
number of subreddits, and the main ``targets'' change over time.
However, intergroup contact is different from intergroup conflict as 
it may help improve intergroup attitudes and 
reduce intergroup tensions and conflicts.
Approaching this topic from a CSCW lens raises additional questions about how socio-technical design
decisions can influence the outcomes reported in traditional offline settings.

\subsection{\added{Fan Behavior in Online Sports Communities}}
Despite the important role of sports in modern life, 
sports fan behavior in online communities for 
professional sports remains understudied.
\citet{yu2015world} collect real-time tweets from US soccer fans during
five 2014 FIFA World Cup games to examine soccer fans' emotional
responses in their tweets. 
The quantitative analyses show that fear and anger were the most
common negative emotions and in general increased when the opponent team
scored and decreased when the US team scored. 
\citet{zhang+tan+lv:18} investigate the connection 
between online fan behavior and offline team performance, 
and \citet{Leung2017Effect} study the effect of NFL game outcomes 
on content contribution to Wikipedia.
It is also worth noting that fan behavior 
can differ depending on the environment. \citet{cottingham2012interaction}
demonstrates the difference in emotional energy between fans in sports bars and those attending
the game in the stadium. 
We believe that there exist exciting opportunities in online sports discussion forums
for understanding human behavior, including intergroup contact.

\section{Professional sports as a testbed}
\label{sec:setup}
We focus on the professional sports context derived from NBA-related discussion forums 
(\communityname{/r/NBA} and 30 team subreddits) on Reddit, an active
community-driven platform where 
users can submit posts and make comments. 
These user-created discussion forums are also called ``subreddits''.
Each subreddit has multiple moderators to make sure that posts are 
relevant to the subreddit's theme. 
Over the years, basketball fans all over the world have flocked
to \communityname{/r/NBA}, the site's professional basketball subreddit,
to discuss games in progress, seek meaning in the latest trade rumors,
and debate the legality of calls by the referees.
In fact, \communityname{/r/NBA} has become 
the largest single-sport subreddit with more than 1.9M subscribers~\cite{redditlist}
and one of the most active subreddits on Reddit~\cite{topreddits}.
NBA-related subreddits represent an ideal testbed 
for understanding how intergroup contact relates to intragroup behavior 
because their structure allows us to identify 
many users' team affiliation.
Moreover, \communityname{/r/NBA} is the place for 
all basketball fans to congregate, 
where intergroup contact between fans of different teams happens.

\begin{figure}
    \centering
    \includegraphics[width=0.7\textwidth]{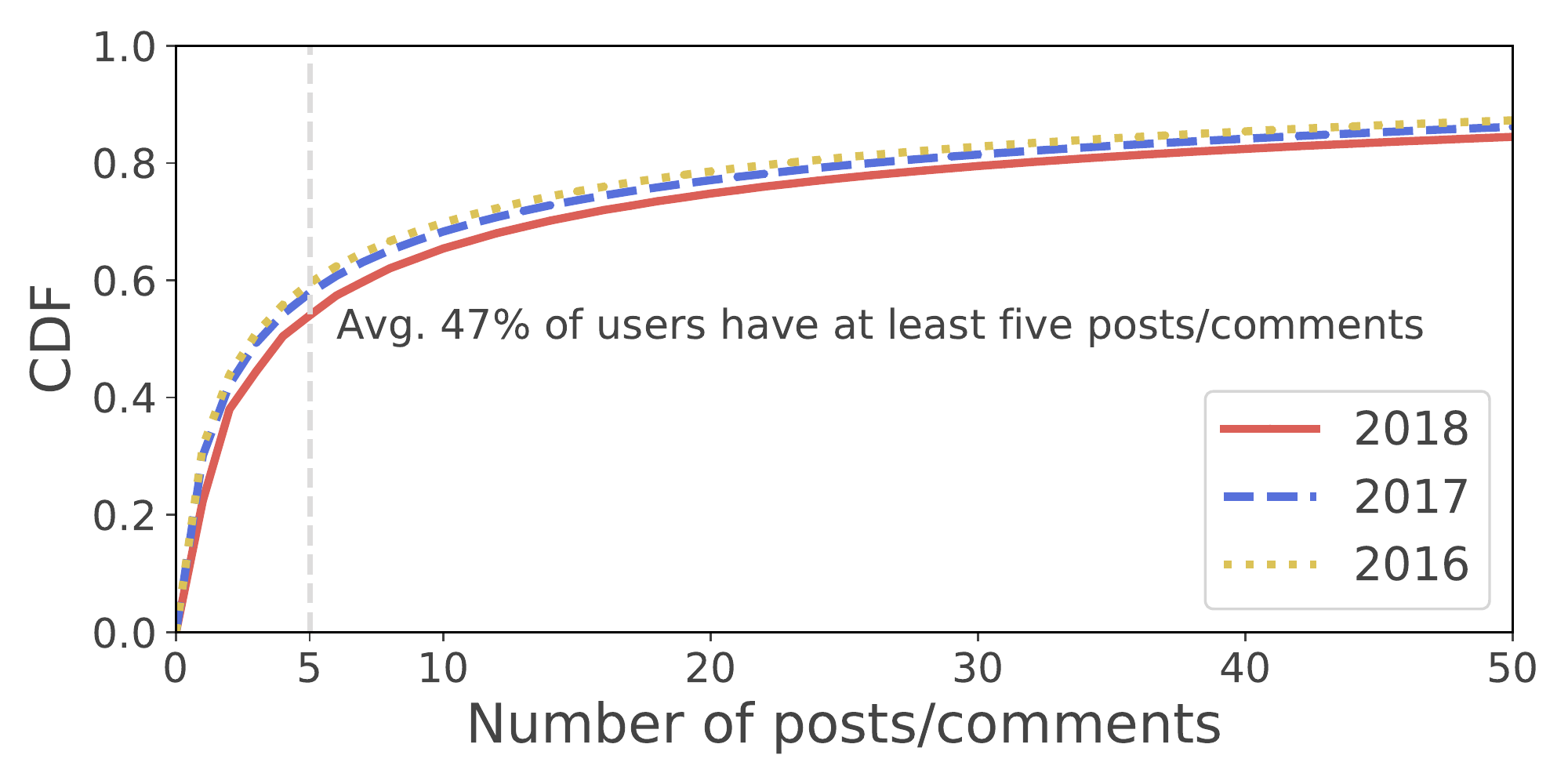}
    \caption{The distribution of the number of activities made by users 
    in NBA-related subreddits in the 2018, 2017, and 2016 seasons.
    }
    \label{fig:CDFNumofPosting}
\end{figure}

\subsection{Dataset and NBA Seasons}
We obtain 2.1M posts and 61M comments in NBA-related subreddits from 
\url{https://pushshift.io} \cite{pushshift}.
As pointed out in \citet{zhang+tan+lv:18}, 
offline NBA seasons are reflected in user behavior 
in these NBA-related subreddits.
We organize our dataset according to the timeline of NBA seasons 
and focus on the most recent three seasons, i.e., from July 2015 to June 2018.
For simplicity and clarity, we refer to a specific season by the calendar
year when it ends. For instance, the official 2017-2018 NBA season is referred to 
as \textit{the 2018 season} or \textit{2018} in this paper.

\subsection{Identifying Team Affiliation and Intergroup Contact}
To identify the team affiliation of users in a season, 
we first define active users in NBA-related subreddits 
in a season as those who have at least five activities, 
where an activity refers to either submitting a post or making a comment.
\figref{fig:CDFNumofPosting} shows the distribution of the number of activities by a user in NBA-related subreddits.
These active users contribute over 95\% of all the activities in NBA-related subreddits.

\begin{figure}
    \center
    \begin{subfigure}[t]{0.48\textwidth}
        \includegraphics[width=\textwidth]{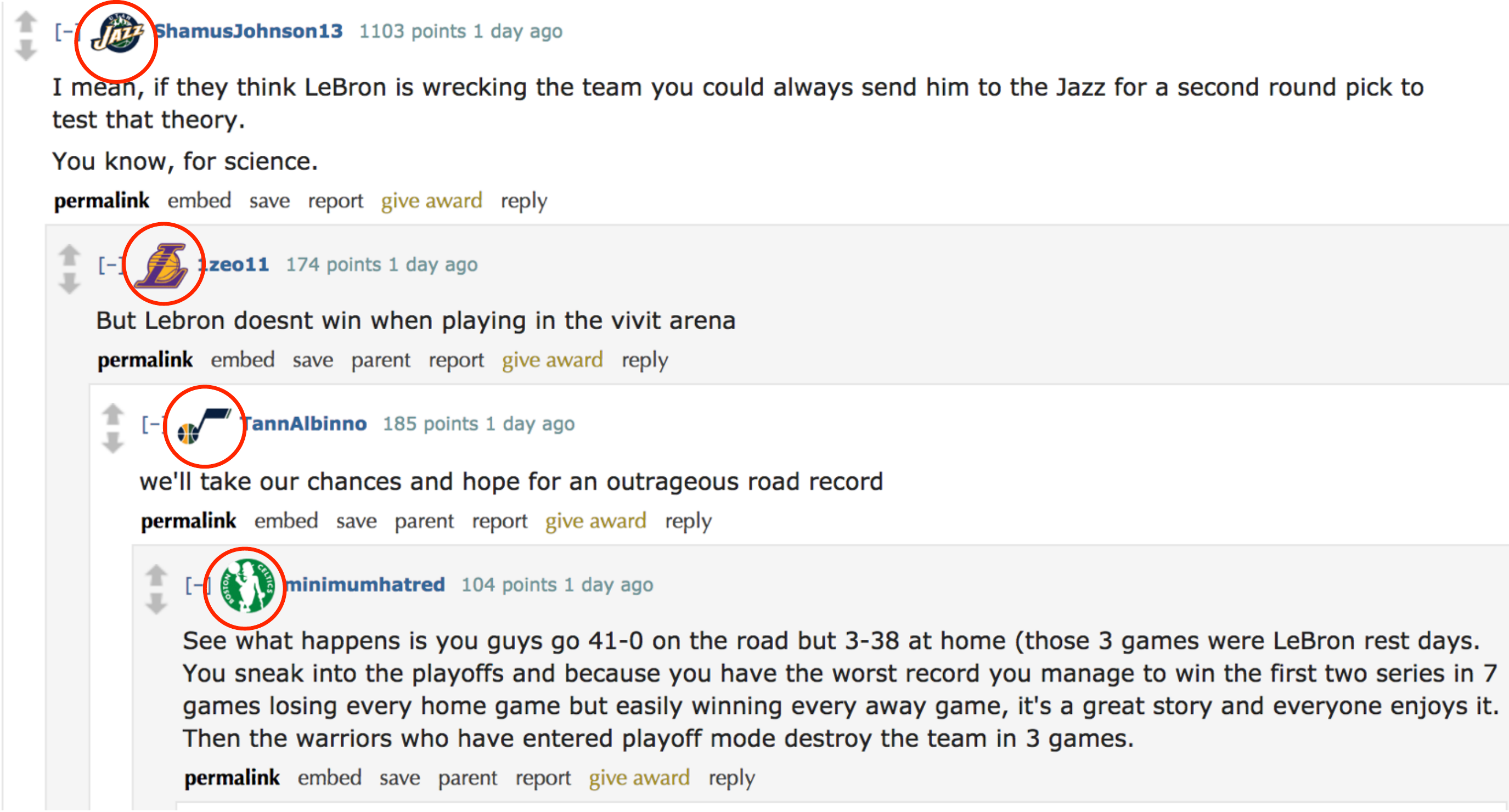}
        \label{fig:designbefore}
    \end{subfigure}
    \hfill
    \begin{subfigure}[t]{0.48\textwidth}
        \includegraphics[width=\textwidth]{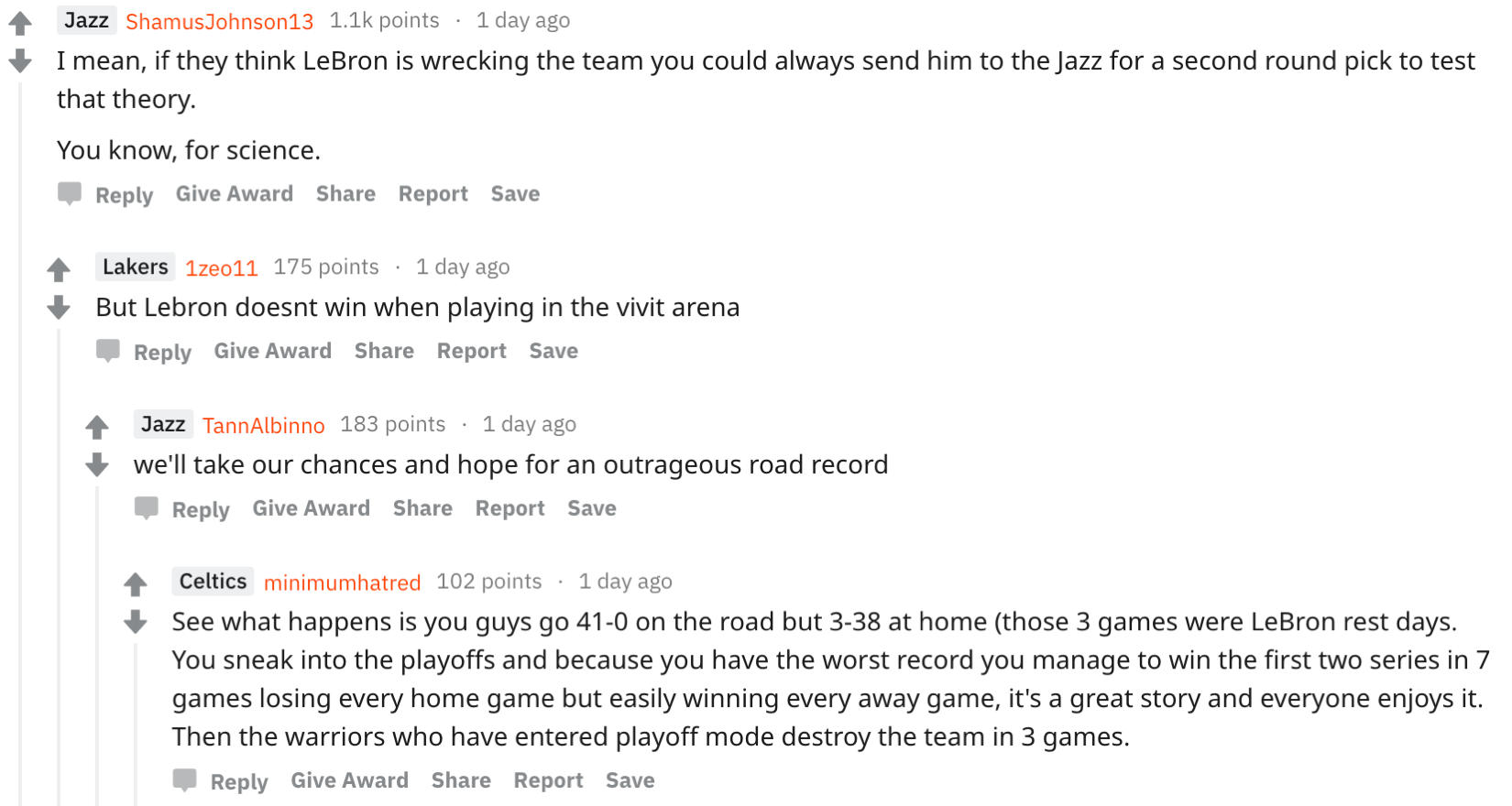}
        \label{fig:designafter}
    \end{subfigure}
    \caption{An example of the flair usage in /r/NBA 
    before and after the design change.
    Before the design change, flairs are represented 
    by team logos while after the change,
    flairs are represented by the team name.  
    }
    \label{fig:FlairChange}
\end{figure}

We identify the team affiliation of active users based on 
where their activities occur and by using a special mechanism on Reddit, known as flair.
Flair appears as an icon next to the username in posts and comments.
Every comment can have at most one flair.
Before April 2018, flairs are represented by team logos. After that, Reddit adopted a new
design to the entire platform, and the flairs are represented by team names in \communityname{/r/NBA}.
An example is shown in \figref{fig:FlairChange}.
In \communityname{/r/NBA}, fans can use flairs to indicate support of a team.
$\sim$80\% of the comments/posts in our \communityname{/r/NBA} dataset
have been made with flairs even though flairs are optional.
We 
use all the flairs that fans used in \communityname{/r/NBA} 
for the inference of their team affiliation 
\footnote{In our \communityname{/r/NBA} dataset, every comment's JSON format has 
the ``author\_flair\_css\_class'' key, and
the corresponding value represents a unique flair this comment uses.
The value is a string with the team's name and the flair id.
For example, the flair values of the Log Angeles Lakers are phrased as
``Lakers1'', ``Lakers2'', ``Lakers3'', and ``Lakers4''.}.

We view posting/commenting in a team subreddit and using a team's flair 
in \communityname{/r/NBA} as an indication of support towards that team.
An active user is defined as a fan of a team 
if the user indicates support only for that team and 
such support sustains over all activities in an entire NBA season.
In other words, all activities of a fan indicate support towards his/her affiliated team.
It follows that not every active user in NBA-related subreddits is identified as a fan of some team.

We further determine whether a fan of a team is exposed to intergroup contact based on 
his/her (lack of) behavior in \communityname{/r/NBA}, which we refer to as intergroup status.
To summarize, we categorize fans of a team into the following two categories:
\begin{itemize}
\item \textbf{\Intergroup}: 
Fans of a team who posted in both the affiliated team subreddit and 
\communityname{/r/NBA} in the season.

\item \textbf{\Teamonly}: 
Fans of a team who had no activity in \communityname{/r/NBA} 
throughout the season.

\end{itemize}

\begin{table}
\centering
\caption{The number of \intergroup and \teamonlyusers in the 2018, 2017, and 2016 seasons.}
\begin{tabular}{lccc}
\toprule
         & 2018 & 2017  & 2016 \\
\midrule
\Teamonly    & 6,023     & 5,941     &   4,843  \\ 
\Intergroup & 28,296      & 24,528    &   20,467 \\
\bottomrule
\end{tabular}
\label{tab:stats}
\end{table}

\tableref{tab:stats} shows the number of members in each category. 
Since our study is concerned with intragroup behavior, i.e., behavior in the 
affiliated team subreddit, %
we view these \intergroup and \teamonly fans as \intergroup and \teamonly members of the affiliated team 
and study their behavior in the affiliated team subreddit.

\begin{figure}
    \center
    \includegraphics[width=0.7\textwidth]
    {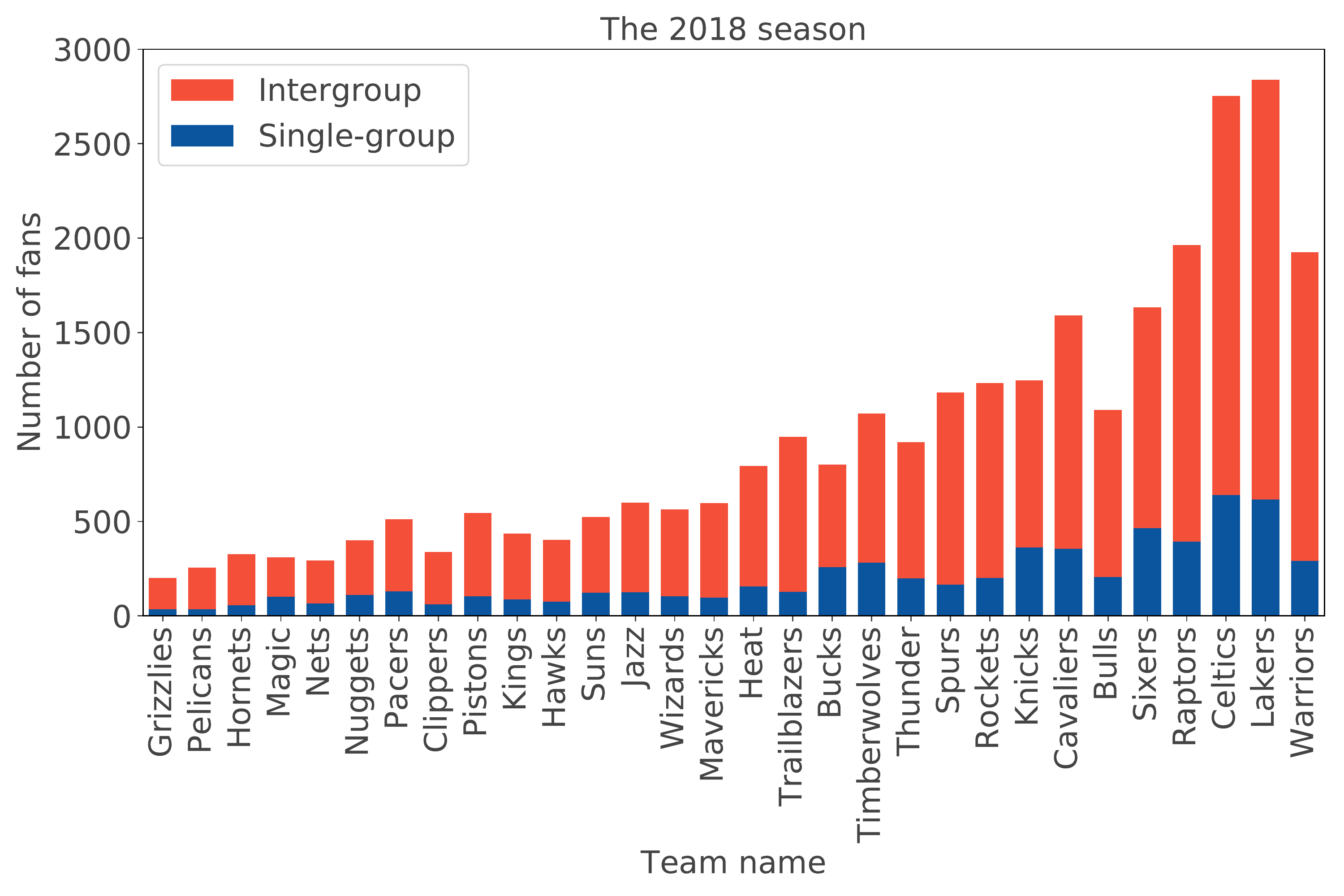}
    \caption{The number of \intergroup and \teamonlyusers 
    affiliated with each NBA team in the 2018 season.
    We rank 30 team subreddits by the number of subscribers 
    each team has by the end of the 2018 season. 
    (see \figref{fig:appoverallnumofusers} for the data statistics in the 2017 and 2016 seasons).
    }
    \label{fig:overallnumofusers}
\end{figure}

\figref{fig:overallnumofusers} presents the number of
\intergroup and \teamonlyusers
in all 30 team subreddits in the 2018 season 
(see \figref{fig:appoverallnumofusers} for the numbers in the 2017 and 2016 seasons).
In every team subreddit, there are many more \intergroupusers than \teamonlyusers.
Our definitions are based on user behavior in a single NBA season, 
and the label of a user can change across seasons. 
However, a single-group member rarely becomes an \intergroupuser 
in the next season in our dataset (6.0\% of \teamonlyusers become \intergroupusers
from 2016 to 2017, and 8.5\% of \teamonlyusers become \intergroupusers
from 2017 to 2018). 
This also confirms the tendency of \teamonlyusers to avoid intergroup contact.

We use both posting and commenting behavior to identify fans' team affiliation, but we 
focus on analyzing comments in the rest of the paper since the posts are usually much longer
and more formal, and are thus not comparable to comments.

\section{Methods}
\label{sec:methods}
We study the behavioral differences between \intergroup and \teamonlyusers
in their affiliated team subreddit
by examining the expression of emotions in their comments in team subreddits
for two reasons:
(1) emotion is a central theme in understanding sports fans and their opinions;
(2) textual content constitutes the main observed behavior 
in NBA-related discussion forums~\cite{zhang+tan+lv:18,lehr2019outgroup,sutton1997creating}.

\subsection{RQ1: Intragroup Behavior Differences}
\label{sec:RQ1}

\noindent{\textbf{Matched \intergroupusers.}}
A na\"ive way to compare these two categories of users is to directly examine all users in each category.
However, such an approach does not take into account other important confounding factors,
such as how active a member is in the group.
We thus seek to ensure that \intergroup and \teamonlyusers are a priori balanced 
on any observable features in the affiliated team subreddit,
which indicates similar loyalty to the team.
To achieve this, we adopt matching techniques: for each \teamonlyuser, we match him/her with 
the most similar unmatched \intergroupuser
from the same affiliated team, where similarity is based on all the
observed features. 
Due to the observational nature, whether a member has intergroup contact or not is not randomly assigned.
In other words, our study reflects the behavioral differences between those who engage in intergroup contact and those who do not.

Following prior studies on factors 
associated with fan behavior in
online sports communities~\cite{zhang+tan+lv:18,mann1989sports,psysportsfan,Leung2017Effect}, 
we consider the following observable feature set for matching:
(1) the number of comments in the affiliated team subreddit, 
(2) the average time gap between comments,
(3) the average length of comments,
(4) the proportion of comments in the playoff season, and
(5) the proportion of comments in the game threads (these game threads are created for discussions during a game).
All the comments examined here are in the members' affiliated team subreddit.
We collect all of these feature values for each season.
The similarities between fans are estimated using the 
nearest neighbor matching technique~\cite{stuart2010matching}.
Min-max normalization is applied to each feature before feeding it into 
the matching model so that no single feature dominates the matching.
We do not include the feedback (upvotes/downvotes) that members received from the NBA subreddits for matching because 
it can be endogenous with the language used in the comments 
(e.g., comments with hate speeches may not get 
many upvotes). 

To evaluate the outcome of our matching procedure,
for each observable feature, we check 
distributional differences between the treatment group (intergroup members) %
and the control group (single-group members). We compare their empirical cumulative distributions before and after
matching using the Mann-Whitney U test~\cite{mann1947test}. 
The results of the 2018, 2017, and 2016 seasons are summarized in 
\figref{fig:matching}. 
A small p-value here indicates that there exists a significant difference 
between the treatment group and the control group.
Prior to matching, the p-value for each feature is close to 0, 
implying that the distributions do differ between groups. 
After matching, we find no 
difference between the treatment group and the matched control group
for any observable feature at the 5\% significance level ($\alpha=0.05$) in all three seasons,
indicating that the data is balanced across all the covariates after matching.

\smallskip

\noindent{\textbf{Language usage analysis.}}
The proportion of emotional words (i.e., positive emotions, negative emotions, and swear words) in members' comments are analyzed
using the Linguistic Inquiry and Word Count software (LIWC \cite{pennebaker2007linguistic}),
a word frequency-based text analysis tool
(see \tableref{tab:negativeexamples} for examples of emotional words detected 
using this software).
The hate speech comments are identified 
using an automated hate speech detection model \cite{hateoffensive}.
It is a multi-class classifier that can reliably separate 
hate speech from other offensive language
(see \tableref{tab:hatespeechexamples} for examples of hate speech comments detected 
using this detection model).
According to \citet{hateoffensive}, the model achieved an overall precision of 0.91,
recall of 0.90, and F1 score of 0.90 on detecting hate speech tweets.

\smallskip

\noindent{\textbf{\fightin model.}}
To identify a list of distinguishing keywords 
that are over-used by \intergroup or \teamonlyusers,
we apply the \fightin algorithm \cite{monroe2008fightin} to compare the 
word frequencies to the background frequencies found in the other fan group's corpora
using the informative Dirichlet prior model.
This method estimates the log-odds ratio of each word $w$ between
two corpora $\alpha$ and $\beta$ given the frequencies obtained from the background corpus $\mathcal{D}$.
Then the log-odds ratio $\delta_w^{(\alpha\text{-}\beta)}$ for word $w$ can be estimated as:
\begin{equation}
\delta_w^{(\alpha\text{-}\beta)} = log\frac{c_w^{\alpha} + c_w^{\mathcal{D}}}
{c^{\alpha} + c^{\mathcal{D}} - c_w^{\alpha} + c_w^{\mathcal{D}}} \
- log\frac{c_w^{\beta} + c_w^{\mathcal{D}}}{c^{\beta} + c^{\mathcal{D}} 
- c_w^{\beta} + c_w^{\mathcal{D}}},
\end{equation}
where $c_w^{\alpha}$ and $c_w^{\beta}$ are the counts of word $w$ 
in corpora $\alpha$ and $\beta$,
$c^{\alpha}$ and $c^{\beta}$ are the counts of all words in corpora $\alpha$ and $\beta$,
$c_w^{\mathcal{D}}$ is the count of word $w$ in the background corpus $\mathcal{D}$,
and $c^{\mathcal{D}}$ is the count of all words in corpus $\mathcal{D}$.
The \fightin algorithm also provides an estimation for the variance of the
log-odds ratio,
\begin{equation}
\sigma^2(\delta_w^{(\alpha\text{-}\beta)}) 
\backsim \frac{1}{c_w^{\alpha} + c_w^{\mathcal{D}}} \
+ \frac{1}{c_w^{\beta} + c_w^{\mathcal{D}}},
\end{equation}
and the corresponding $z$-score can be calculated as follows: 
\begin{equation}
Z = \frac{\delta_w^{(\alpha\text{-}\beta)}}
{\sqrt{\sigma^2(\delta_w^{(\alpha\text{-}\beta)})}}.
\end{equation}

The \fightin model is known to outperform other traditional methods 
in detecting word usage differences between corpora, such as
PMI (pointwise mutual information) \cite{manning1999foundations} and 
TF-IDF \cite{salton1986introduction}, by not over-emphasizing
fluctuations of rare words \cite{monroe2008fightin}.
We use the comments made by \intergroupusers as the background corpus 
for \teamonlyusers and vice versa to identify 
differences in language usage by each fan group 
\footnote{Part of our code is borrowed from Jack Hessel's \fightin model implementation \cite{fightinwords}.}. 
We rank each word by averaging its z-scores calculated by the \fightin model
across all 30 teams. 
A higher positive z-score indicates this word is over-used by \teamonlyusers,
and a higher negative z-score means this word is over-used 
by \intergroupusers.

\subsection{RQ2: Different Levels of Intergroup Contact}
\label{sec:level}
\noindent{\textbf{Matched \intergroupusers with different levels of intergroup contact.}}
We define a user's level of intergroup contact based on 
the fraction of comments in the 
intergroup setting (\communityname{/r/NBA}).
The fraction is calculated as 
the proportion of the number of comments the user made in the
\communityname{/r/NBA} versus the total number of comments
in NBA-related subreddits.
Specifically, for each \teamonly member, we again apply the nearest neighbor
matching technique to find five closest \intergroup members in the
same affiliated team and 
assign a label of 1, 2, 3, 4, or 5 to them based on their 
fraction of comments
in \communityname{/r/NBA} in a complete NBA season.
Different from pairing \intergroup members and \teamonly members before,
we do it with replacement because there are not enough \intergroup members
to conduct this matching uniquely.
As such, an \intergroup member can be matched to
multiple \teamonly members.
We compare the empirical cumulative distributions before and after
matching for each level using the Mann-Whitney U test [35]. 
The results of the 2018, 2017, and 2016 seasons are
presented in Figure \ref{fig:matchinglevel2018}, \ref{fig:matchinglevel2017}, 
and \ref{fig:matchinglevel2016}, respectively.
We also assign a label of 0 to \teamonly members.
A larger label indicates a higher level of intergroup contact that the member
has in \communityname{/r/NBA}. 
We aggregate \intergroupusers at each level across all 30 team subreddits to 
compare their intragroup behavior.
Note that the number of members at each level is the same, but some 
\intergroupusers may be counted more than once.

\smallskip

\noindent{\textbf{Regression analyses of the relationship between different levels of 
intergroup contact and language usage.}}
To understand the relationship between members' intergroup contact level and language usage,
we also conduct  OLS regression analyses after the above matching procedure. 
The independent variables considered in the regression model are the same set of features 
used in matching intergroup and \teamonlyusers (\secref{sec:RQ1}).
We standardize all independent variables before feeding into the regression model.
Our full linear regression model to test each language 
usage pattern is shown below:

\begin{align*}
\variablename{Proportion of language usage} \sim & 
\beta_0 + \beta_1\variablename{number of comments} + 
\beta_2\variablename{average comment hours gap} \\ 
&\quad + \beta_3\variablename{average comment length} 
+ \beta_4\variablename{proportion of playoff comments}  \\
&\quad + \beta_5\variablename{proportion of game thread comments}\\
&\quad + \beta_6\variablename{fraction} + 
\beta_7\variablename{fraction}\times \variablename{level1}
+ \beta_8\variablename{fraction}\times \variablename{level2}\\
&\quad + \beta_9\variablename{fraction}\times \variablename{level3} + 
\beta_{10}\variablename{fraction}\times \variablename{level4} +
\beta_{11}\variablename{fraction}\times \variablename{level5}\\
\end{align*}

The fraction in the linear regression model refers to 
the proportion of the number of comments the user made in the
intergroup setting (/r/NBA) versus the total number of comments
in NBA-related subreddits. 
All the control variables for matching \intergroupusers and \teamonlyusers
are included. 
There are repeated measures in our regression model as 
an \intergroupuser can be matched to more than one \teamonlyuser.
The average number of times for an \intergroupuser to be matched is 1.77
(excluding the \intergroupusers who never get matched).
Among the \intergroupusers who are matched more than once, 
the average variance of their intergroup contact levels in different matches
is 0.27. 
The small variance shows the consistency of our matching technique.

\section{Results}
\label{sec:results}

In this section, we examine intragroup language differences between
\intergroup and \teamonlyusers (RQ1). We further discuss how different levels 
of intergroup contact relate to intragroup behavior (RQ2).

\subsection{RQ1: Intragroup Language Differences}
\label{sec:languagedifference}

\begin{figure}
    \center
    \begin{subfigure}[t]{0.32\textwidth}
        \includegraphics[width=\textwidth]
        {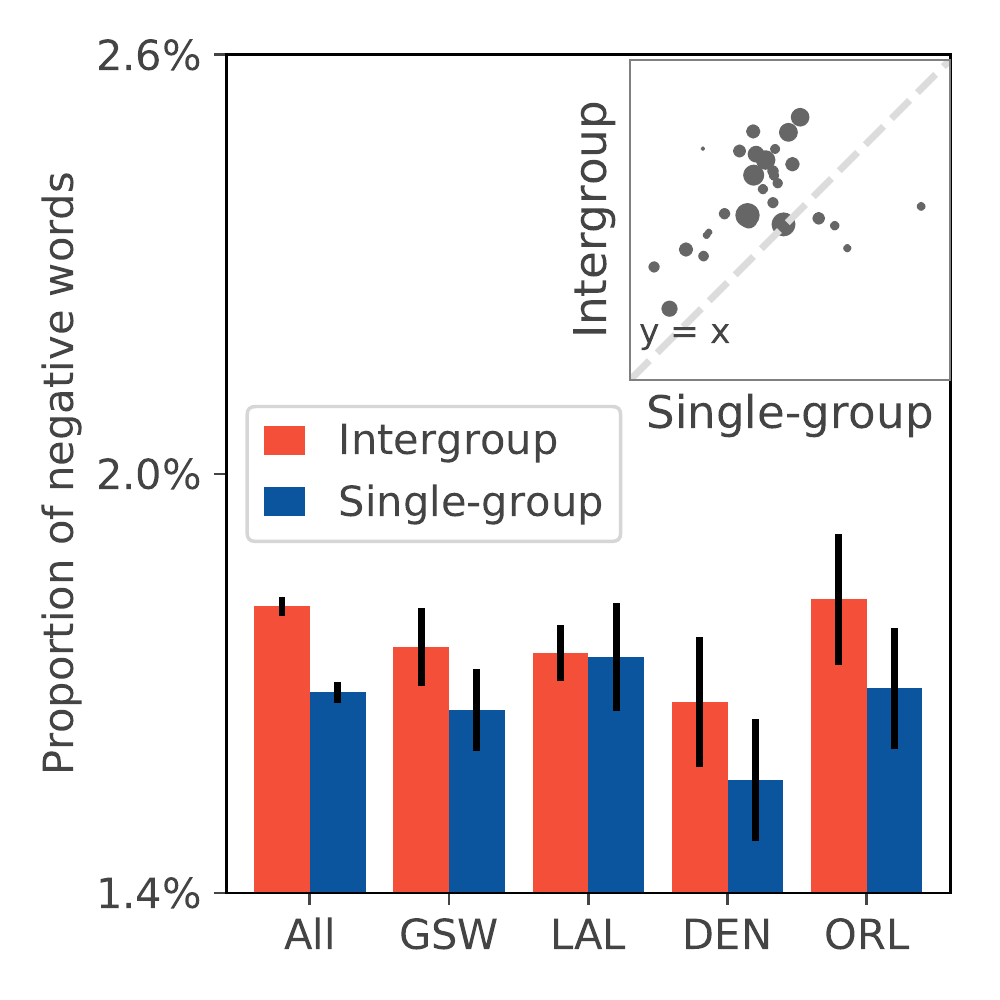}
        \caption{Proportion of negative words.}
        \label{fig:effectneg}
    \end{subfigure}
    \hfill
    \begin{subfigure}[t]{0.32\textwidth}
        \includegraphics[width=\textwidth]
        {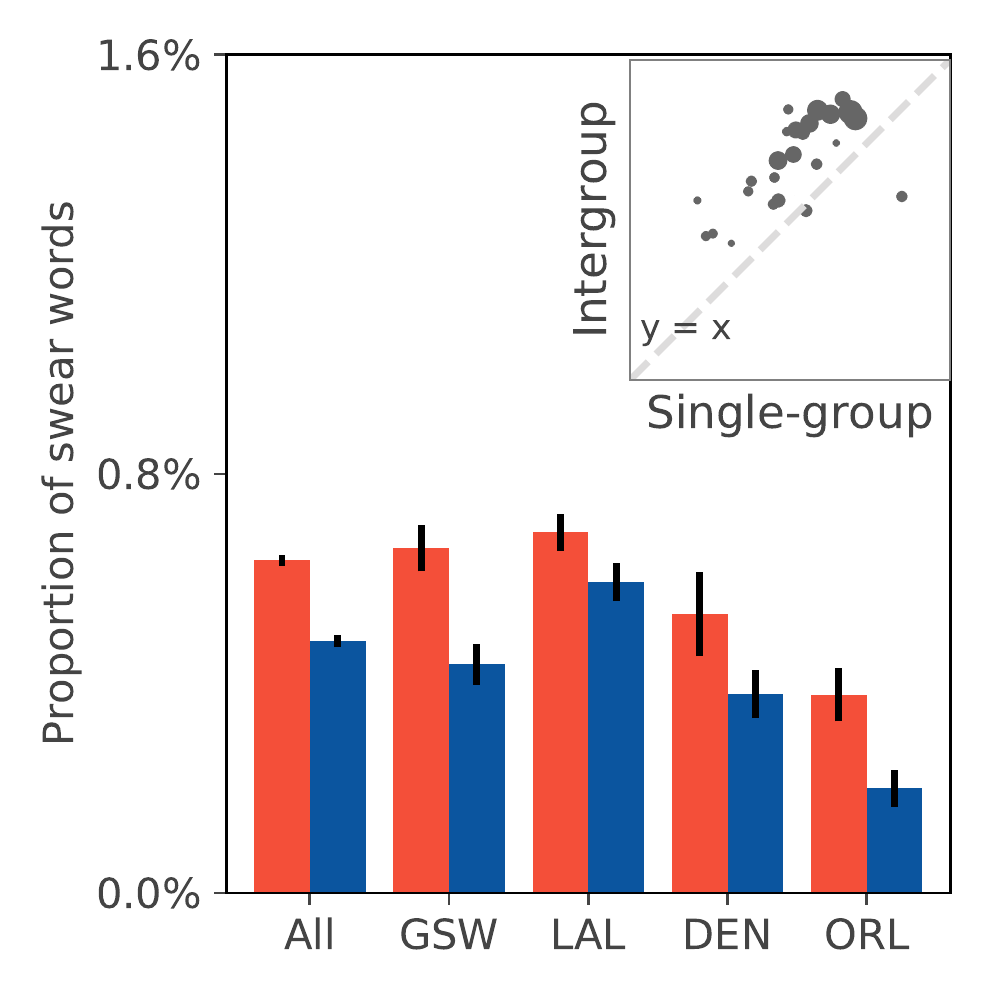}
        \caption{Proportion of swear words.}
        \label{fig:effectswear}
    \end{subfigure}
    \hfill
    \begin{subfigure}[t]{0.32\textwidth}
        \includegraphics[width=\textwidth]
        {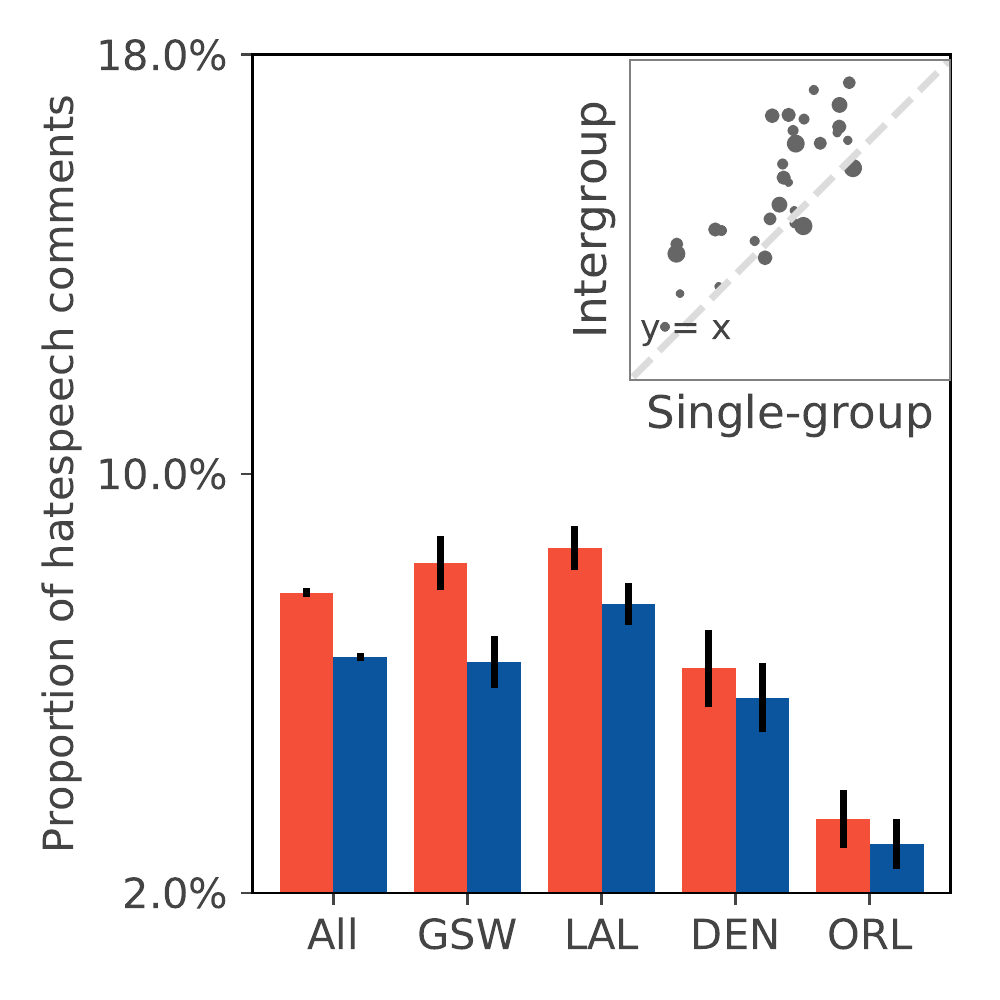}
        \caption{Proportion of hate speech comments.}
        \label{fig:effecthate}
    \end{subfigure}
    \caption{The comparison of language usage between \intergroup and \teamonlyusers
    in the 2018 season.
    \Intergroupusers use more negative words (\figref{fig:effectneg};  
    two-tailed t-test, $t=6.23$, $p<0.001$,
    95\% CI=0.08\% to 0.16\%; 26 out of 30 teams, two-tailed binomial test $p<0.001$) 
    and swear words (\figref{fig:effectswear}; 
    two-tailed t-test, $t=3.51$, $p<0.001$, 95\% CI=0.02\% to 0.08\%; %
    28 out of 30 teams, two-tailed binomial test $p<0.001$), 
    and generate more hate speech comments (\figref{fig:effecthate};
    two-tailed t-test, $t=10.44$, $p<0.001$, 95\% CI=1.00\% to 1.46\%; 
    26 out of 30 teams, two-tailed binomial test $p<0.001$).
    ``All'' is based on concatenating the samples from all 30 NBA team subreddits, 
    and we also show the top two and bottom two teams ranked by the number of subscribers 
    that have at least 100 \teamonlyusers.
    We further show the scatter plot of all 30 teams 
    in the top right to illustrate 
    that the findings are robust across teams (the size of the dot is proportional to the number of subscribers). 
    Error bars represent standard errors.
    The results are consistent in the 2017 and 2016 seasons
    (see \figref{fig:intrasentiment2017} 
    and \figref{fig:intrasentiment2016}).
    }
    \label{fig:intrasentiment2018}
\end{figure}

\figref{fig:intrasentiment2018} compares negative language usage between matched \intergroup and \teamonlyusers.
\Intergroupusers tend to use more negative language
than \teamonlyusers, which is indicated by the use of more negative words 
(two-tailed t-test, $t=6.23$, $p<0.001$, 
95\% CI=0.08\% to 0.16\%; 26 out of 30 teams, 
two-tailed binomial test $p<0.001$)
and swear words 
(two-tailed t-test, $t=3.51$, $p<0.001$, 
95\% CI=0.02\% to 0.08\%; 
28 out of 30 teams, two-tailed binomial test $p<0.001$) 
based on lexicon analysis.
We further compute the proportion of hate speech with an automated hate speech
and offensive language detection model~\cite{hateoffensive}. 
It is consistent that \intergroupusers also generate more hate speech 
(two-tailed t-test, $t=10.44$, $p<0.001$, 
95\% CI=1.00\% to 1.46\%; 
26 out of 30 teams, two-tailed binomial test $p<0.001$).
These results indicate that \intergroupusers are more emotionally charged in their intragroup behavior compared with the \teamonlyusers and
are somewhat different from the hypothesis that intergroup
contact enhances empathy and perspective 
thinking~\cite{batson1997empathy,pettigrew2008does,stephan1999role}.
Our results are consistent when excluding the NBA playoffs (see \secref{sec:regular}).
We also compare positive language usage between \intergroupusers and \teamonlyusers
and do not find a consistent trend at the 5\% significance level 
(see \figref{fig:positive}).

\begin{figure}
    \center
    \includegraphics[width=0.7\textwidth]
    {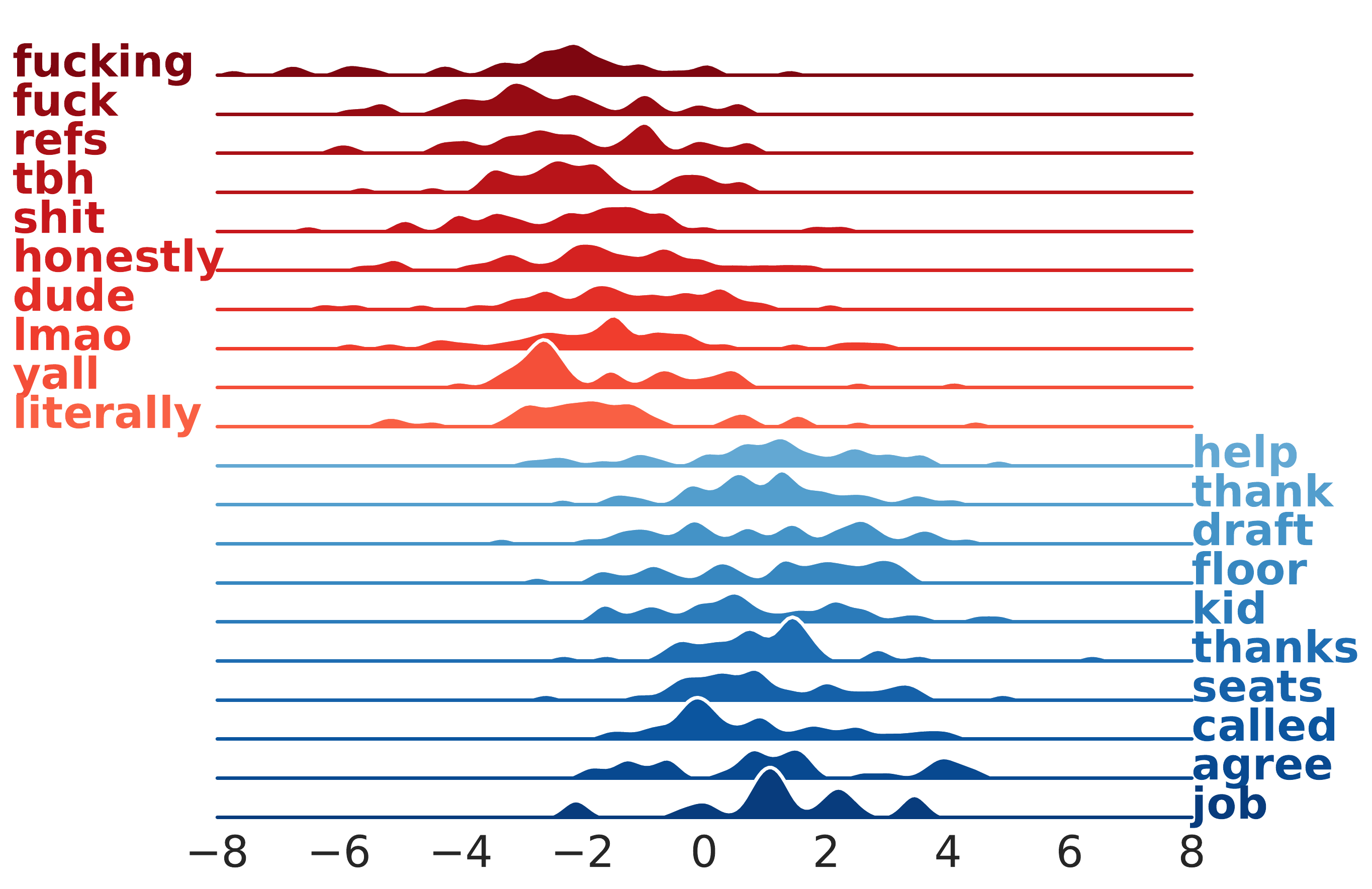}
    \caption{
    The top-10 over-represented words used 
    by \intergroup (red) and \teamonly (blue) members in the 2018 season. 
    For each word, we show the distribution of the z-scores for 
    all 30 teams calculated by the \fightin 
    algorithm \cite{monroe2008fightin}.
    }
    \label{fig:fightinwords}
\end{figure}

To further understand the difference between \intergroup and \teamonlyusers in language usage,
we identify a list of distinguishing words 
that are more likely to be used by \intergroup or by \teamonlyusers, 
using the \fightin algorithm 
with the informative Dirichlet prior model \cite{monroe2008fightin}.
Figure~\ref{fig:fightinwords} lists the top-10 over-represented words used
by \intergroup and \teamonlyusers in the 2018 season.
We rank each word by its average z-score calculated by the \fightin algorithm
across all 30 teams.
A positive z-score indicates that this word is over-used by \teamonlyusers,
while a negative z-score suggests that this word is over-used 
by \intergroupusers.
Our results show that \teamonlyusers are more friendly and calm
when commenting in the affiliated team subreddit and use more polite words, such as ``agree'', ``thanks'', and ``help''.
Also, ``seats'' suggest that some \teamonlyusers
are local fans, as they frequently discuss information about attending live games.
In comparison, 
\intergroupusers use more swear words and talk more about the referees (likely complaining).

\subsection{RQ2: Different Levels of Intergroup Contact}
\label{sec:differentlevel}
In addition to identifying the intragroup behavioral differences, 
our observational study allows us to quantify different levels of intergroup contact,
which can be difficult to operationalize in experimental studies. 
Here, we examine the mechanisms of how increased levels of 
intergroup contact relate to differences in intragroup behavior.

\begin{figure}
    \center
    \begin{subfigure}[t]{0.32\textwidth}
        \includegraphics[width=\textwidth]
        {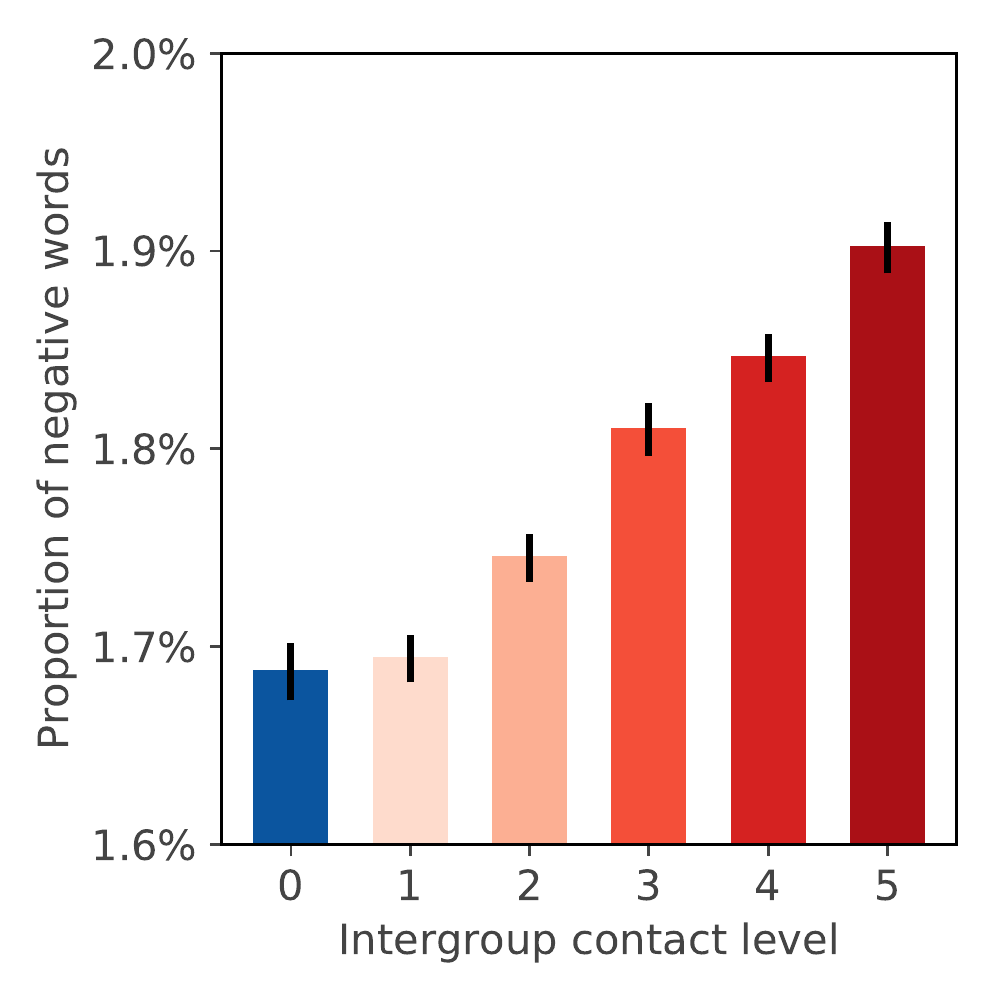}
        \caption{Proportion of negative words.}
        \label{fig:levelneg}
    \end{subfigure}
    \hfill
    \begin{subfigure}[t]{0.32\textwidth}
        \includegraphics[width=\textwidth]
        {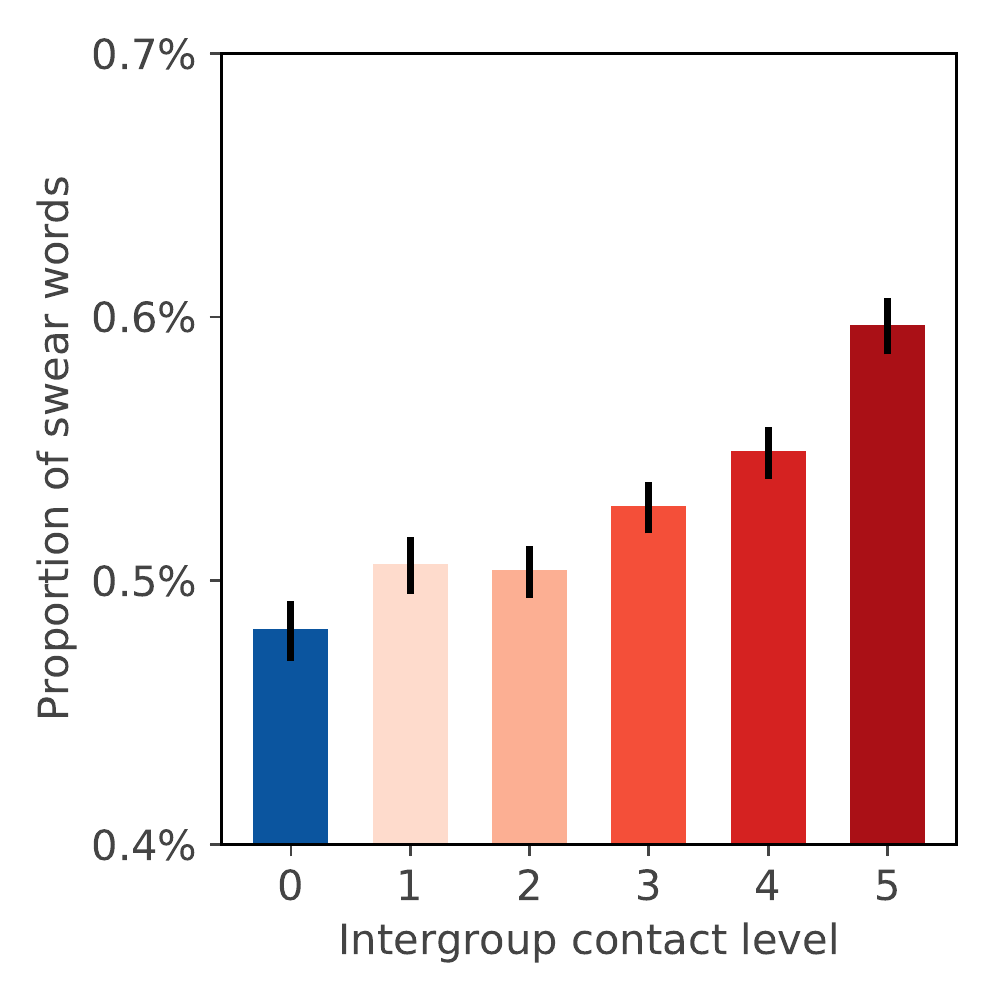}
        \caption{Proportion of swear words.}
        \label{fig:levelpos}
    \end{subfigure}
    \hfill
    \begin{subfigure}[t]{0.32\textwidth}
        \includegraphics[width=\textwidth]
        {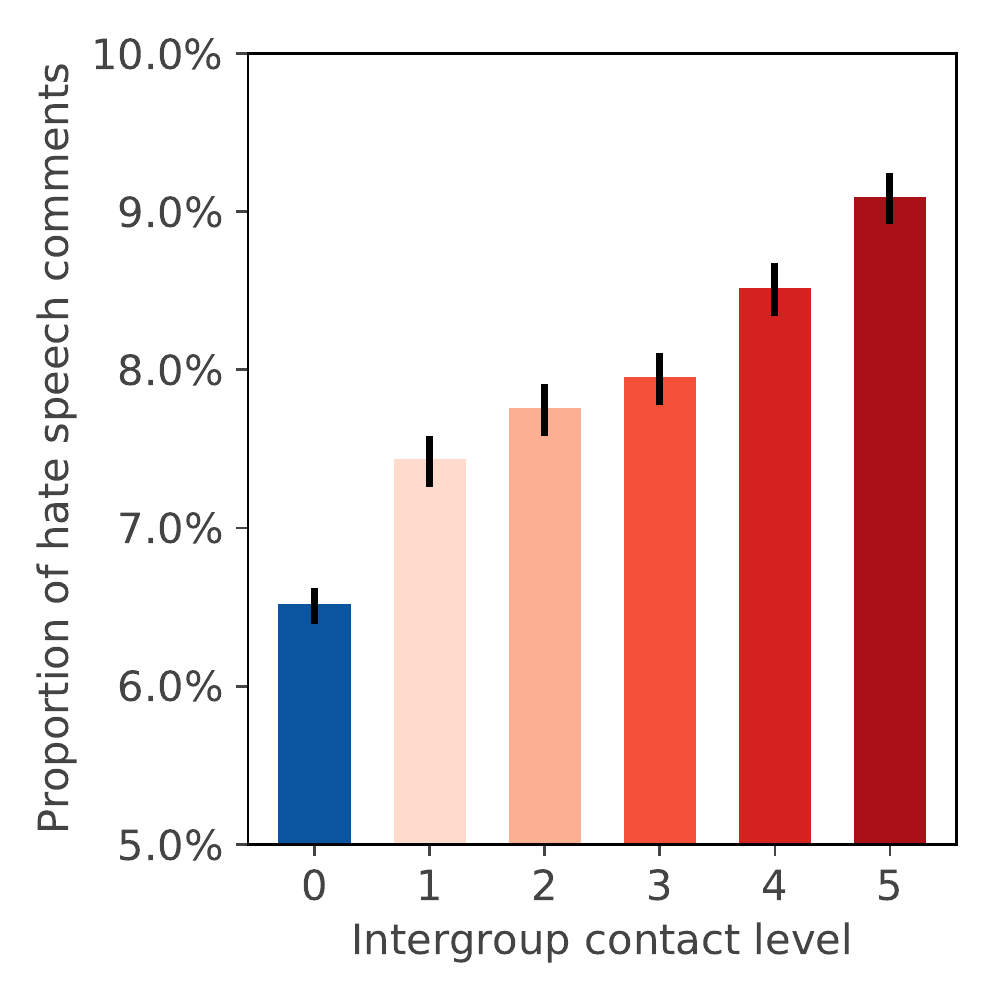}
        \caption{Proportion of hate speech comments.}
        \label{fig:levelhate}
    \end{subfigure}
    \caption{Intragroup language usage differences of members with different
    intergroup contact levels in the 2018 season.
    x-axis represents intergroup levels determined by the fraction of 
    comments in \communityname{/r/NBA}. 
    We observe a consistent monotonic pattern in the proportion of 
    negative words (mean = 1.69\%, 1.69\%, 1.74\%,
    1.81\%, 1.85\%, and 1.90\%, respectively for labels from 0 to 6;), 
    swear words (mean = 0.48\%, 0.51\%, 0.50\%, 0.53\%, 0.55\%, and 
    0.60\%, respectively for labels from 0 to 6), 
    and hate speech comments
    (mean = 6.51\%, 7.42\%, 7.75\%, 7.94\%, 8.51\%, and 9.08\%,  
    respectively for labels from 0 to 6).
    The monotonic trend is consistent in the 2017 and 2016 season
    (see \figref{fig:levellanguage2017} and \figref{fig:levellanguage2016}).
    Error bars represent standard errors.
    }
    \label{fig:levellanguage2018}
\end{figure}

Figure~\ref{fig:levellanguage2018} shows language usage differences between
members with different intergroup contact levels. 
Members of higher intergroup contact levels are generally more negative in language usage:
They tend to use more negative words
(mean = 1.69\%, 1.69\%, 
1.76\%, 1.80\%, 1.85\%, and
1.90\%,  
respectively for labels from 0 to 6%
)
and swear words 
(mean = 0.48\%, 0.50\%, 
0.50\%, 0.53\%, 0.55\%, and 
0.59\%,  
respectively for labels from 0 to 6
),
and generate more hate speech comments 
(mean = 6.51\%, 7.61\%, 7.72\%, 7.77\%, 8.43\%, and 9.15\%,  
respectively for labels from 0 to 6
)
in the affiliated team subreddit.
However, the trends are not necessarily linear.
For instance, \intergroupusers at level 1 do not show significant differences from \teamonlyusers in negative word usage,
while \intergroupusers at level 5 present a significant jump from previous levels 
in negative words, swear words, and the use of hate speech.
\tableref{tab:reg} shows the results of regression analyses.
The fraction of intergroup contact has a statistically significant positive coefficient in regressions for
the proportion of negative words, swear words, and hate speech comments.
Moreover, the coefficients for some interaction terms with levels are also statistically significant (e.g., level 5, $\beta_{11}$, is statistically significant in regressions for
the proportion of negative words, swear words, and hate speech comments), indicating that nonlinear corrections are required.
Note that the BIC score is consistently better by incorporating the interaction terms, although adjusted $R^2$ remains the same due to the fact that this is a very challenging regression task.

\begin{table}[t]
\small
\centering
\caption{Regression analyses for the proportion of negative words, swear words, and
hate speech comments. The variables used for matching \intergroup
and \teamonlyusers are also included for control. For each of the analyses, 
the fraction of intergroup contact has a positive coefficient. 
The number of stars indicates p-values, ***: 
$p<0.001$, **: $p<0.01$ *: $p<0.05$.
}
\begin{tabular}{l|LL|LL|LL}
\toprule
\multicolumn{1}{l|}{}  & \multicolumn{2}{c|}{Prop. of negative words}  & \multicolumn{2}{c|}{Prop. of swear words} & \multicolumn{2}{c}{Prop. of hate speech}\\
\multicolumn{1}{c|}{Variable}   & \multicolumn{1}{c}{Reg. 1}  & \multicolumn{1}{c|}{Reg. 2} 
        & \multicolumn{1}{c}{Reg. 1}                 & \multicolumn{1}{c|}{Reg. 2}    
        &\multicolumn{1}{c}{Reg. 1}                 & \multicolumn{1}{c}{Reg. 2}     
        \\ \midrule
\textit{Control}  &&&&&&\\
number of comments    & 0.003* & 0.003* & 0.023***  & 0.022*** & 0.002* & 0.002*  \\
average comment hours gap &-0.004***&0.004***&-0.017***& -0.017***&-0.003***&-0.003***  \\
average comment length  & -0.028***&-0.028***&-0.000&-0.000&-0.042***&-0.042***  \\
Prop. of playoff comments & 0.006***&0.006***&0.019***& 0.018***&0.004***&0.004***  \\
Prop. of game thread comments & 0.053***&0.053***&0.090***&0.089***&0.028***&0.028***  \\[5pt]
\textit{Fraction}    &&&&&&\\
fraction              & 0.007*** & 0.005***& 0.027***&0.019***& 0.004***&0.003***\\[5pt]
\textit{Levels} &&&&&&\\
fraction $\times$ level1  &  & -0.002  &           & -0.001  & & 0.001\\
fraction $\times$ level2  &  & 0.001  &           & -0.001  & &-0.001\\
fraction $\times$ level3  &  & 0.002**  &           & 0.003  & &0.000\\
fraction $\times$ level4  &  & 0.002**  &           & 0.005*  & &0.001\\
fraction $\times$ level5  &  & 0.002***  &           & 0.010*** & &0.002*** \\[5pt]
 \midrule
intercept        & 0.052***  & 0.053*** & 0.040***  & 0.042***  & 0.028***&0.028***  \\  
Adjusted $R^2$   & 0.172     & 0.172 & 0.033     & 0.033     & 0.138 & 0.138    \\
BIC & \multicolumn{1}{c}{-147675} & \multicolumn{1}{c|}{-147641}   
& \multicolumn{1}{c}{-45838} & \multicolumn{1}{c|}{-45803}
&\multicolumn{1}{c}{-164859} &  \multicolumn{1}{c}{-164835}  \\
\bottomrule
\end{tabular}
\label{tab:reg}
\end{table}

\section{Intragroup Behavior vs. Intergroup Behavior of the Same User}
\label{sec:twofaces}
\begin{figure}
    \center
    \begin{subfigure}[t]{0.32\textwidth}
        \includegraphics[width=\textwidth]
        {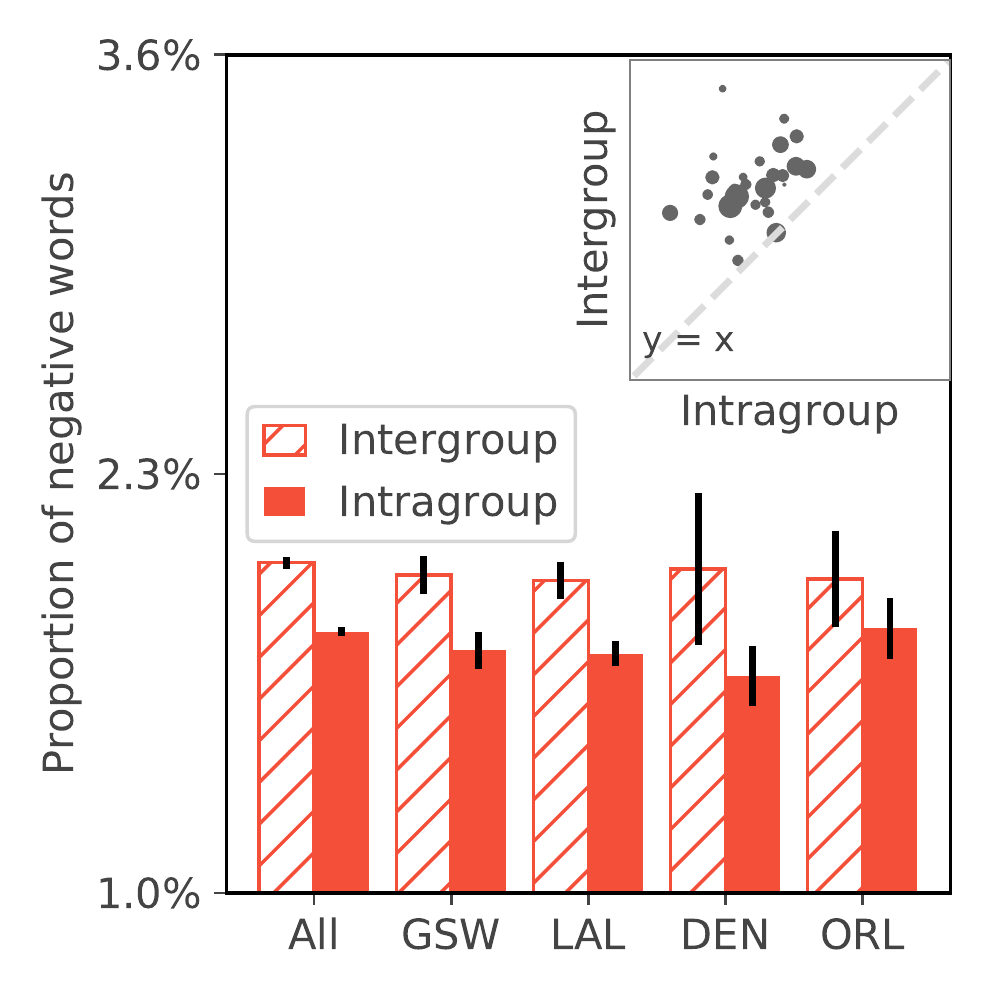}
        \caption{Proportion of negative words.}
        \label{fig:interneg}
    \end{subfigure}
    \hfill
    \begin{subfigure}[t]{0.32\textwidth}
        \includegraphics[width=\textwidth]
        {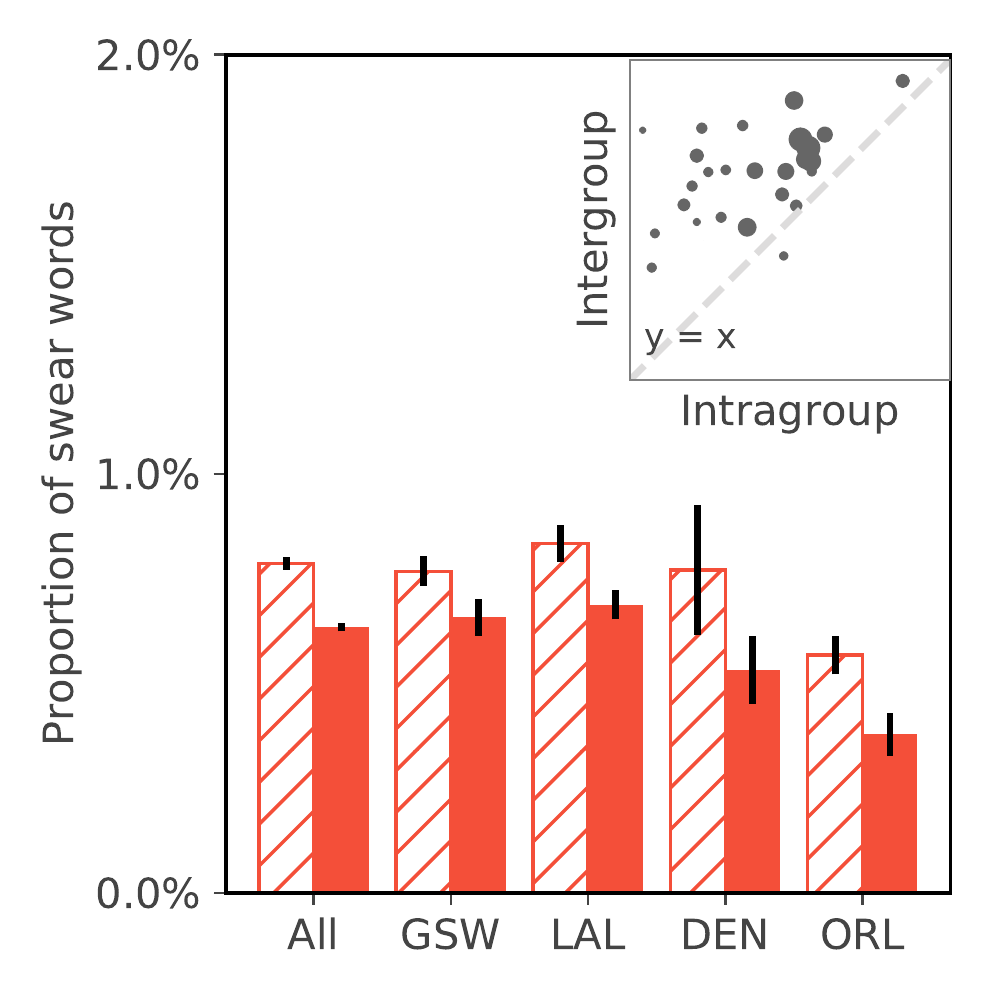}
        \caption{Proportion of swear words.}
        \label{fig:interswear}
    \end{subfigure}
    \hfill
    \begin{subfigure}[t]{0.32\textwidth}
        \includegraphics[width=\textwidth]
        {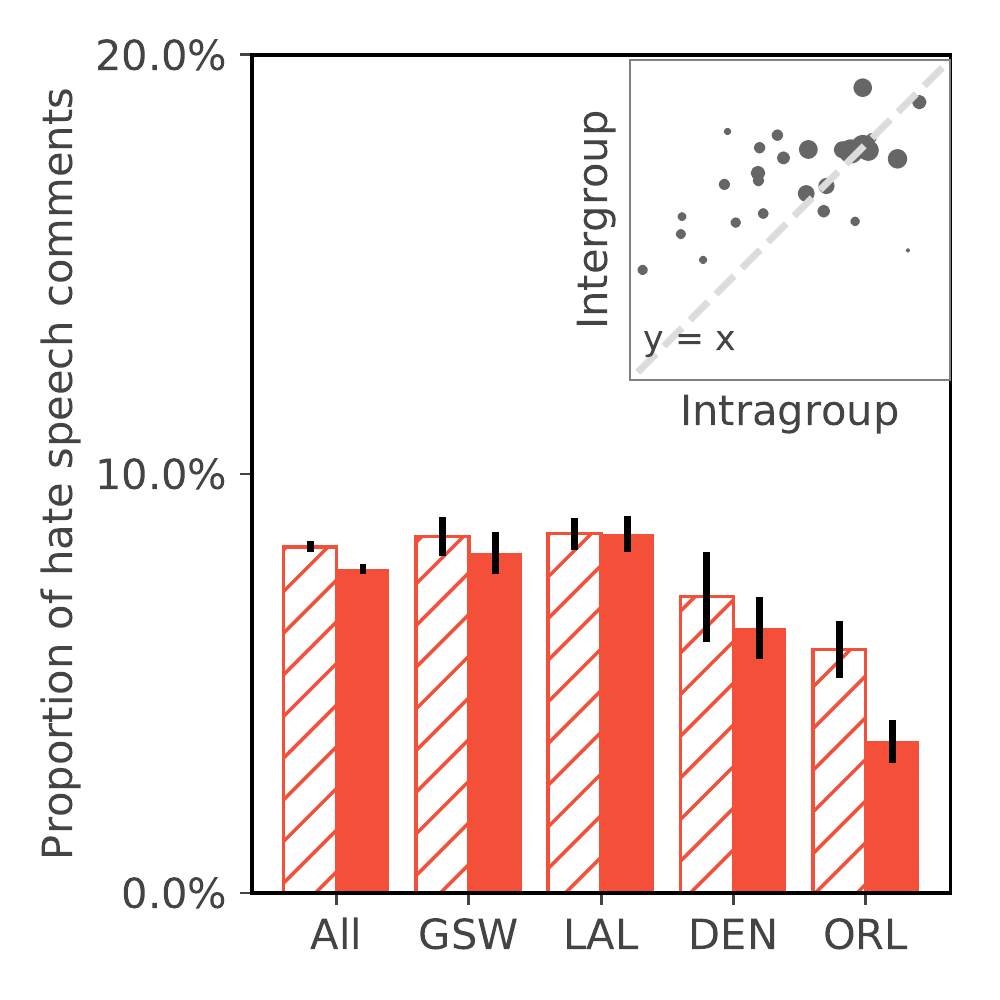}
        \caption{Proportion of hate speech comments.}
        \label{fig:interhate}
    \end{subfigure}
    \caption{\Intergroupusers use more negative language in the intergroup setting
    than in the intragroup setting in the 2018 season. 
    Here we only consider the matched \intergroupusers in \figref{fig:intrasentiment2018} (i.e., the solid red bars in this figure are identical to the red bars in \figref{fig:intrasentiment2018}%
    ).
    They use more negative words
    (two-tailed t-test, $t=9.39$, $p<0.001$, 
    95\% CI=0.17\% to 0.26\%; 
    30 out of 30 teams, two-tailed binomial test $p=0.001$)
    and swear words 
    (two-tailed t-test, $t=13.69$, $p<0.001$, 95\% CI=0.22\% to 0.29\%; 
    29 out of 30 teams, two-tailed binomial test $p<0.001$), 
    and generate more hate speech comments
    (two-tailed t-test, $t=2.97$, $p=0.003$, 
    95\% CI=0.18\% to 0.88\%; 
    23 out of 30 teams, two-tailed binomial test $p=0.005$).
    ``All'' is based on concatenating the samples from all 30 NBA team subreddits, 
    and we also show the top two and bottom two teams ranked by the number of subscribers 
    that have at least 100 \teamonlyusers.
    We further show the scatter plot of all 30 teams 
    in the top right to illustrate 
    that the findings are robust across teams (the size of the dot is proportional to the number of subscribers).
    Error bars represent standard errors. 
    The results are consistent in the 2017 and 2016 seasons
    (see \figref{fig:intersentiment2017} and \figref{fig:intersentiment2016}).
    }
    \label{fig:intersentiment2018}
\end{figure}

Given the clear intragroup behavioral differences in language usage between \intergroup and \teamonlyusers,
we end our study by exploring the potential reasons behind them. 
We study the differences in language usage of the same user in his/her affiliated team subreddit vs. in \communityname{/r/NBA}.
We compare the same person in two different contexts and naturally control for most of the confounding factors, which is also connected with the personality vs. situation debate \citep{kenrick1988profiting}.

\figref{fig:intersentiment2018} shows that \intergroupusers 
use even more negative language
in the intergroup setting, as they use more negative words
(two-tailed t-test, $t=9.39$, $p<0.001$, 95\% CI=0.17\% to 0.26\%; 
30 out of 30 teams, two-tailed binomial test $p=0.001$)
and swear words 
(two-tailed t-test, $t=13.69$, $p<0.001$, 95\% CI=0.22\% to 0.29\%; 
29 out of 30 teams, two-tailed binomial test $p<0.001$), 
and generate more
hate speech comments than in the intragroup setting 
(two-tailed t-test, $t=2.97$, $p=0.003$, 
95\% CI=0.18\% to 0.88\%; 
23 out of 30 teams, two-tailed binomial test $p=0.005$).
This indicates that fans are more hostile when facing fans from 
other teams than from the same team. 
This observation is robust after controlling for topics of discussion 
by only considering game threads (see \secref{sec:appgamethreads} and \figref{fig:gamethread}).
Our observation suggests that although \intergroupusers
are more emotional than \teamonlyusers in the affiliated subreddit, 
they are not as ``outrageous'' as they are in the intergroup setting.
In comparison, when going to the intergroup setting and confronting fans from other team groups, 
they tend to have more negative interactions and troll each other.

These observations may provide explanations for 
the characteristics of \intergroup fans in intragroup behavior.
Prior studies suggest that negative intergroup contact
is more influential in shaping people's attitudes and may curb 
the contact's ability to reduce 
prejudice~\citep{graf2014negative,paolini2010negative,stark2013generalization}.
The emotionally charged intergroup contact from the intergroup setting
may 
connect to \intergroup fans' more sentimental attitudes 
in their affiliated team subreddit.
It requires further research to establish the causal link here, but the fact that we are able to observe these contrasts demonstrates the importance of such observational studies based on real interactions over substantial time periods.

\section{Discussion}
\label{sec:discussion}

Although most previous studies have focused on the role of intergroup contact 
in changing attitudes of individual members, our study highlights the fact 
that users selectively become \intergroupusers, 
and \intergroup and \teamonly members in turn interact with each other in their affiliated group.
Such interaction can potentially influence members' language usage and shape the entire group.
Moreover, we demonstrate a variety of ways in which intergroup contact levels can moderate intragroup behavior.
This indicates that observational studies can provide important complementary evidence to experimental studies on this topic
because interventions can hardly result in deep and regular contact.
Novel methodologies are required to further bridge the gap between observational studies and experimental studies.

\para{Could social media be driving polarization?}
Twitter, Reddit, and Facebook have become 
important platforms for political discussions as well as misinformation \citep{vosoughi2018spread,grinberg2019fake}.
Service providers are designing new features that would actively expose people to opposing views.
For example, Twitter recently experimented with new algorithms that would promote alternative viewpoints 
in Twitter's timeline to address misinformation and reduce the effect of echo chambers \cite{jackdorsey}.
However, the proposed solution may increase polarization.
Unlike decades of offline experiments which mostly indicate intimate contact between
members of rival groups across an extended period can produce positive effects,
the results in \citet{Bail201804840} and our paper suggest that 
encountering views from opposing groups online may make them 
even more wedded to their own views.
There are several possible explanations of this contrast
by examining the possible mechanisms that intergroup contact affects individual attitudes.
First, the comments created on social media are usually brief. 
These short messages without enough context may not
enhance knowledge about opposing groups. 
Several studies suggest that people interpret short text-based messages inconsistently, 
which creates significant potential for miscommunication
\cite{wu2014short,miller2017understanding,kelly2012s}.
Second, 
the discussion structure may facilitate the spread of negative interaction.
\citet{cheng2017anyone} examines the evolution of discussions on CNN.com and 
show that existing trolling comments in a discussion thread significantly increase 
the likelihood of future trolling comments.
The spread of negativity will 
increase rather than reduce people's anxiety levels when facing opposing groups.
Third, the anonymous, spontaneous nature of communications on social media 
may not be conducive to cultivating empathy.
In an experiment designed to examine the relationship between the presence
of mobile devices and the quality of social interactions, results show that 
participants who have conversations in the absence of mobile devices
report high levels of empathetic concern \cite{misra2016iphone}.
In summary, 
intergroup contact may lead to diverging outcomes depending on the environment and the nature of the contact.
Further research is required to examine these possibilities and understand how social and technical 
design decisions can influence the outcomes.

\para{Can we design better online discussion forums for different groups?}
The findings in this work indicate that
social platforms designers should 
consider 
strategies to shape intergroup contact online.
As hinted above, it is insufficient to recommend users to follow members of opposing groups or opposing views.
Better design strategies need to be experimented for encouraging civil and extended intergroup contact.
It would also be useful to take into account how different levels of intergroup contact may moderate individual opinions differently.
Content moderation can be a promising area for future studies in the context of intergroup contact \citep{kiesler+kraut+resnick+kittur:2011}.
For instance, 
\citet{matias2019preventing} shows the displaying community rules can prevent harassment, but how to reduce negative intergroup contact remains an open question. 
Similarly, a powerful way of spreading online information is through social consensus cues
and online endorsement (e.g., upvotes, likes). 
However, promoting content with the highest popularity can sometimes be problematic.
Earlier research suggests that tweets with more sentiment-laden 
words are likely 
to be favorited or retweeted, and politicians may intentionally use this strategy to 
maximize impacts on Twitter \cite{brady2018ideological,tan+lee+pang:14,tan+etal:16b}. 
Our study also finds that \intergroupusers receive better feedback from their affiliated team subreddit
even though they use more negative language (\figref{fig:feedback}). 
This type of behavior can generate negative reactions from opposing groups and push
the whole discussion to cycle towards more emotionally-laden and potentially polarizing content.
It is thus important to develop comment ranking systems that are cognizant of intergroup contact 
and prioritizes constructive interactions.

\smallskip
\noindent\textbf{Limitations.}
Our findings are subject to the following limitations.
First, the causal relationship between intergroup contact and negative language usage is
not entirely clear.
Due to the nature of our observational study, 
whether a member has intergroup contact is not randomly assigned. 
Though we match users based on a series of activity features, an important confounding factor
could be that people who seek intergroup contact are inherently different from those who do not.

Second, our definition of intergroup contact entails that we focus on relatively
active users. Thus, we cannot observe indirect intergroup contact,  
such as browsing \communityname{/r/NBA}. Prior studies have shown that 
indirect contact, such as imagining oneself interacting with an out-group
member and observing an in-group member interacting with an out-group 
member~\cite{turner2010imagining,dovidio2011improving}, 
may also shape human behavior.
It also follows that \intergroupusers have more activities on NBA-related discussion forums 
as a whole than \teamonlyusers.
We want to note that 
the nature of intergroup contact is that given the same amount of time in life, 
individuals with intergroup contact put more effort into intergroup contact than those without such contact.

Third, 
we use a coarse proxy to consider any users who have posted in our intergroup setting (\communityname{/r/NBA}) as \intergroupusers, and study the language differences in the intragroup setting at the user level instead of at the dyad level.
However, some comments created by the fans in /r/NBA may be
replies to the fans who are from the same team
or do not have a team affiliation.
More in-depth characterization of different types of discussions happened in the intergroup setting
is required to further understand the differences observed in this study.

Fourth, the observations made in this study are limited to Reddit NBA fan groups. 
The sports context might be a strong case for understanding intergroup relations, as all the teams 
are created to compete with each other for the final championship.
The expression of hostile attitudes towards opposing sides are culturally acceptable
and even encouraged \cite{cikara2011usversusthem}. 
We should expect 
less negative intergroup contact between groups that 
do not contend for the same resources (e.g., music fans of different musicians may not have conflicts with each other at all). 
However, 
politics, especially in a polarized bipartisan situation, share common properties with the sports context.
Examining the
generalization of our results in other contexts is a promising avenue for future work.

Finally, the negative language observed in our study may not necessarily bring negative effects 
to the community. 
Prior studies suggest the main reason people use swearing words on the online platform
is to express some strong emotions, such as anger 
and frustration \cite{wang2014cursing,diakopoulos2011towards,cheng2017anyone}.
\citet{heath2001emotional} 
examine users' emotional selection in memes when emotion is 
manipulated and observe that people prefer the version of the story that produced the highest
levels of disgust and evoke strong sentiment.
\citet{jay2009offensive} further argues that only when cursing occurs in the form of insults
toward others, such as name-calling, harassment, and hate speech, it becomes harmful.
In addition, earlier literature 
suggests that the reason people use swearing words on the online platform may
relate to Internet humor, such as jokes and memes. 
Posting humorous content on the Internet has the potential to engage other users
in art activities that are closely connected to their lives and receive 
online endorsements \cite{yoon2016not,shifman2014memes}. 
Attempting to be funny could be another reason that \intergroupusers adopt a
more negative language style than \teamonlyusers. 
However, as pointed out by \citet{lockyer2005beyond},
a significant proportion of Internet humor has offensive, sexism, and racism content,
and its consequences are often overlooked.

\section{Conclusion}
\label{sec:conclusion}

In this paper, by applying our computational framework to NBA-related discussion forums on Reddit, 
we identify clear language differences between intergroup and \teamonlyusers
in their affiliated group (the intragroup setting). 
We find that in the affiliated team subreddit, \intergroupusers tend to use more negative and swear words, and  generate more hate speech comments compared 
with single-group members.
Moreover, we quantify different levels of intergroup contact 
for each intergroup member based on the fraction of their comments in the intergroup setting (\communityname{/r/NBA}).
Interestingly, the level of intergroup contact can relate to
differences in language usage in different ways, though the relationship is 
mostly monotonic.
To further shed light on the behavior of intergroup members, we also compare 
the language usage of intergroup members between the intragroup setting 
and the intergroup setting. This setup naturally controls for the subject because 
we compare the same person across two different environments. 
We observe that intergroup members are even more negative and more likely to swear
in the intergroup setting. 
As intergroup contact in online platforms becomes increasingly common and can play an important role in opinion formation,
our work 
demonstrates how observational studies can provide
complementary evidence to experimental studies on this topic.

\bibliographystyle{ACM-Reference-Format}
\bibliography{references}

\setcounter{table}{0}
\renewcommand{\thetable}{A\arabic{table}}
\setcounter{figure}{0}
\renewcommand{\thefigure}{A\arabic{figure}}

\appendix

\section{Appendix}

\begin{table}[]
\centering
\small
\caption{A sample of positive, negative and swear words in  
the Linguistic Inquiry and Word Count dictionary 
(LIWC \cite{pennebaker2007linguistic}).
Words ending with ``*'' match any string with the same prefix.}
\begin{tabular}{l|p{0.7\textwidth}}
\toprule
Positive & credit*, graced, attract*, graceful*, terrific*, bonus*, affection*, humour*, delicious*, love, openness, sweetheart*, bless*, bold*, madly, fine, friend*, hurra*, ready, trust*, secur*, won, improving, fiesta*, dynam*, toleran*, sunniest, optimal*, helpful*, neat*, enthus*, joking, favour*, giving, agreeab*, easiness, supportive*, frees*, graces, gentler\\ \midrule
Negative & ignor*, aggravat*, unattractive, scary, attack*, offend*, grief, fright*, domina*, unfriendly, violat*, grave*, nast*, suck, shock*, sucker*, impatien*, wept, heartless*, shake*, battl*, moron*, vanity, aggress*, masochis*, unsure*, screw*, lost, losing, mocker*, envie*, sadness, nag*, timid*, afraid, hateful*, turmoil, agoniz*, obnoxious*, pain  \\ \midrule
Swear & prick*, dyke*, tit, cock, dicks, butt, bloody, dick, sob, asshole*, pussy*, screw*, suck, wanker*, mofo, fucks, shit*, bastard*, arse, butts, darn, sucked, jeez, nigger*, fucker*, arses, ass, hell, crappy, dang, motherf*, dumb*, heck, crap, tits, queer*, bitch*, sonofa*, titty, fuckin*\\  
\bottomrule 
\end{tabular}
\label{tab:negativeexamples}
\end{table}
\begin{table}[]
\centering
\small
\caption{Examples of hate speech comments detected 
with the automated detection model~\cite{hateoffensive}.}
\begin{tabular}{p{0.8\textwidth}}
\toprule
Kevin Sorbo's Hercules was such a pussy magnet                                                                                                  \\ \midrule
Because I want losers like you to fuck off                                                                                                      \\ \midrule
Same reason KAT ass rapes our team everytime we play.                                                                                           \\ \midrule
Holy shit Sabonis is stuntin' like his daddy right now with these passes                                                          \\ \midrule
That's like me saying you're a dumbass because your team is currently shit. Sorry pal, don't talk about basketball until you make the playoffs. \\ \midrule
Melo gets NO calls ever - I've always said he should bitch more                                                                                 \\ \midrule
No defense, no rebounding. Same old shit. Embarrassing                                                                                          \\ \midrule
Fuck the Celtics!                                                                                                                               \\ \midrule
now since your dumb ass sees that it doesnt make difference like ive been saying what is your excuse?                                           \\ \midrule
Tyler Ennis you bum. Worst player in the league.\\ 
\bottomrule 
\end{tabular}
\label{tab:hatespeechexamples}
\end{table}

\subsection{\added{The language differences between \intergroupusers and \teamonlyusers 
are consistent when we exclude the NBA playoffs}}
\label{sec:regular}
\begin{figure}
    \centering
    \begin{subfigure}[t]{0.32\textwidth}
        \includegraphics[width=\textwidth]
        {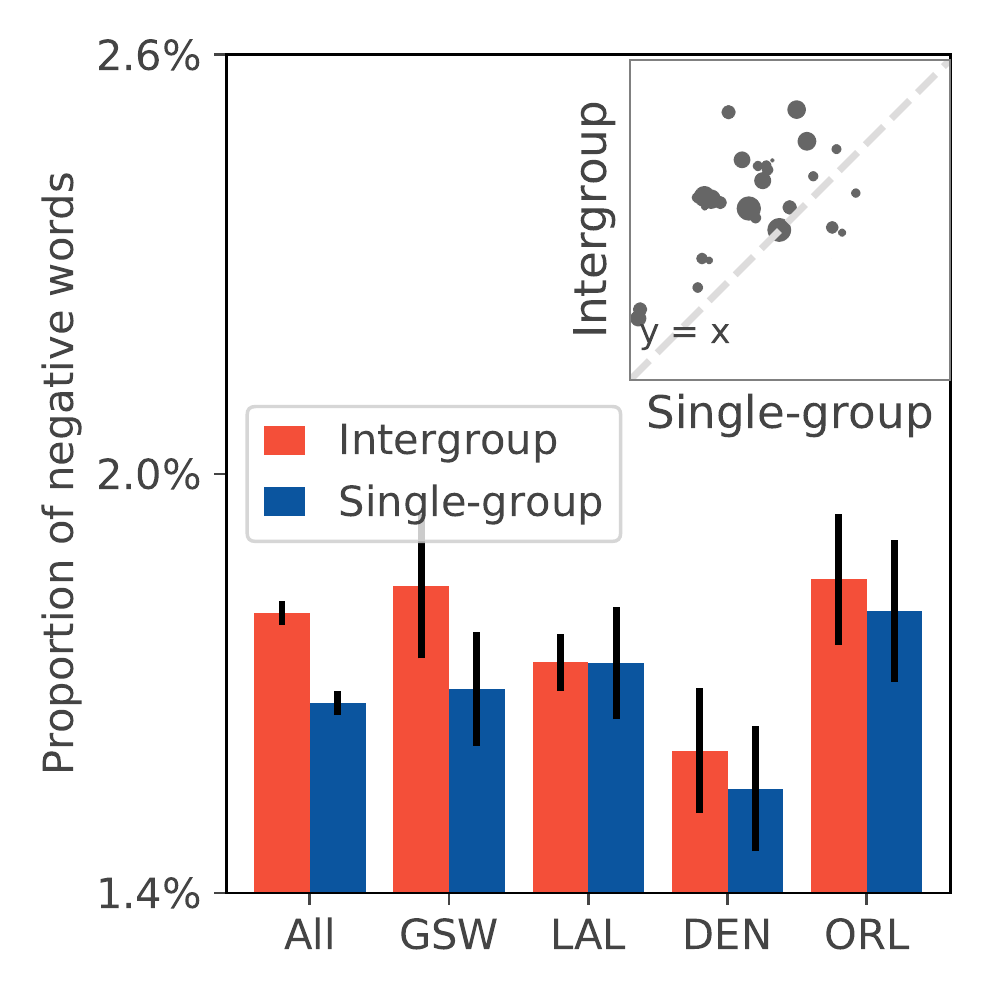}
        \caption{Proportion of negative words.}
        \label{fig:effectnegregular2018}
    \end{subfigure}
    \hfill
    \begin{subfigure}[t]{0.32\textwidth}
        \includegraphics[width=\textwidth]
        {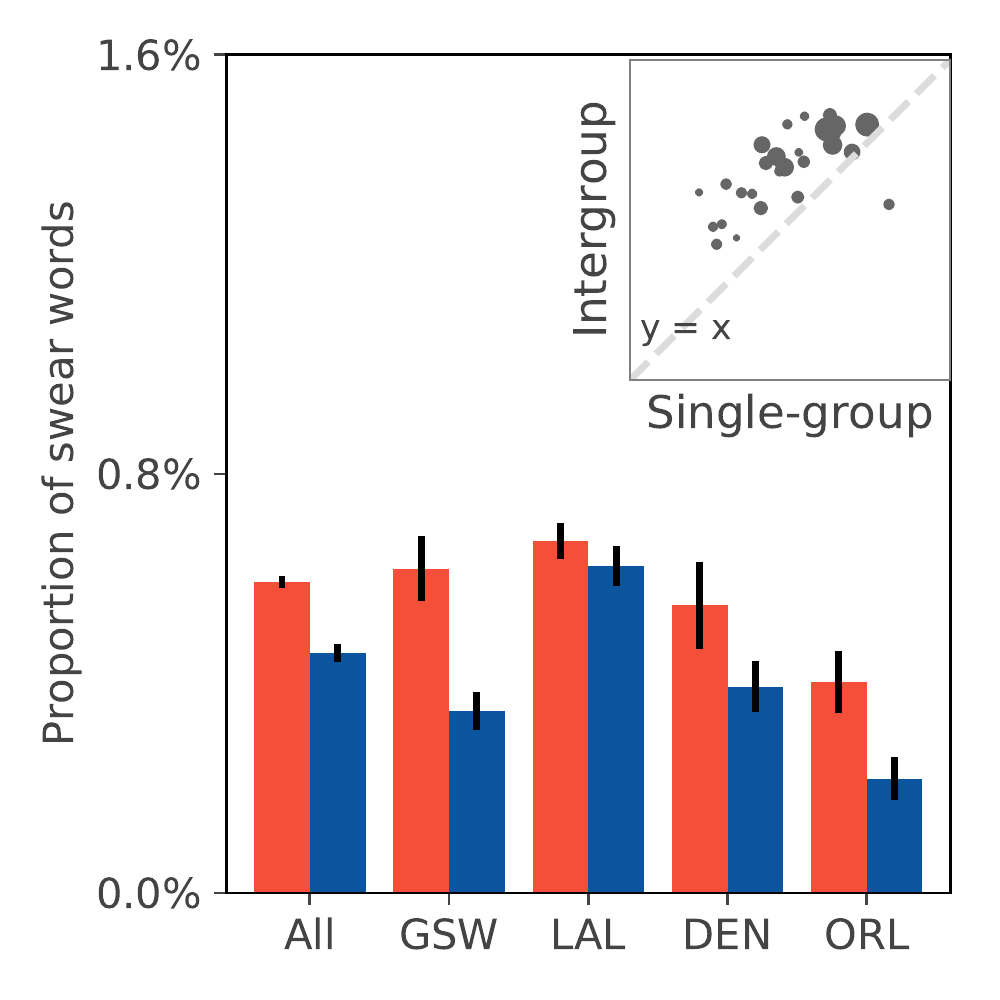}
        \caption{Proportion of swear words.}
        \label{fig:effectswearregular2018}
    \end{subfigure}
    \hfill
    \begin{subfigure}[t]{0.32\textwidth}
        \includegraphics[width=\textwidth]
        {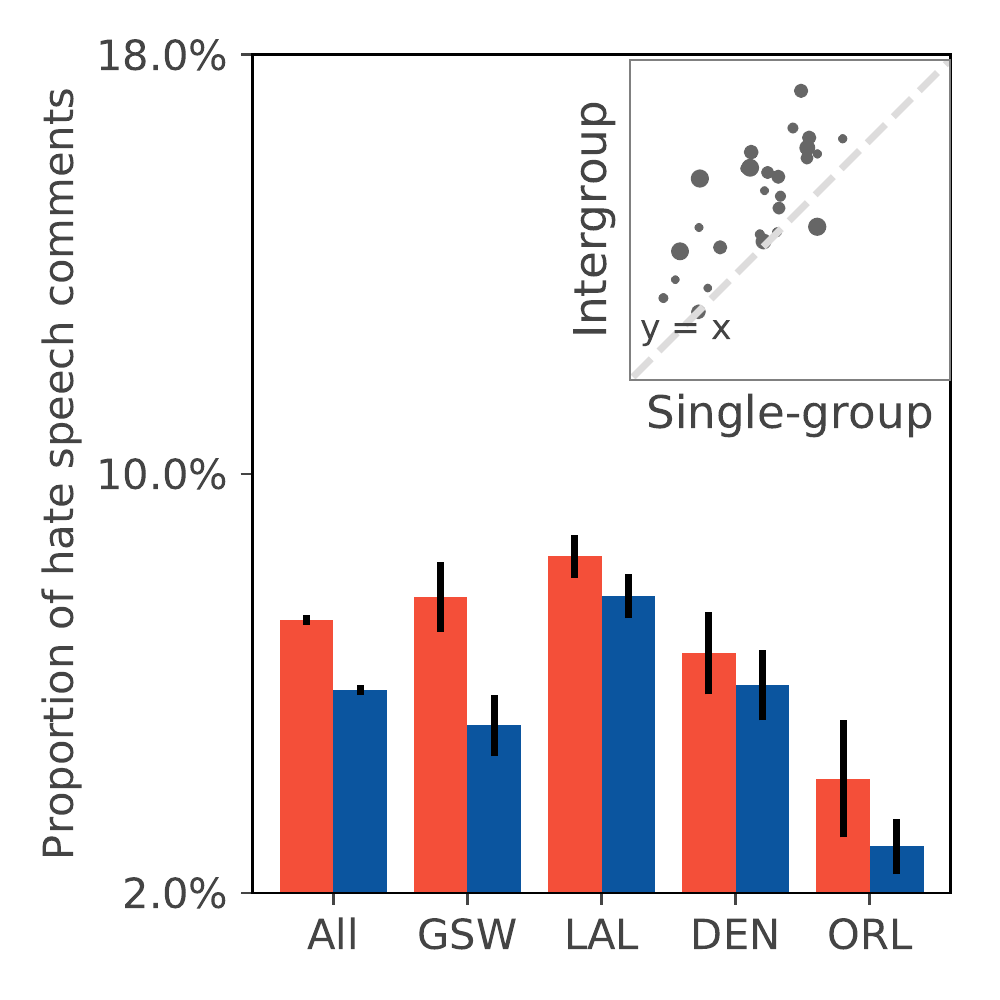}
        \caption{Proportion of hate speech comments.}
        \label{fig:effecthateregular2018}
    \end{subfigure}
    \caption{
    \added{The comparison of language usage between intergroup and \teamonlyusers in the 2018 season
    when excluding the NBA playoffs.
    \Intergroupusers use more negative words 
    (two-tailed t-test, $t=4.31$, $p<0.001$,  
    95\% CI=0.07\% to 0.19\%; 24 out of 30 teams, two-tailed binomial test $p=0.001$) 
    and swear words 
    (two-tailed t-test, $t=3.43$, $p<0.001$, 
    95\% CI=0.04\% to 0.14\%; 28 out of 30 teams, two-tailed binomial test $p<0.001$) 
    and generate more hate speech comments 
    (two-tailed t-test, $t=10.05$, $p<0.001$, 95\% CI=1.38\% to 2.05\%; 
    24 out of 30 teams, two-tailed binomial test $p=0.001$).
    Error bars represent standard errors.}}
    \label{fig:intrasentimentregular2018}
\end{figure}

\begin{figure}
    \centering
    \begin{subfigure}[t]{0.32\textwidth}
        \includegraphics[width=\textwidth]
        {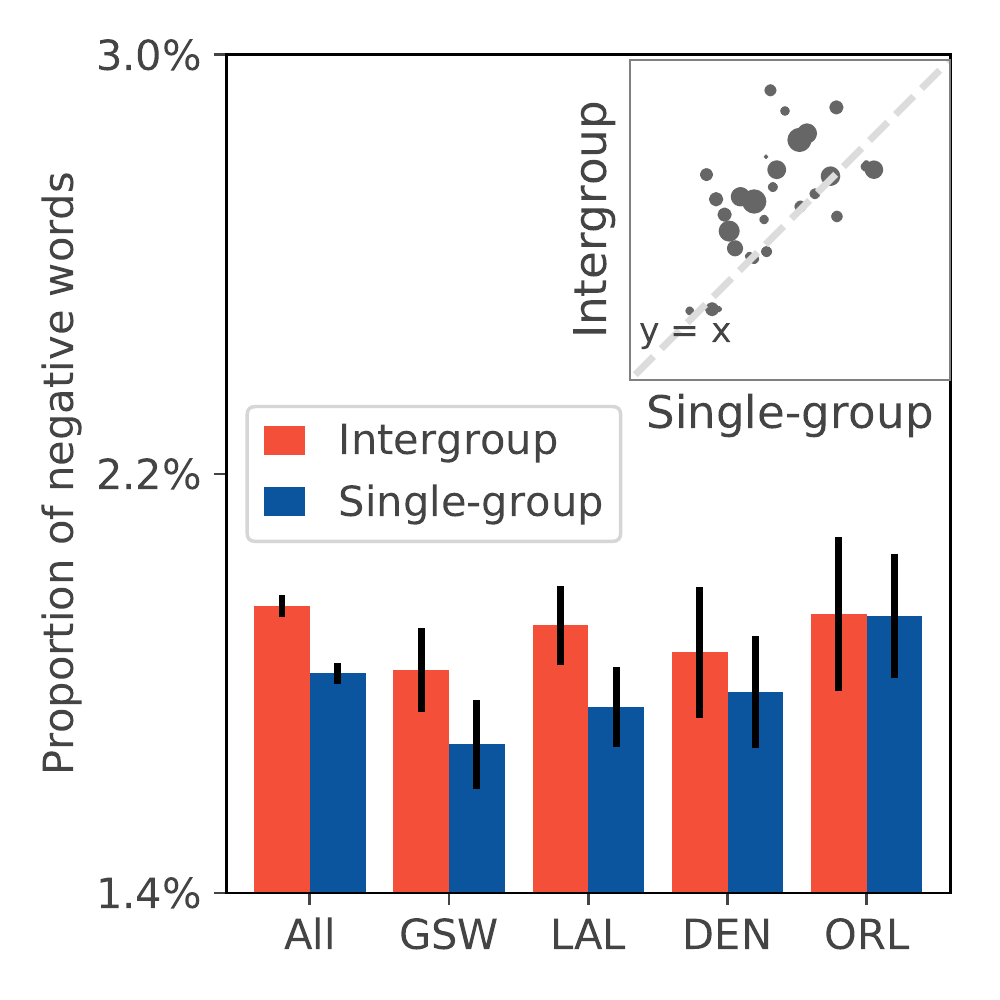}
        \caption{Proportion of negative words.}
        \label{fig:effectnegregular2017}
    \end{subfigure}
    \hfill
    \begin{subfigure}[t]{0.32\textwidth}
        \includegraphics[width=\textwidth]
        {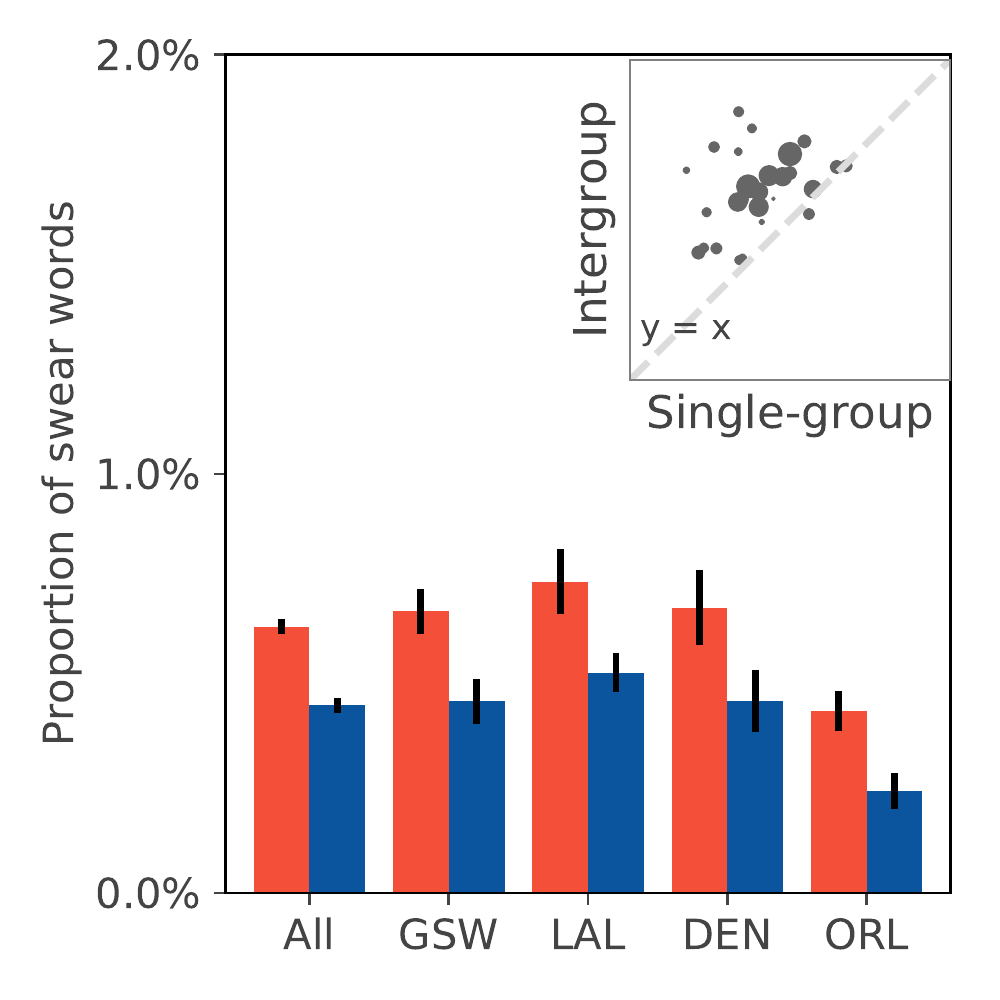}
        \caption{Proportion of swear words.}
        \label{fig:effectswearregular2017}
    \end{subfigure}
    \hfill
    \begin{subfigure}[t]{0.32\textwidth}
        \includegraphics[width=\textwidth]
        {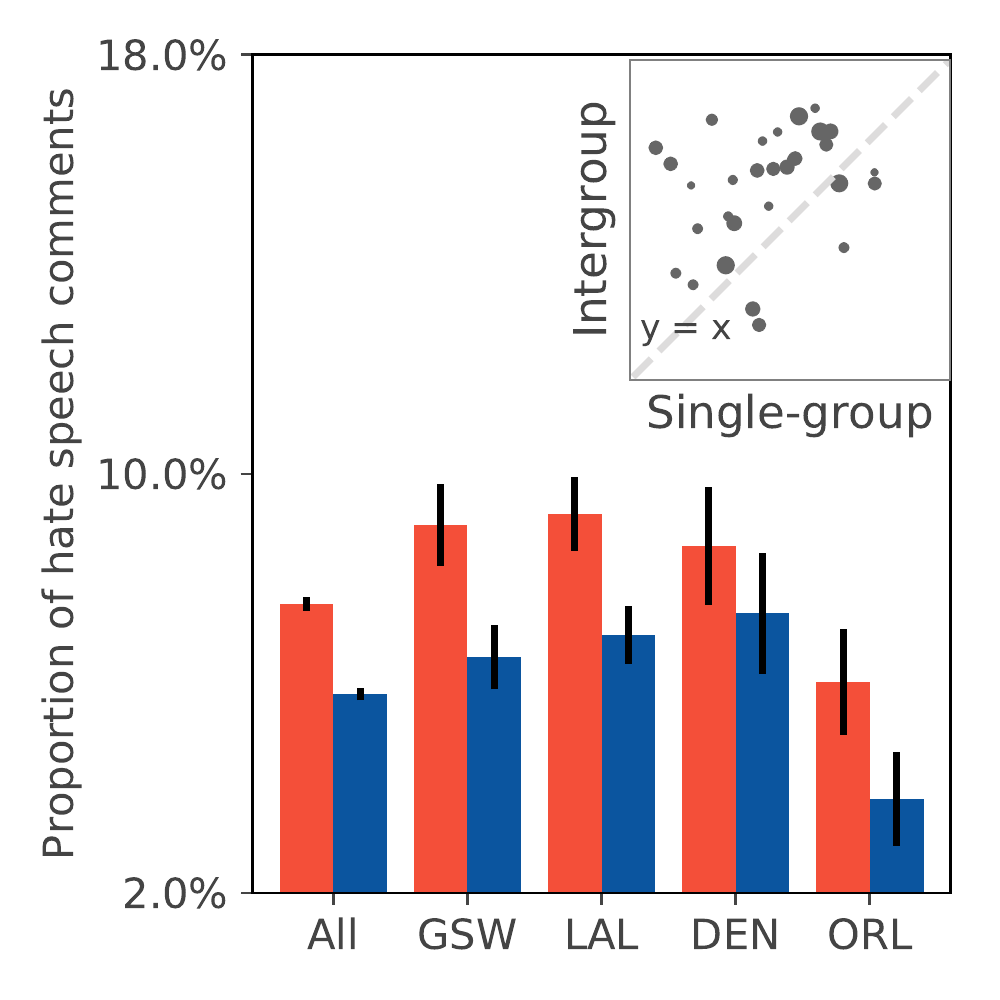}
        \caption{Proportion of hate speech comments.}
        \label{fig:effecthateregular2017}
    \end{subfigure}
    \caption{\added{The comparison of language usage between intergroup and \teamonlyusers in the 2017 season
    when excluding the NBA playoffs.
    \Intergroupusers use more negative words 
    (two-tailed t-test, $t=4.31$, $p<0.001$,  
    95\% CI=0.07\% to 0.19\%; 24 out of 30 teams, two-tailed binomial test $p=0.001$) 
    and swear words 
    (two-tailed t-test, $t=3.43$, $p=0.002$, 
    95\% CI=0.04\% to 0.14\%; 28 out of 30 teams, two-tailed binomial test $p<0.001$) 
    and generate more hate speech comments 
    (two-tailed t-test, $t=10.05$, $p<0.001$, 95\% CI=1.38\% to 2.05\%; 
    23 out of 30 teams, two-tailed binomial test $p<0.005$).
    Error bars represent standard errors.}}
    \label{fig:intrasentimentregular2017}
\end{figure}

\begin{figure}
    \centering
    \begin{subfigure}[t]{0.32\textwidth}
        \includegraphics[width=\textwidth]
        {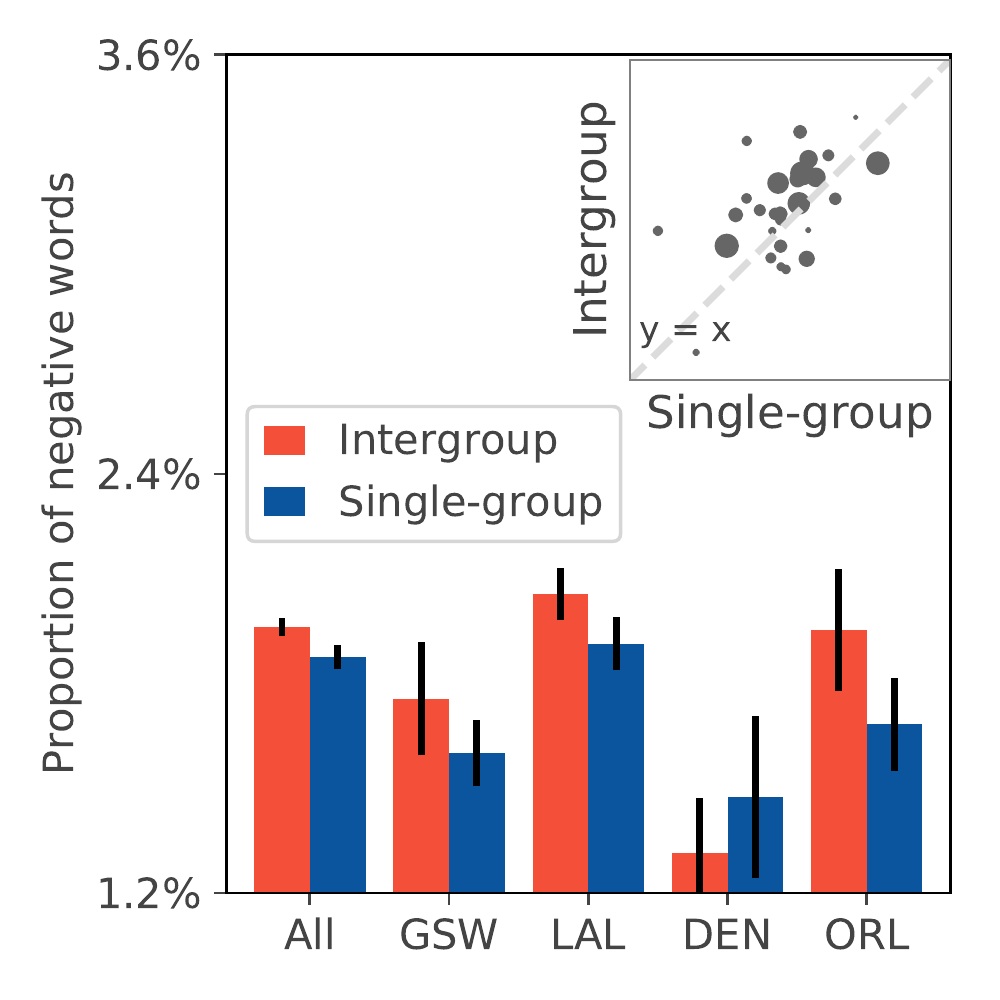}
        \caption{Proportion of negative words.}
        \label{fig:effectnegregular2016}
    \end{subfigure}
    \hfill
    \begin{subfigure}[t]{0.32\textwidth}
        \includegraphics[width=\textwidth]
        {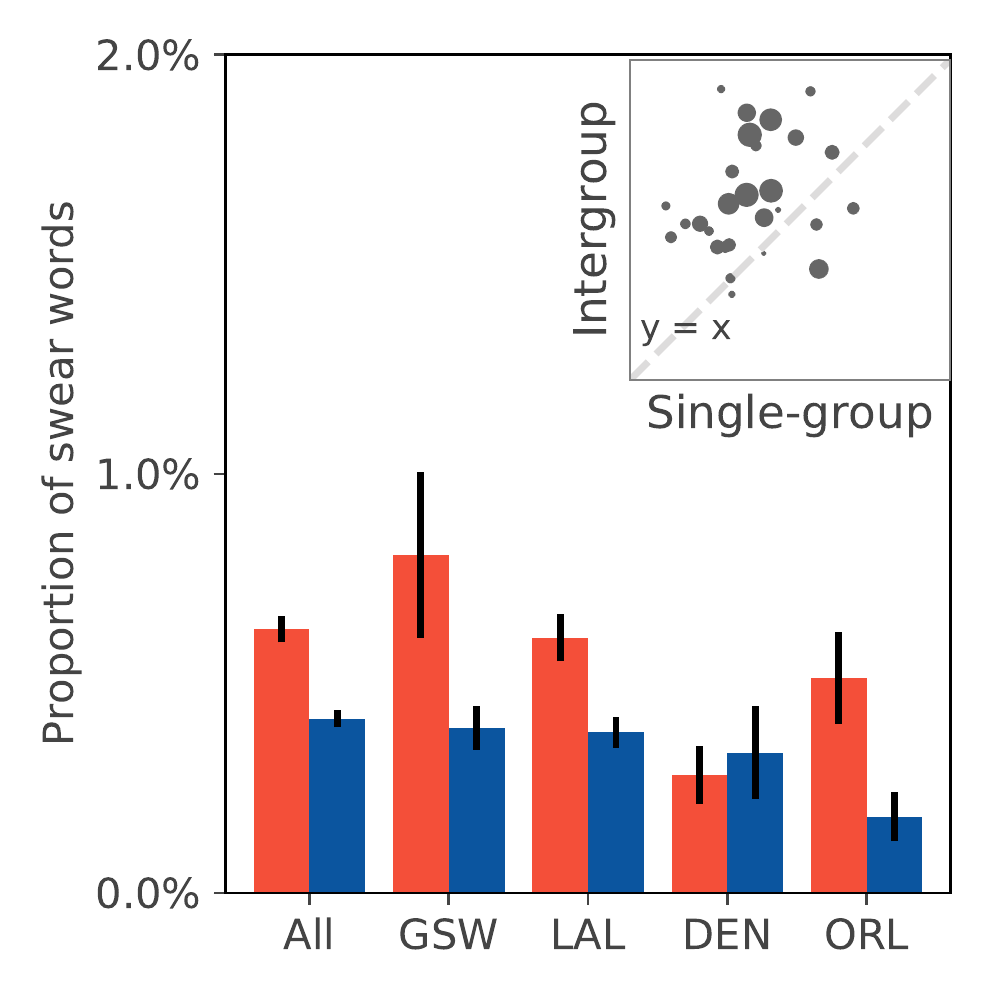}
        \caption{Proportion of swear words.}
        \label{fig:effectswearregular2016}
    \end{subfigure}
    \hfill
    \begin{subfigure}[t]{0.32\textwidth}
        \includegraphics[width=\textwidth]
        {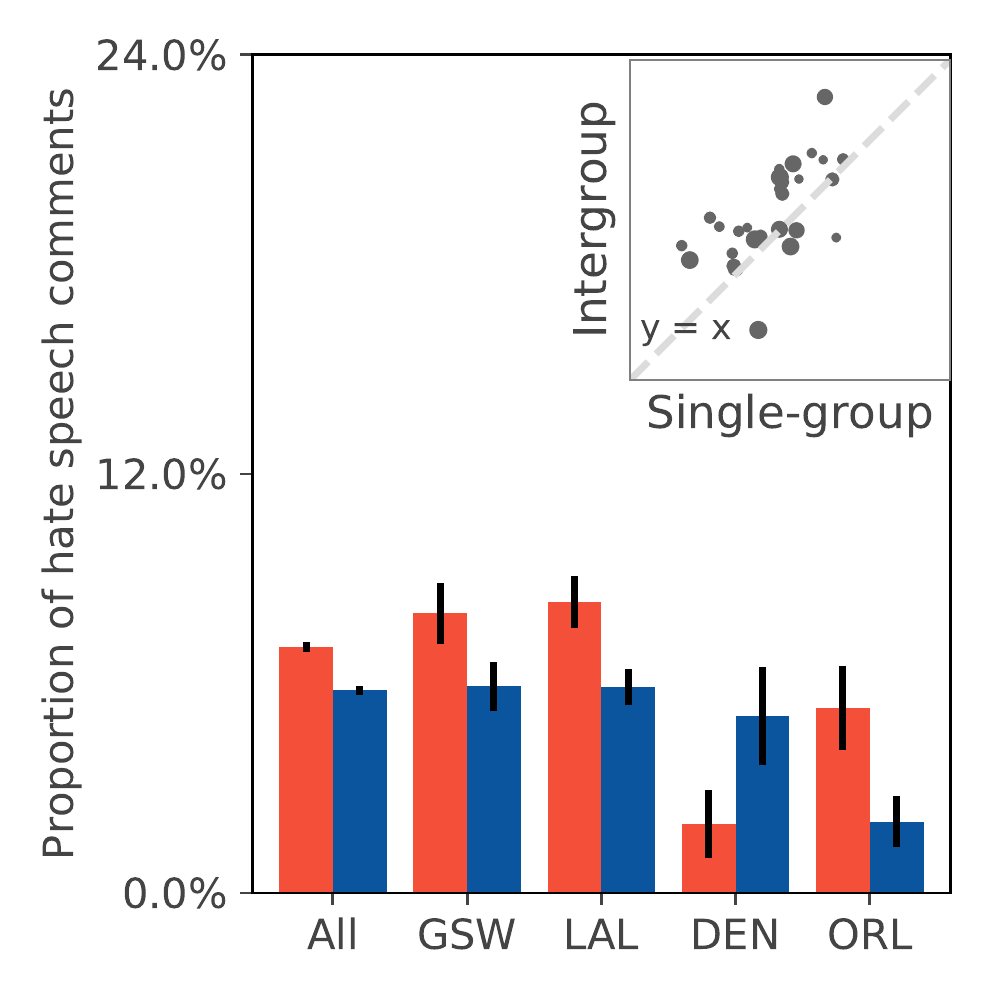}
        \caption{Proportion of hate speech comments.}
        \label{fig:effecthateregular2016}
    \end{subfigure}
    \caption{\added{The comparison of language usage between intergroup 
    and \teamonlyusers in the 2016 season when excluding the NBA playoffs.
    \Intergroupusers use more negative words 
    (two-tailed t-test, $t=1.99$, $p=0.04$,  
    95\% CI=0.01\% to 0.17\%; 21 out of 30 teams, two-tailed binomial test $p=0.04$) 
    and swear words (two-tailed t-test, $t=3.06$, $p=0.002$, 95\% CI=0.04\% to 0.18\%; 
    24 out of 30 teams, two-tailed binomial test $p=0.001$) 
    and generate more hate speech comments (two-tailed t-test, $t=7.95$, $p<0.001$, 
    95\% CI=0.92\% to 1.53\%; 25 out of 30 teams, two-tailed binomial test $p<0.001$).
    Error bars represent standard errors.}}
    \label{fig:intrasentimentregular2016}
\end{figure}

\added{In \secref{sec:languagedifference}, we find that \intergroupusers use more
negative and swear words and generate more hate speech comments compared to
\teamonlyusers in the intragroup setting. 
It is possible that comments posted during the playoff season play a significant
factor, 
as 
the activity in \communityname{/r/NBA} peaks in the playoff season \citep{zhang+tan+lv:18}.
Figure \ref{fig:intrasentimentregular2018}, \ref{fig:intrasentimentregular2017}, 
and \ref{fig:intrasentimentregular2016} show that our results are consistent 
in all three seasons when we exclude the playoff season.}

\subsection{\Intergroupusers receive better feedback than 
\teamonlyusers}
\begin{figure}
    \centering
    \begin{subfigure}[t]{0.32\textwidth}
        \includegraphics[width=\textwidth]
        {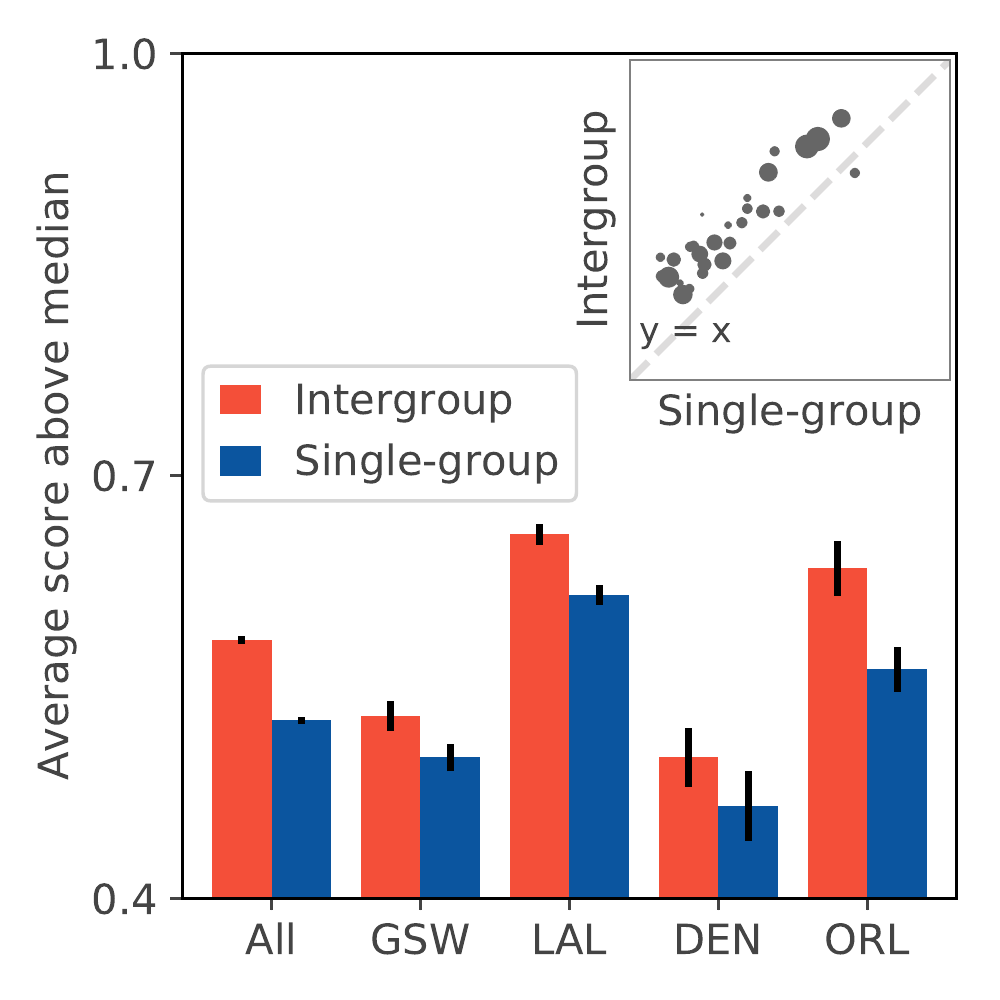}
        \caption{The 2018 season.}
        \label{fig:feedback2018}
    \end{subfigure}
    \hfill
    \begin{subfigure}[t]{0.32\textwidth}
        \includegraphics[width=\textwidth]
        {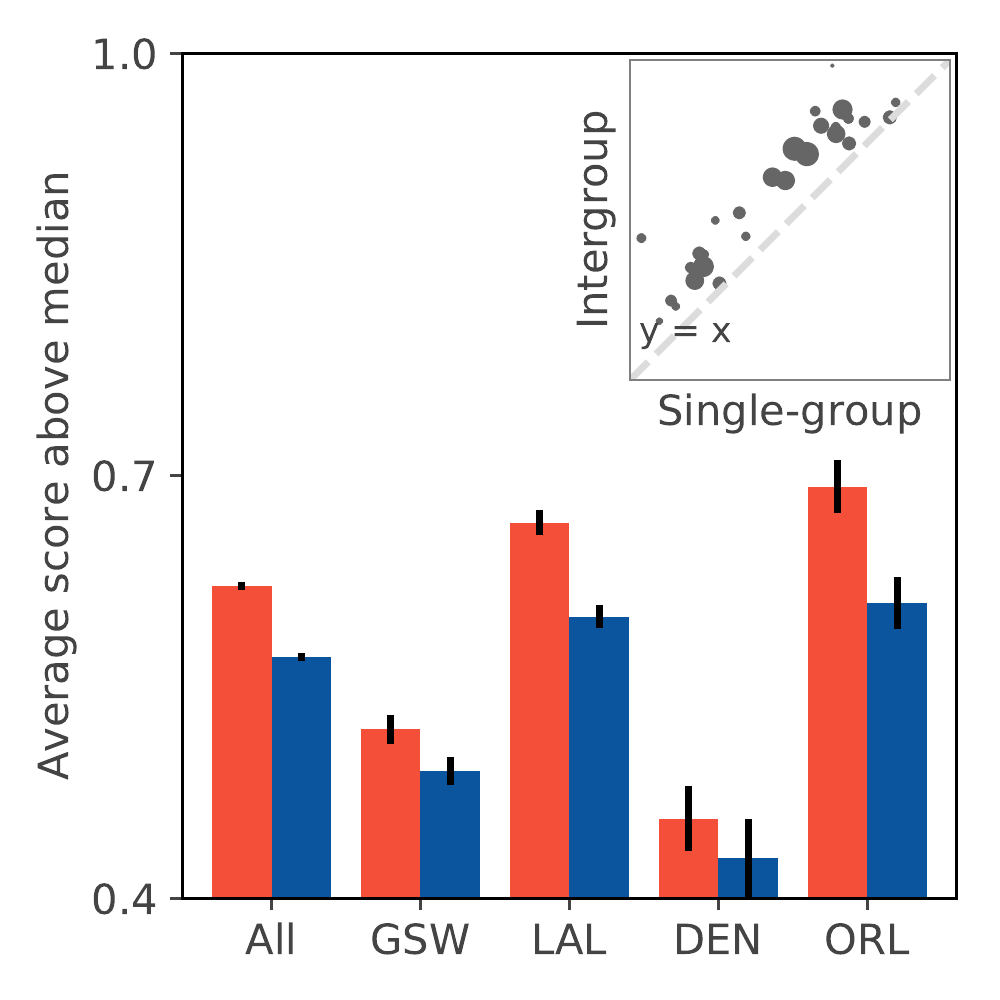}
        \caption{The 2017 season.}
        \label{fig:feedback2017}
    \end{subfigure}
    \hfill
    \begin{subfigure}[t]{0.32\textwidth}
        \includegraphics[width=\textwidth]
        {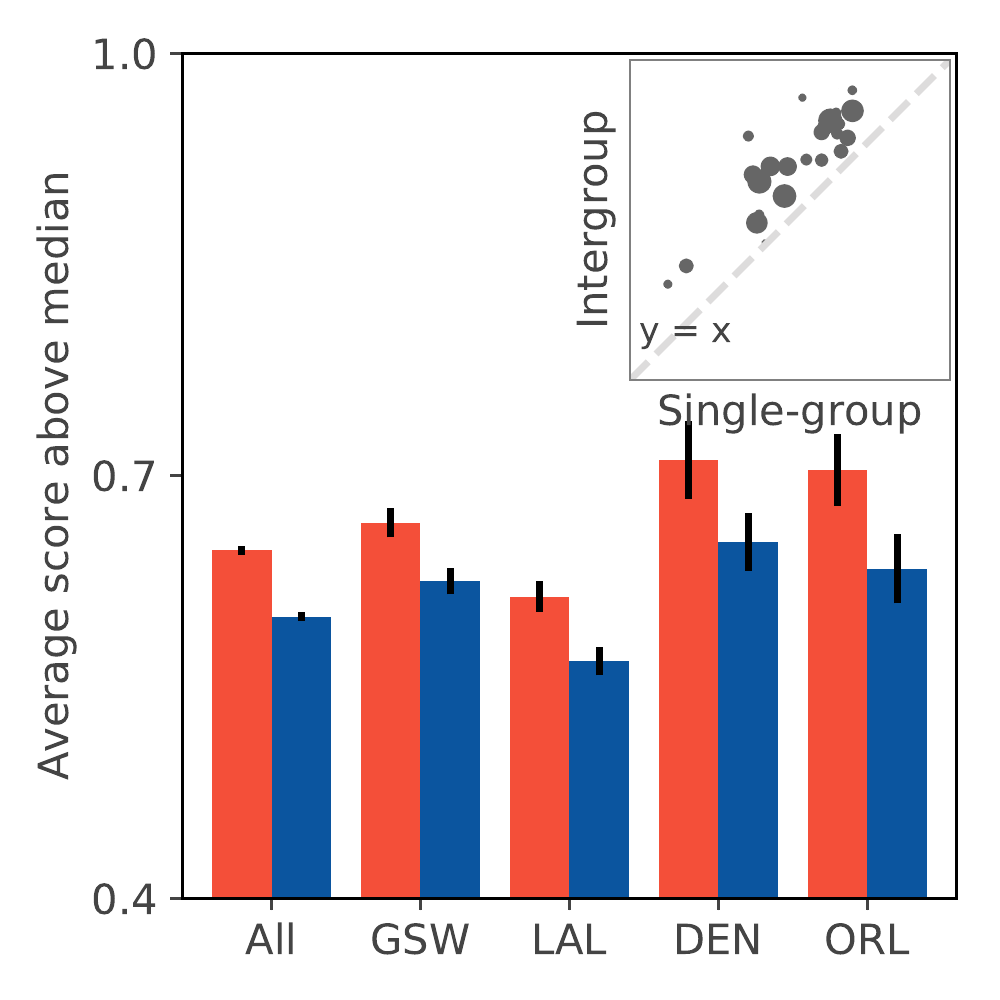}
        \caption{The 2016 season.}
        \label{fig:feedback2016}
    \end{subfigure}
    \caption{The comparison of feedback received from affiliated team subreddit
     between \intergroup and \teamonlyusers in the 2018, 2017, and 2016 seasons.
    \Intergroupusers receive better feedback than 
    \teamonlyusers in all three seasons 
    (two-tailed t-test, $t=15.68$, $p<0.001$, 
    95\% CI=0.048 to 0.062\%, 29 out of 30 teams, 
    two-tailed binomial test $p<0.001$ for the 2018 season;
    two-tailed t-test, $t=12.56$, $p<0.001$, 
    95\% CI=0.043 to 0.058\%, 30 out of 30 teams, 
    two-tailed binomial test $p<0.001$ for the 2017 season;
    two-tailed t-test, $t=12.43$, $p<0.001$,
    95\% CI=0.046 to 0.064\%, 30 out of 30 teams, 
    two-tailed binomial test $p<0.001$ for the 2016 season).
    Error bars represent standard errors.}
    \label{fig:feedback}
\end{figure}

\figref{fig:feedback} shows that \intergroupusers 
receive better feedback in the intragroup setting than 
\teamonly members in all three seasons
(two-tailed t-test, $t=15.68$, $p<0.001$, 
    95\% CI=0.048 to 0.062\%, 29 out of 30 teams, 
    two-tailed binomial test $p<0.001$ for the 2018 season;
    two-tailed t-test, $t=12.56$, $p<0.001$, 
    95\% CI=0.043 to 0.058\%, 30 out of 30 teams, 
    two-tailed binomial test $p<0.001$ for the 2017 season;
    two-tailed t-test, $t=12.43$, $p<0.001$, 
    95\% CI=0.046 to 0.064\%, 30 out of 30 teams, 
    two-tailed binomial test $p<0.001$ for the 2016 season). 
Comment feedback is defined by whether the comment score (\#upvotes-
\#downvotes) is above the median score of that team subreddit in that month, which accounts for the differences across subreddits.
This observation suggests that using negative language is likely to draw attention
in the corresponding team subreddits.

\subsection{\Intergroupusers are more emotional in the intergroup setting than in the intragroup setting when
controlling for the discussion topic}
\label{sec:appgamethreads}
\begin{figure}
    \center
    \begin{subfigure}[t]{0.32\textwidth}
        \includegraphics[width=\textwidth]
        {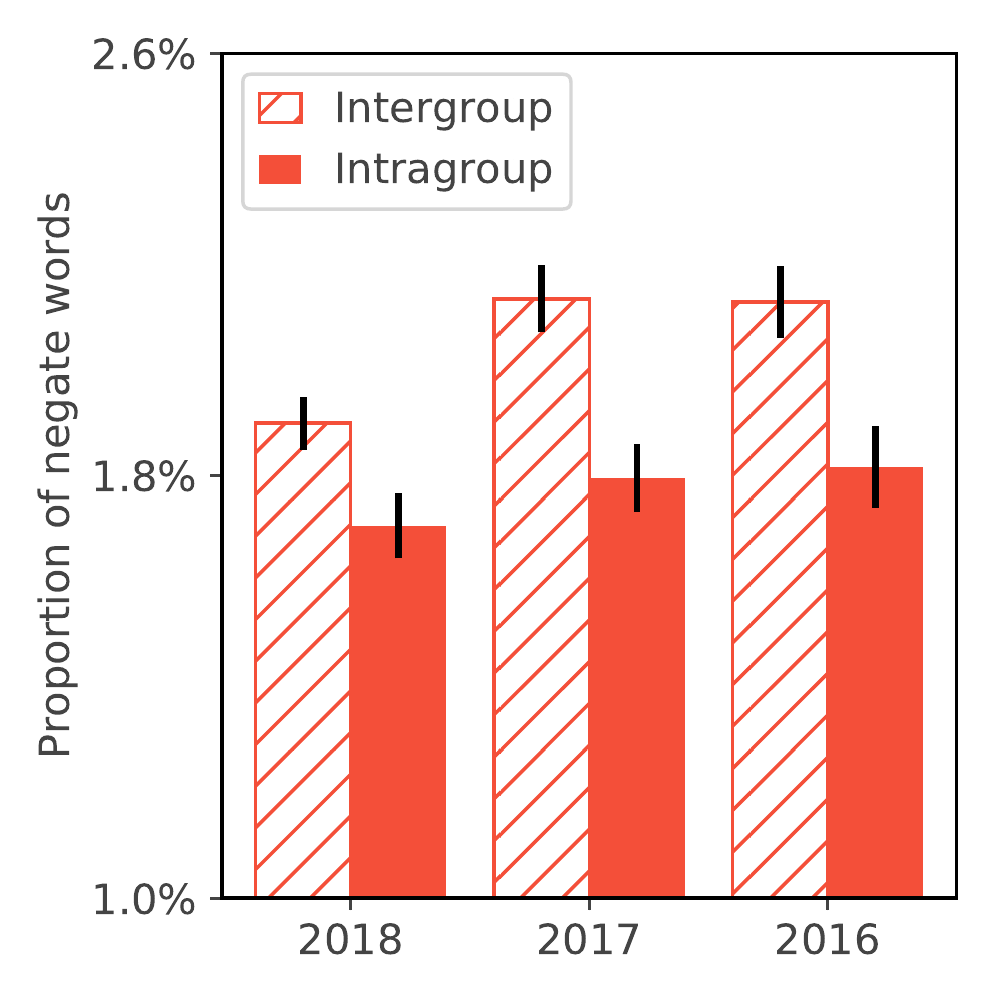}
        \caption{Proportion of negative words.}
        \label{fig:interneg}
    \end{subfigure}
    \hfill
    \begin{subfigure}[t]{0.32\textwidth}
        \includegraphics[width=\textwidth]
        {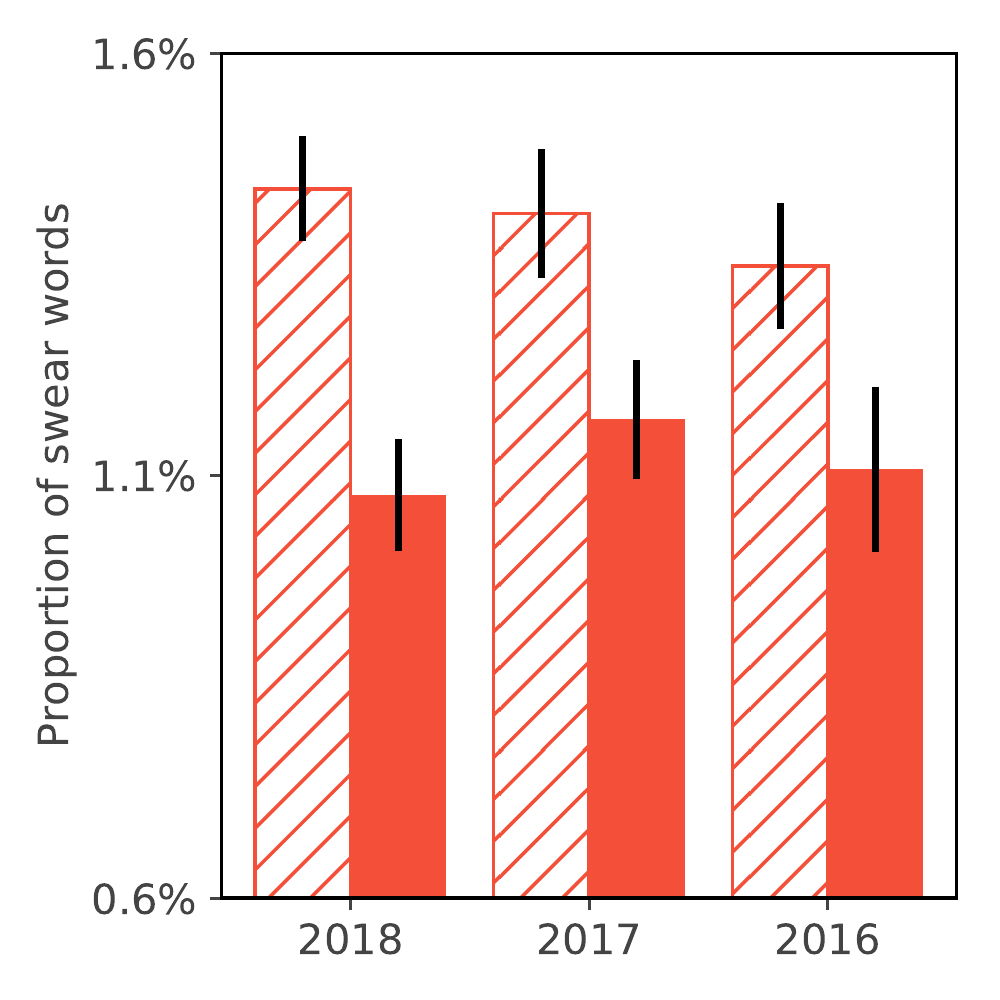}
        \caption{Proportion of swear words.}
        \label{fig:interswear}
    \end{subfigure}
    \hfill
    \begin{subfigure}[t]{0.32\textwidth}
        \includegraphics[width=\textwidth]
        {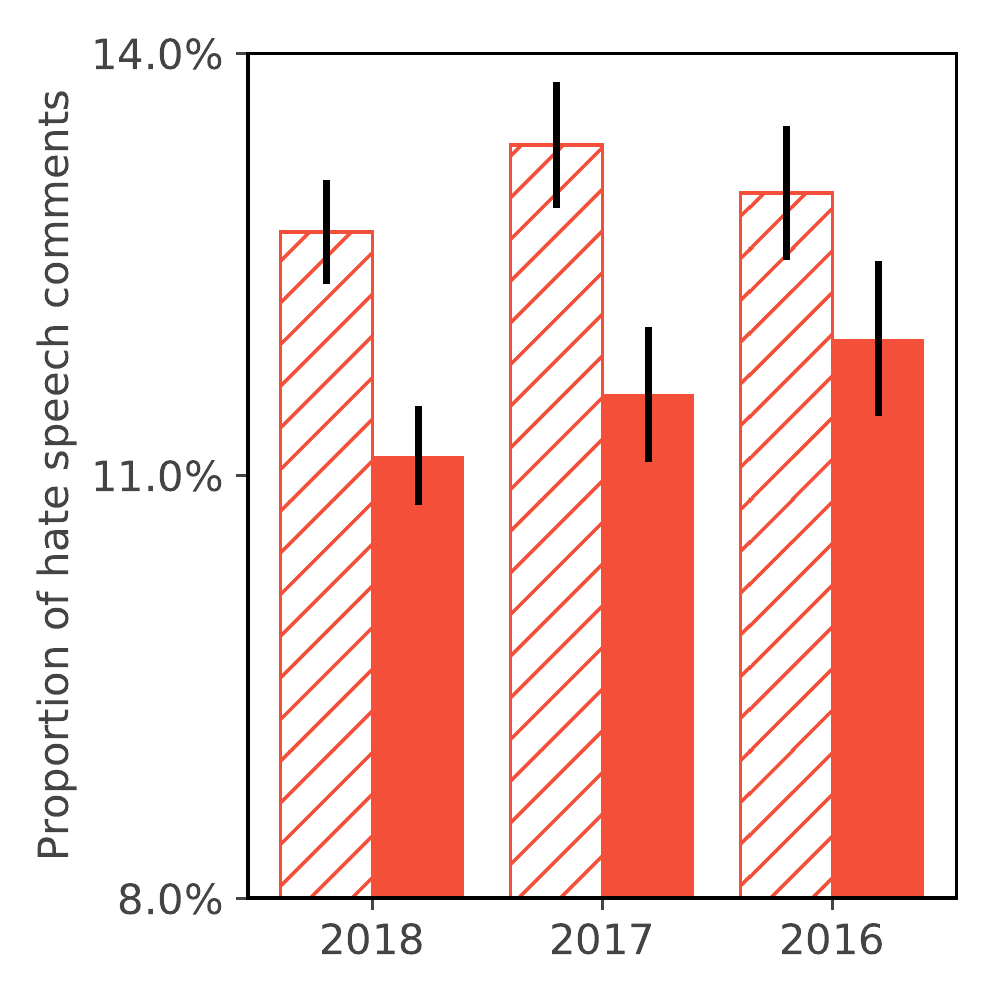}
        \caption{Proportion of hate speech comments.}
        \label{fig:interhate}
    \end{subfigure}
    \caption{The observation that \intergroupusers are more negative in the intergroup setting than in the intragroup setting is robust after controlling 
    for topics of discussion by only considering game threads.
    \Intergroupusers use more negative words 
    (two-tailed t-test, $t=2.43$, $p=0.015$,
    95\% CI=0.04\% to 0.35\% for the 2018 season;
    two-tailed t-test, $t=3.76$, $p<0.001$, 
    95\% CI=0.16\% to 0.52\% for the 2017 season;
    two-tailed t-test, $t=3.04$, $p=0.002$, 
    95\% CI=0.11\% to 0.51\% for the 2016 season)
    and swear words 
    (two-tailed t-test, $t=3.99$, $p<0.001$, 
    95\% CI=0.18\% to 0.54\% for the 2018 season;
    two-tailed t-test, $t=2.34$, $p=0.023$, 
    95\% CI=0.04\% to 0.32\% for the 2017 season;
    two-tailed t-test, $t=1.96$, $p=0.048$,  
    95\% CI=0.00\% to 0.48\% for the 2016 season)
    and generate more hate speech comments 
    (two-tailed t-test, $t=3.10$, $p=0.002$,
    95\% CI=0.06\% to 2.59\% for the 2018 season;
    two-tailed t-test, $t=2.70$, $p=0.007$, 
    95\% CI=0.05\% to 3.05\% for the 2017 season;
    two-tailed t-test, $t=1.42$, $p=0.155$,
    95\% CI=0.00\% to 2.46\% for the 2016 season)
    in the game threads of the intergroup setting than
    that of the intragroup setting. 
    We do not compare the language usage patterns per team in this analysis, as 
    there are teams with less than 20 members after this control. 
    Error bars represent standard errors.
    }
    \label{fig:gamethread}
\end{figure}

In \secref{sec:differentlevel}, 
we find that \intergroupusers use more negative language in the intergroup setting 
than in the intragroup setting.
This difference may occur due to the fact that more heated topics are 
discussed in \communityname{/r/NBA} than in team subreddits. 
To control for this factor, 
we further limit our comparison to the game threads in both settings.
Game threads are important components of NBA-related team subreddits to facilitate game-related discussions
during NBA games.
In practice, each game has a game thread in the home-team subreddit, the away team subreddit, and
the overall \communityname{/r/NBA}.
\figref{fig:gamethread} shows the language usage difference of \intergroup members 
in the game threads of the intergroup and intragroup setting. 
Only members who made comments in the game threads of both settings are included in this analysis
(2118 members for the 2018 season, 1495 members for the 2017 season, 
and 1289 members for the 2016 season).
In all three seasons, it is consistent that \intergroupusers use more negative words 
in the game threads of the intergroup setting than
of the intragroup setting. 
We do not compare the language usage patterns per team in this analysis as 
there are teams with less than 20 members. 

\clearpage

\begin{figure}
    \center
    \includegraphics[width=0.7\textwidth]
    {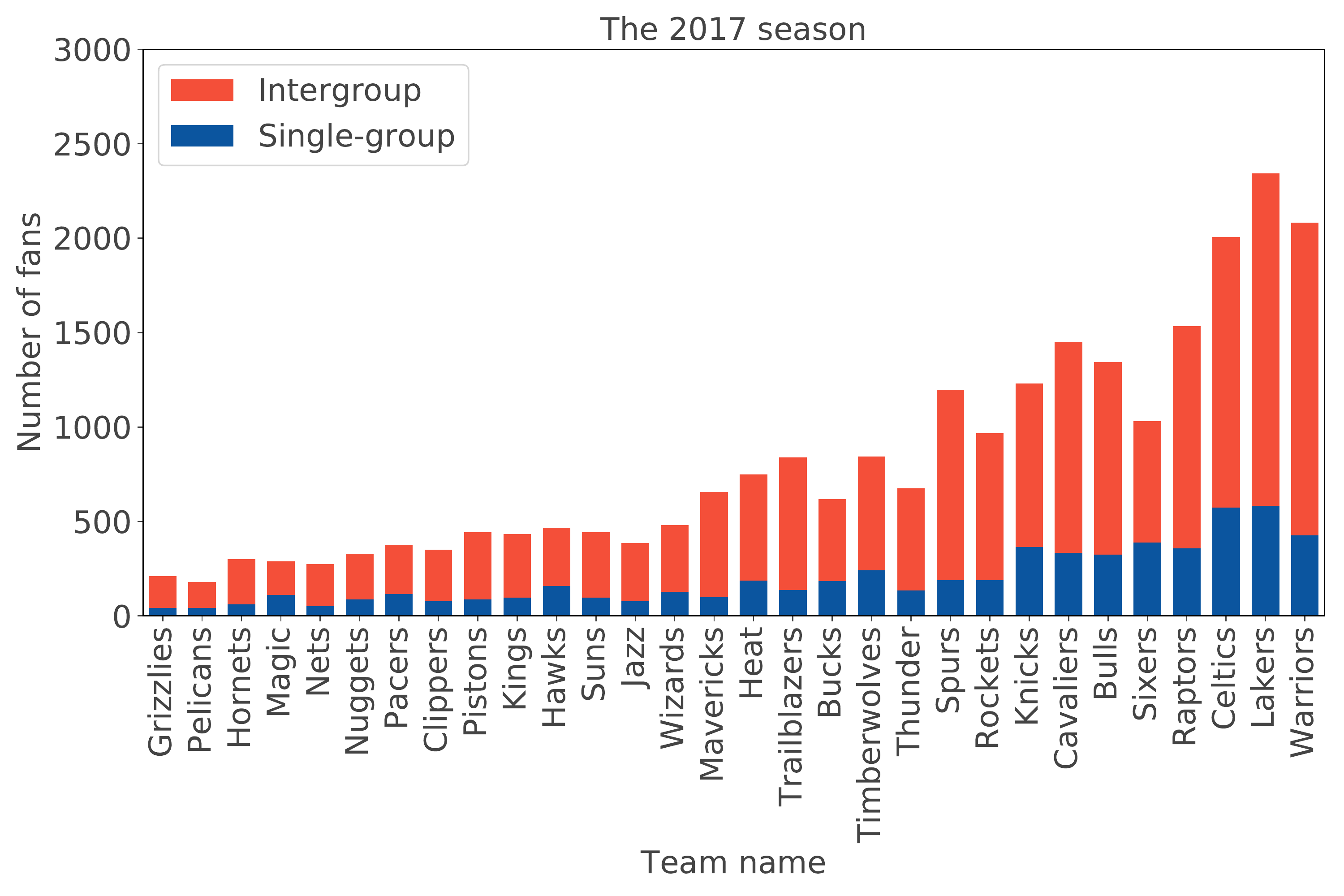}
    \includegraphics[width=0.7\textwidth]
    {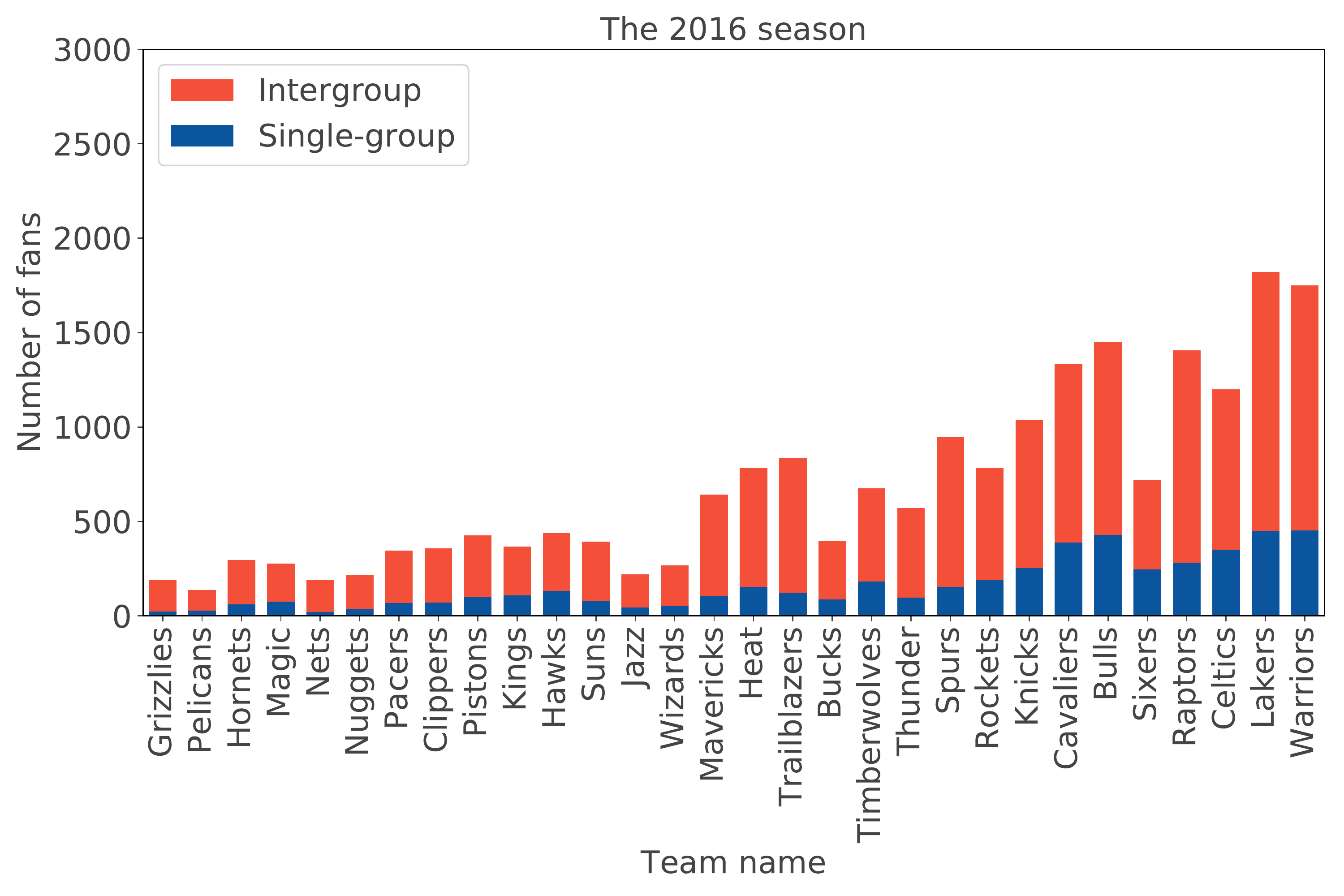}
    \caption{The number of \intergroup and \teamonlyusers 
    affiliated with each NBA team in the 2017 and 2016 seasons.
    We rank 30 team subreddits by the number of subscribers 
    each team has by the end of the 2018 season.
    }
    \label{fig:appoverallnumofusers}
\end{figure}

\begin{figure}
    \center
    \begin{subfigure}[t]{\textwidth}
        \includegraphics[width=0.19\textwidth]
        {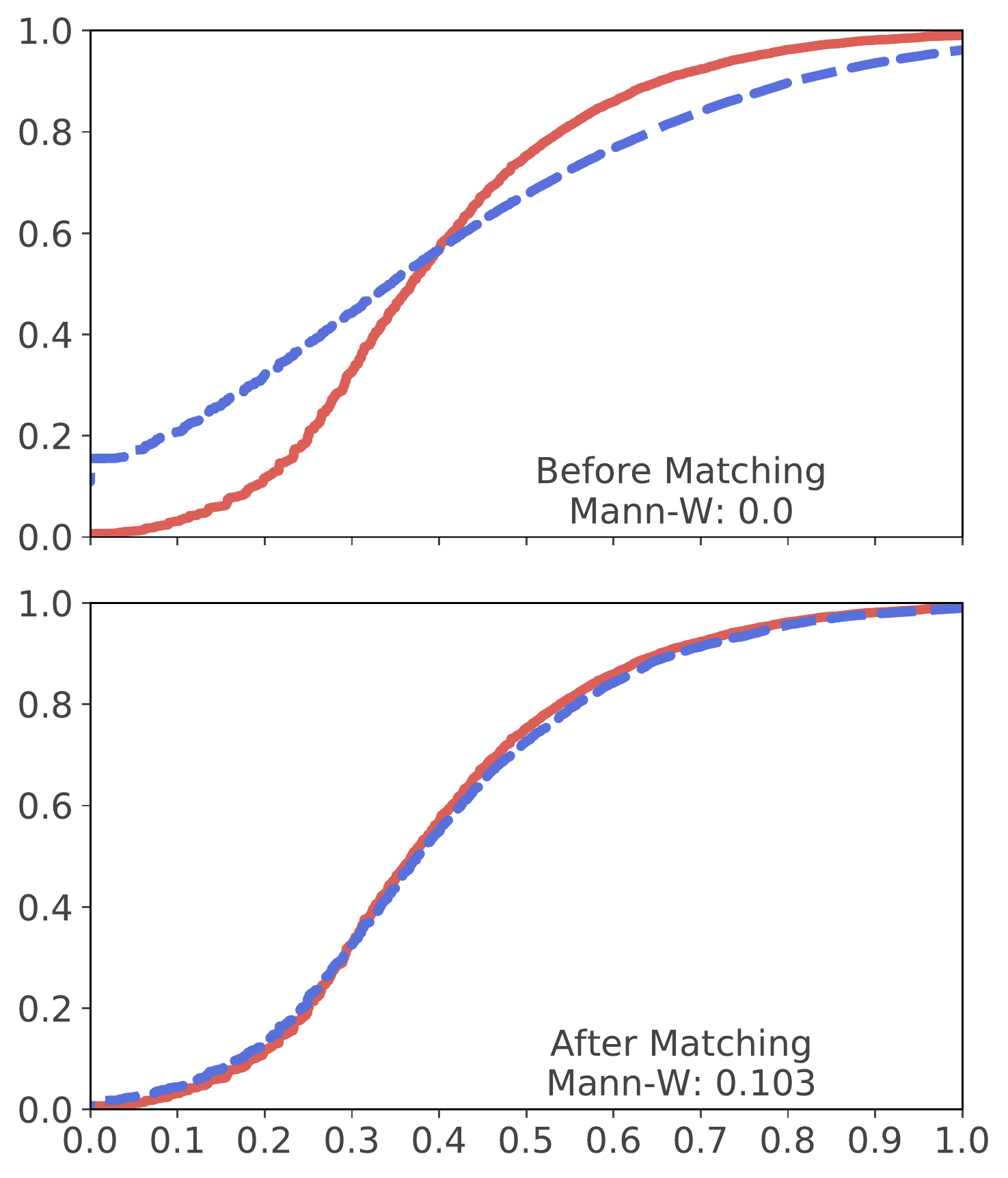}
        \includegraphics[width=0.19\textwidth]
        {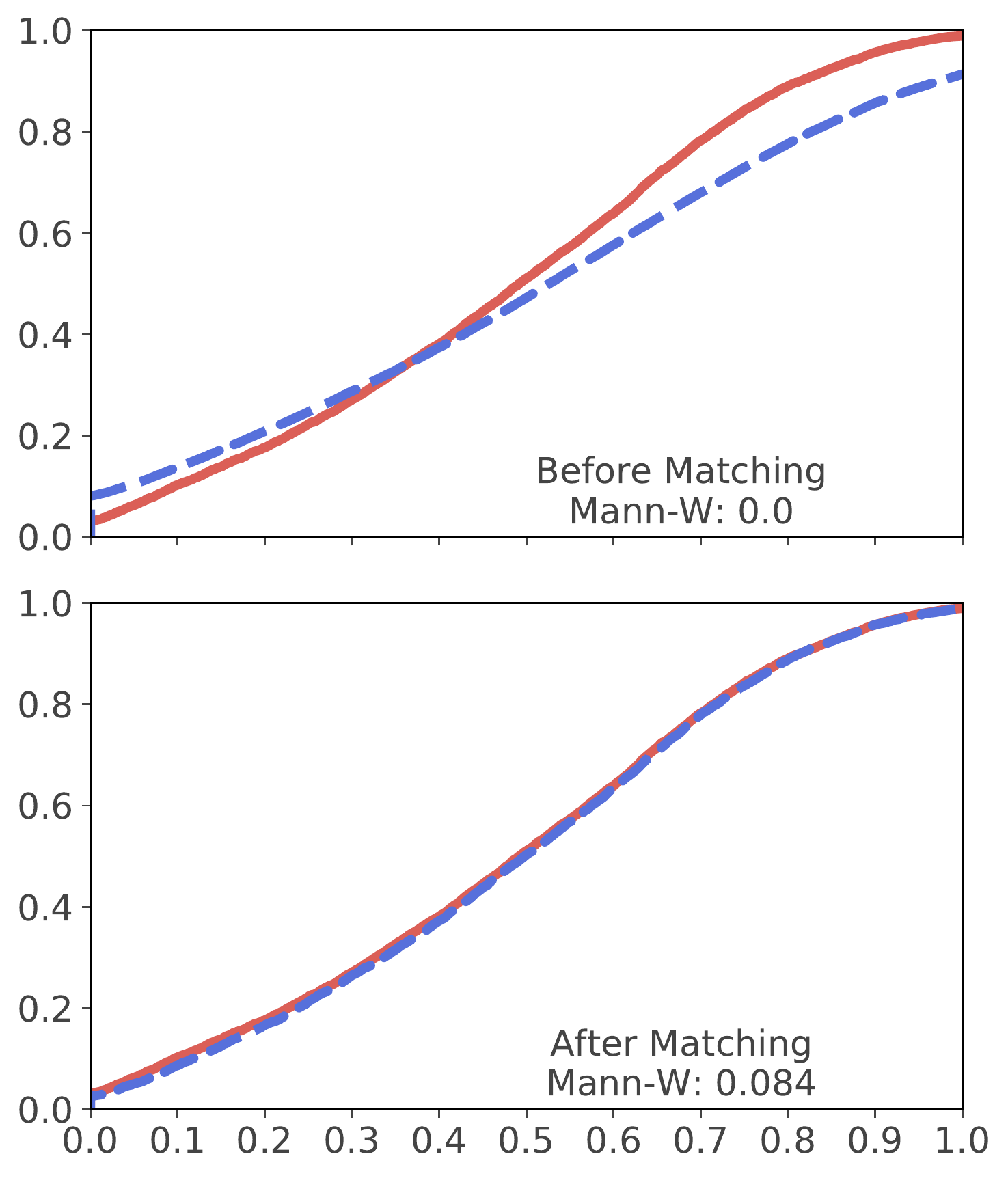}
        \includegraphics[width=0.19\textwidth]
        {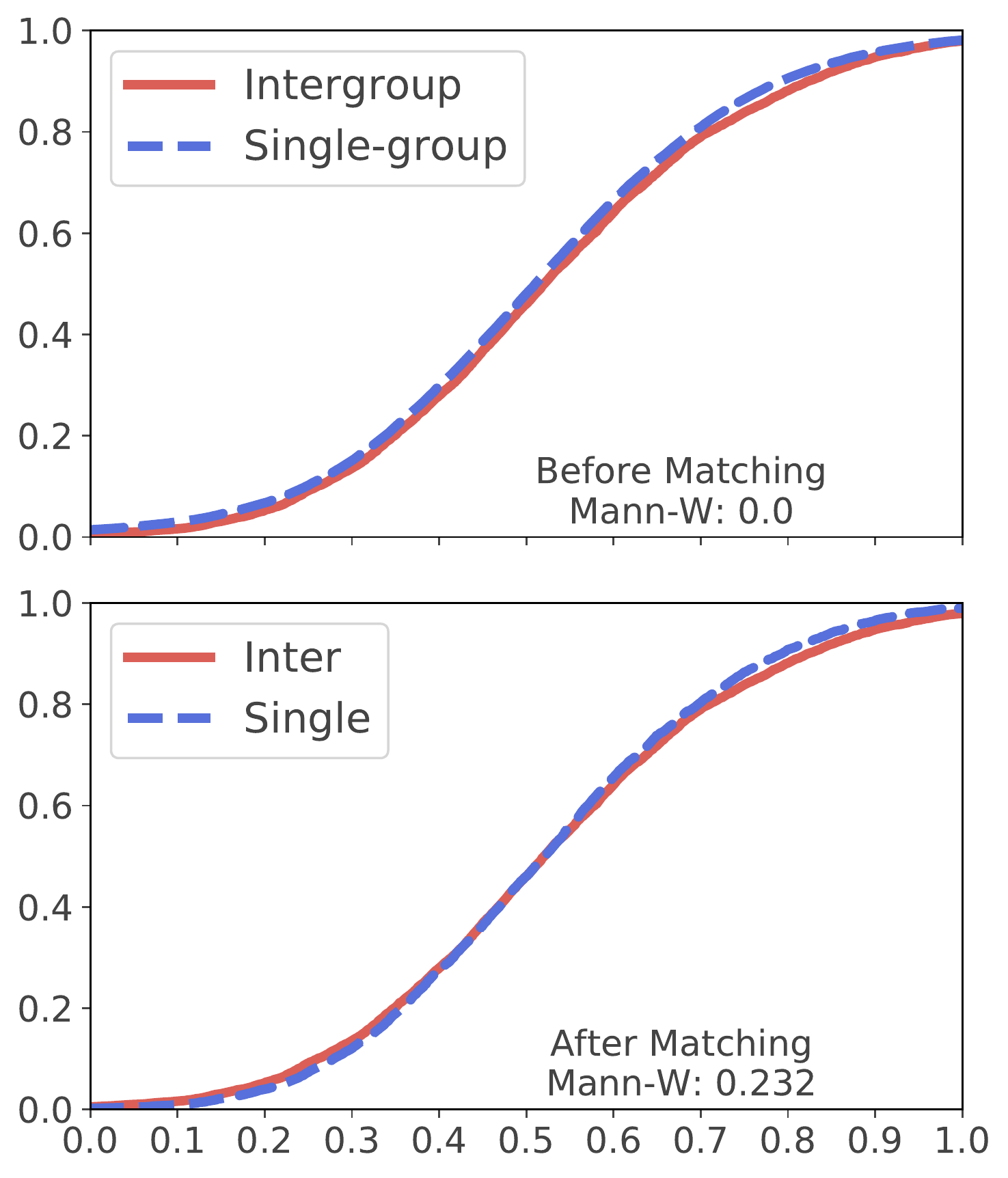}
        \includegraphics[width=0.19\textwidth]
        {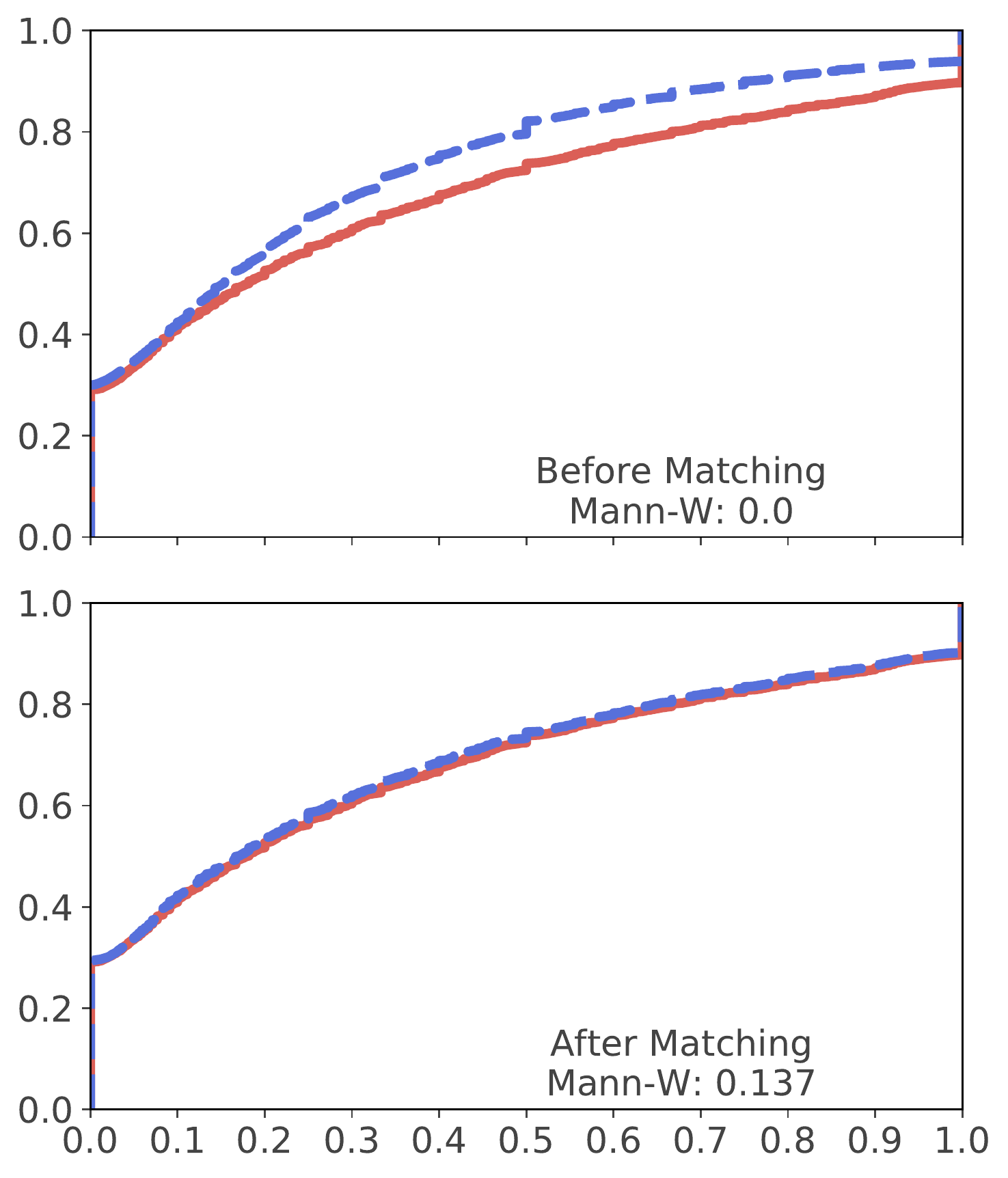}
        \includegraphics[width=0.19\textwidth]
        {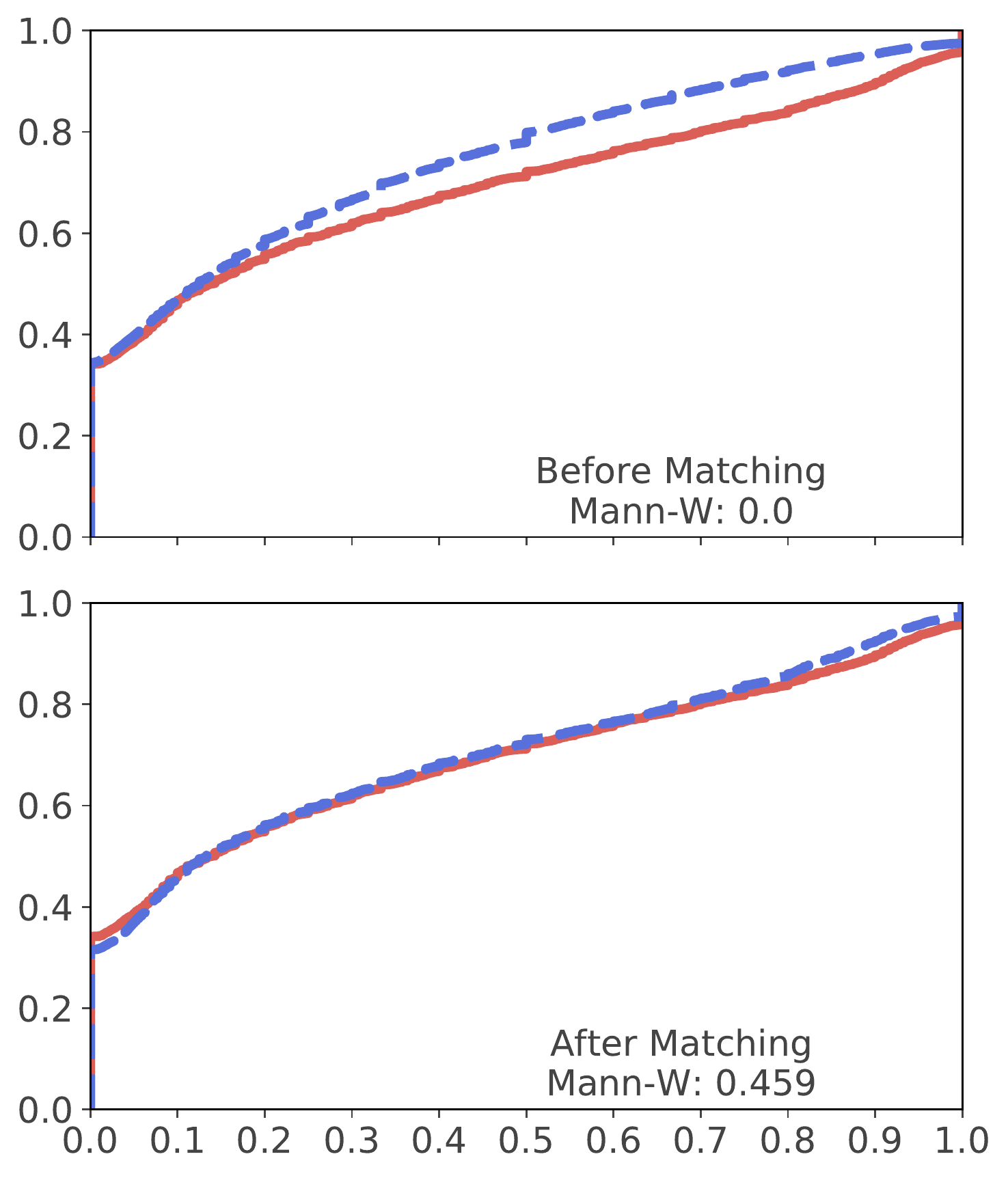}
        \caption{Season 2018}
    \end{subfigure}
    \hfill
    \begin{subfigure}[t]{\textwidth}
        \includegraphics[width=0.19\textwidth]
        {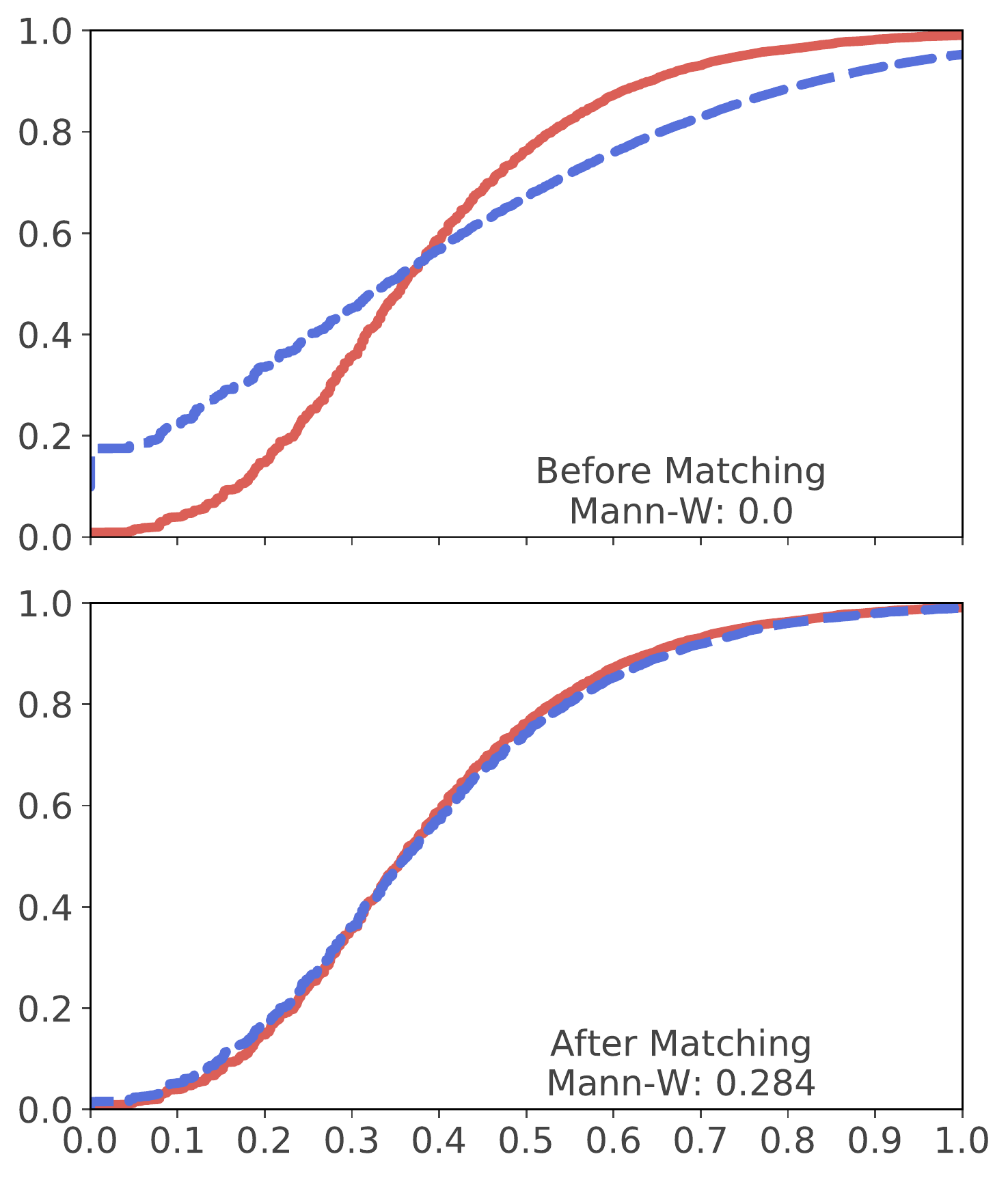}
        \includegraphics[width=0.19\textwidth]
        {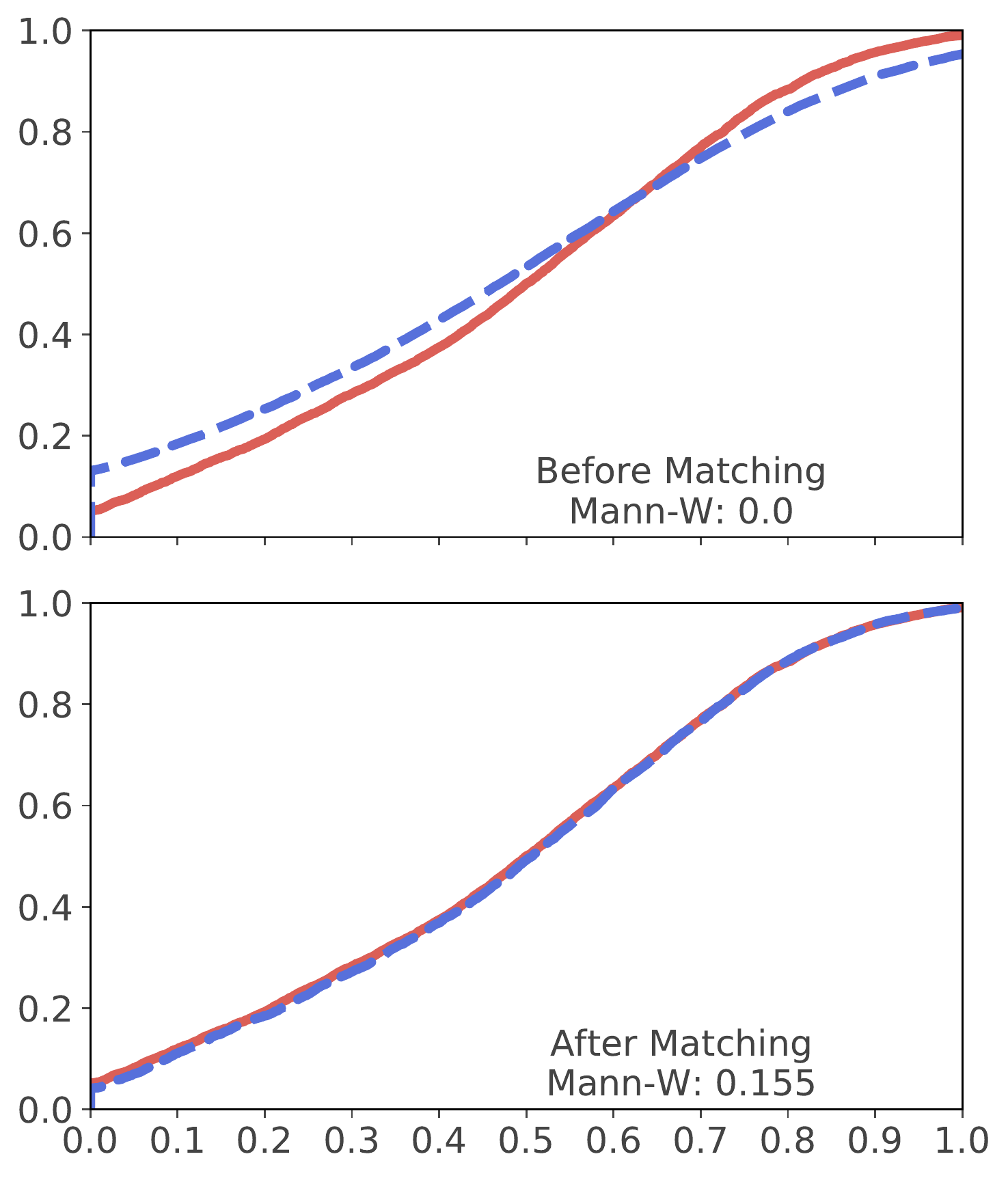}
        \includegraphics[width=0.19\textwidth]
        {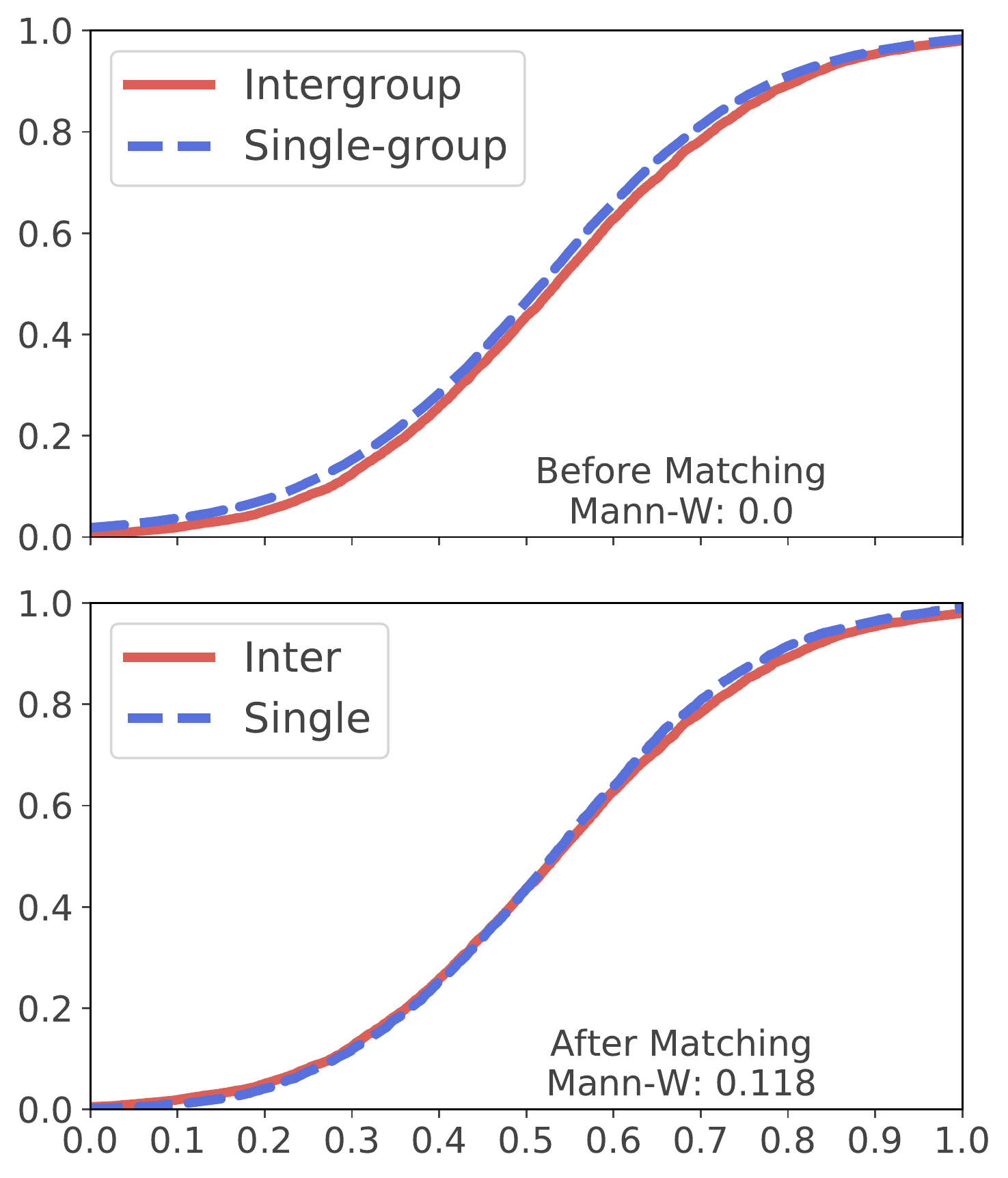}
        \includegraphics[width=0.19\textwidth]
        {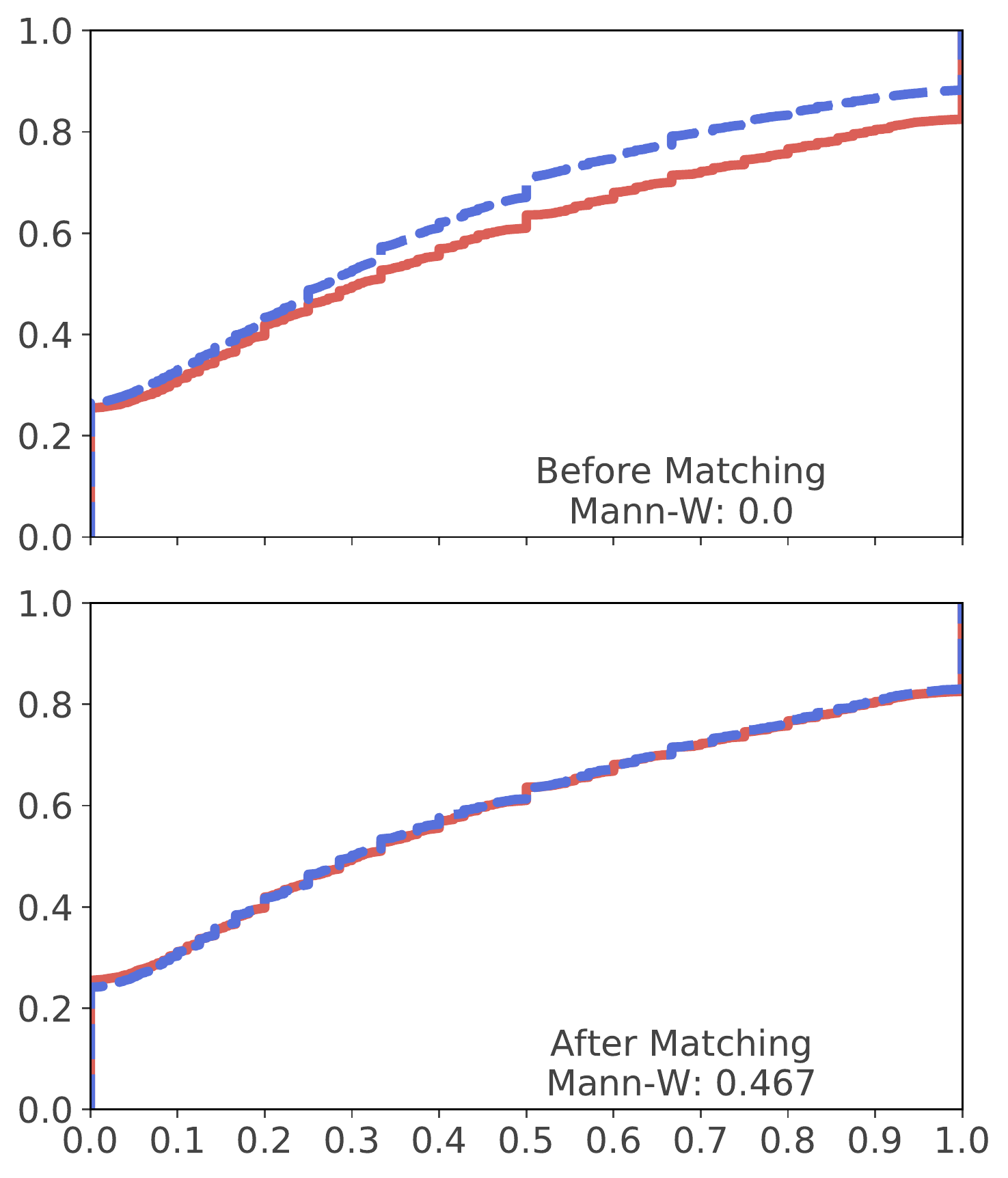}
        \includegraphics[width=0.19\textwidth]
        {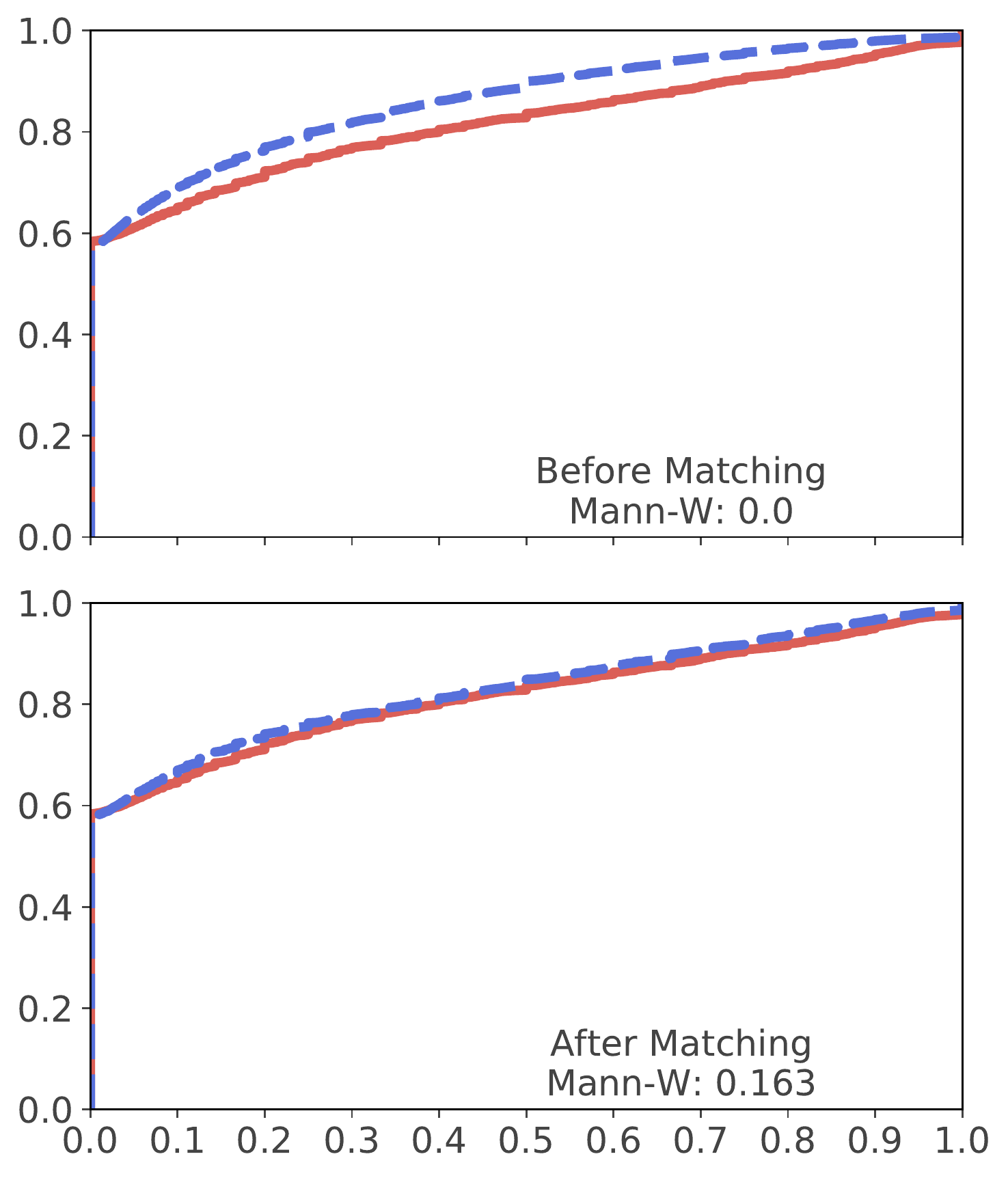}
        \caption{Season 2017}
    \end{subfigure}
    \hfill
    \begin{subfigure}[t]{\textwidth}
        \includegraphics[width=0.19\textwidth]
        {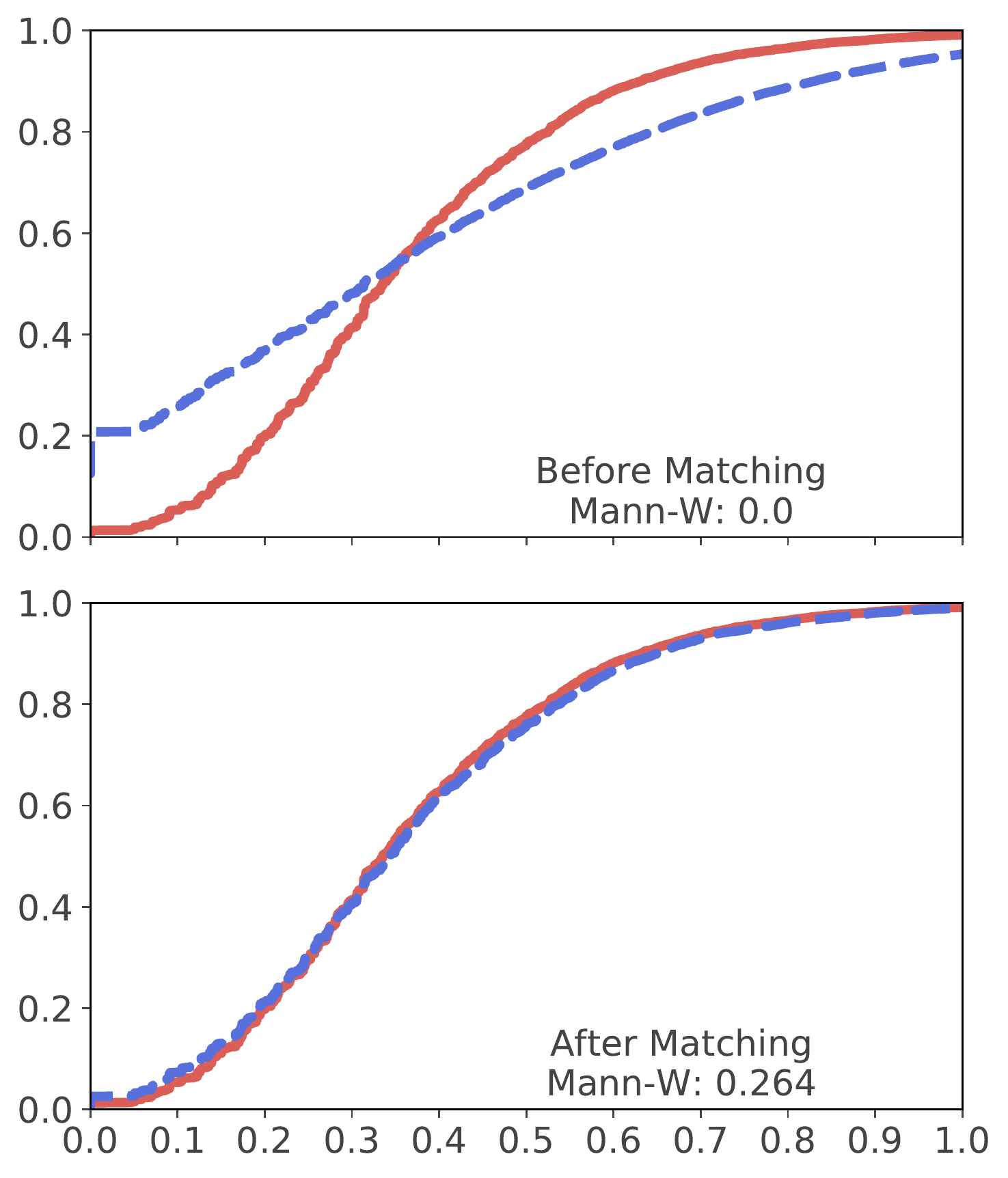}
        \includegraphics[width=0.19\textwidth]
        {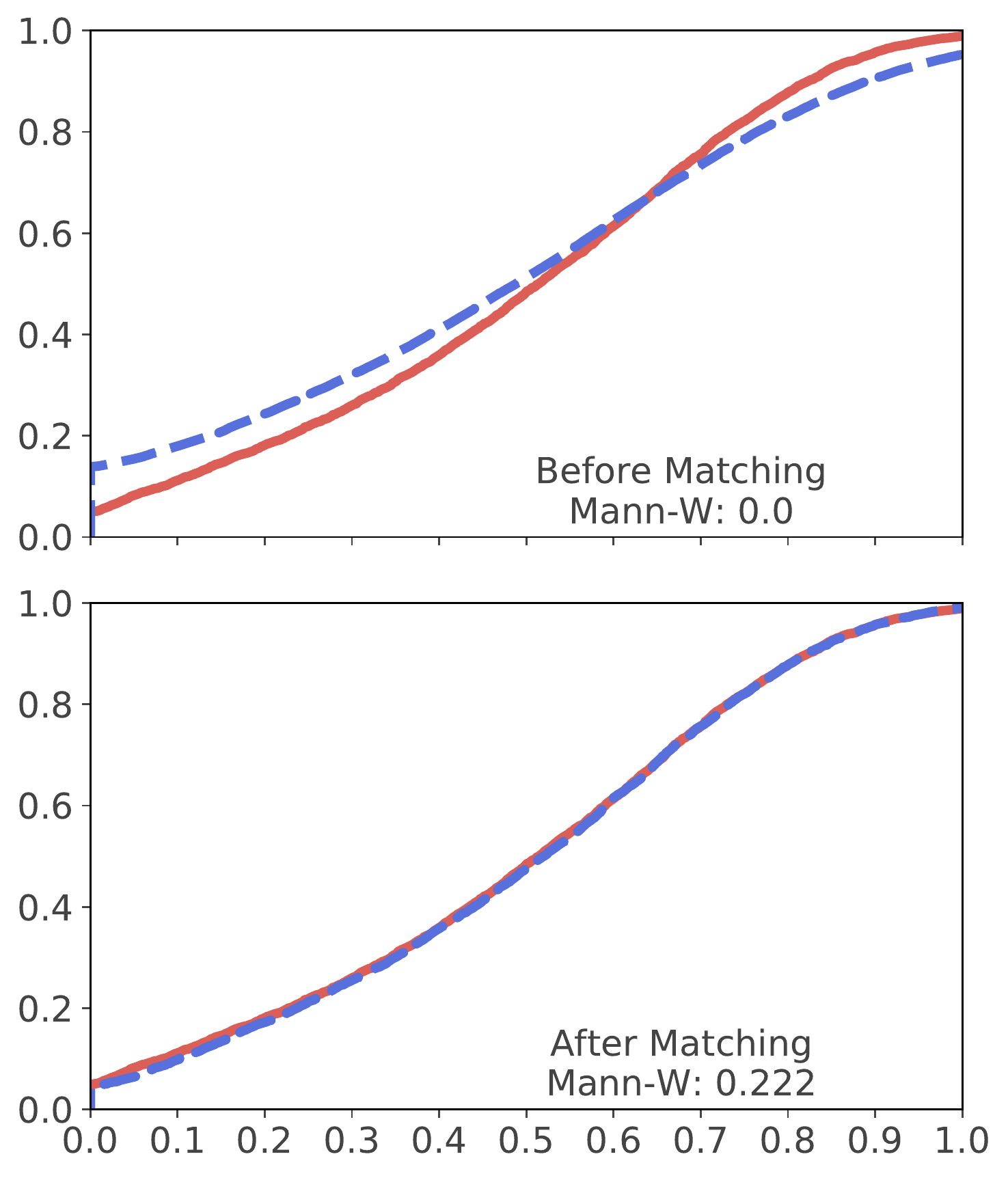}
        \includegraphics[width=0.19\textwidth]
        {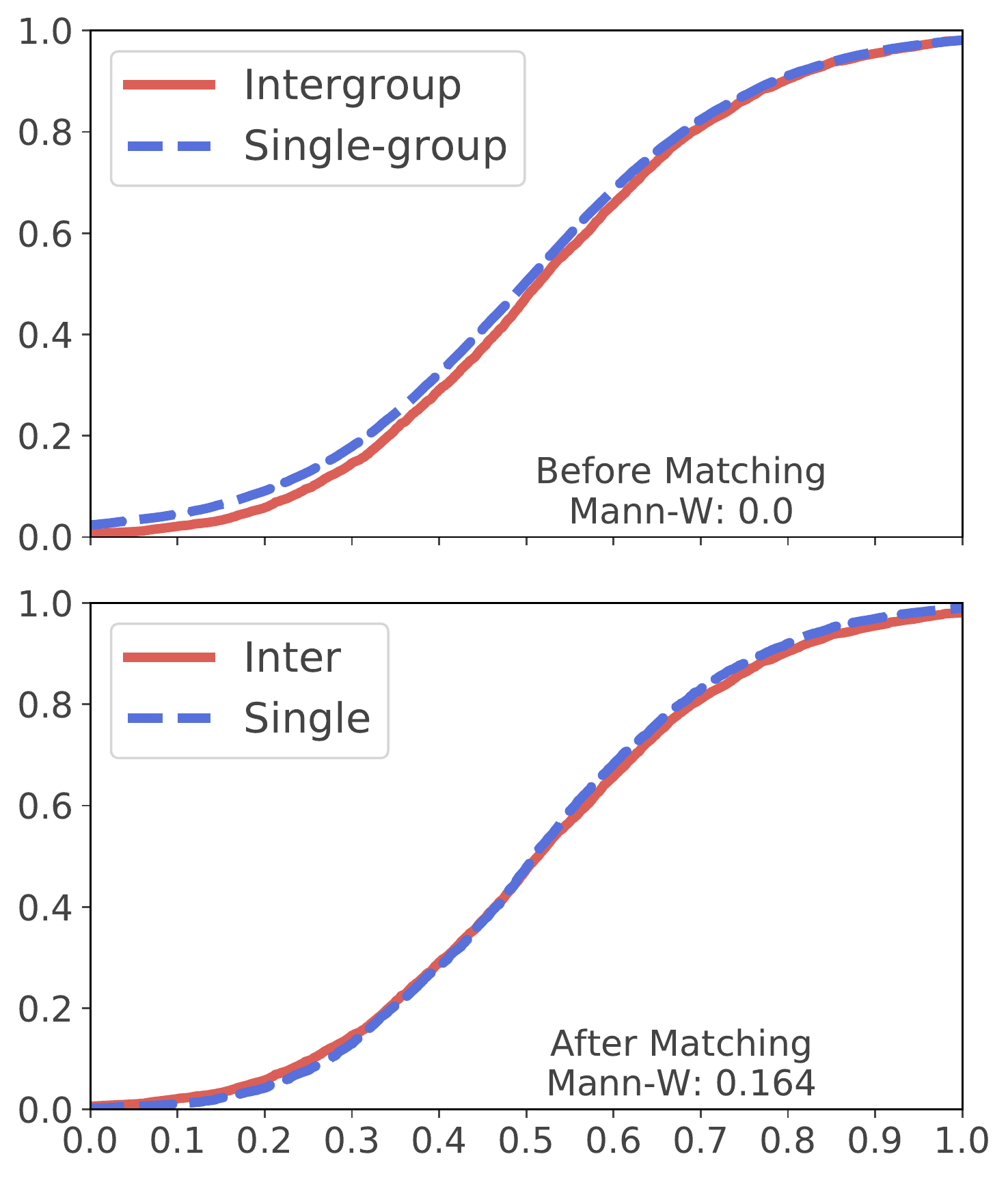}
        \includegraphics[width=0.19\textwidth]
        {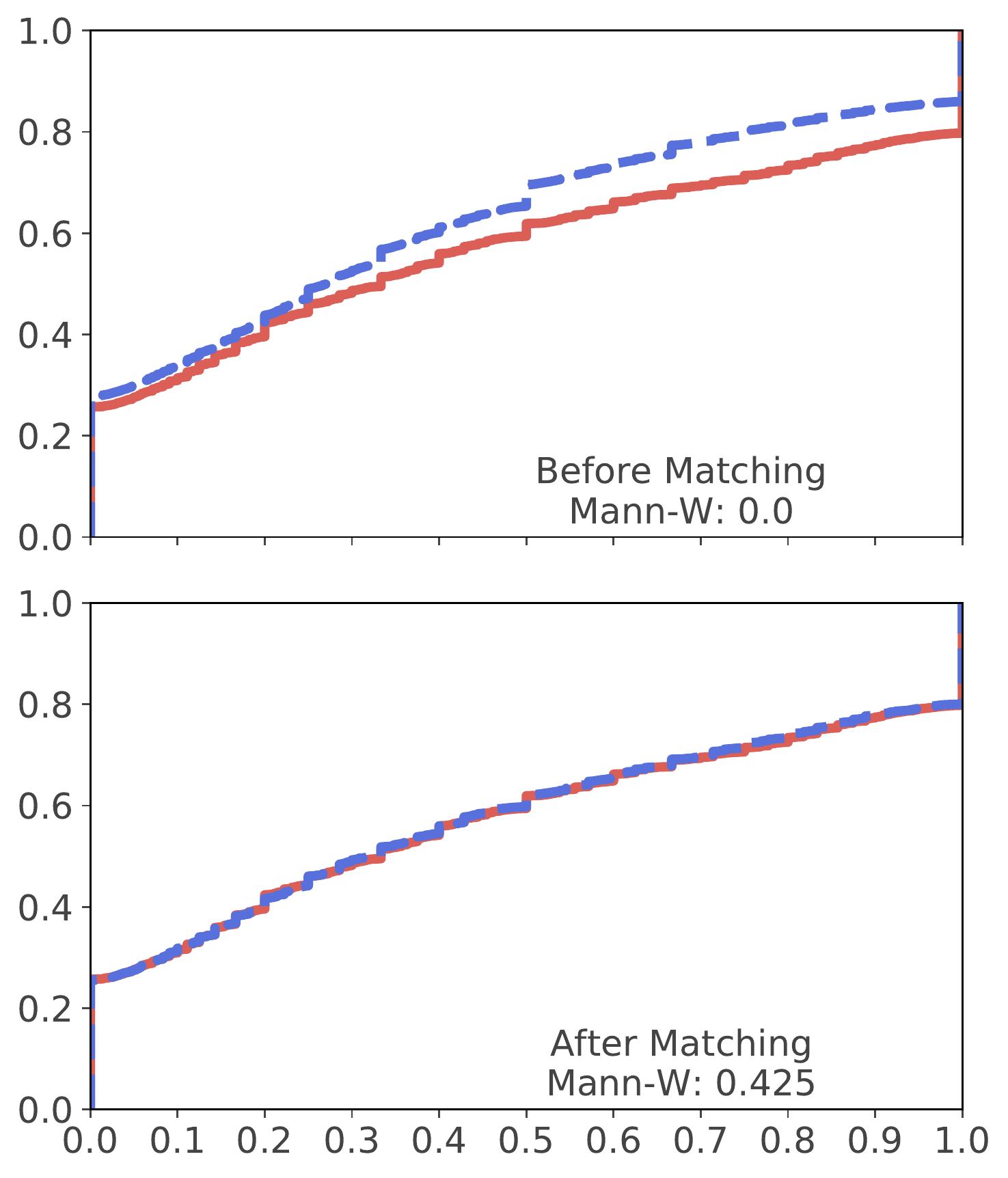}
        \includegraphics[width=0.19\textwidth]
        {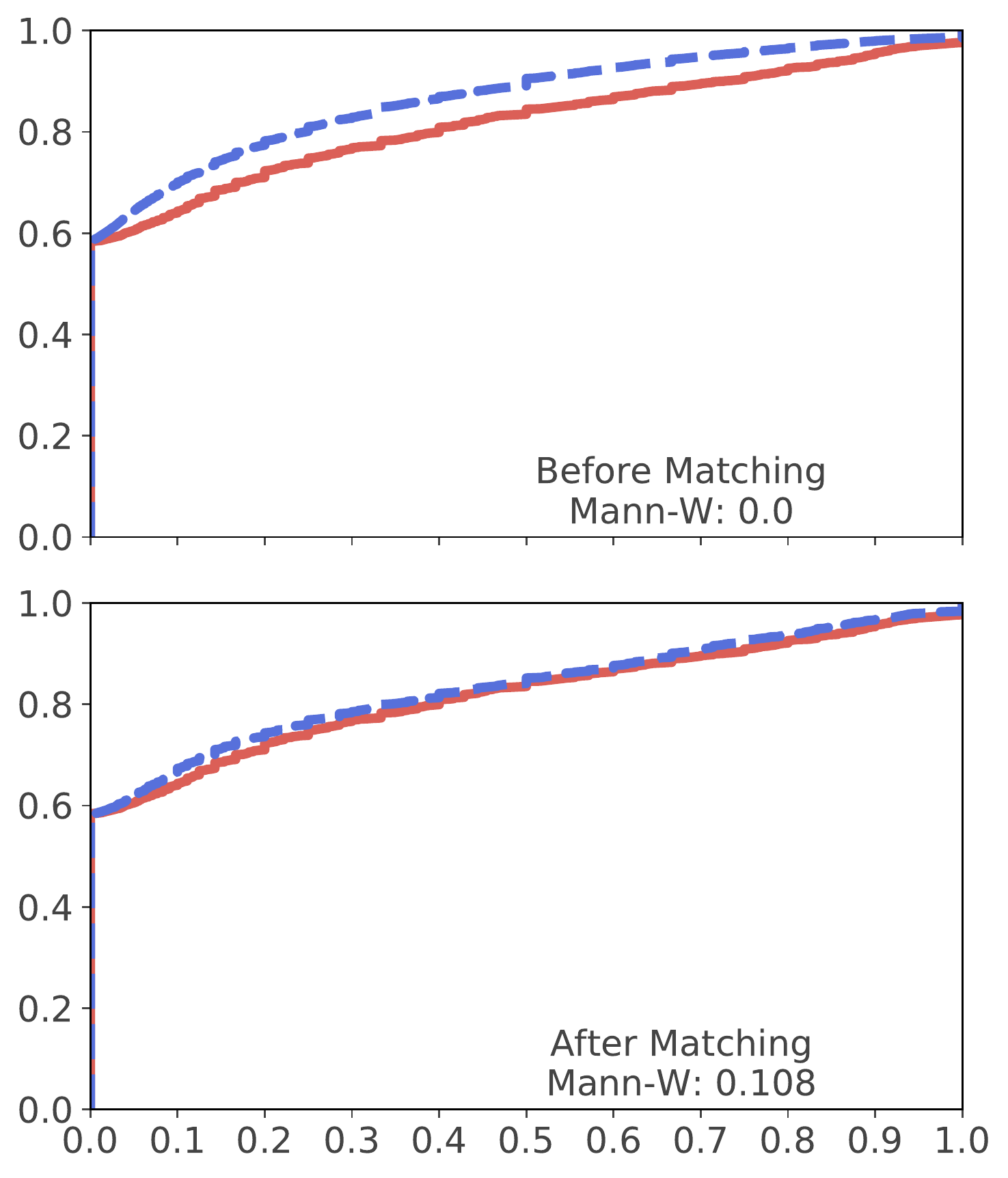}
        \caption{Season 2016}
    \end{subfigure}
    \caption{Empirical cumulative distribution of each activity feature before and 
    after the matching technique in the 2018, 2017, and 2016 seasons. 
    The activity features from left to right are the number of comments,
    the average hour gap between comments, the average comment length, 
    the proportion of playoff comments, and the proportion of game thread comments.
    The corresponding p-values of the Mann-Whitney tests are also reported.
    Recall that a small p-value indicates that there is dependence 
    between the treatment and control groups
    (relative frequencies are different). 
    Prior to matching, each p-value is very close to 0.0. 
    After the matching, at the 0.05 significance level ($\alpha=0.05$), 
    we find no dependence on the group label for any activity feature 
    observed in the matched dataset. These trends are consistent in all three seasons. 
    }
    \label{fig:matching}
\end{figure}

\begin{figure}
    \center
    \begin{subfigure}[t]{\textwidth}
        \includegraphics[width=0.19\textwidth]
        {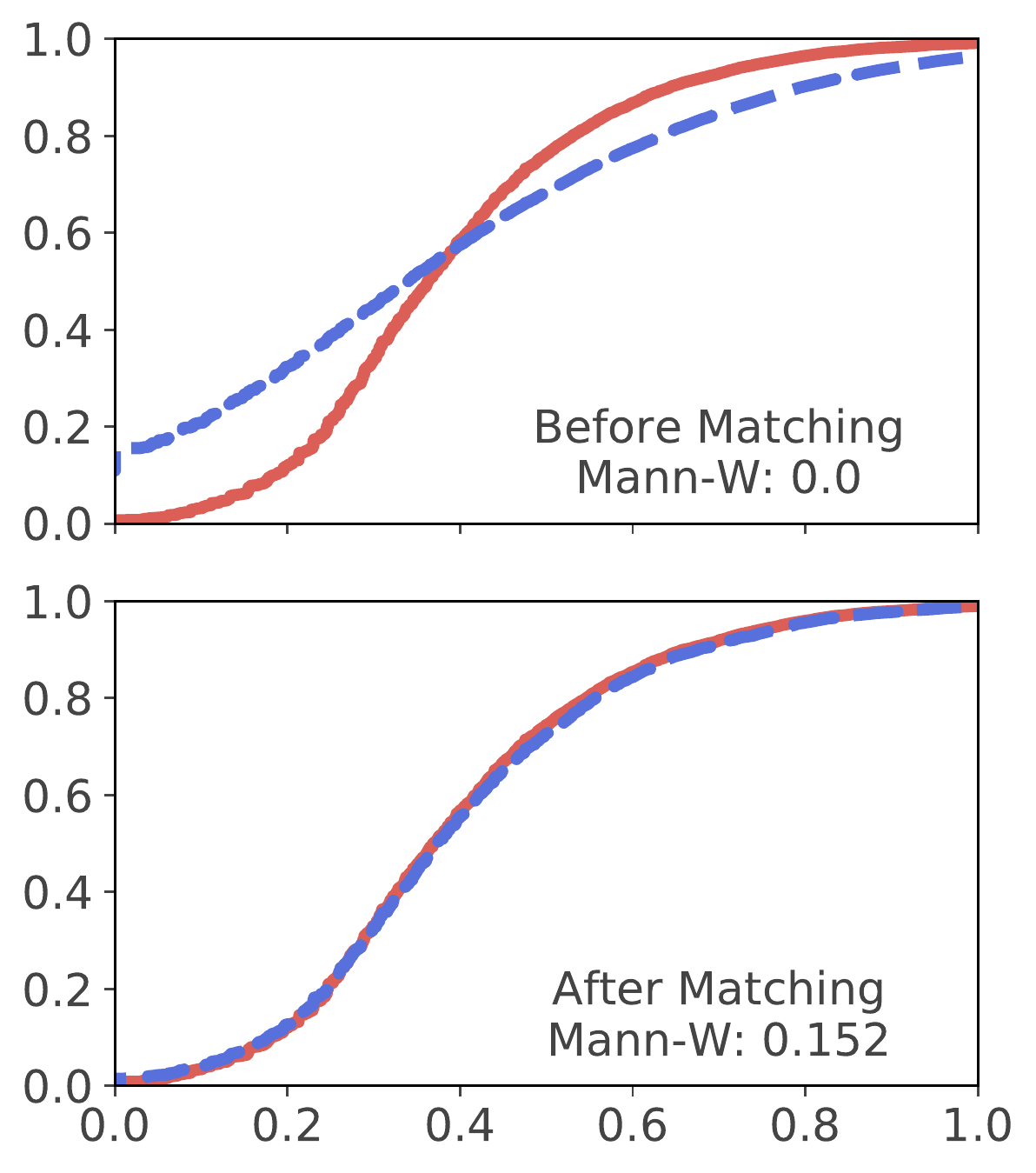}
        \includegraphics[width=0.19\textwidth]
        {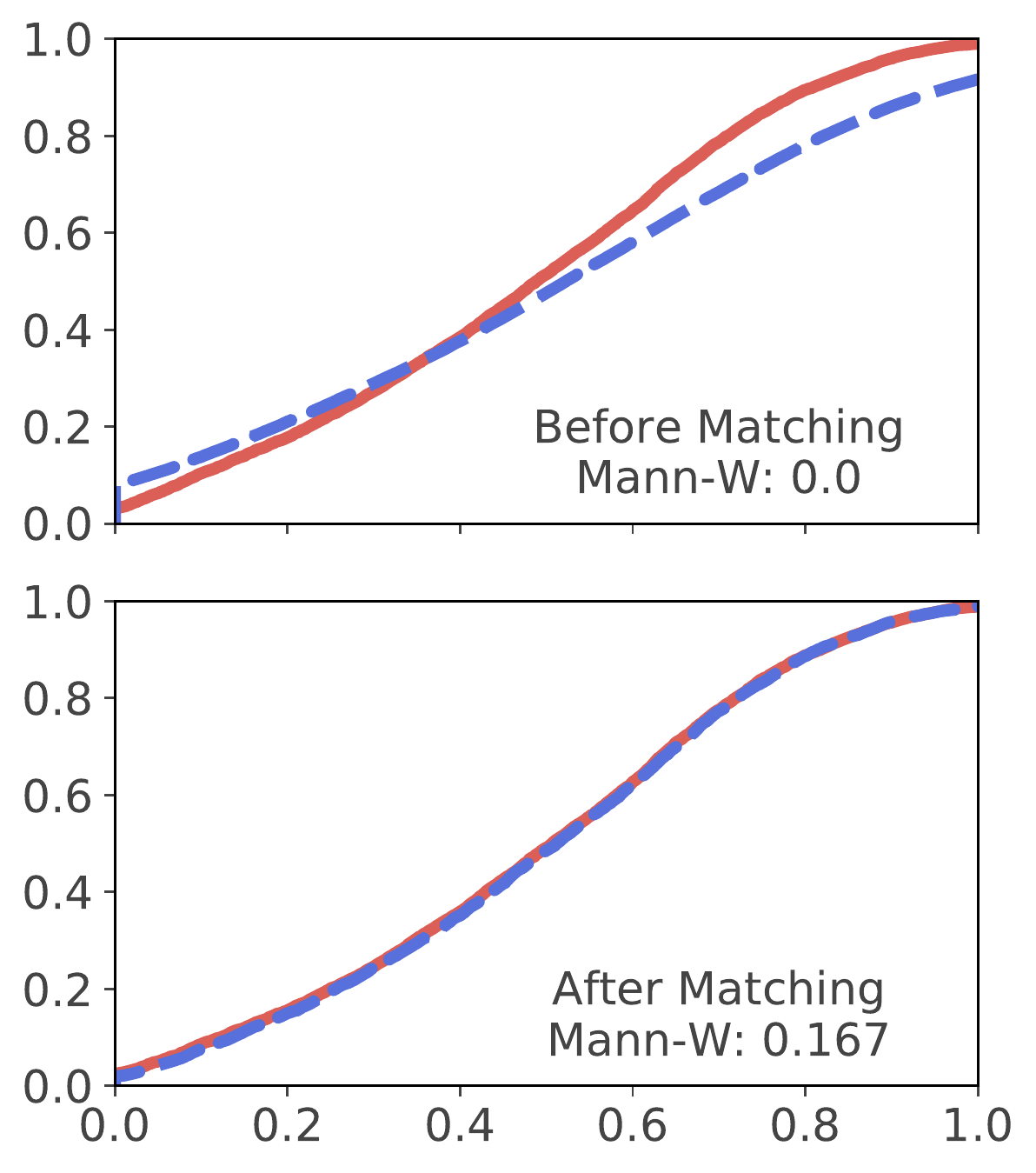}
        \includegraphics[width=0.19\textwidth]
        {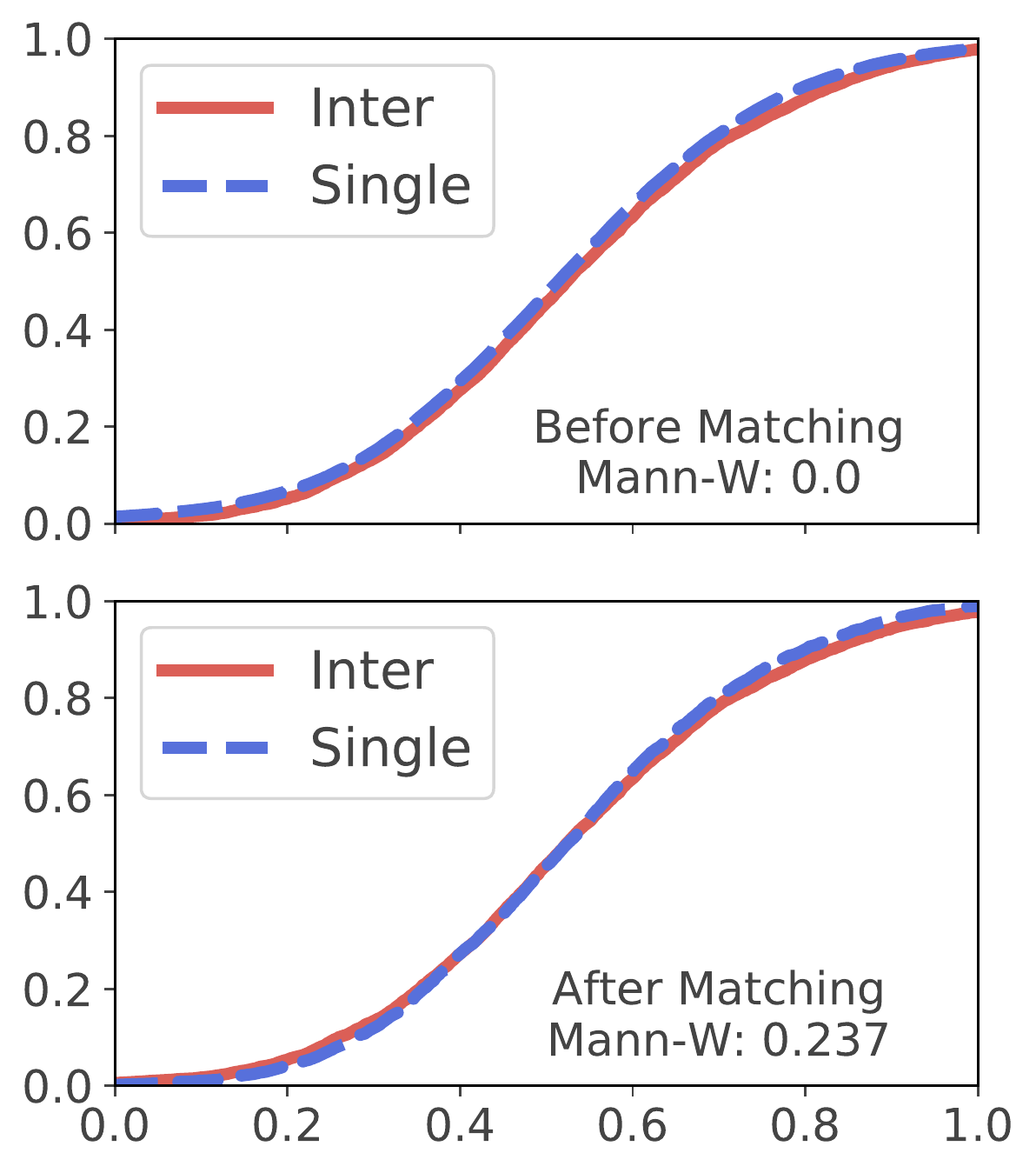}
        \includegraphics[width=0.19\textwidth]
        {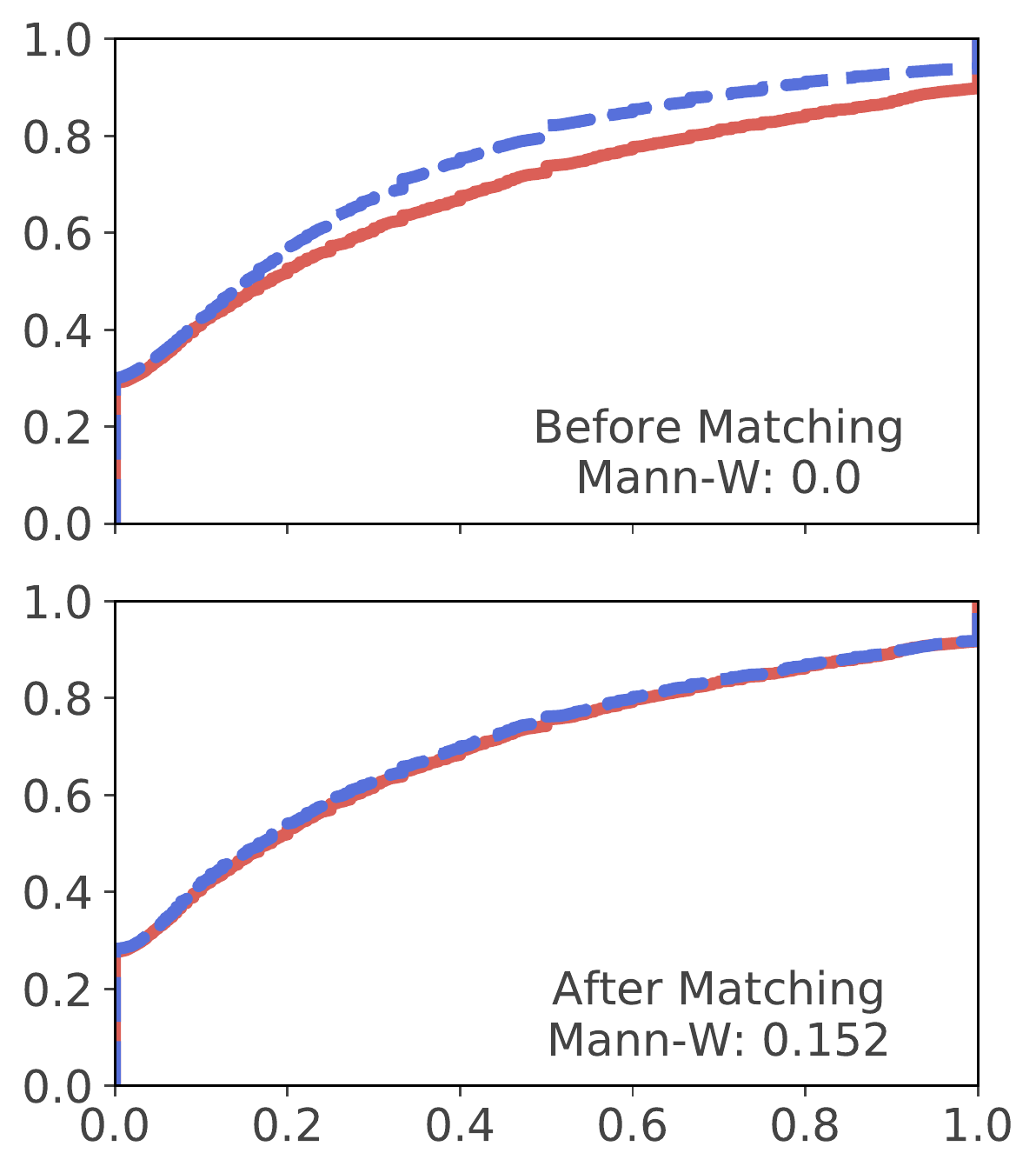}
        \includegraphics[width=0.19\textwidth]
        {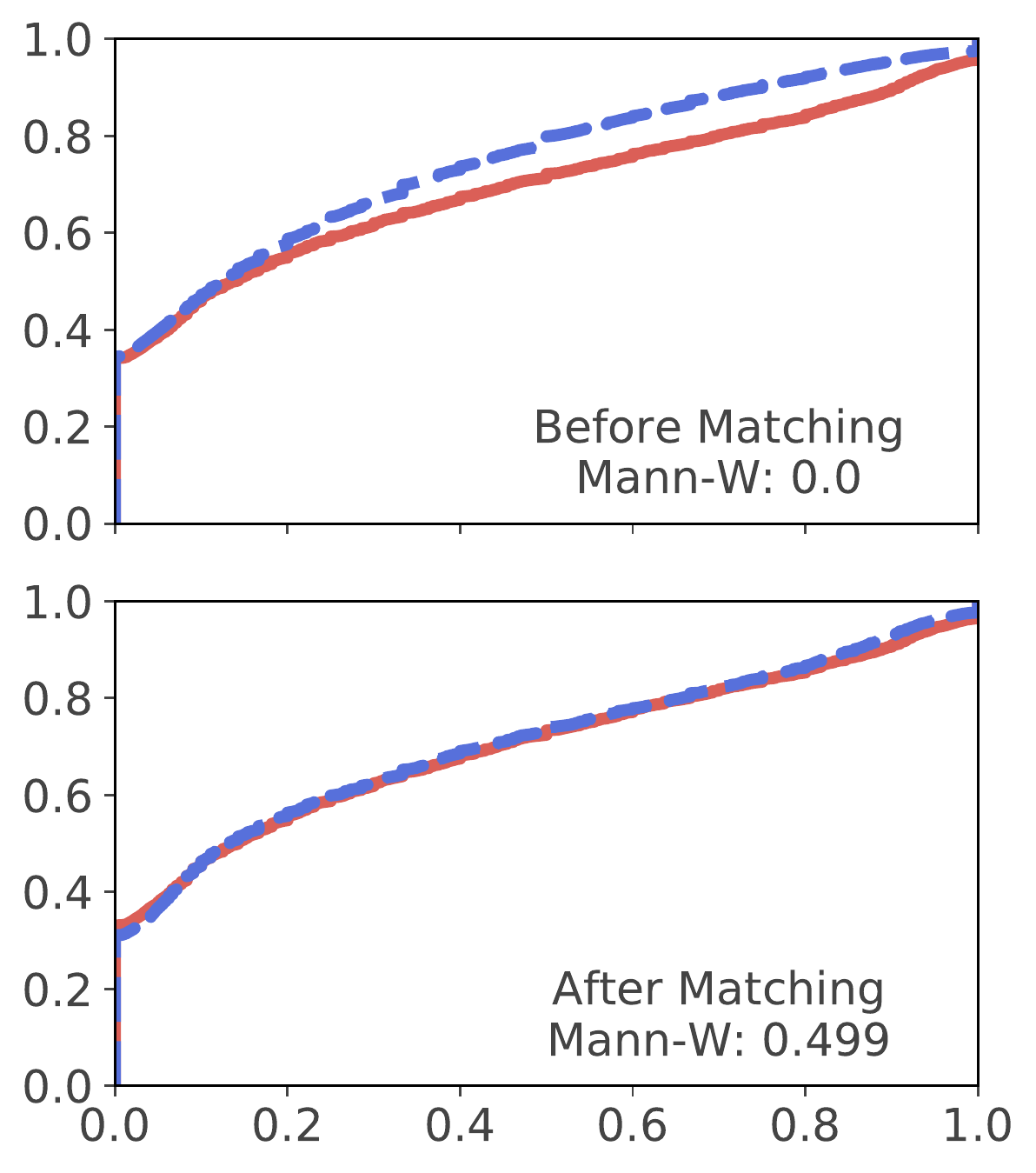}
        \caption{Matching results for level 1.}
    \end{subfigure}
    \hfill
    \begin{subfigure}[t]{\textwidth}
        \includegraphics[width=0.19\textwidth]
        {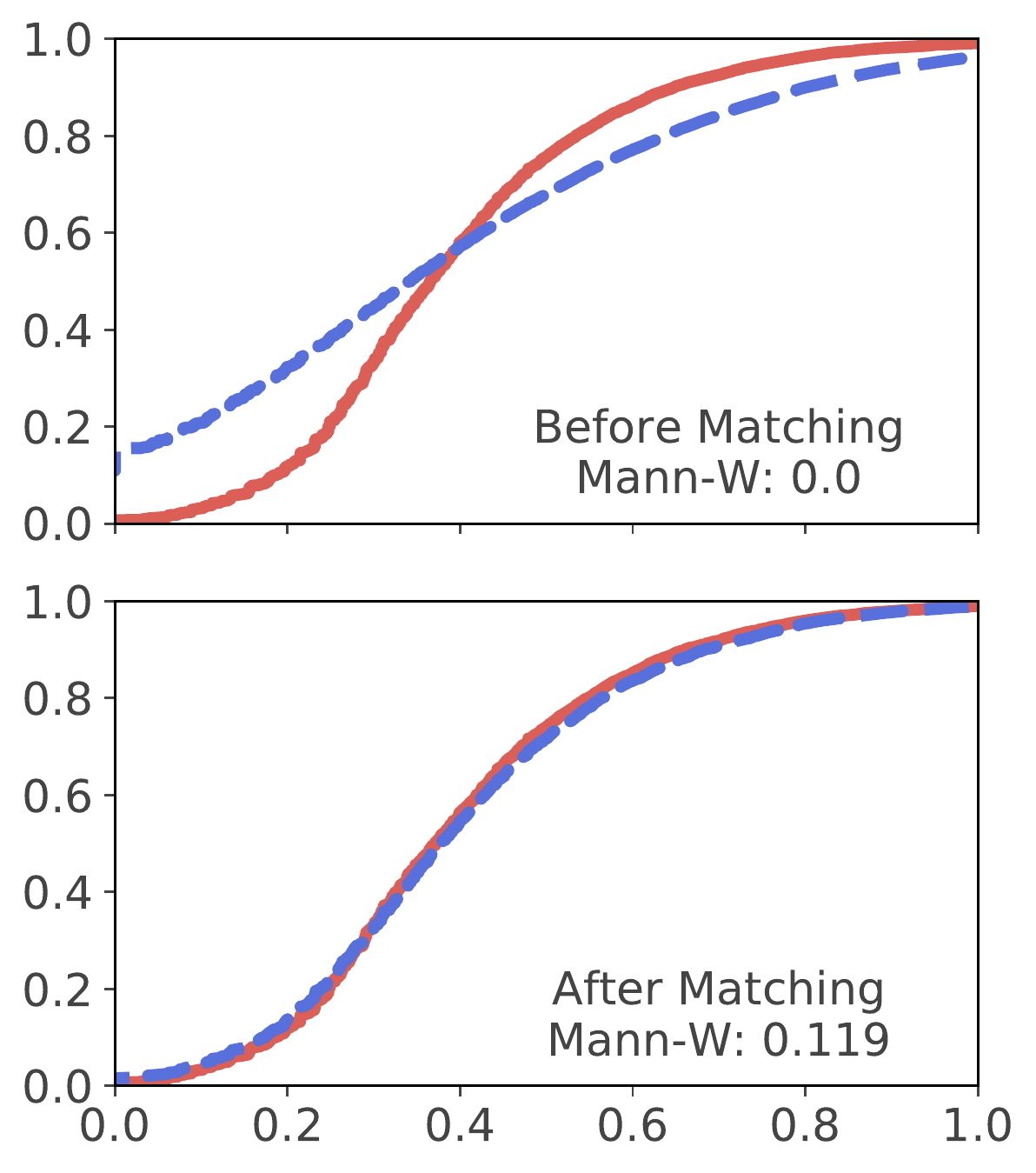}
        \includegraphics[width=0.19\textwidth]
        {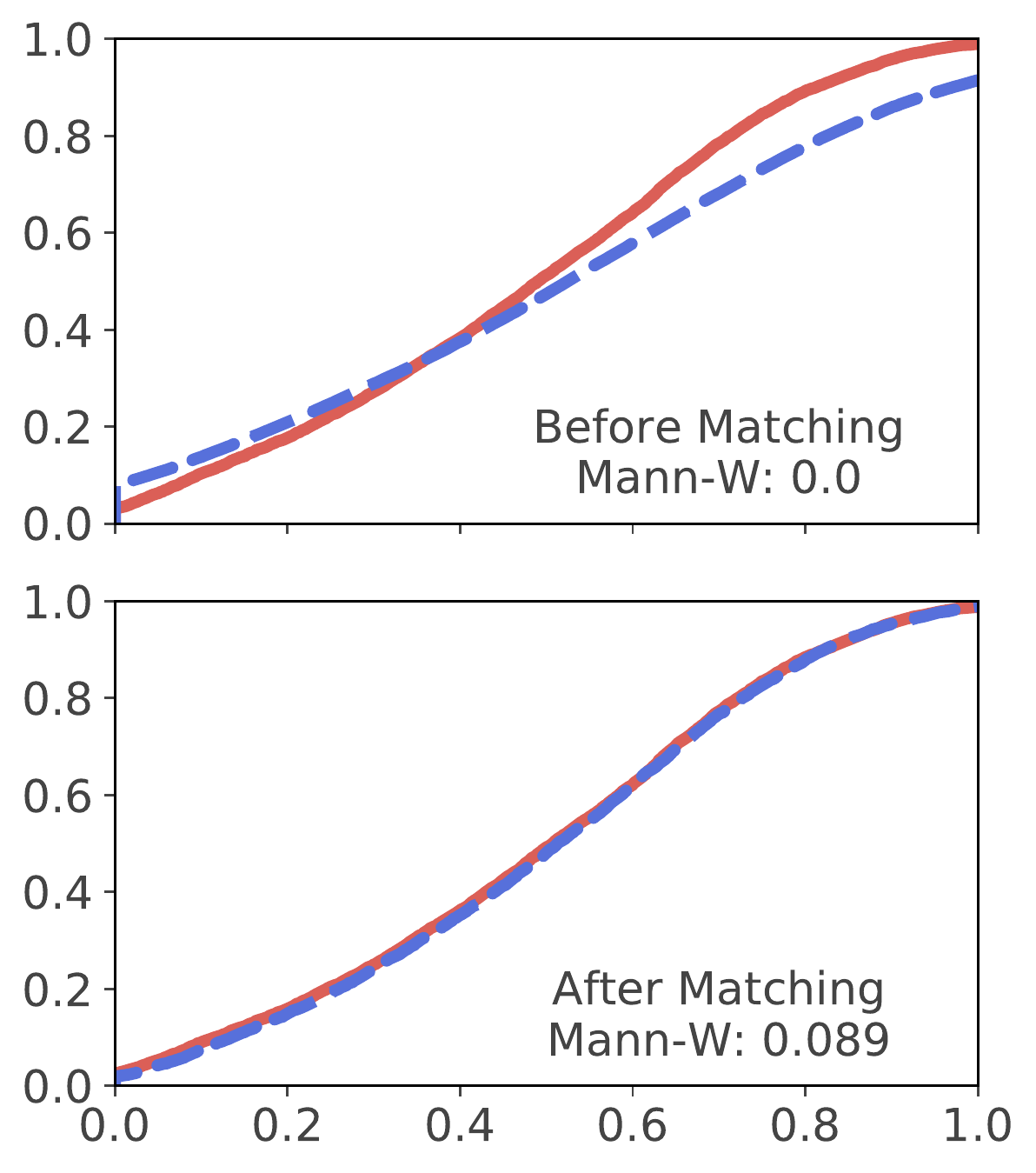}
        \includegraphics[width=0.19\textwidth]
        {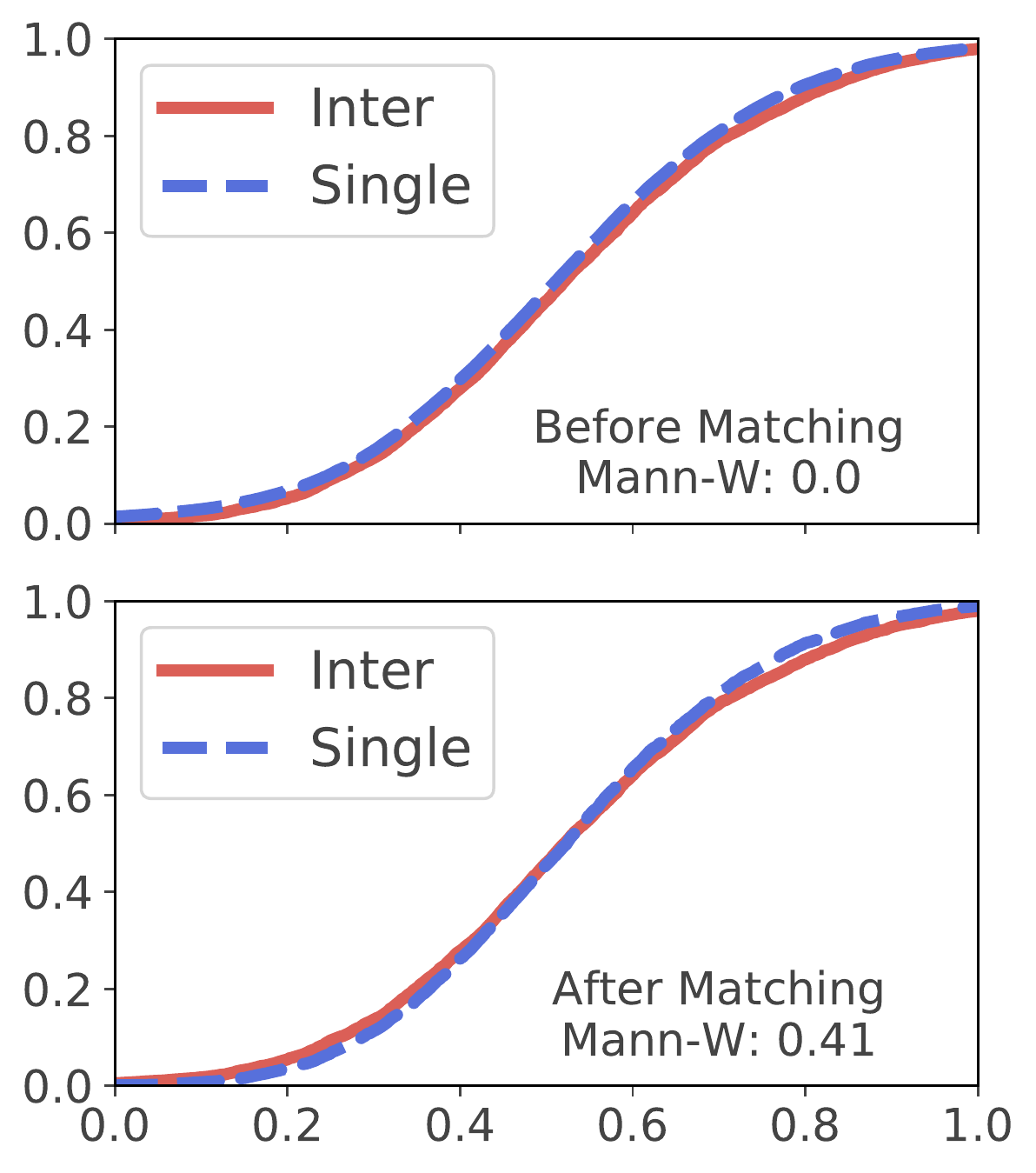}
        \includegraphics[width=0.19\textwidth]
        {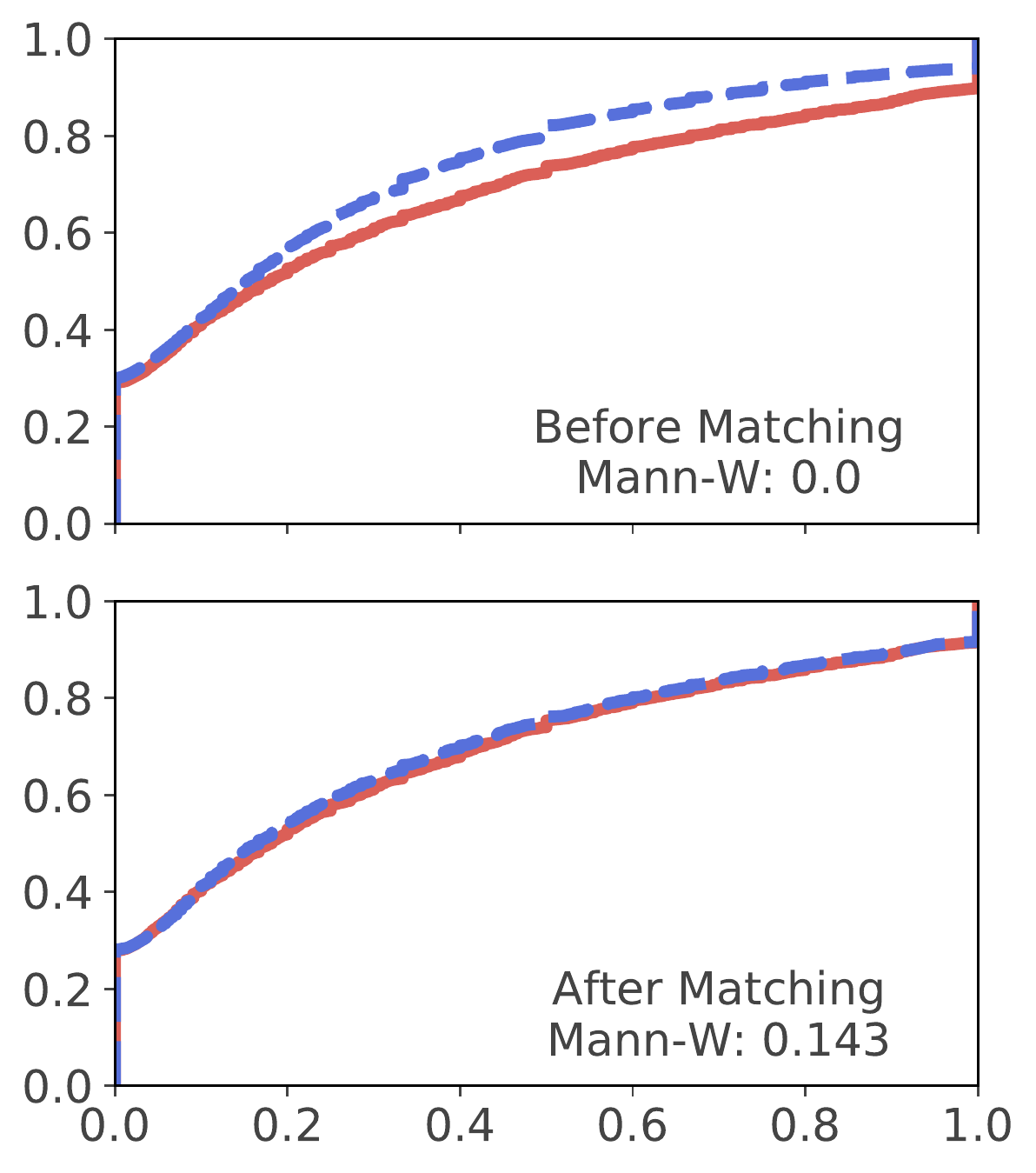}
        \includegraphics[width=0.19\textwidth]
        {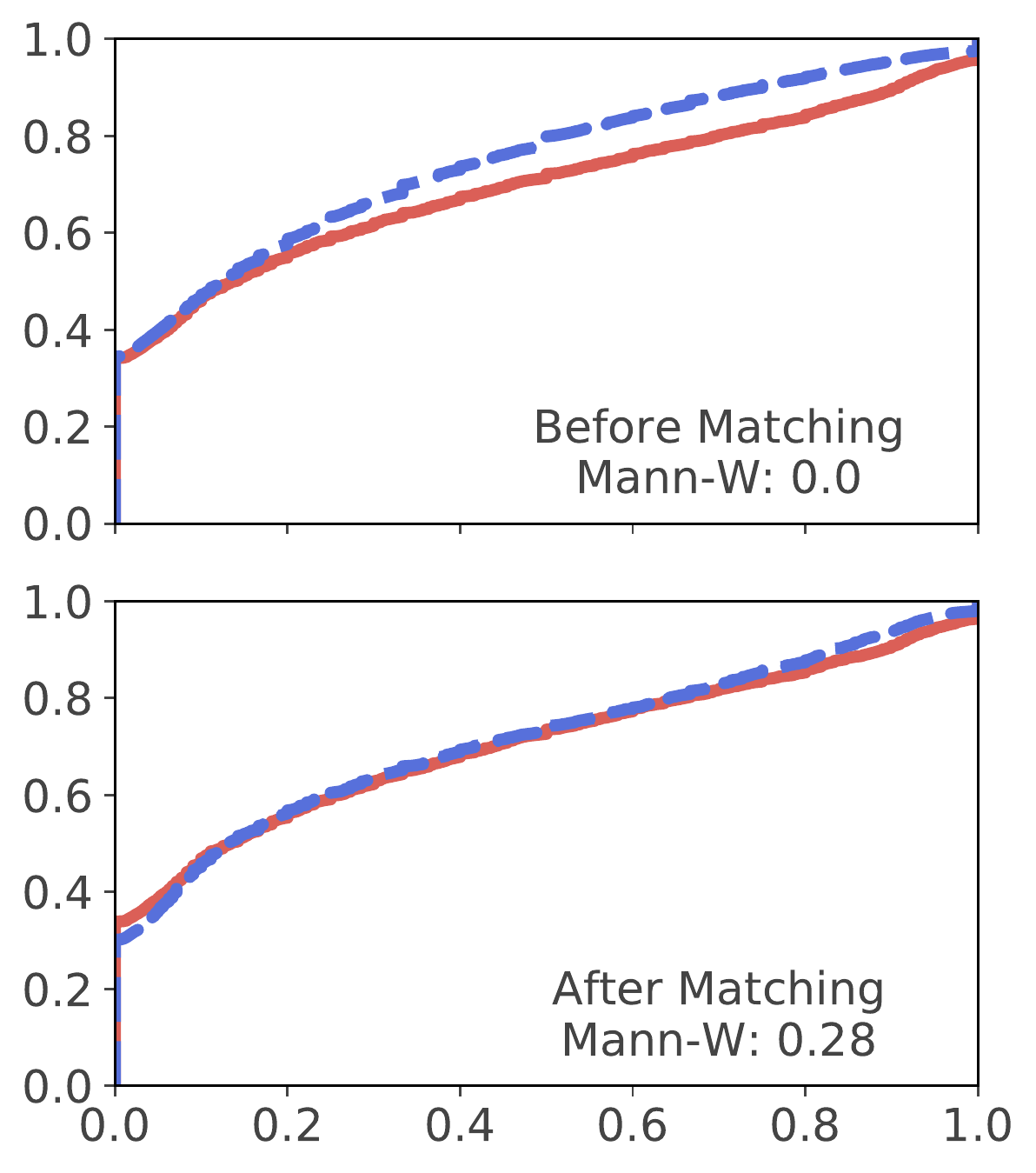}
        \caption{Matching results for level 2.}
    \end{subfigure}
    \hfill
    \begin{subfigure}[t]{\textwidth}
        \includegraphics[width=0.19\textwidth]
        {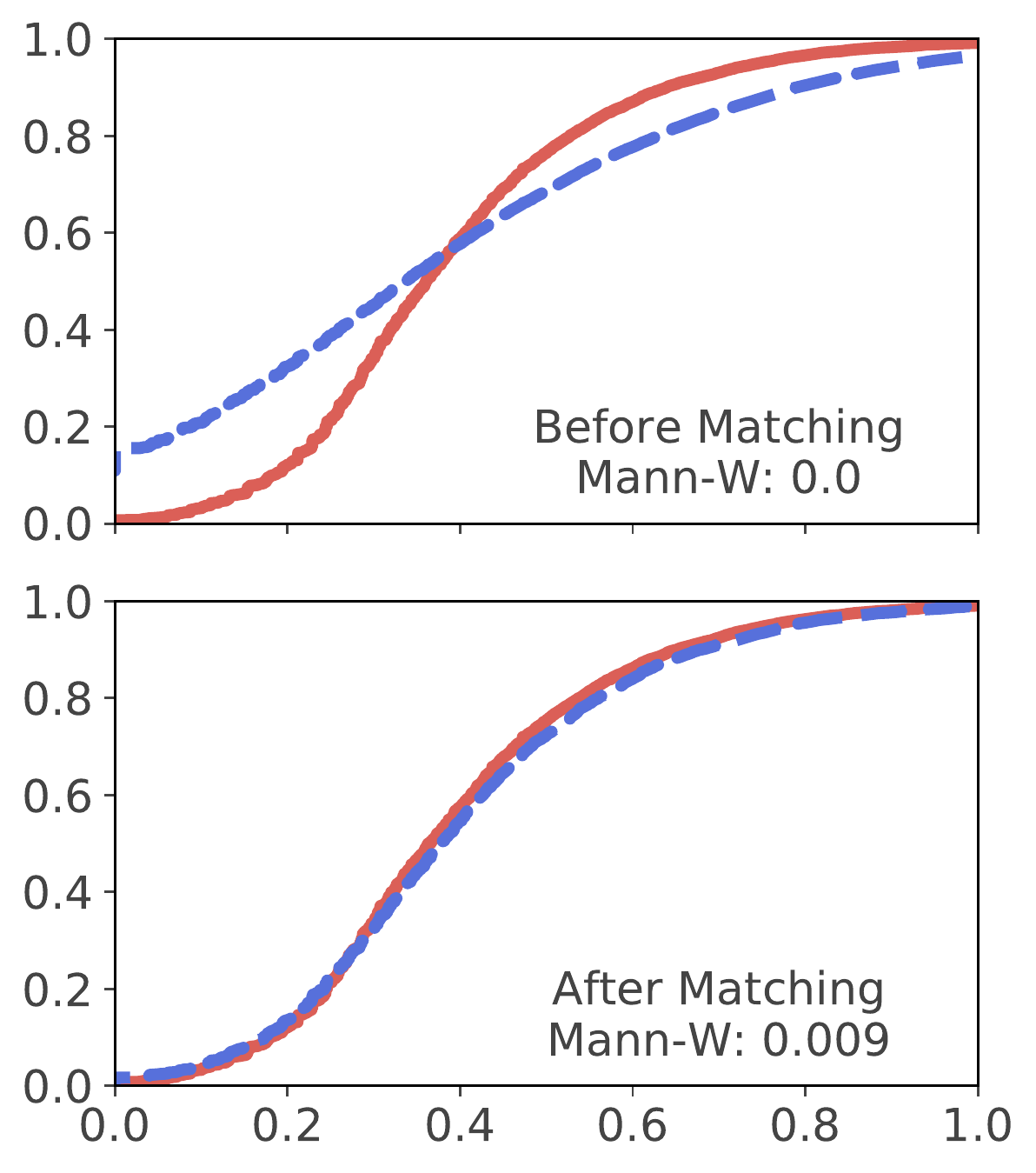}
        \includegraphics[width=0.19\textwidth]
        {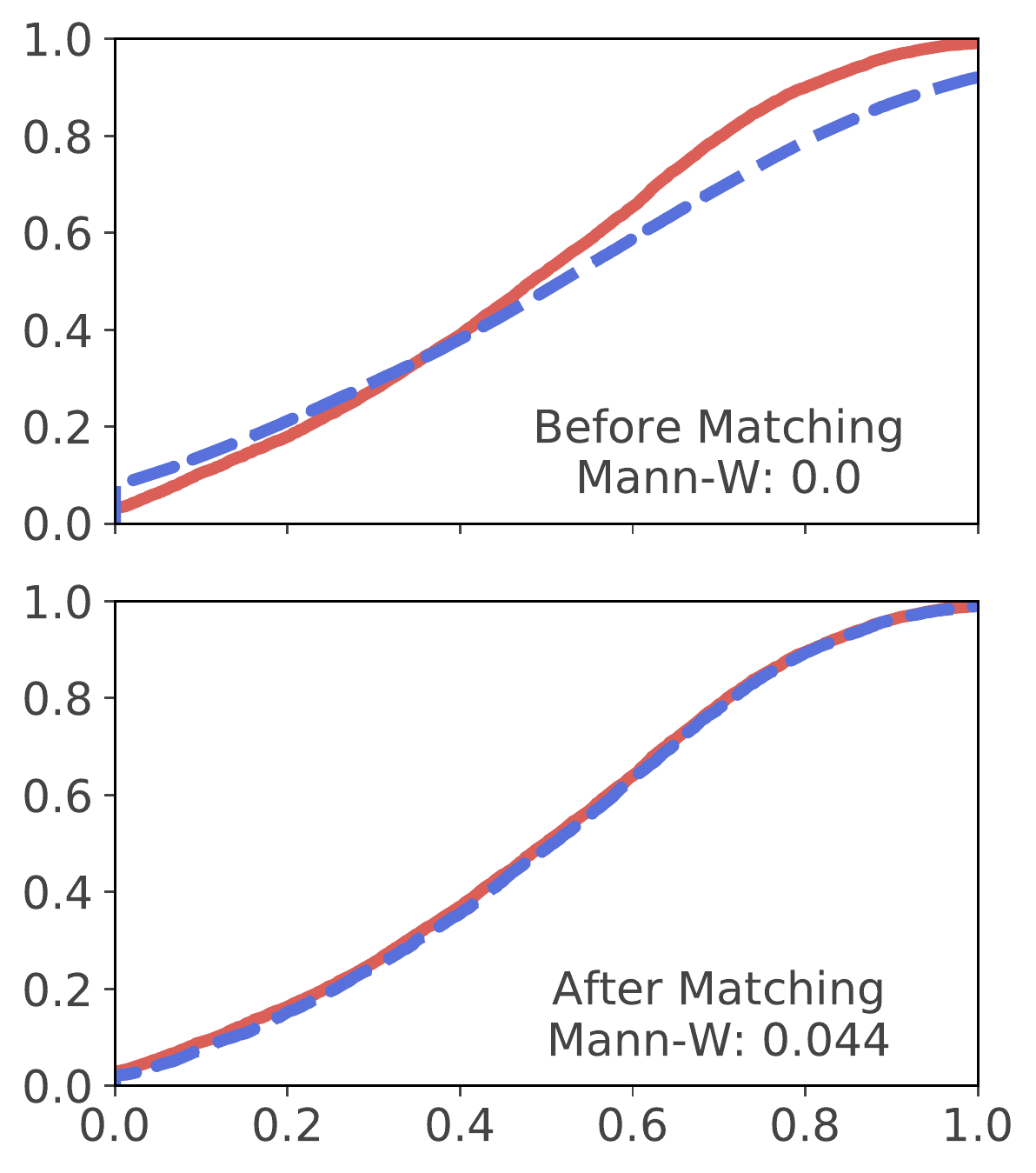}
        \includegraphics[width=0.19\textwidth]
        {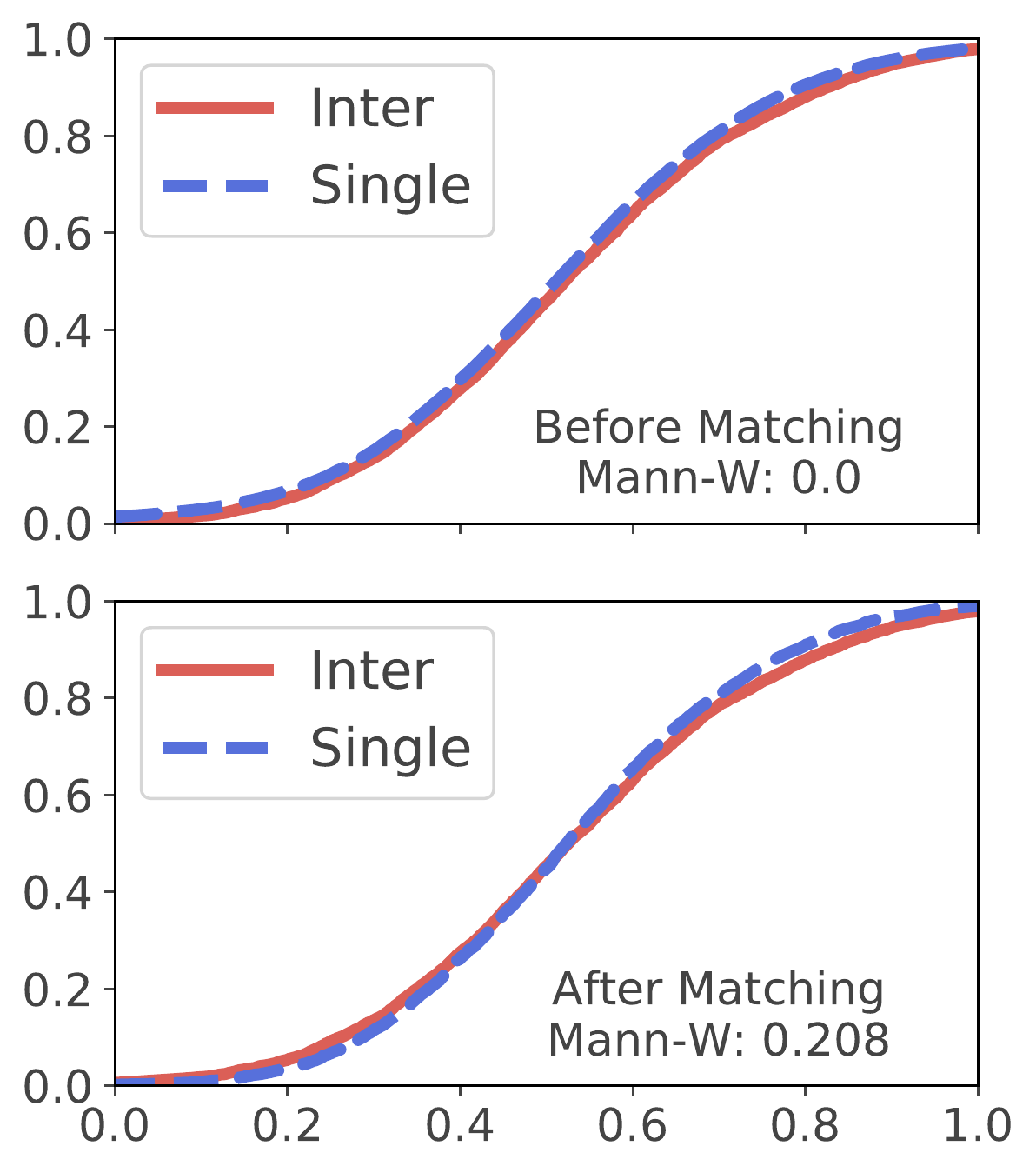}
        \includegraphics[width=0.19\textwidth]
        {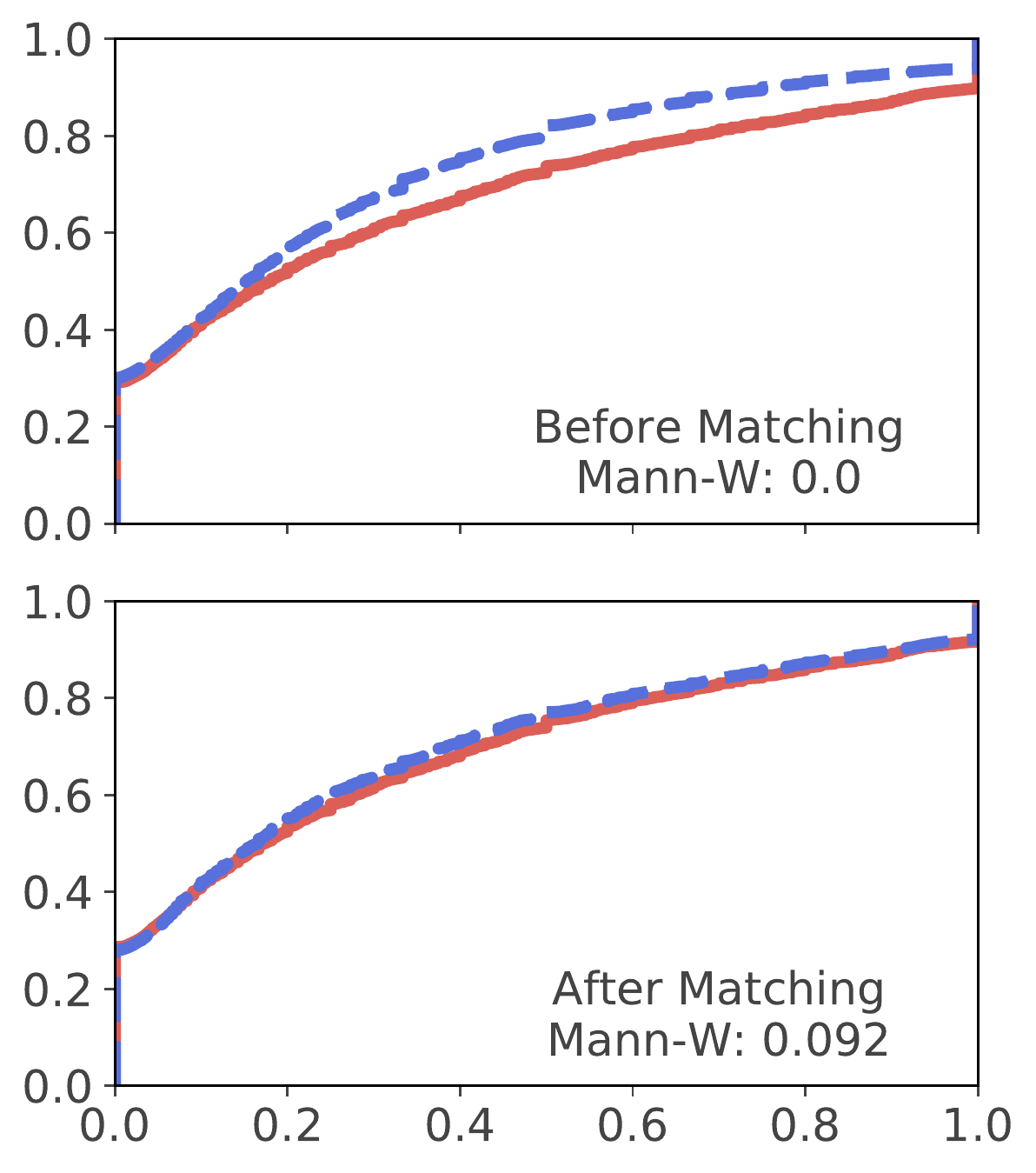}
        \includegraphics[width=0.19\textwidth]
        {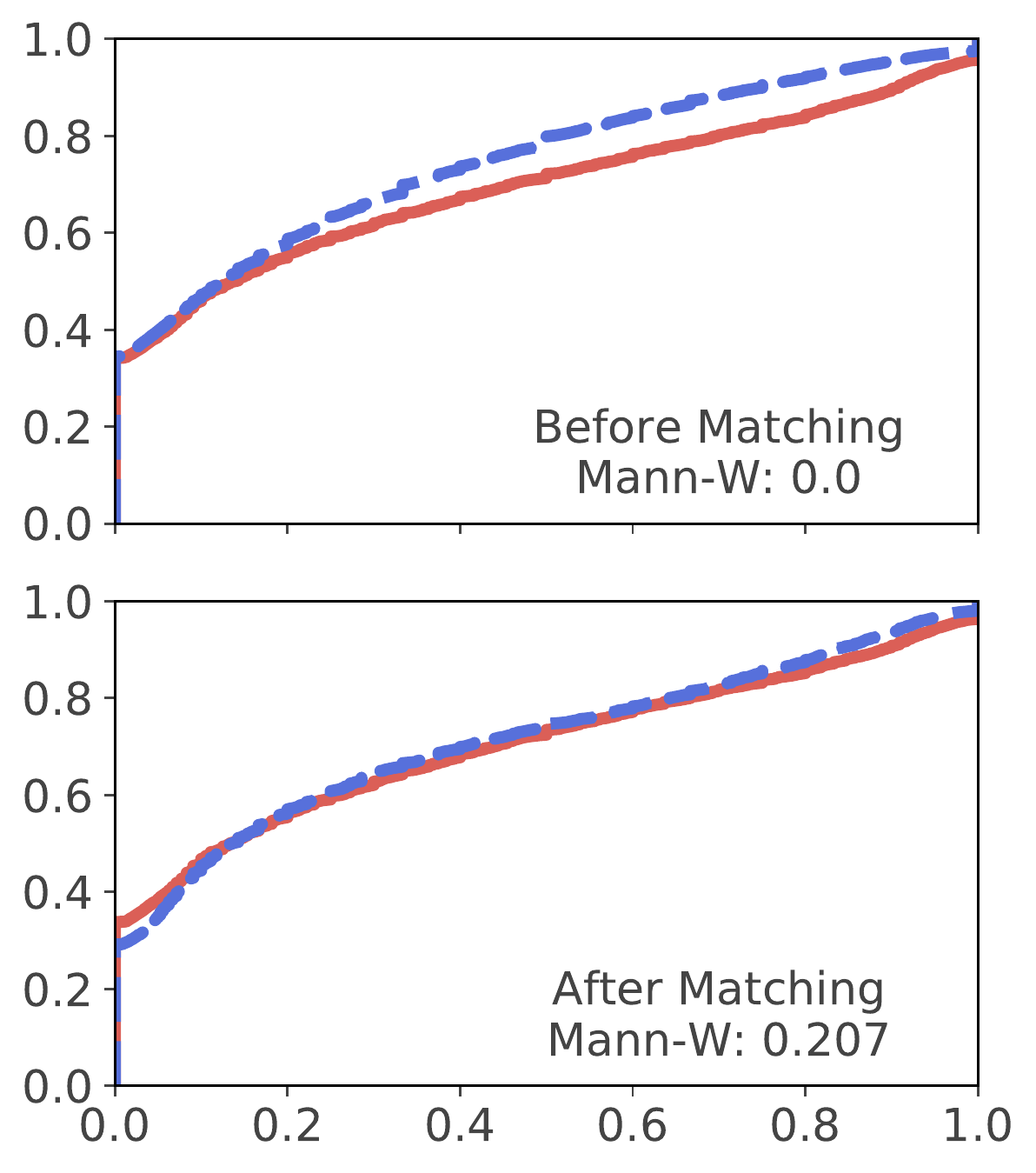}
        \caption{Matching results for level 3.}
    \end{subfigure}
    \hfill
    \begin{subfigure}[t]{\textwidth}
        \includegraphics[width=0.19\textwidth]
        {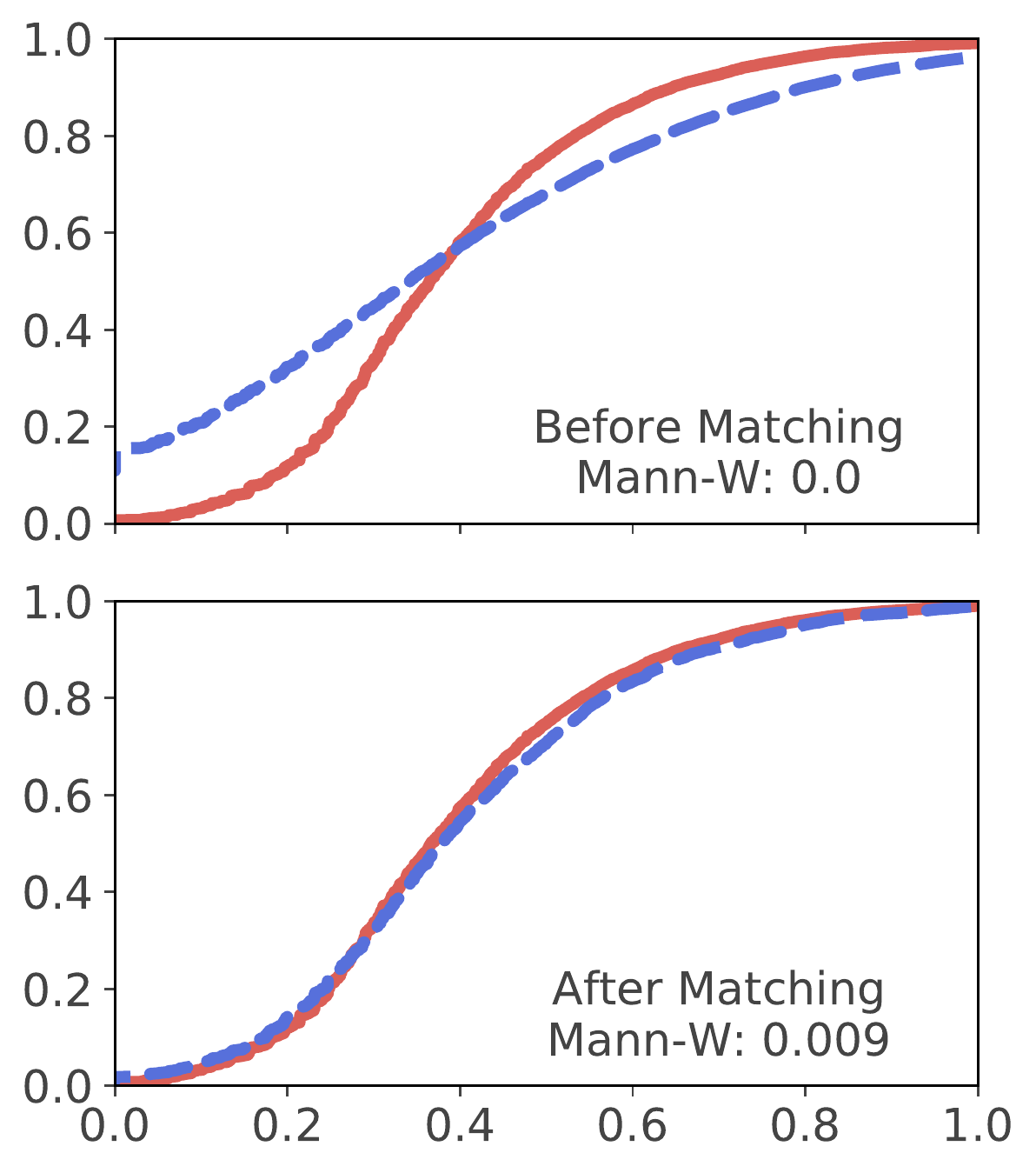}
        \includegraphics[width=0.19\textwidth]
        {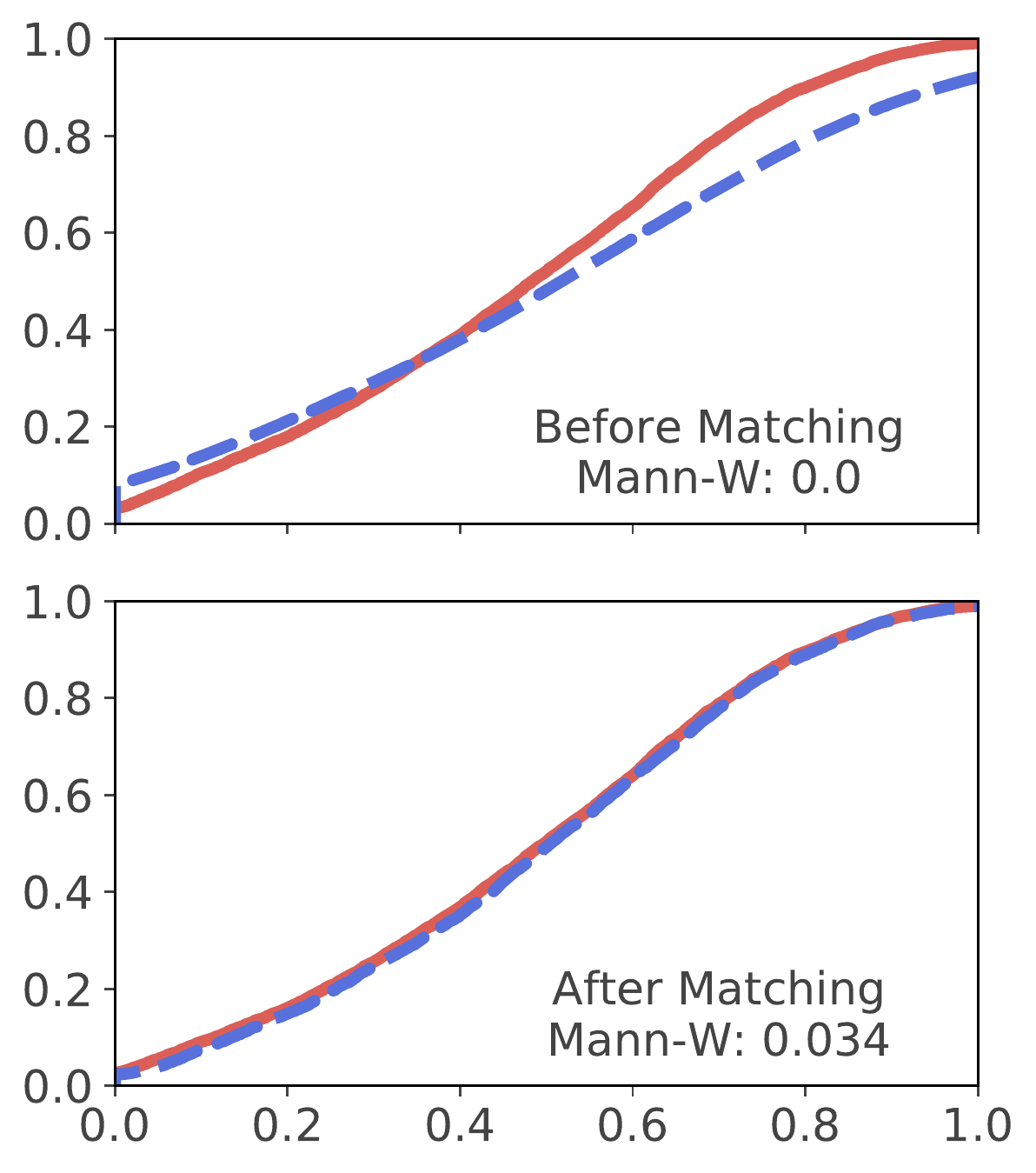}
        \includegraphics[width=0.19\textwidth]
        {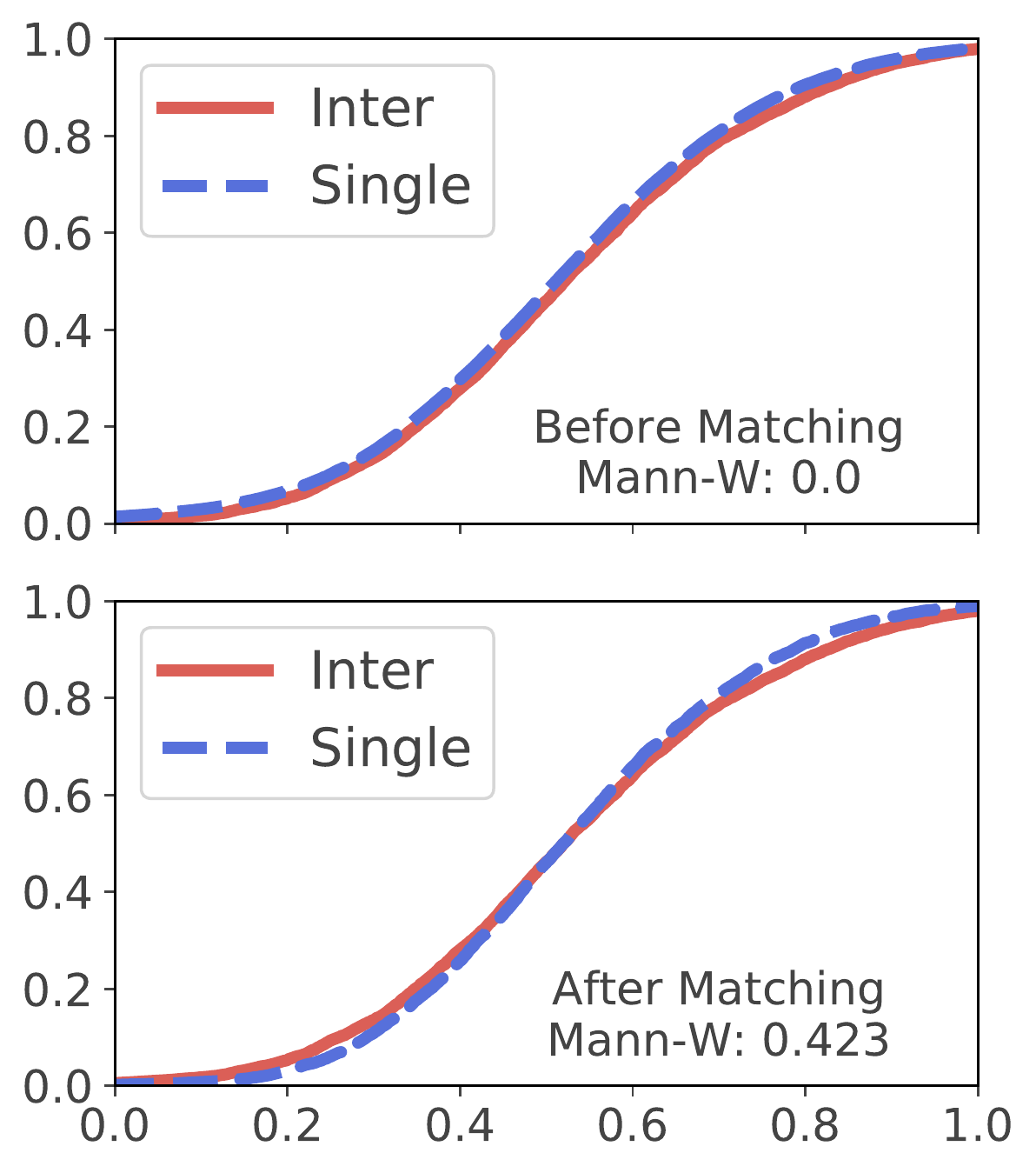}
        \includegraphics[width=0.19\textwidth]
        {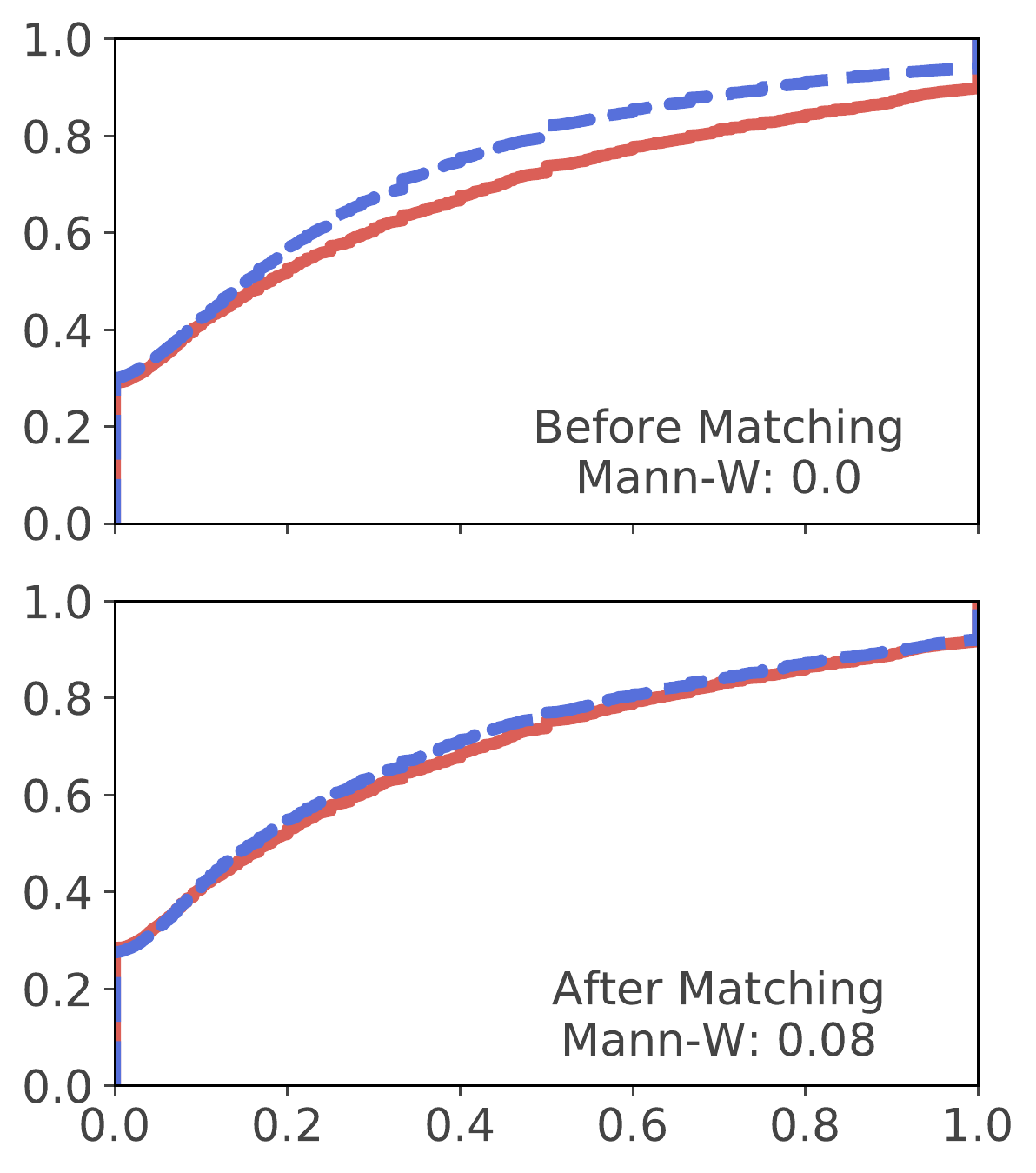}
        \includegraphics[width=0.19\textwidth]
        {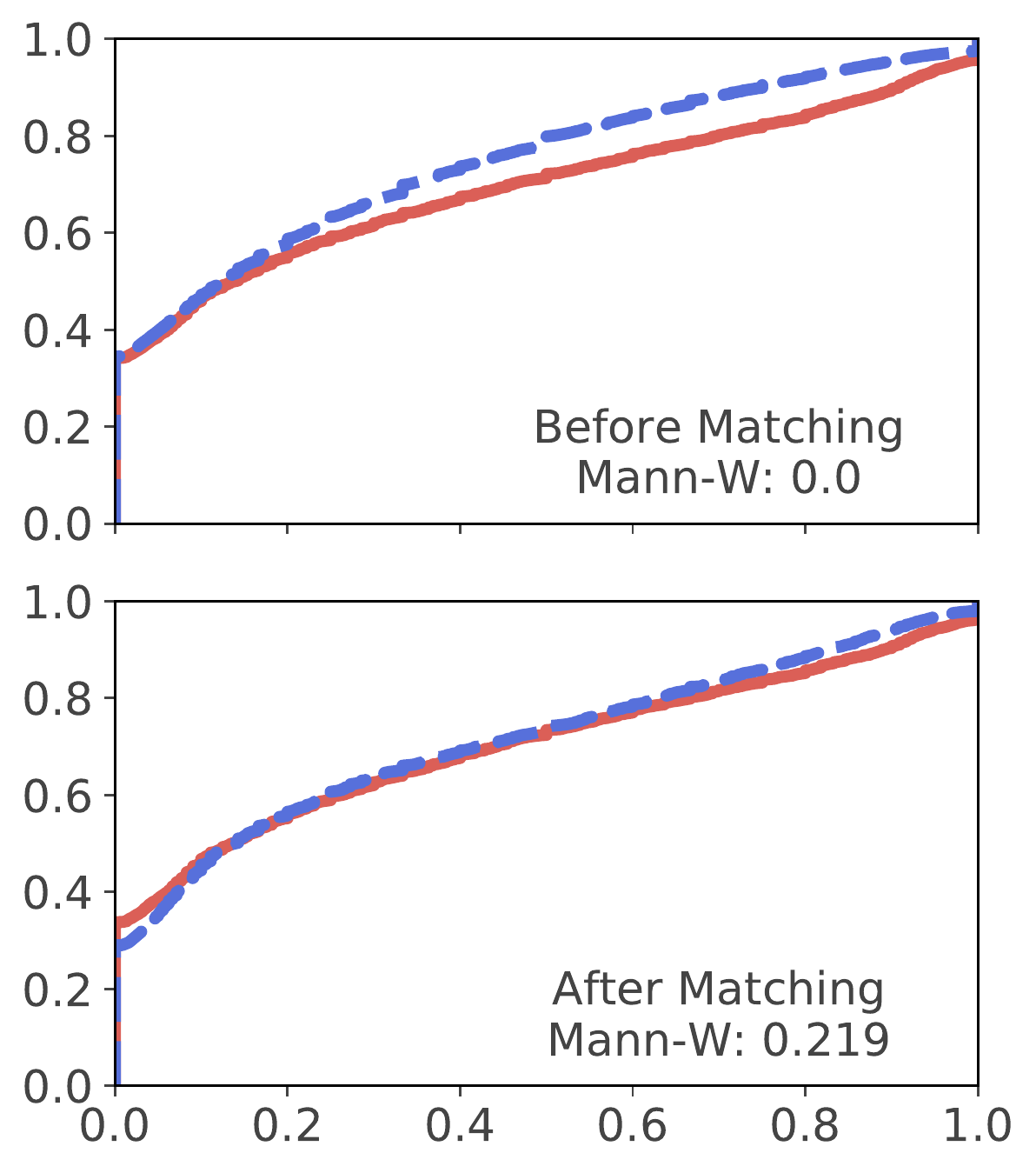}
        \caption{Matching results for level 4.}
    \end{subfigure}
    \hfill
    \begin{subfigure}[t]{\textwidth}
        \includegraphics[width=0.19\textwidth]
        {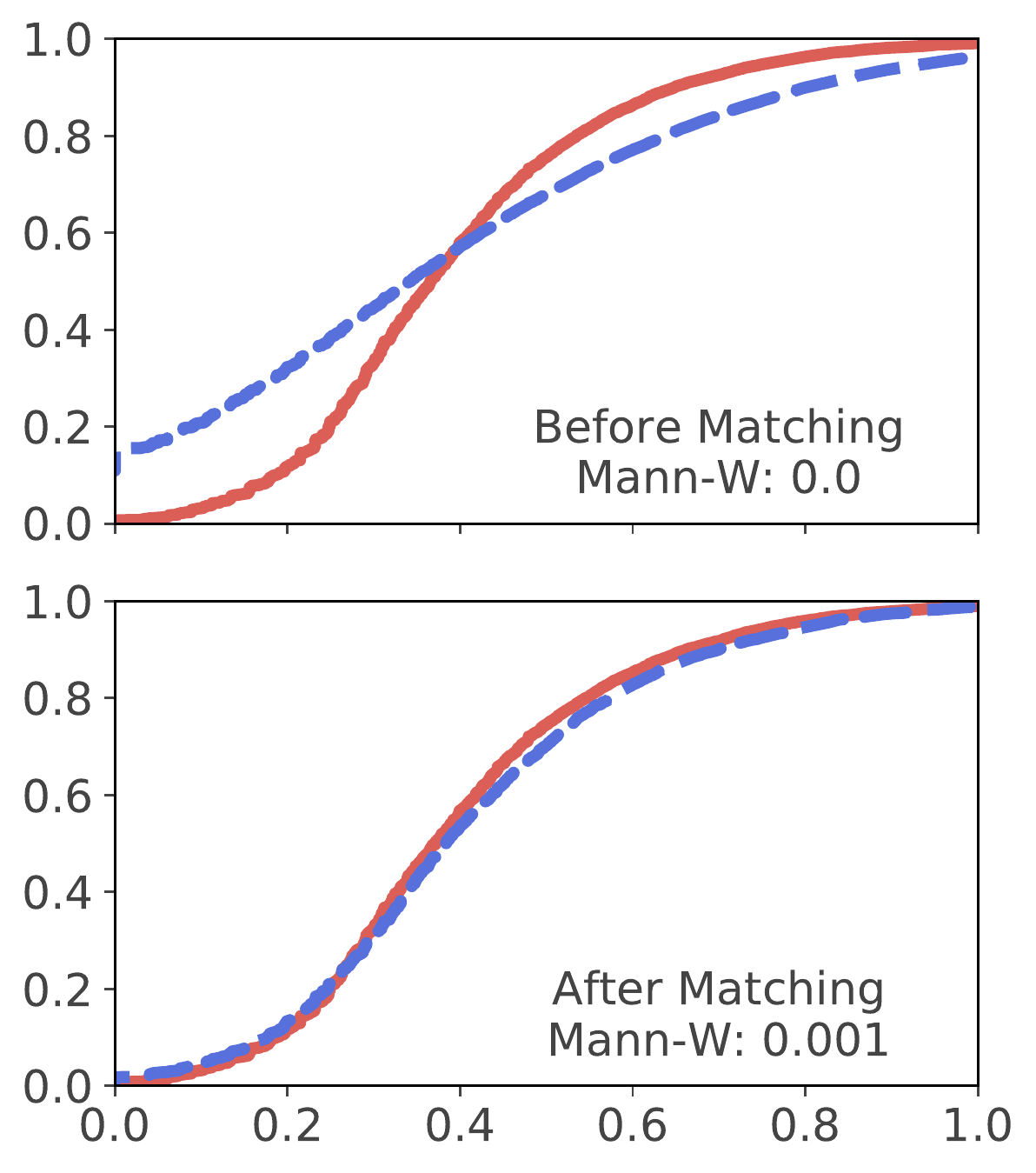}
        \includegraphics[width=0.19\textwidth]
        {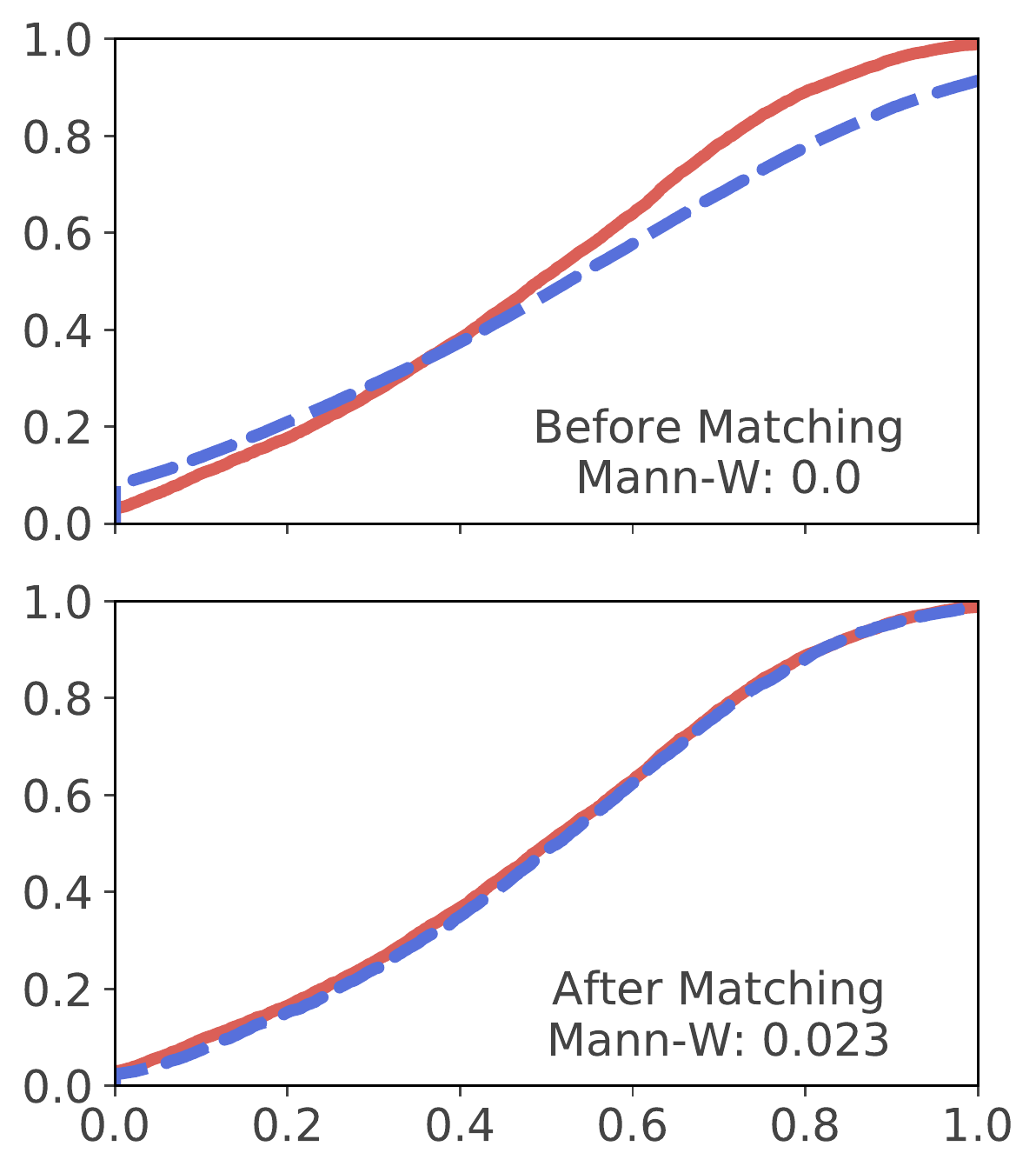}
        \includegraphics[width=0.19\textwidth]
        {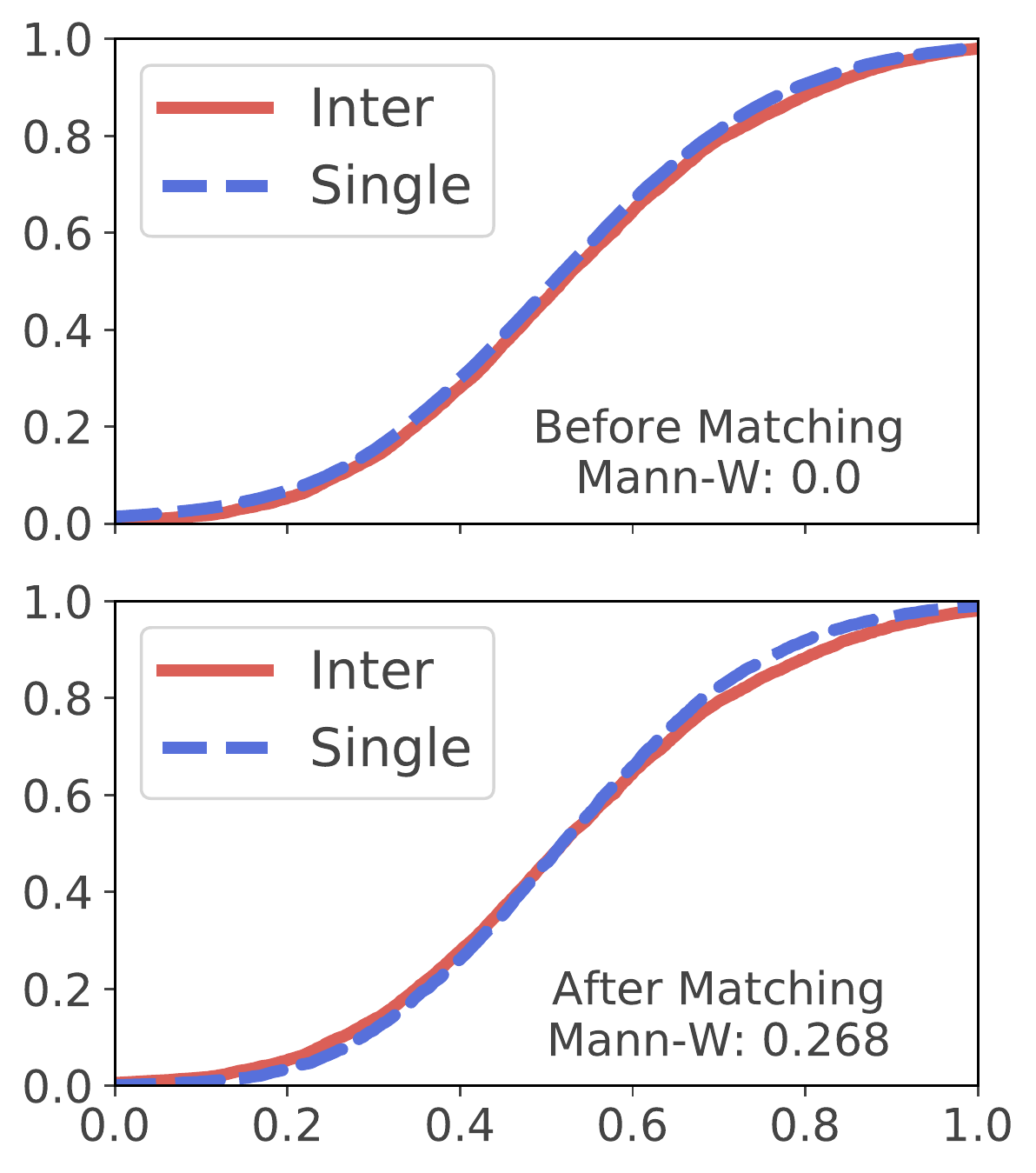}
        \includegraphics[width=0.19\textwidth]
        {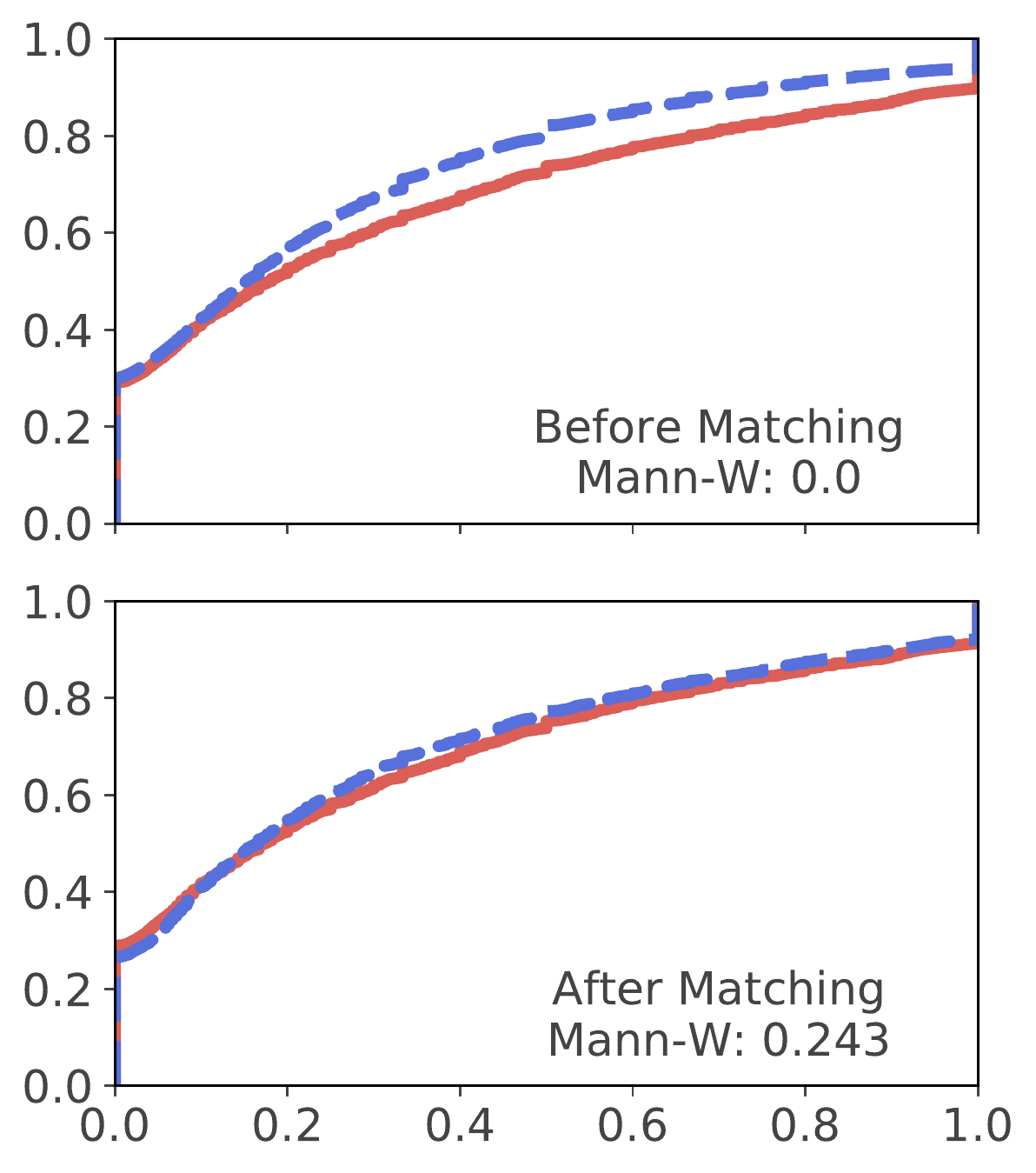}
        \includegraphics[width=0.19\textwidth]
        {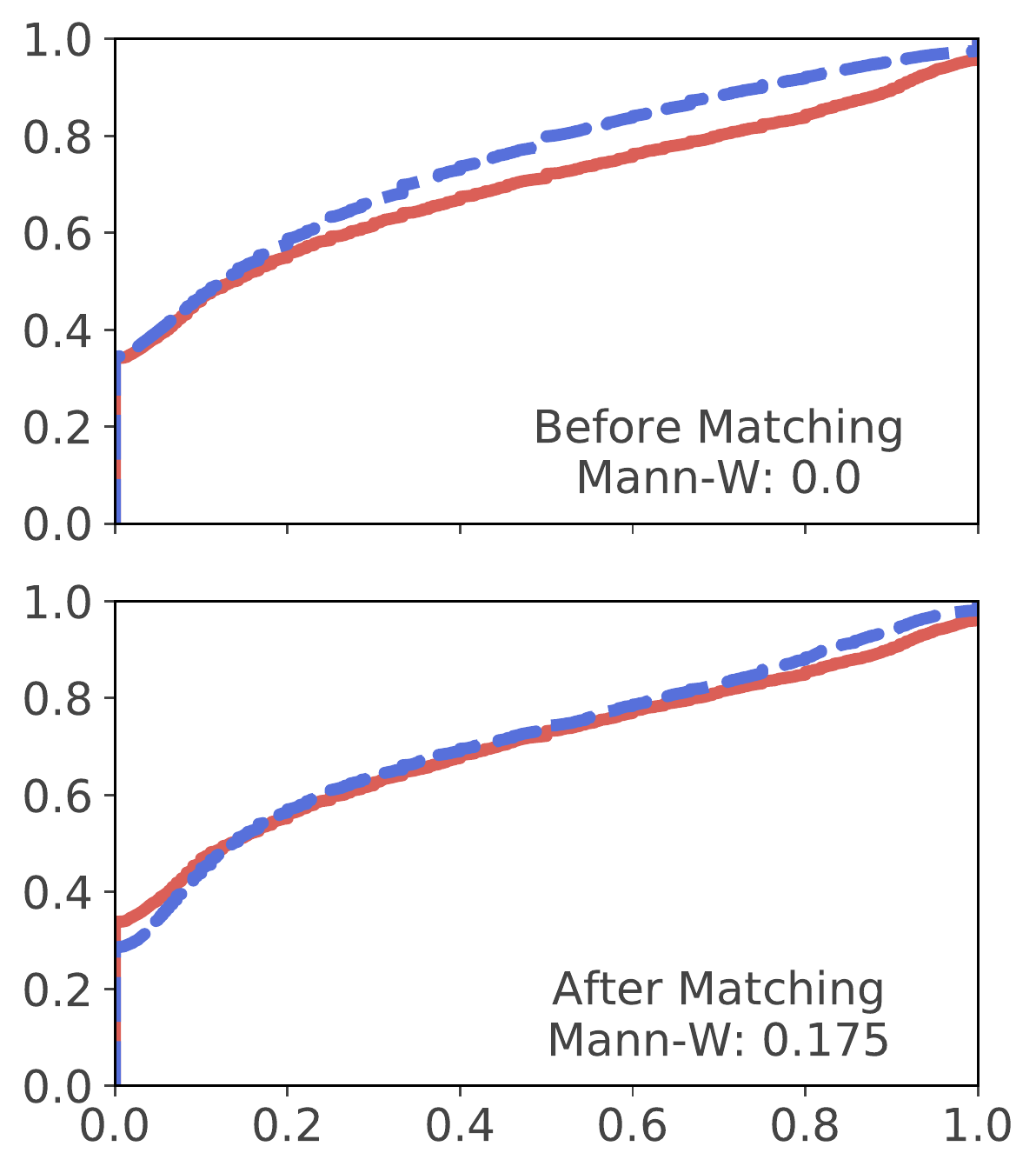}
        \caption{Matching results for level 5.}
    \end{subfigure}
    \caption{Empirical cumulative distribution of each activity feature before and after the matching technique from level 1 to level 5 in the 2018 season.
    The activity features from left to right are the number of comments,
    the average hour gap between comments, the average comment length, 
    the proportion of playoff comments, and the proportion of game thread comments.}
    \label{fig:matchinglevel2018}
\end{figure}

\begin{figure}
    \center
    \begin{subfigure}[t]{\textwidth}
        \includegraphics[width=0.19\textwidth]
        {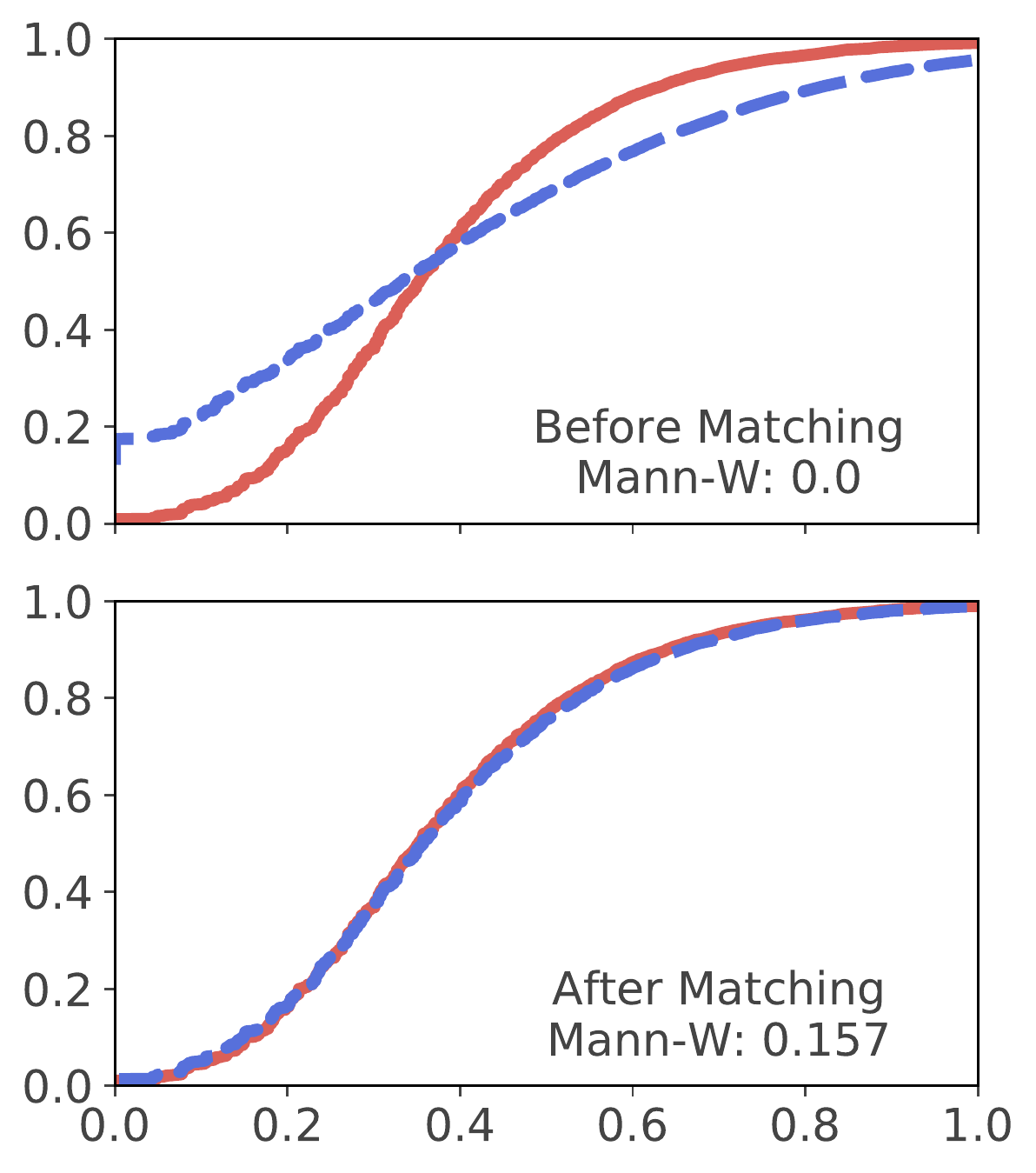}
        \includegraphics[width=0.19\textwidth]
        {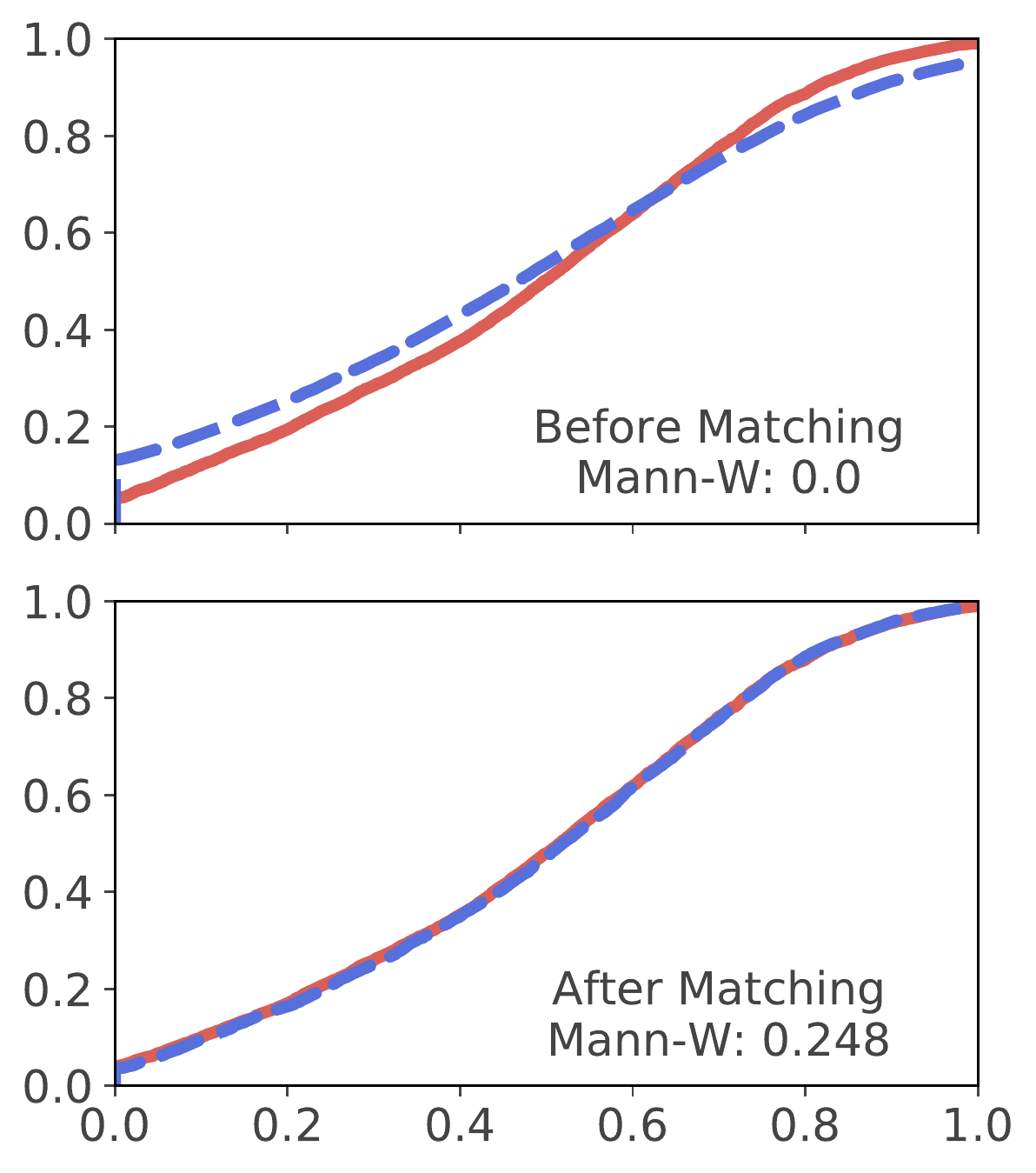}
        \includegraphics[width=0.19\textwidth]
        {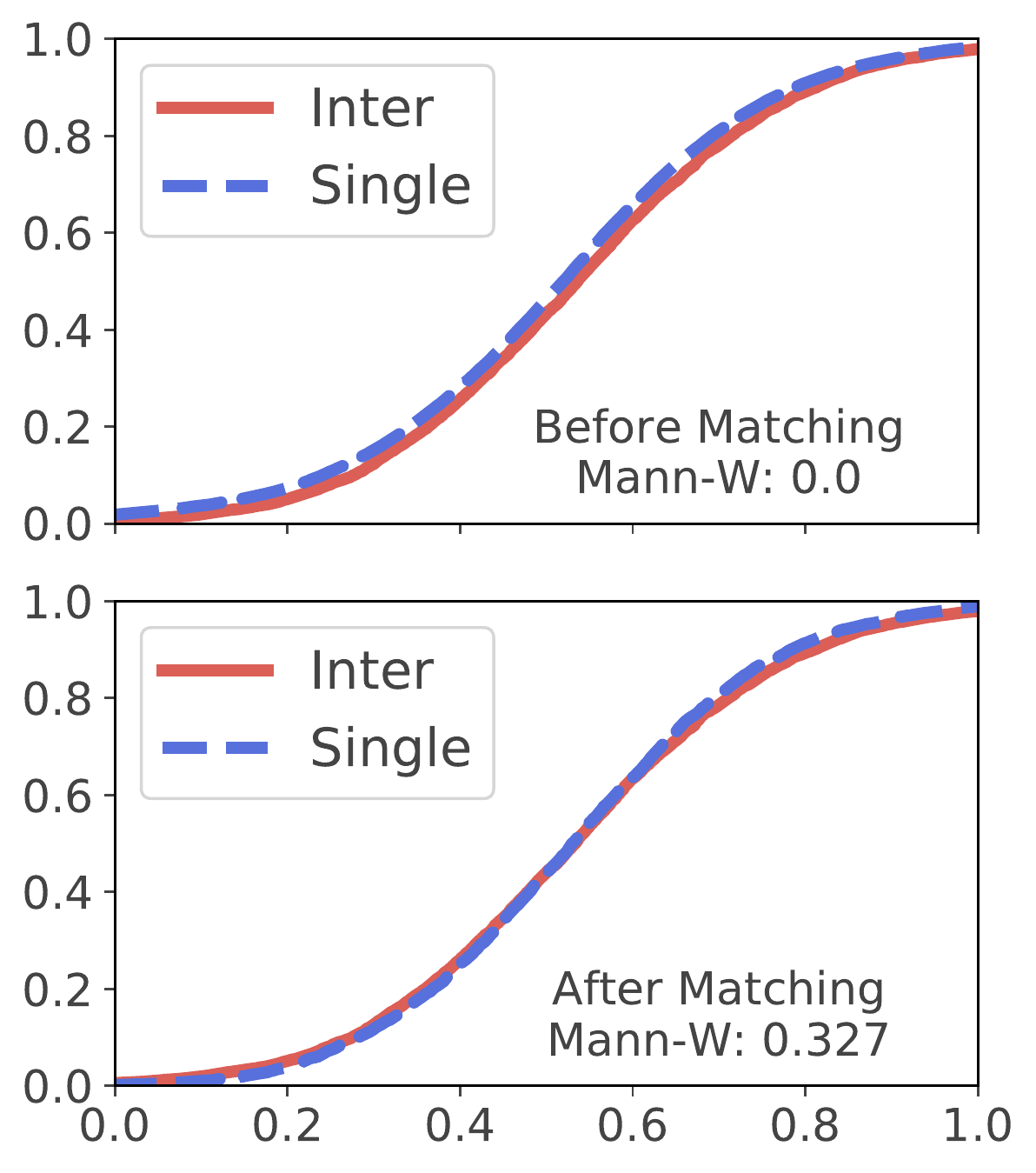}
        \includegraphics[width=0.19\textwidth]
        {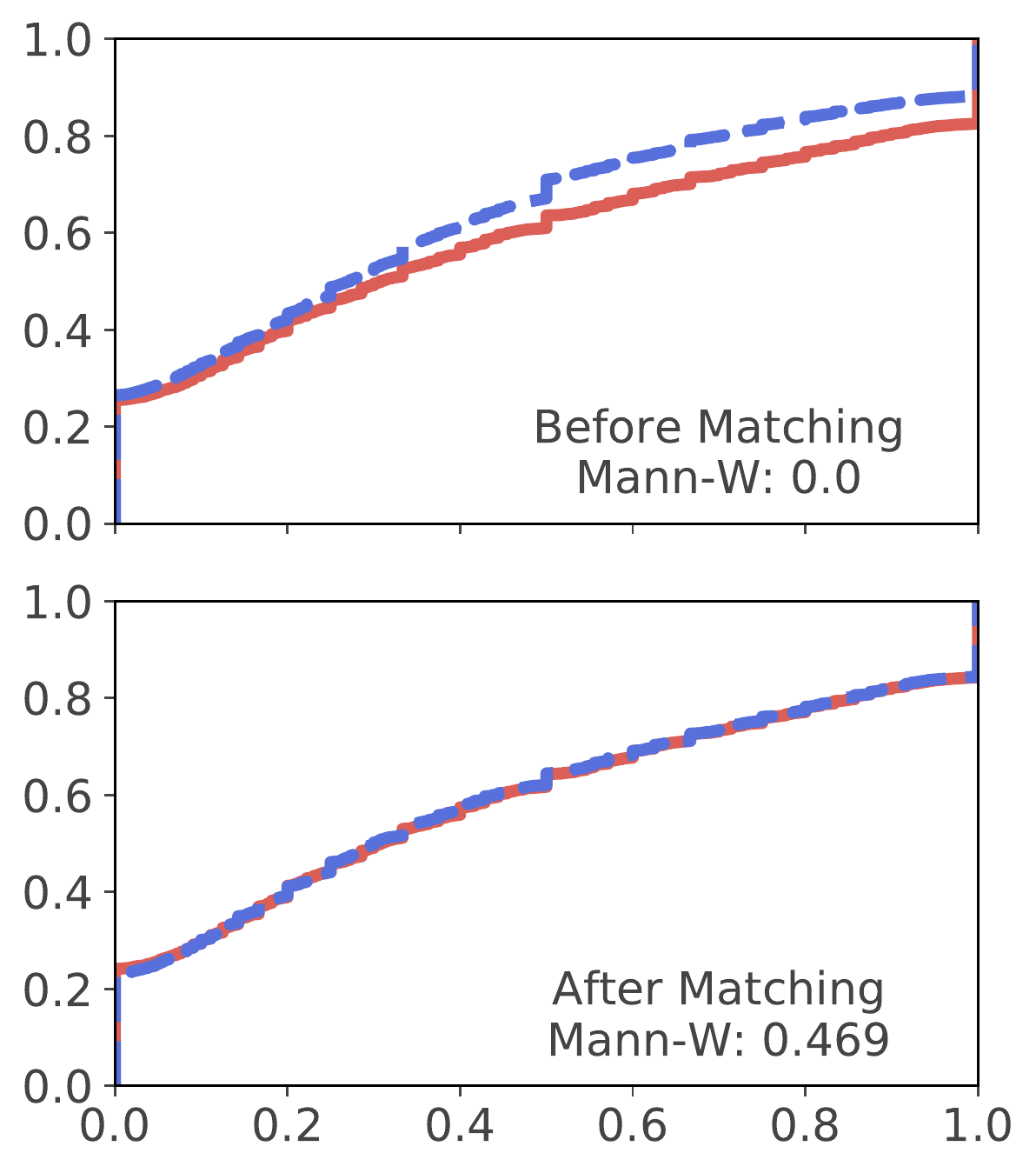}
        \includegraphics[width=0.19\textwidth]
        {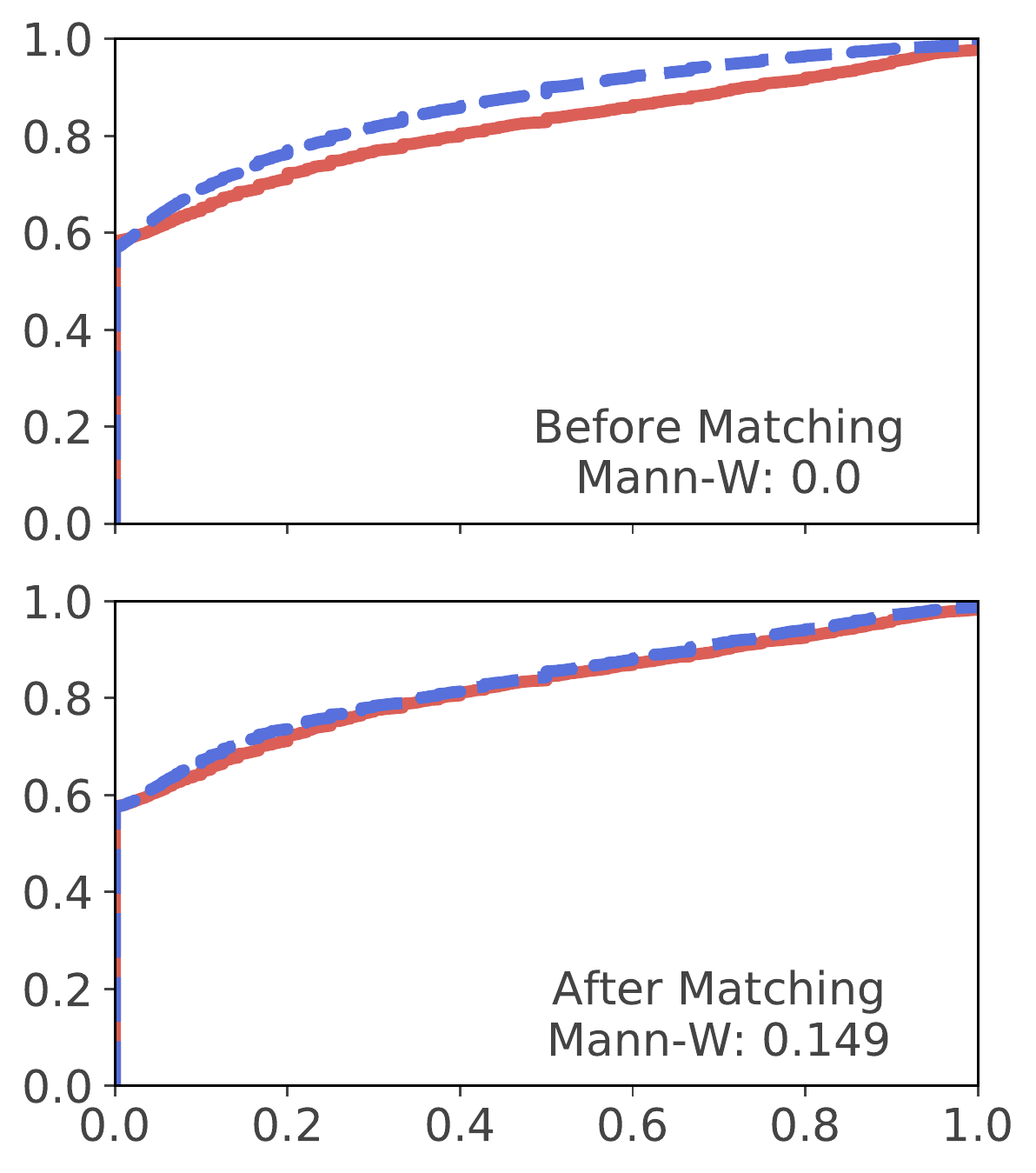}
        \caption{Matching results at level 1.}
    \end{subfigure}
    \hfill
    \begin{subfigure}[t]{\textwidth}
        \includegraphics[width=0.19\textwidth]
        {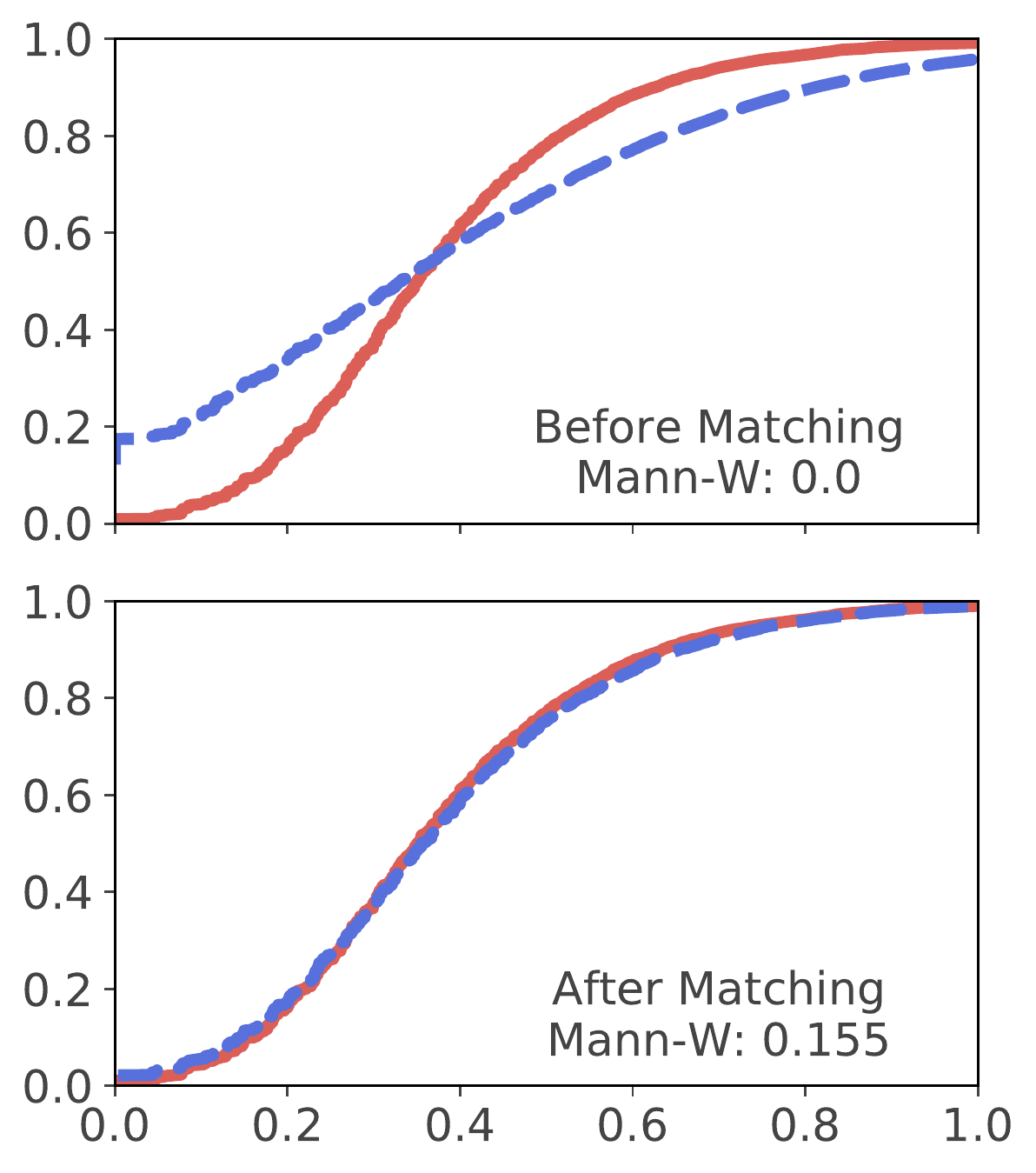}
        \includegraphics[width=0.19\textwidth]
        {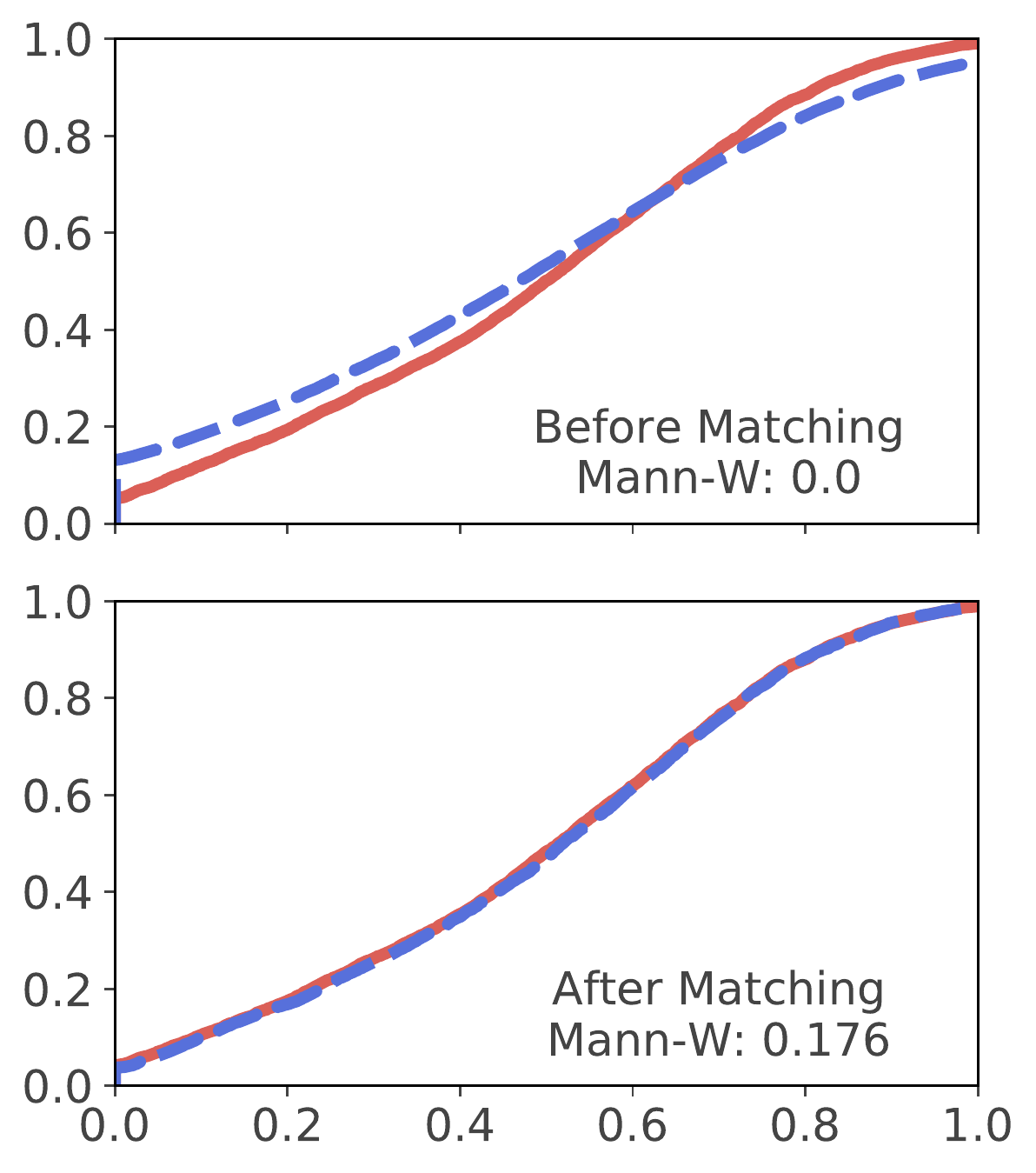}
        \includegraphics[width=0.19\textwidth]
        {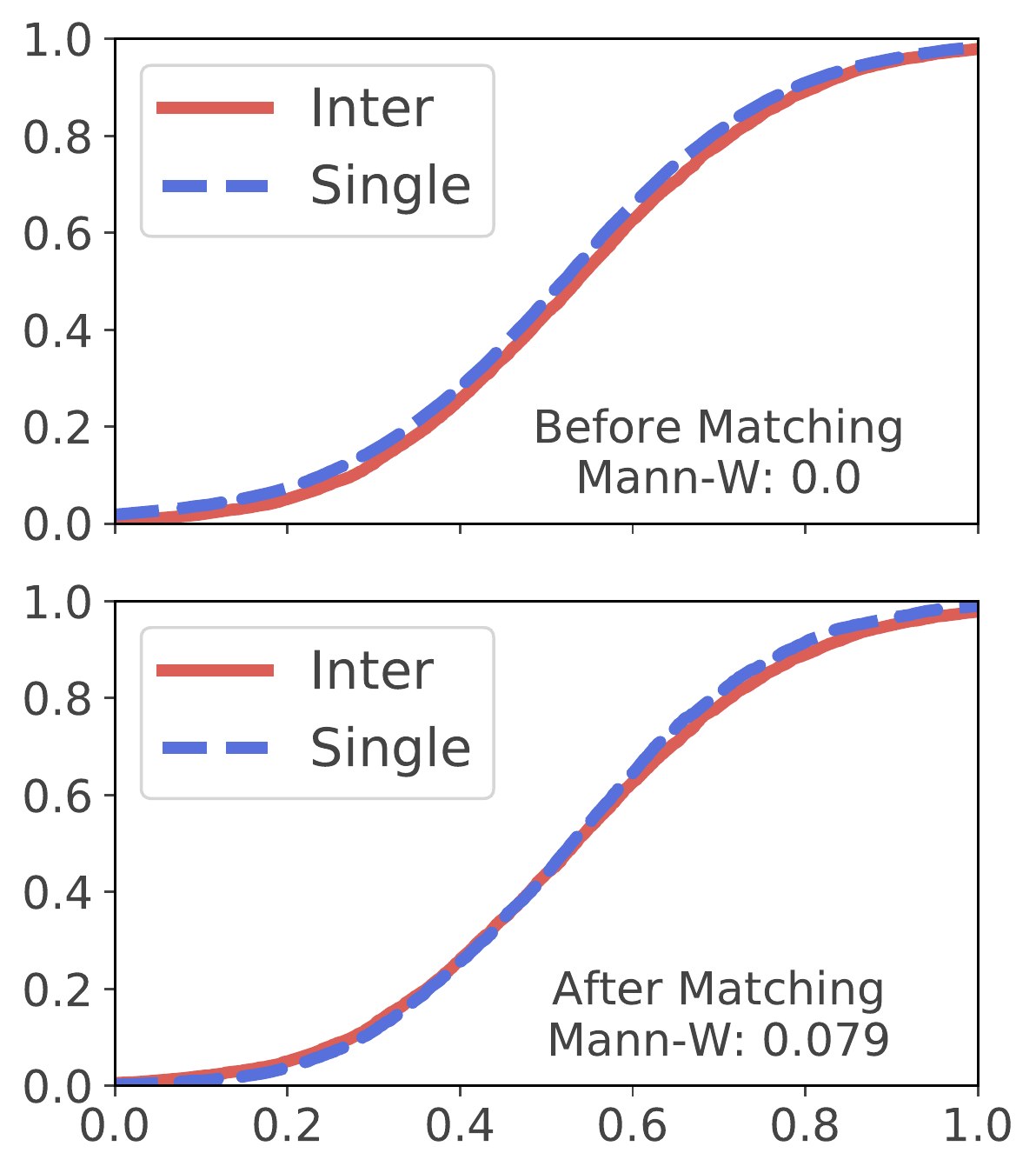}
        \includegraphics[width=0.19\textwidth]
        {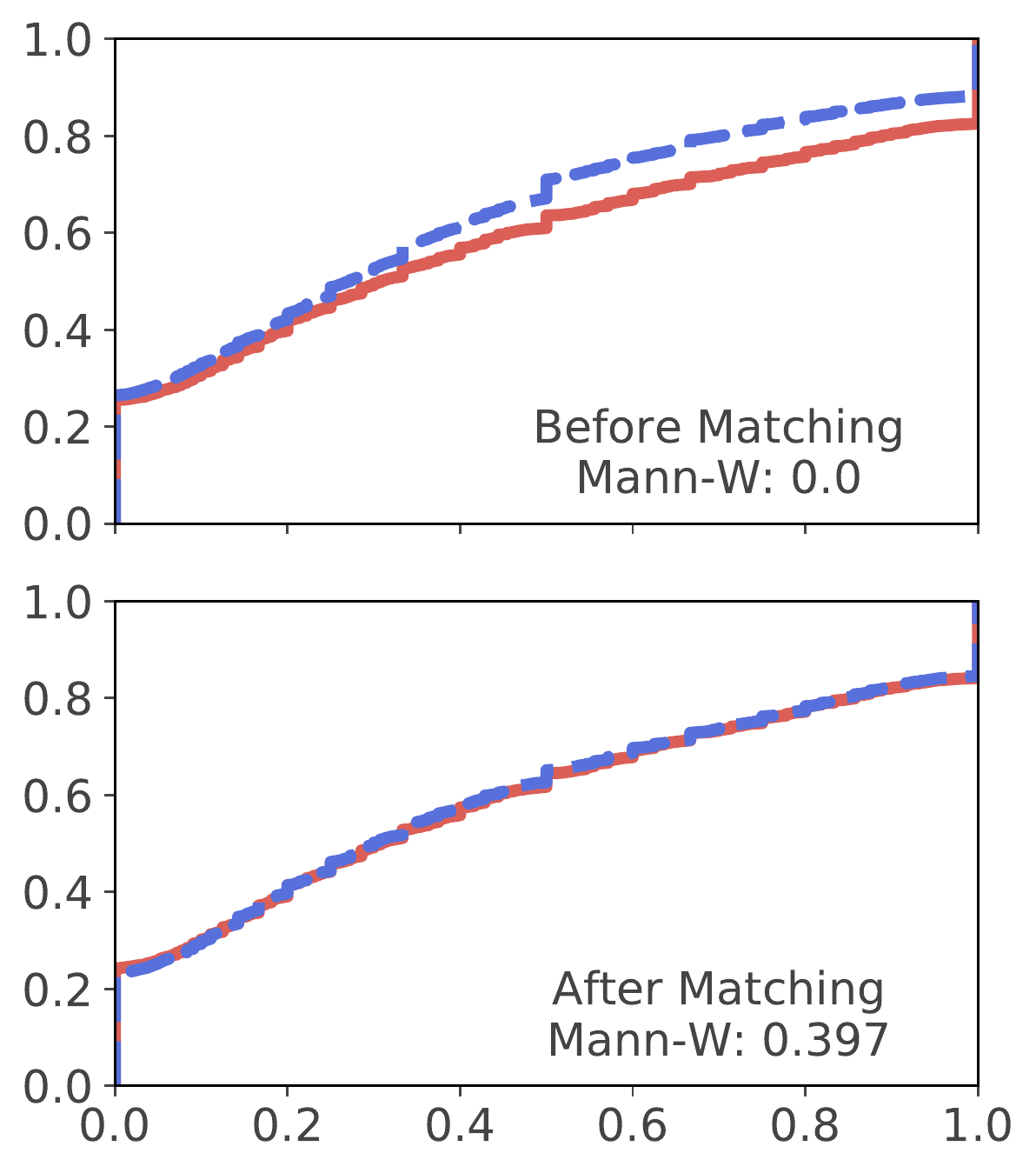}
        \includegraphics[width=0.19\textwidth]
        {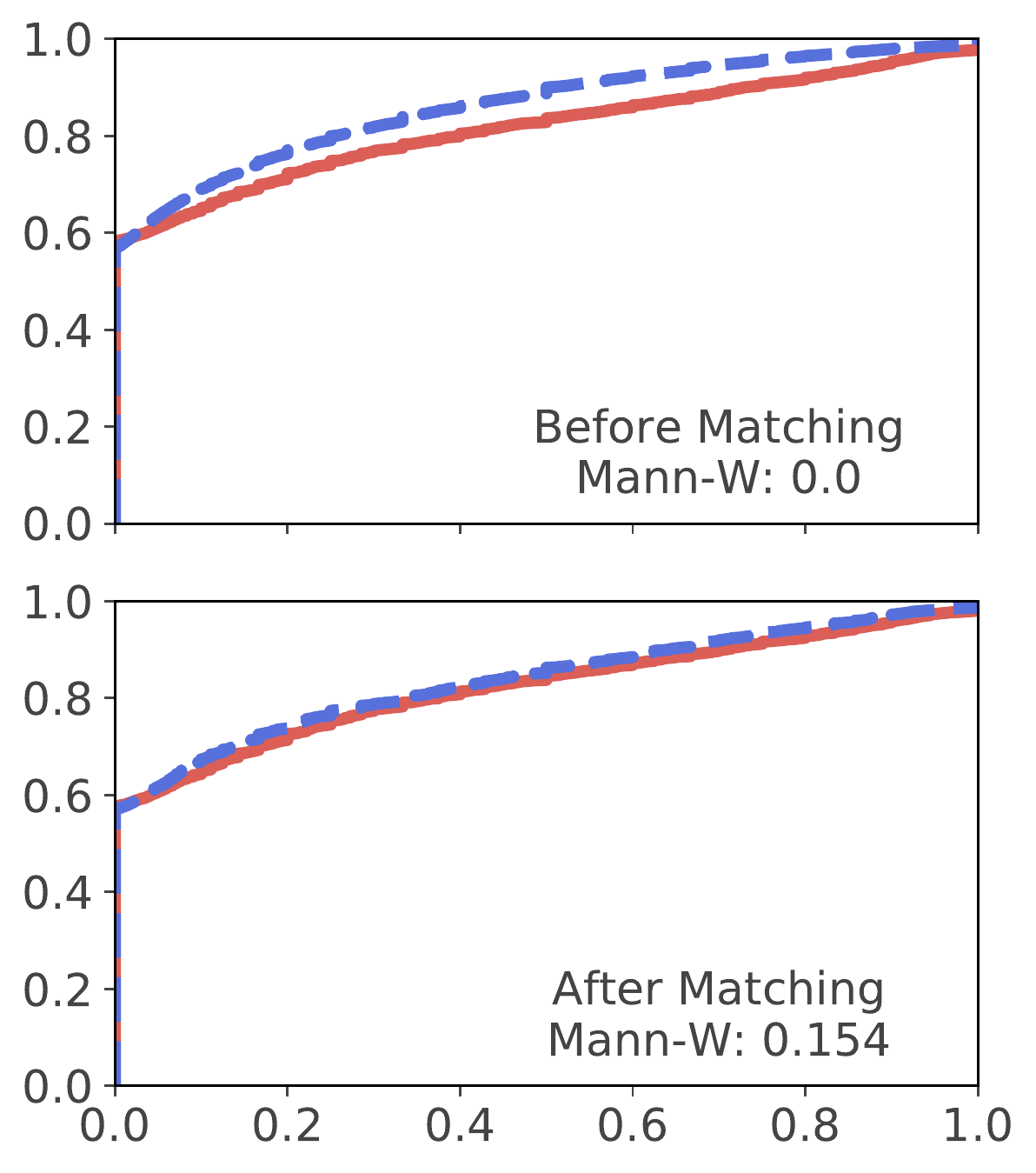}
        \caption{Matching results at level 2.}
    \end{subfigure}
    \hfill
    \begin{subfigure}[t]{\textwidth}
        \includegraphics[width=0.19\textwidth]
        {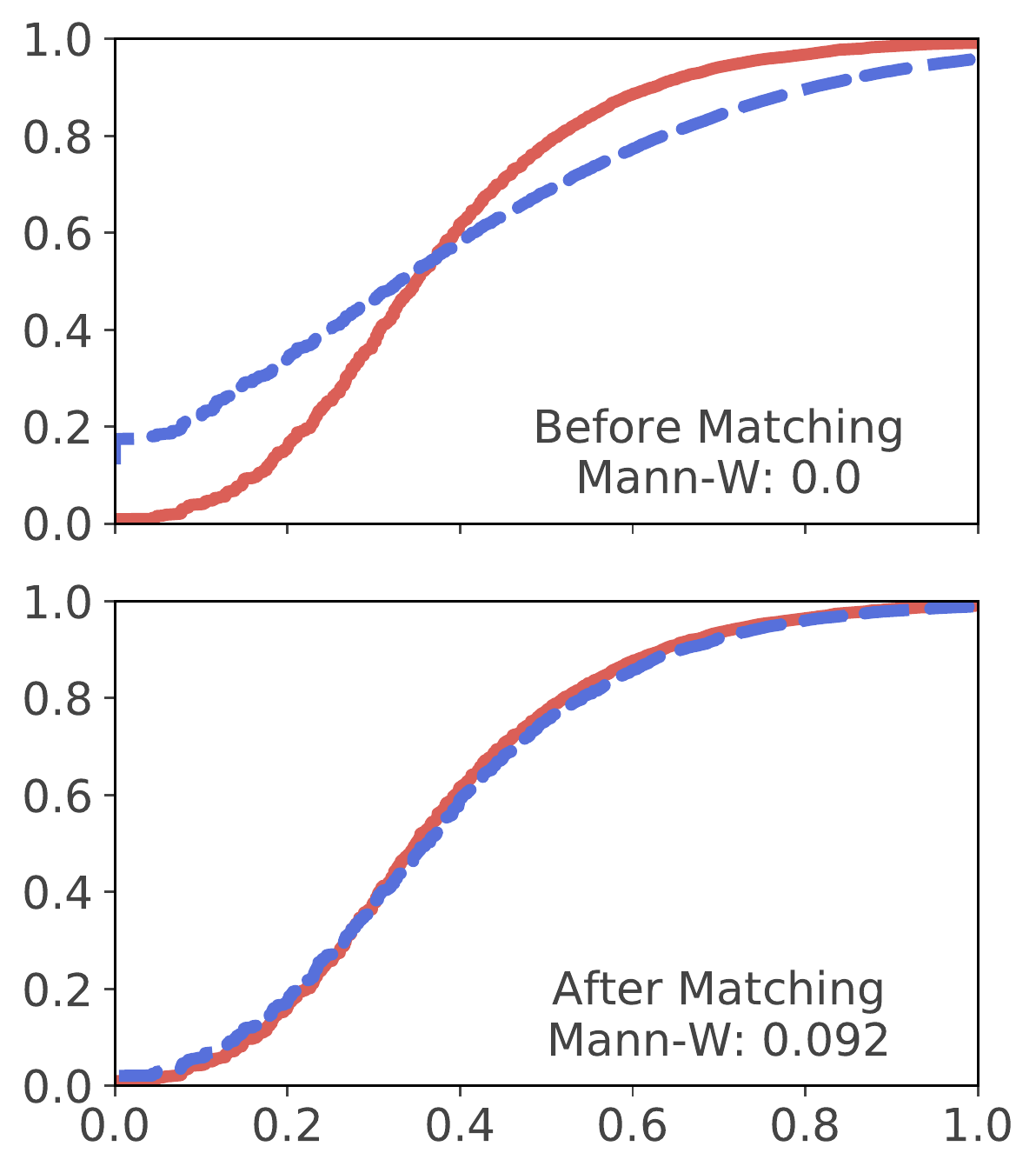}
        \includegraphics[width=0.19\textwidth]
        {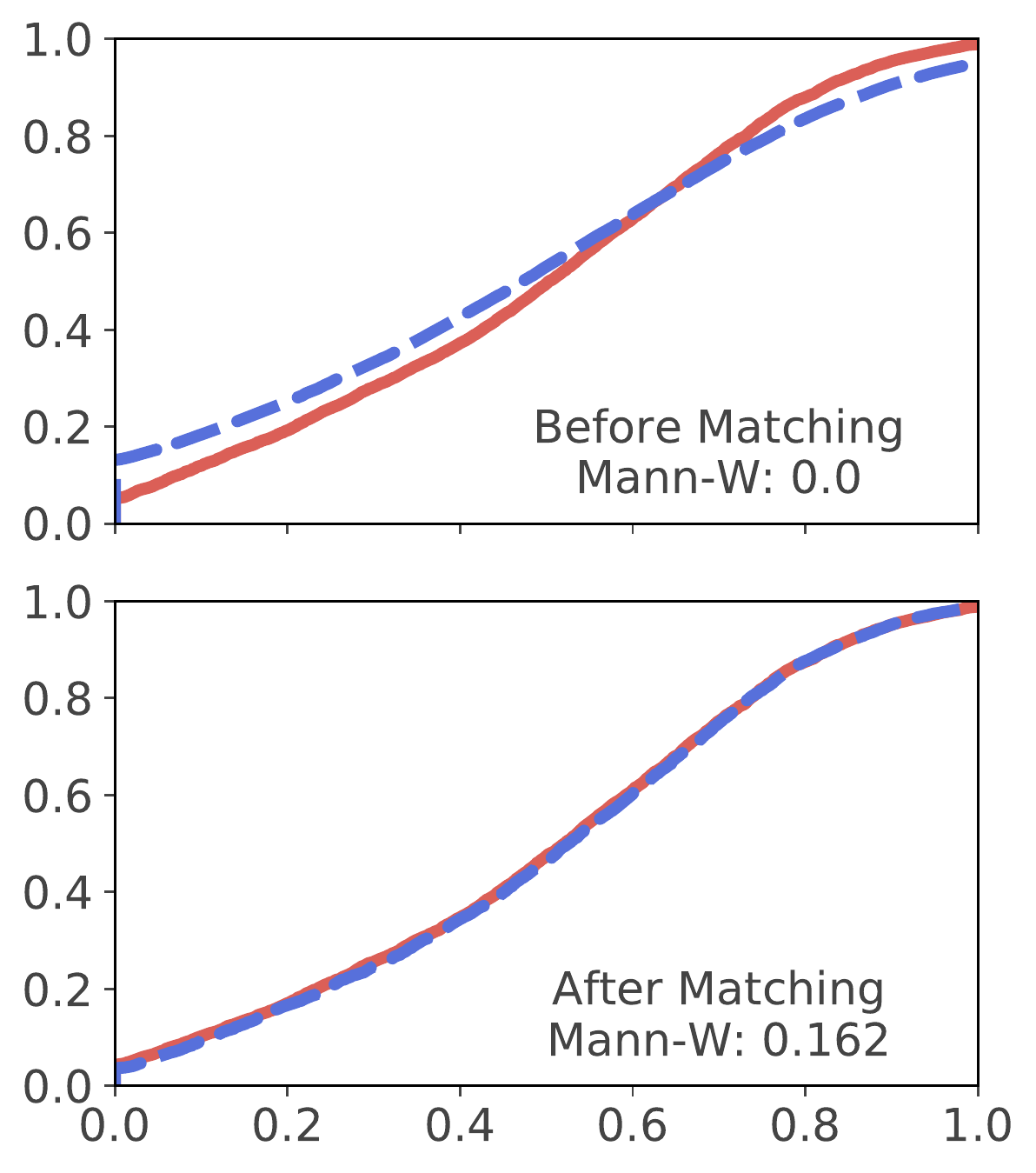}
        \includegraphics[width=0.19\textwidth]
        {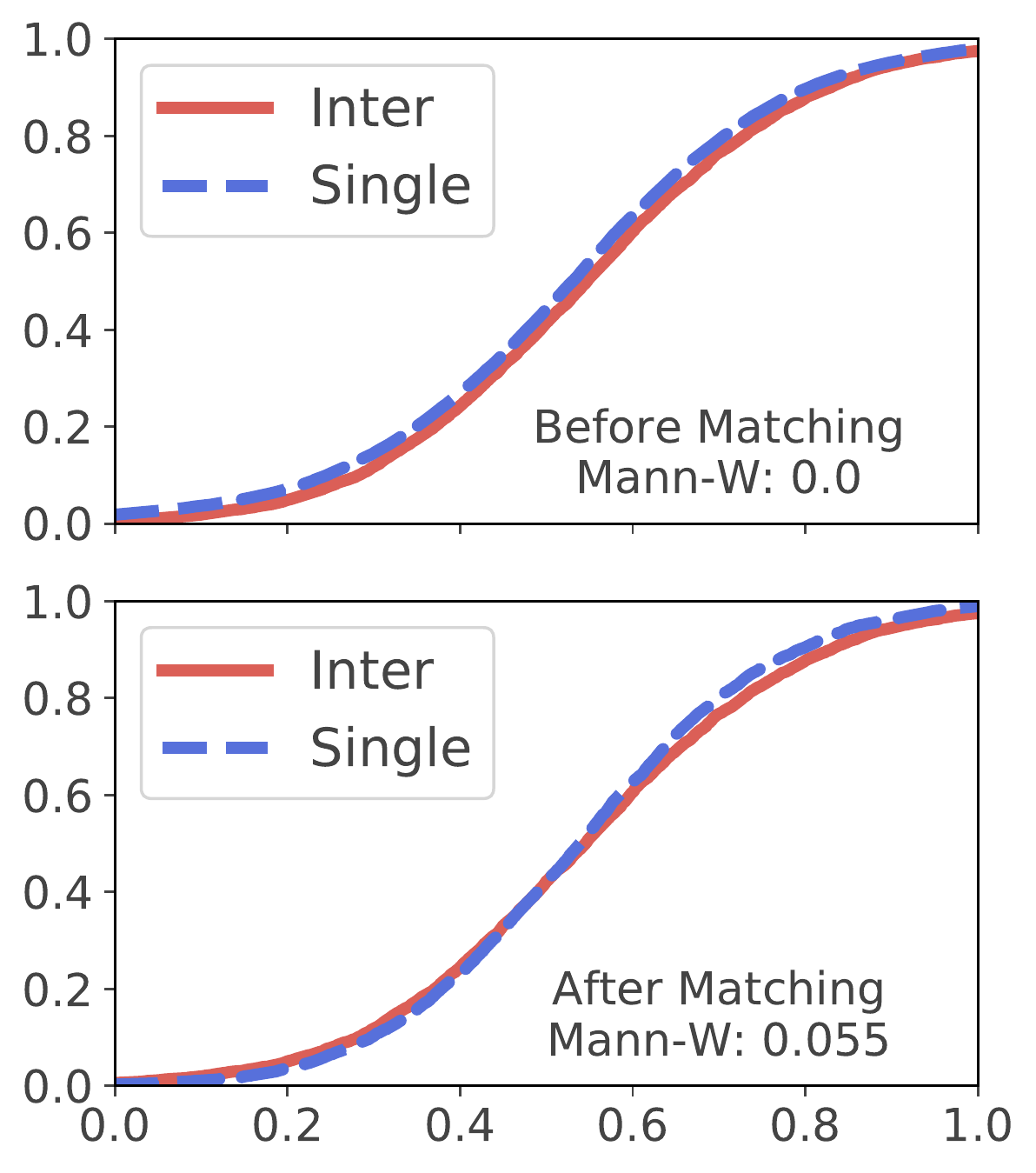}
        \includegraphics[width=0.19\textwidth]
        {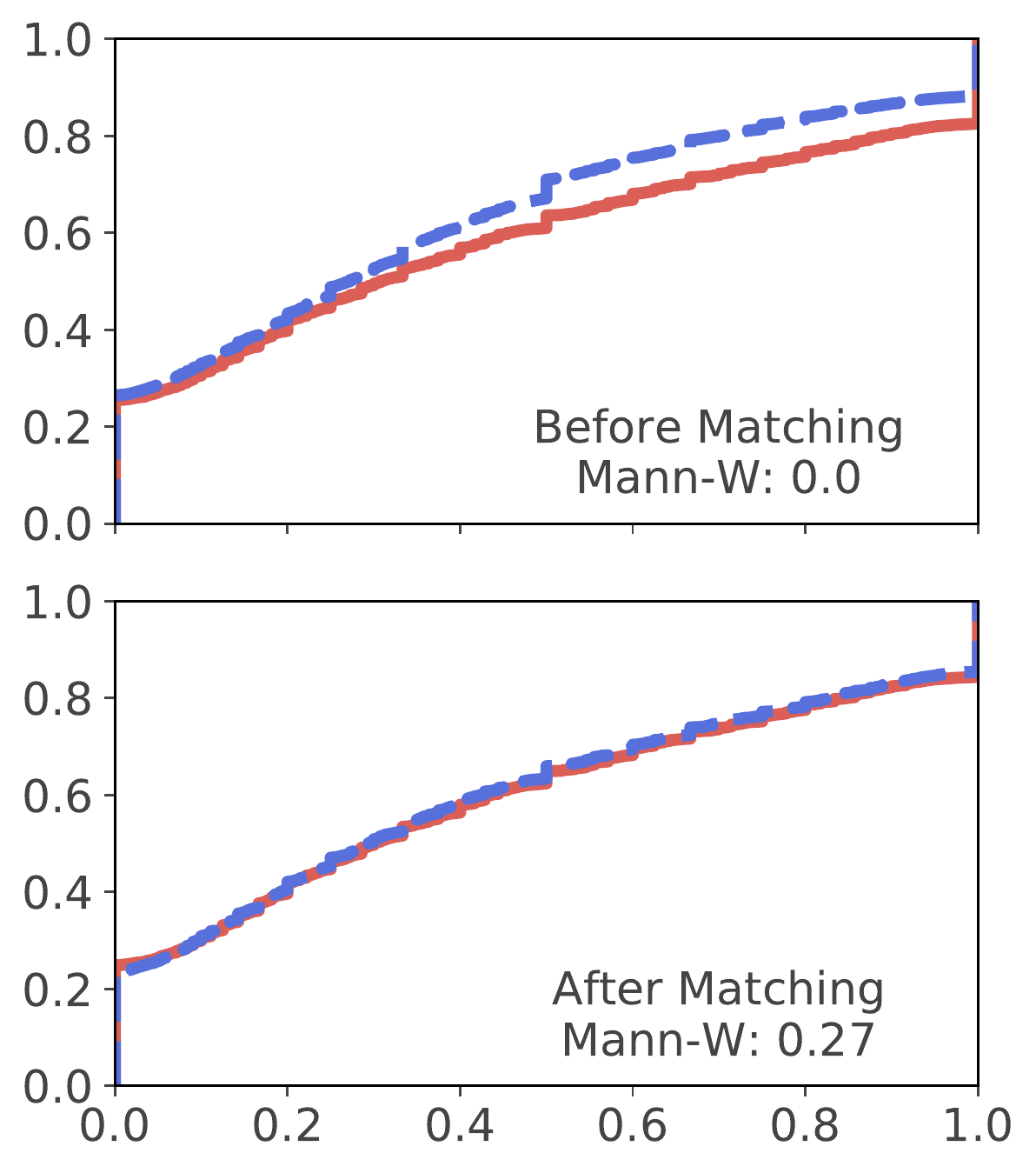}
        \includegraphics[width=0.19\textwidth]
        {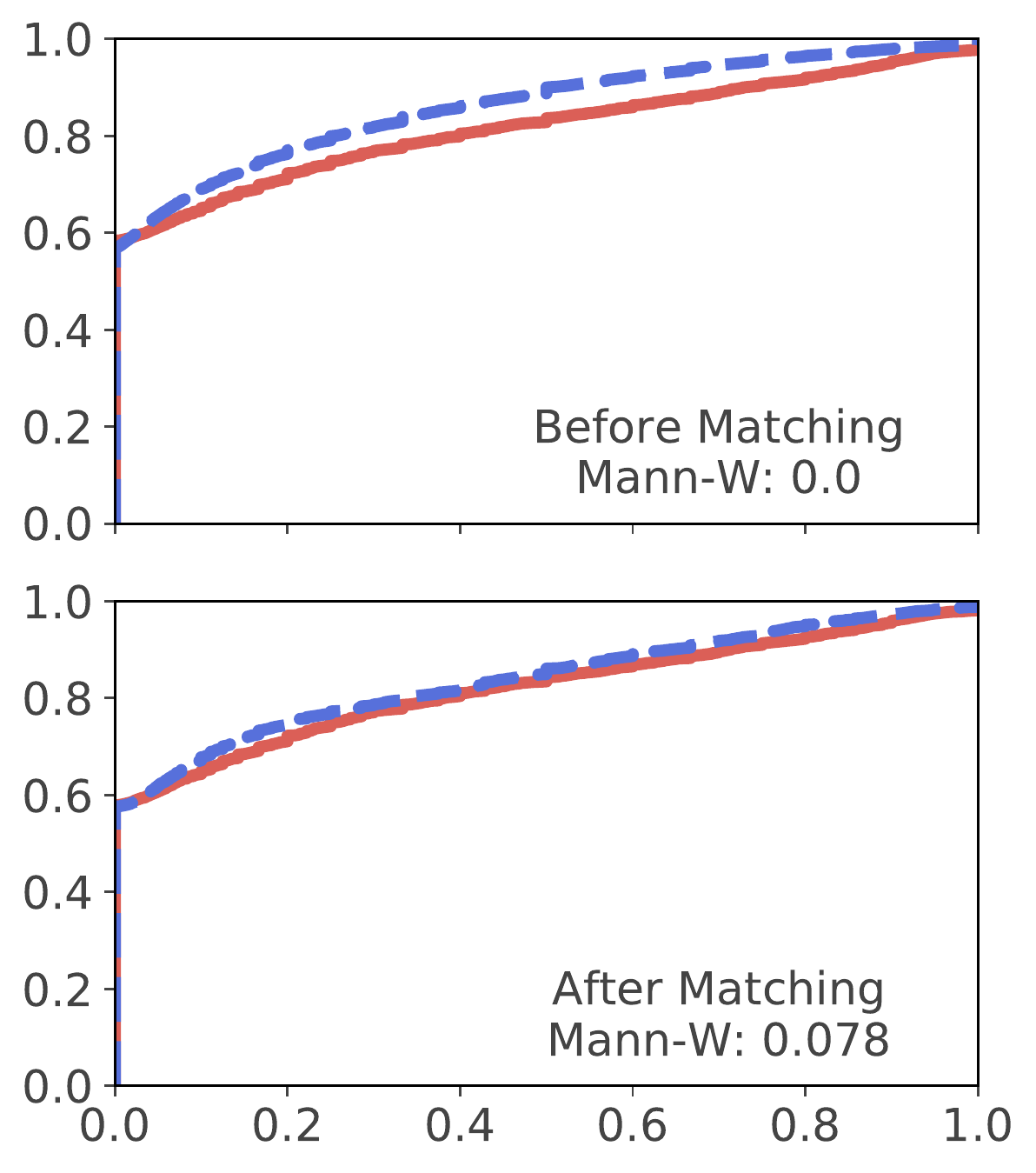}
        \caption{Matching results at level 3.}
    \end{subfigure}
    \hfill
    \begin{subfigure}[t]{\textwidth}
        \includegraphics[width=0.19\textwidth]
        {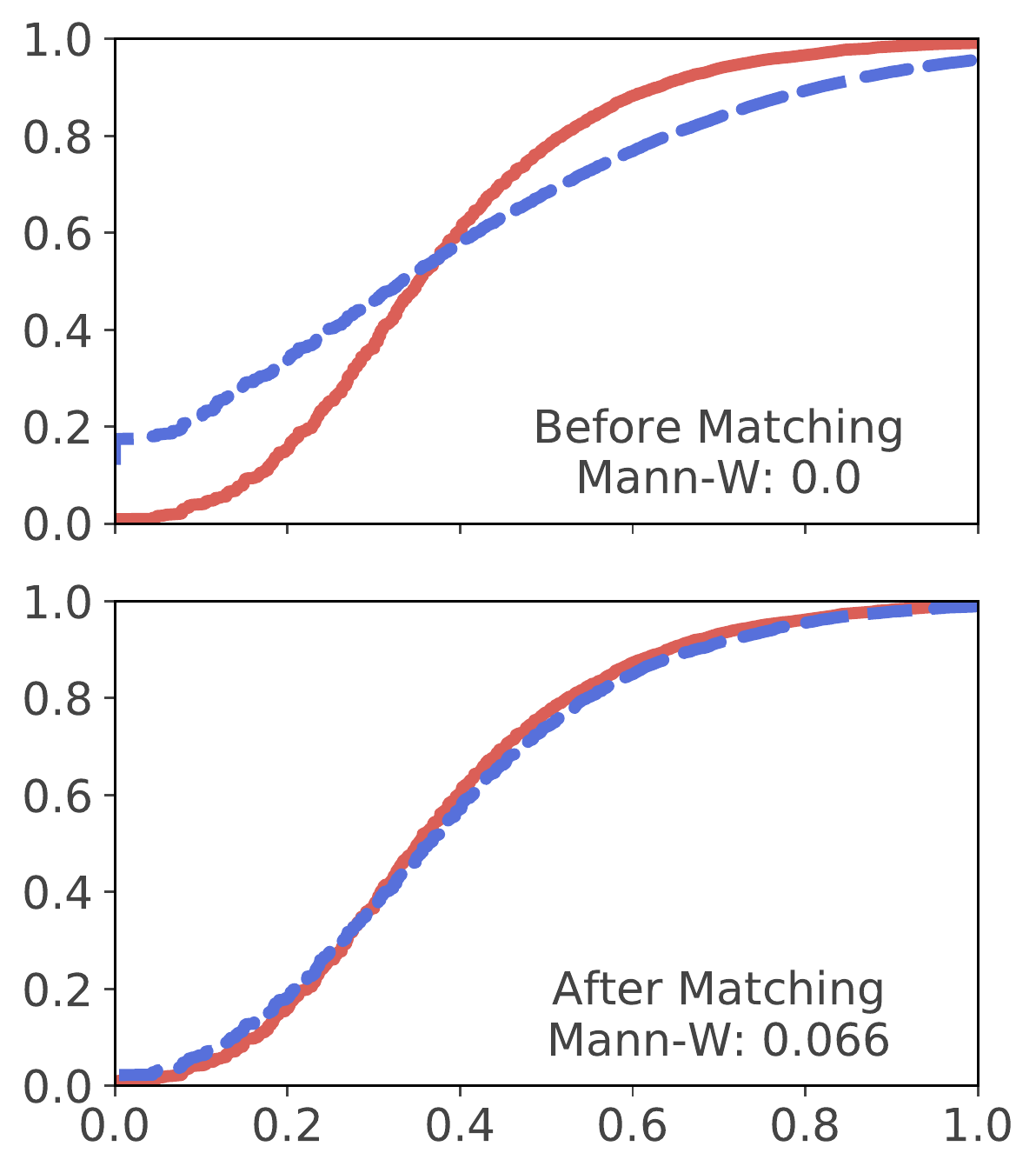}
        \includegraphics[width=0.19\textwidth]
        {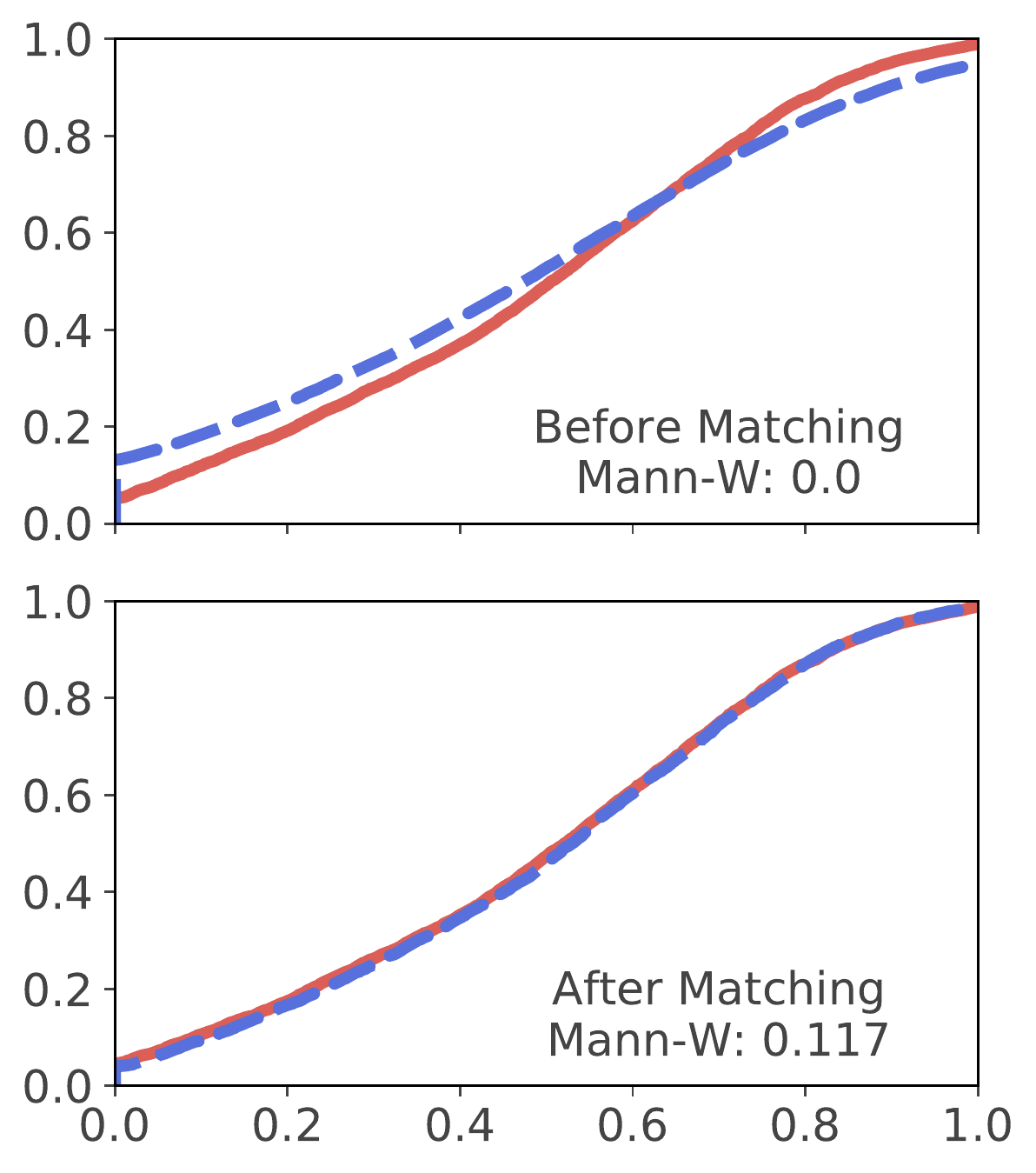}
        \includegraphics[width=0.19\textwidth]
        {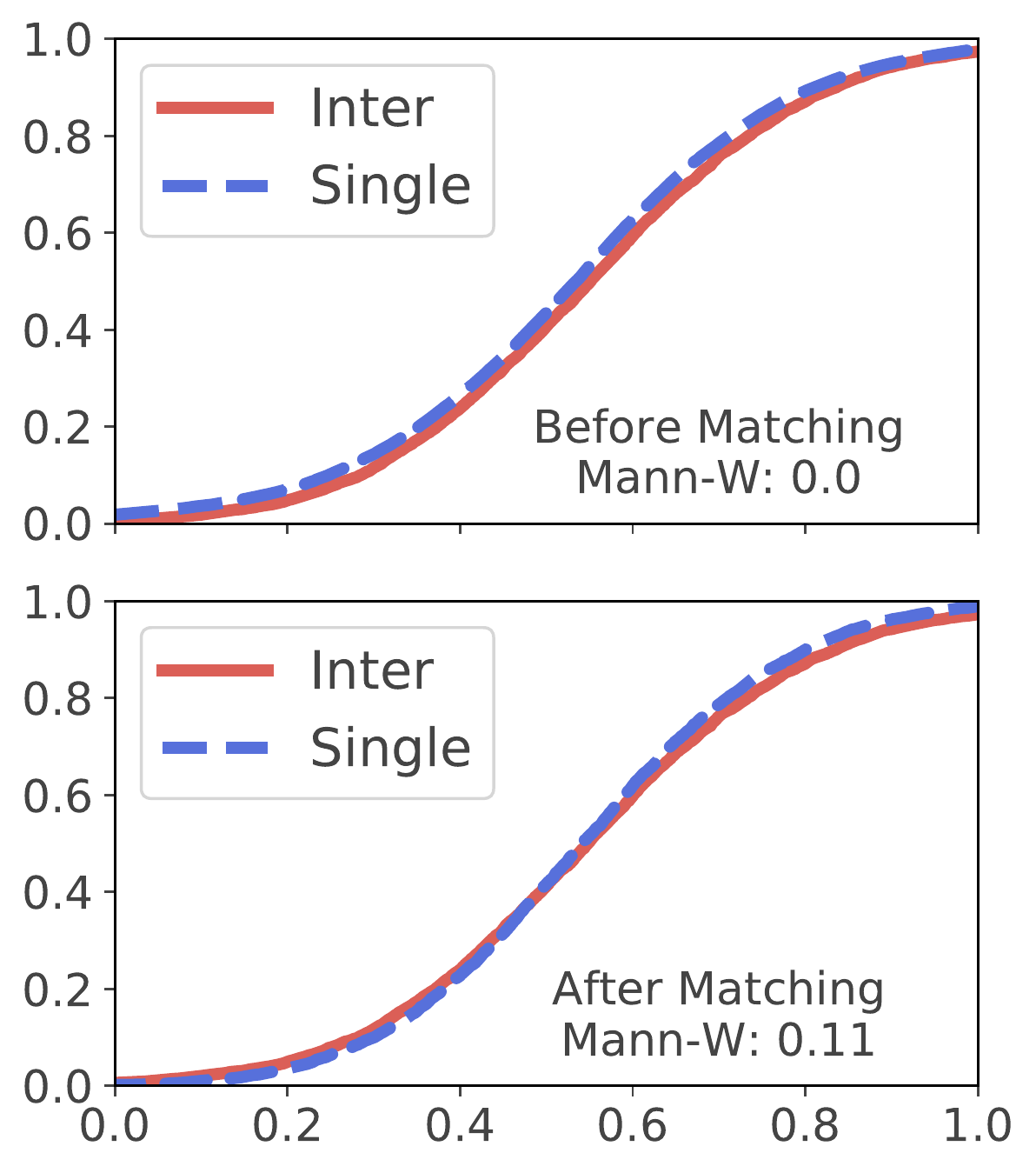}
        \includegraphics[width=0.19\textwidth]
        {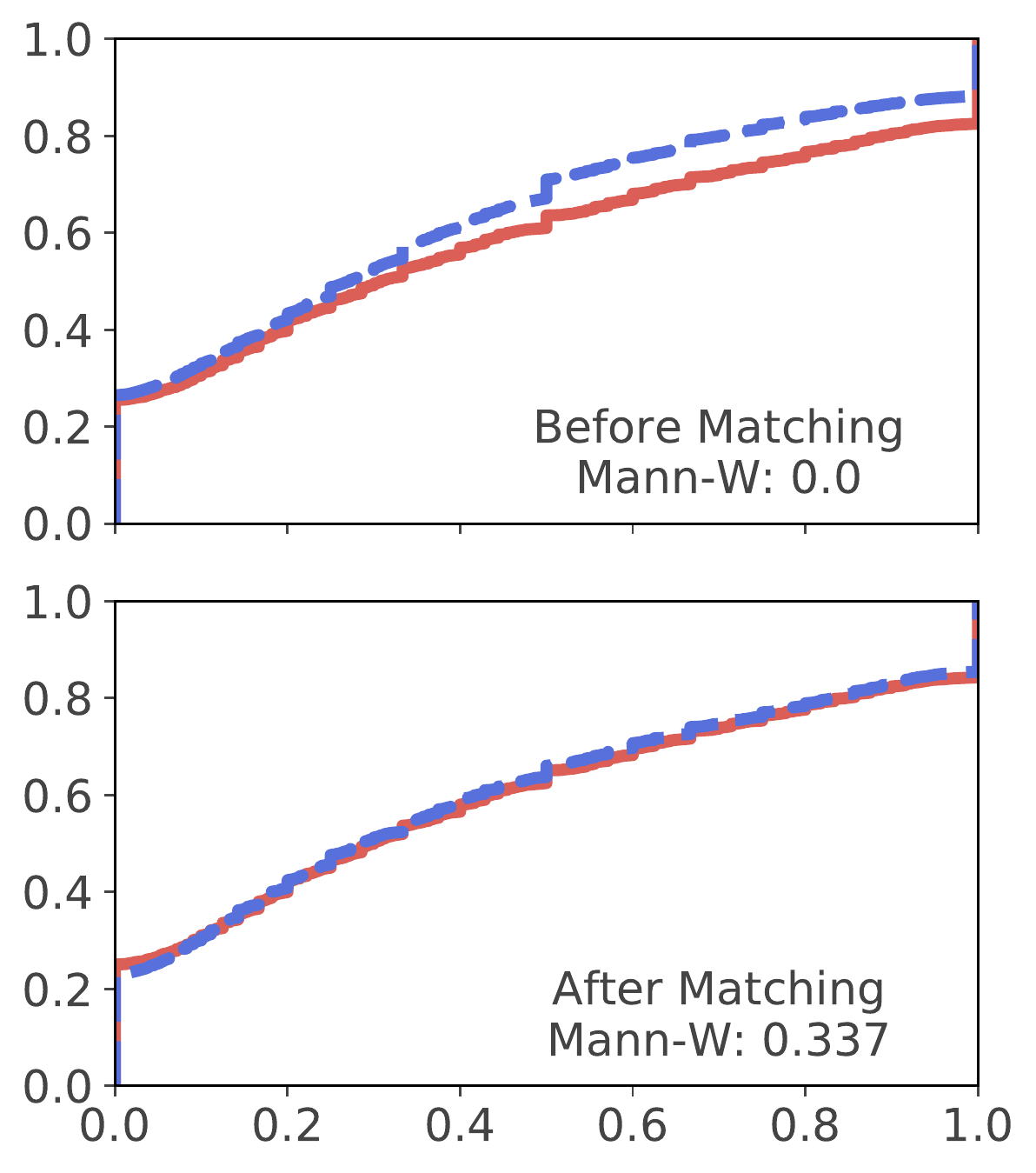}
        \includegraphics[width=0.19\textwidth]
        {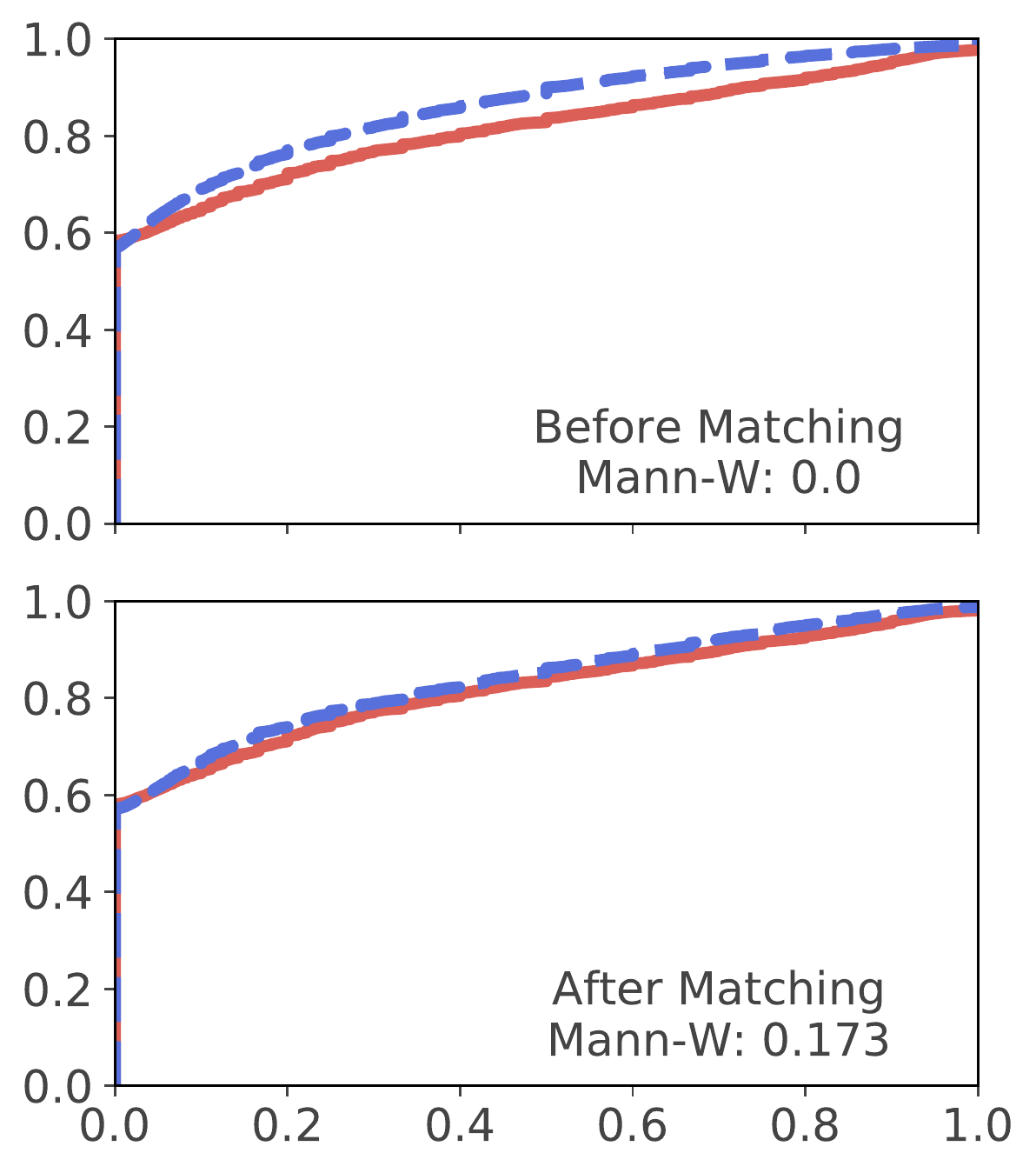}
        \caption{Matching results at level 4.}
    \end{subfigure}
    \hfill
    \begin{subfigure}[t]{\textwidth}
        \includegraphics[width=0.19\textwidth]
        {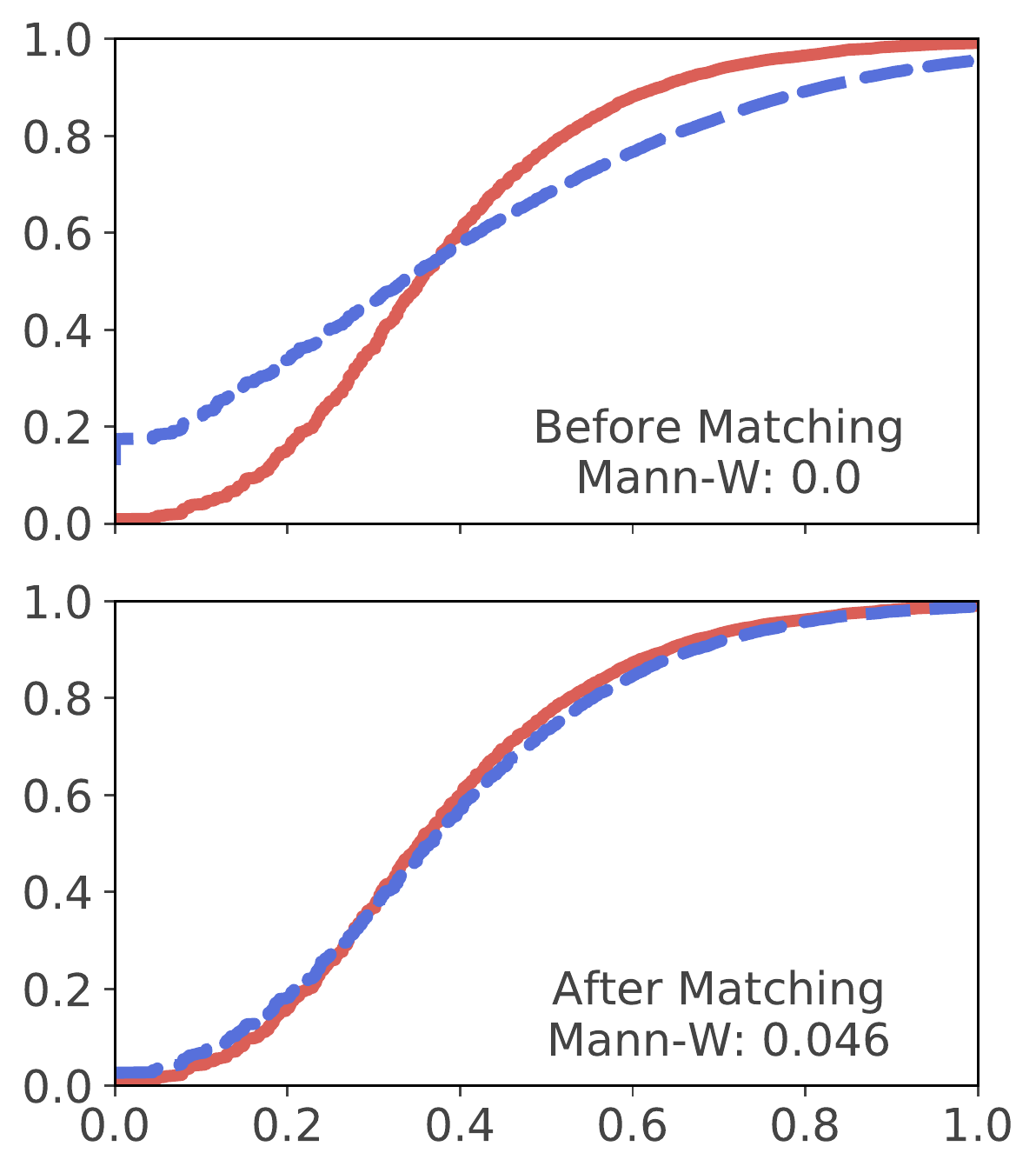}
        \includegraphics[width=0.19\textwidth]
        {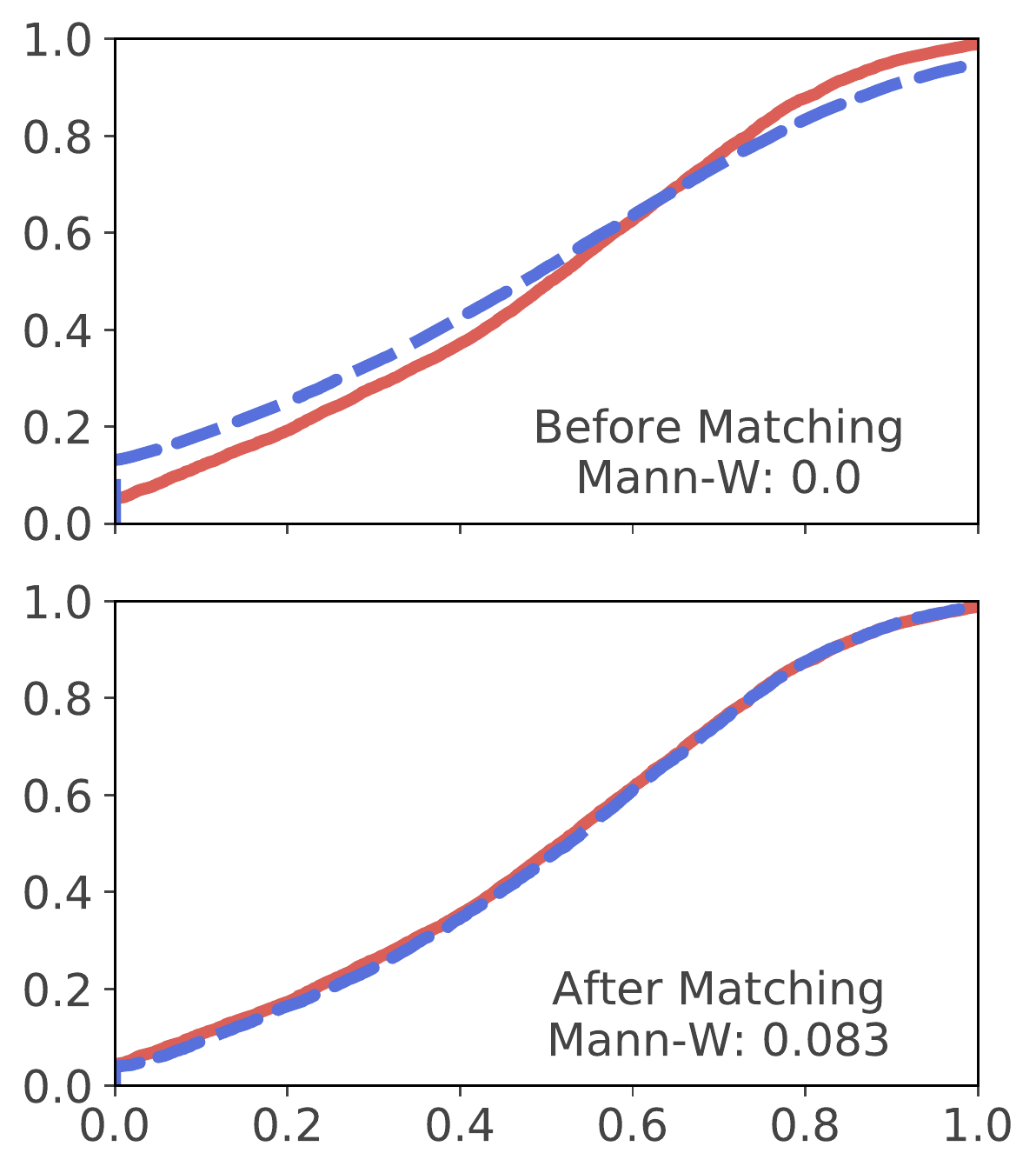}
        \includegraphics[width=0.19\textwidth]
        {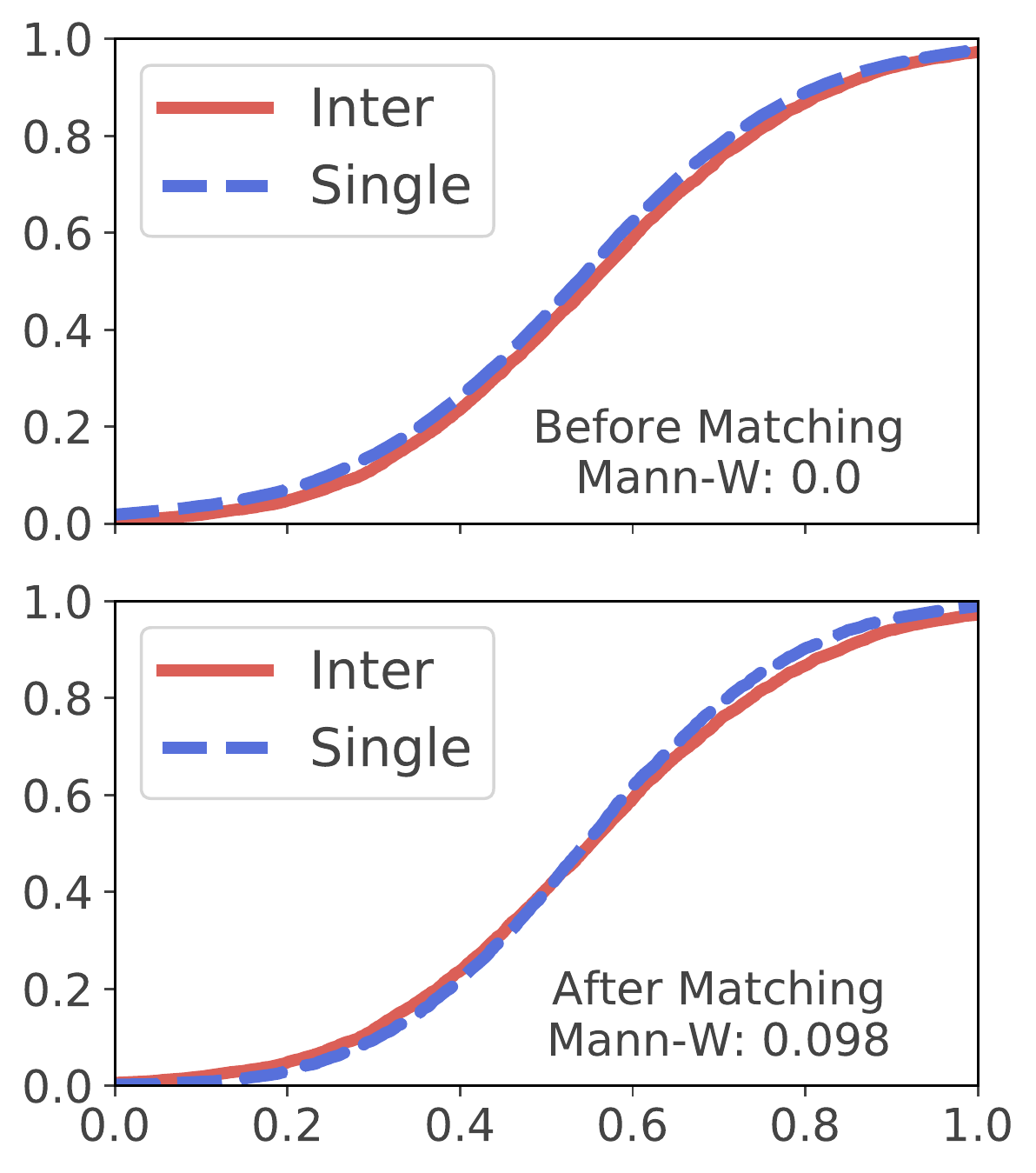}
        \includegraphics[width=0.19\textwidth]
        {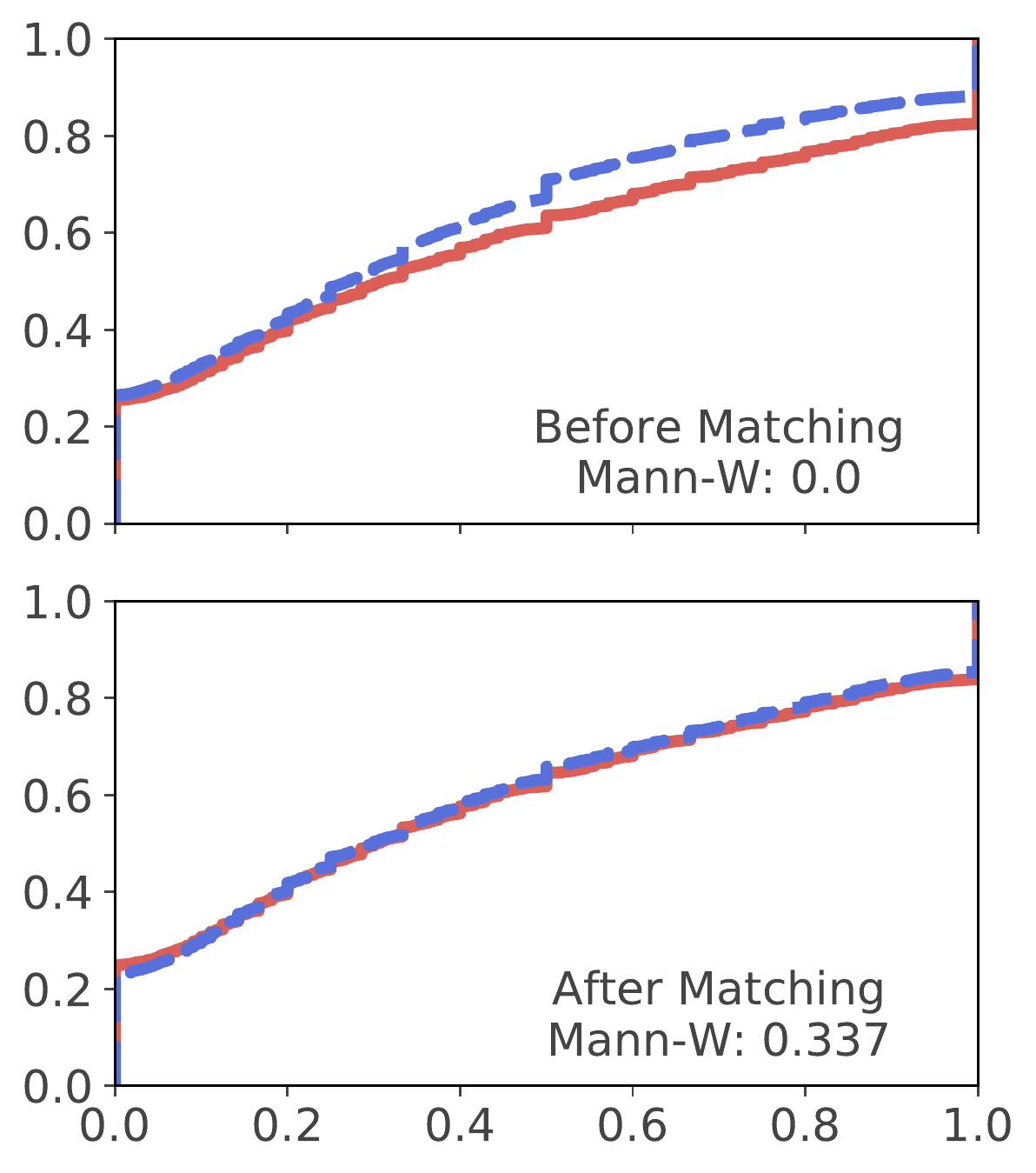}
        \includegraphics[width=0.19\textwidth]
        {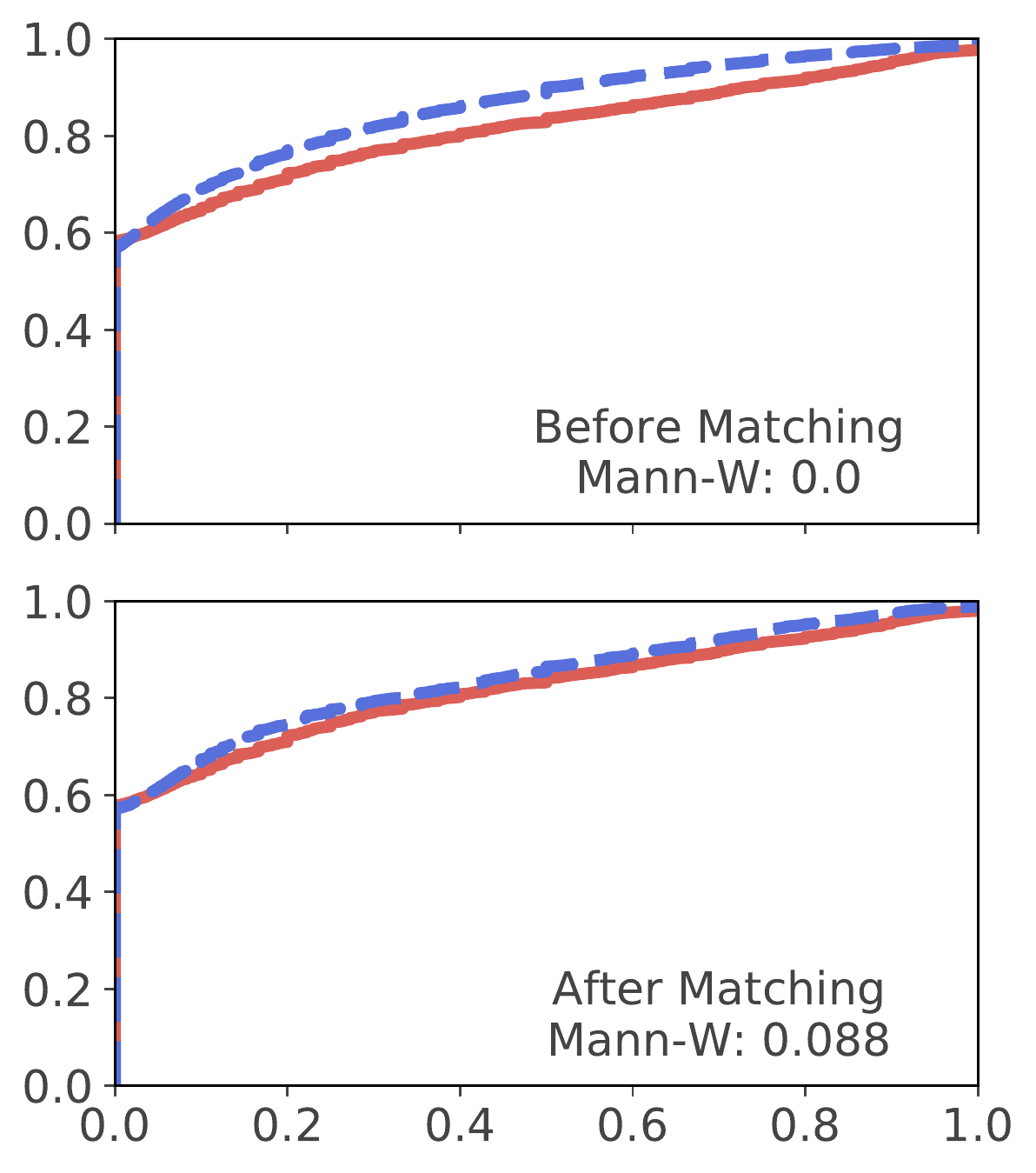}
        \caption{Matching results at level 5.}
    \end{subfigure}
    \caption{Empirical cumulative distribution of each activity feature before and after the matching technique from level 1 to level 5 in the 2017 season.
    The activity features from left to right are the number of comments,
    the average hour gap between comments, the average comment length, 
    the proportion of playoff comments, and the proportion of game thread comments.}
    \label{fig:matchinglevel2017}
\end{figure}

\begin{figure}
    \center
    \begin{subfigure}[t]{\textwidth}
        \includegraphics[width=0.19\textwidth]
        {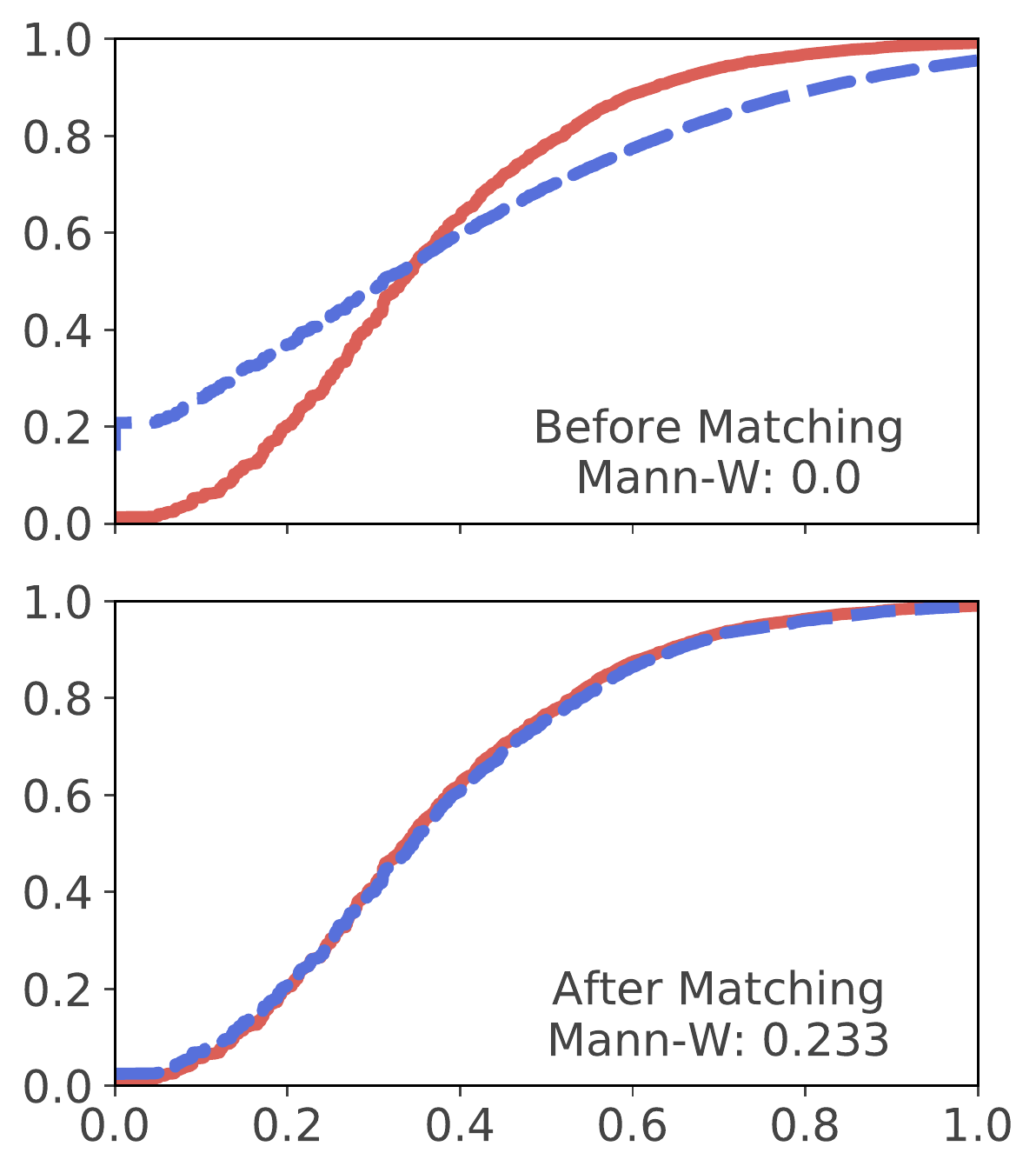}
        \includegraphics[width=0.19\textwidth]
        {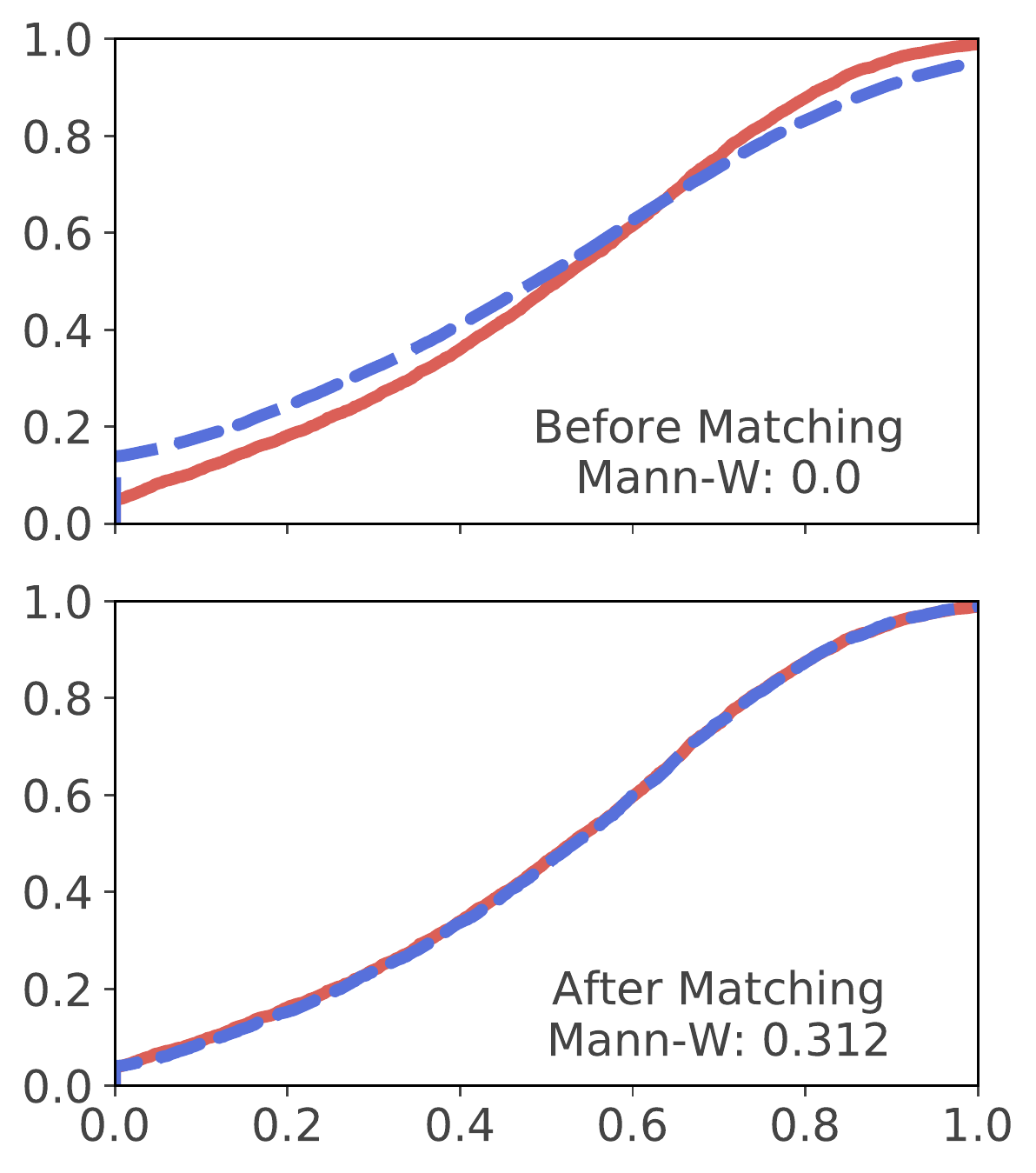}
        \includegraphics[width=0.19\textwidth]
        {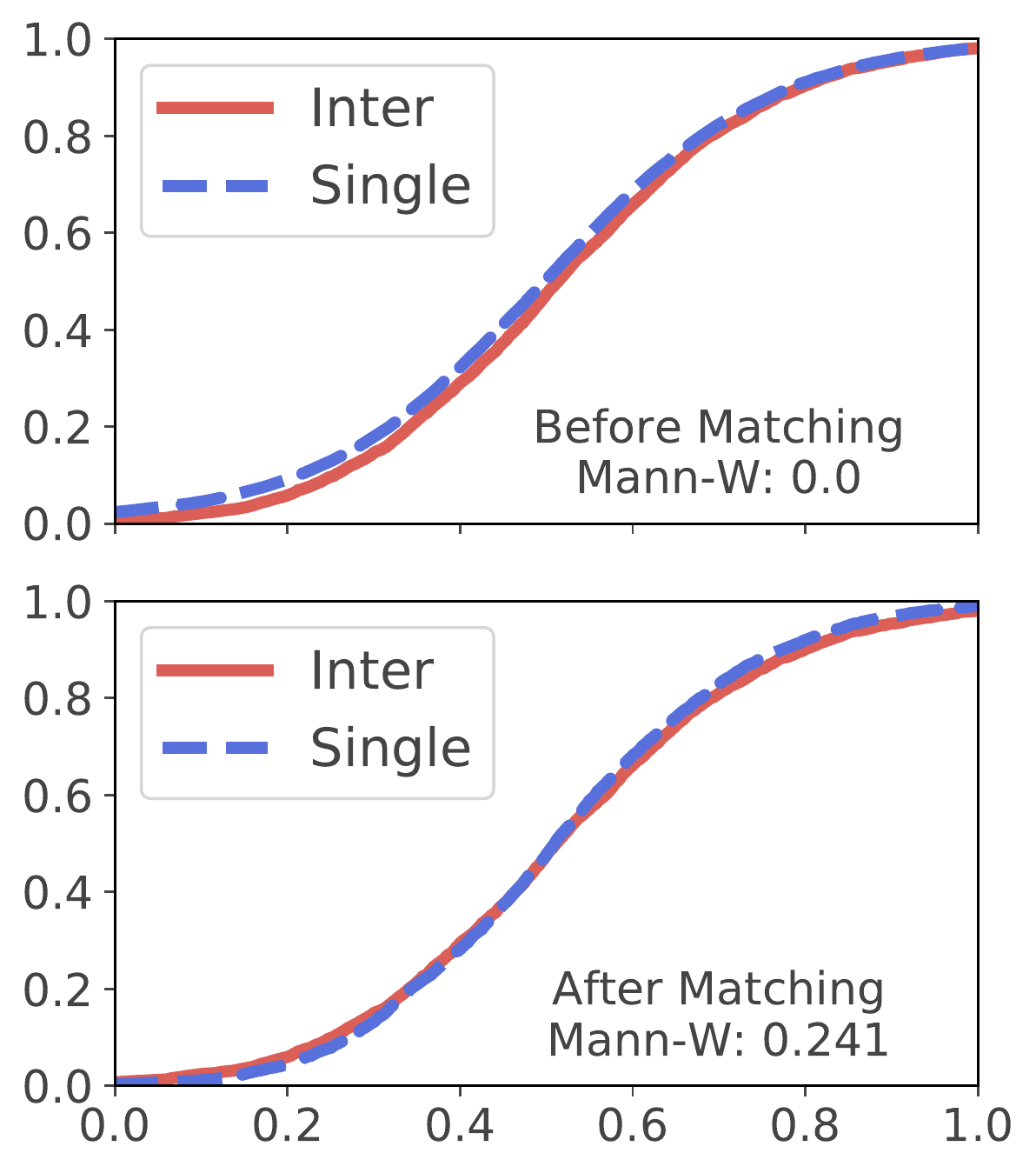}
        \includegraphics[width=0.19\textwidth]
        {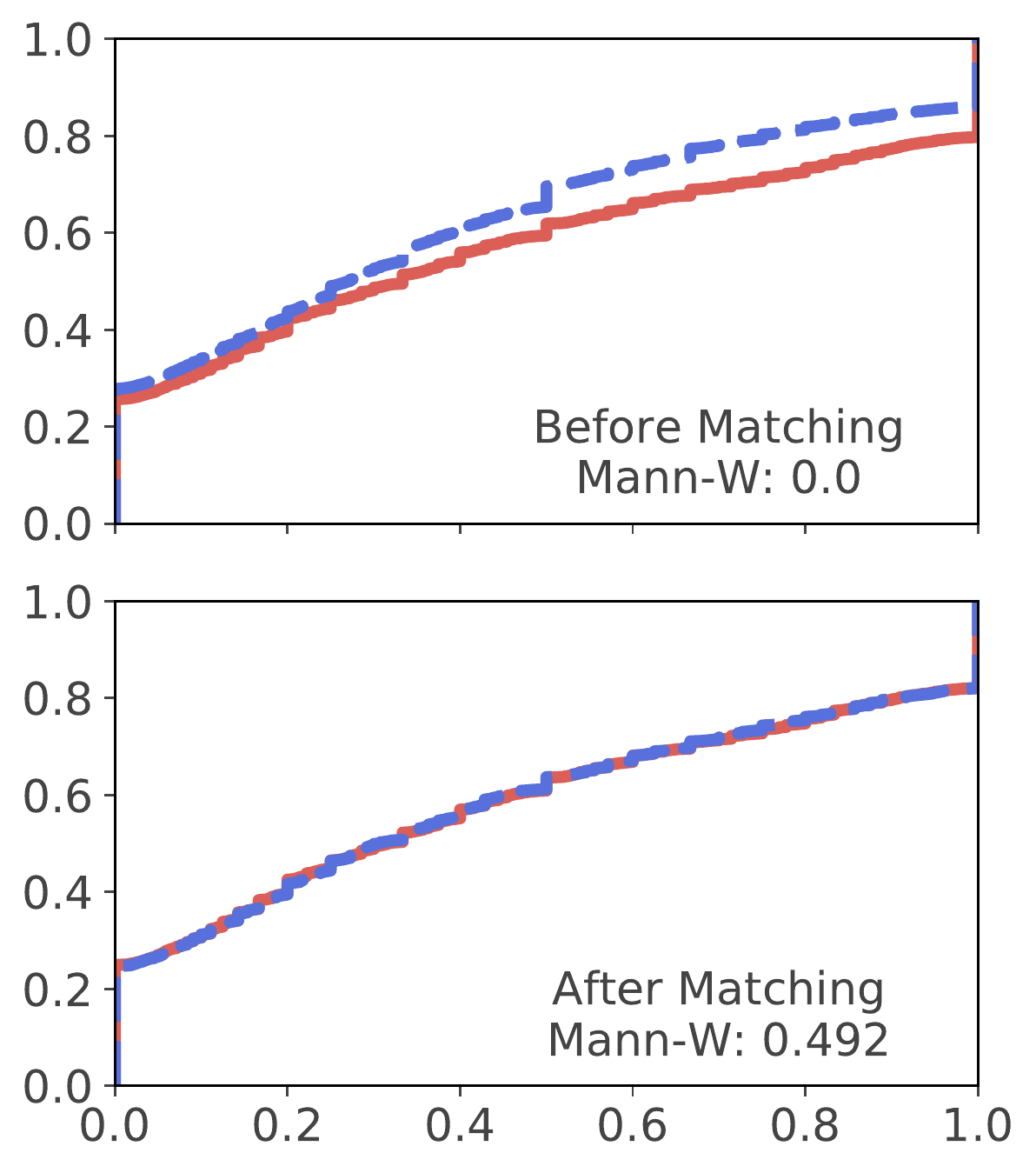}
        \includegraphics[width=0.19\textwidth]
        {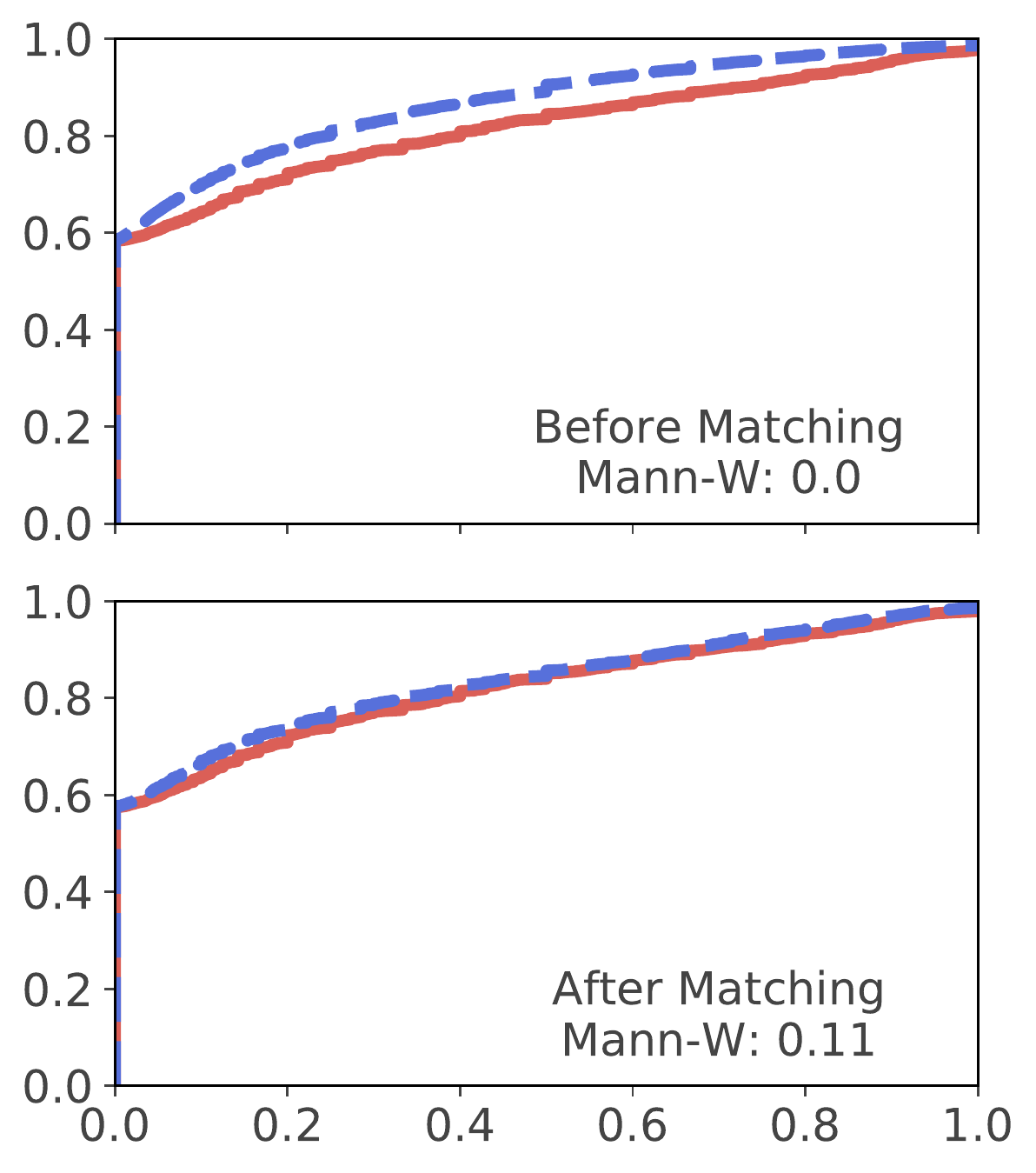}
        \caption{Matching results at level 1.}
    \end{subfigure}
    \hfill
    \begin{subfigure}[t]{\textwidth}
        \includegraphics[width=0.19\textwidth]
        {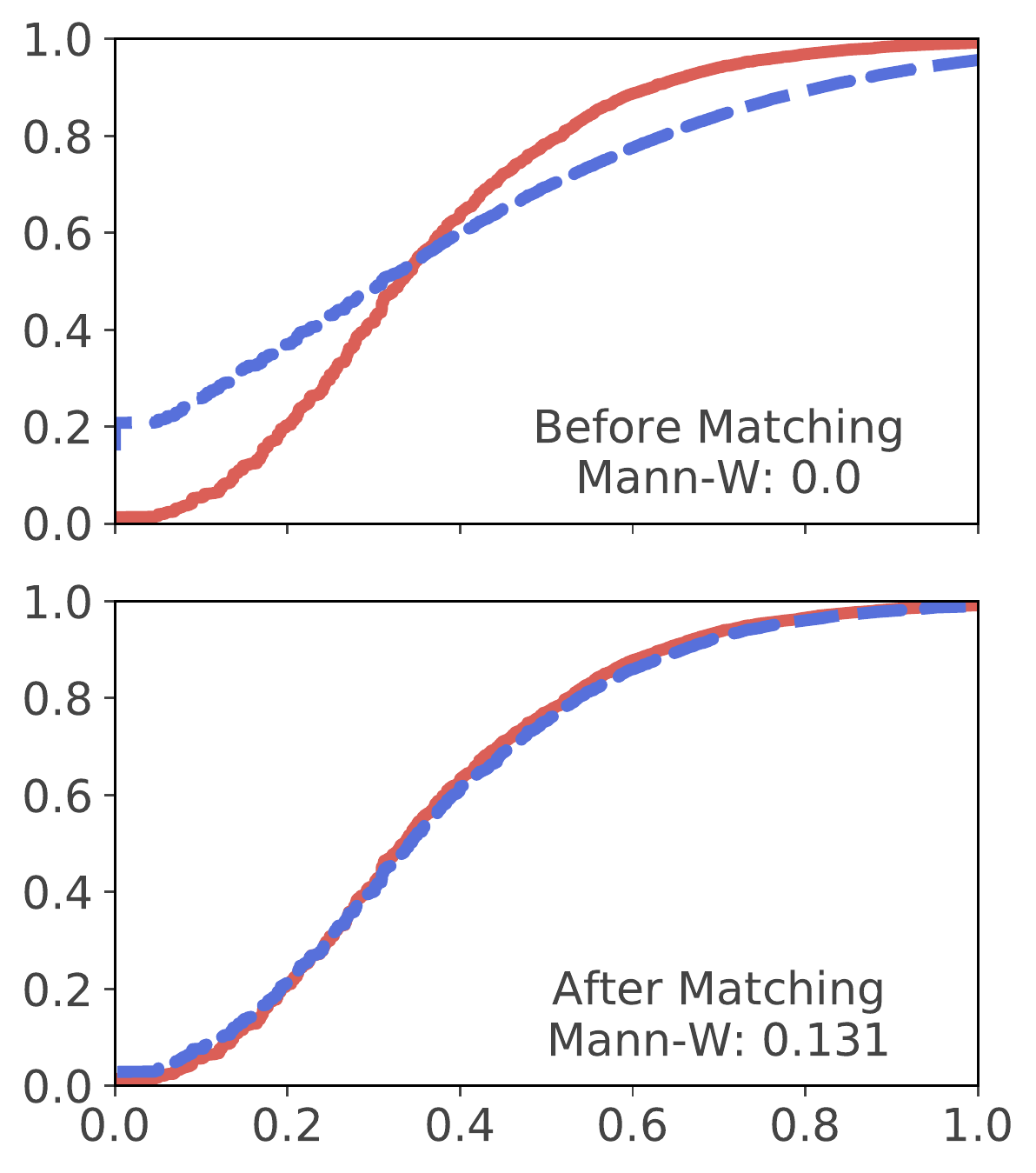}
        \includegraphics[width=0.19\textwidth]
        {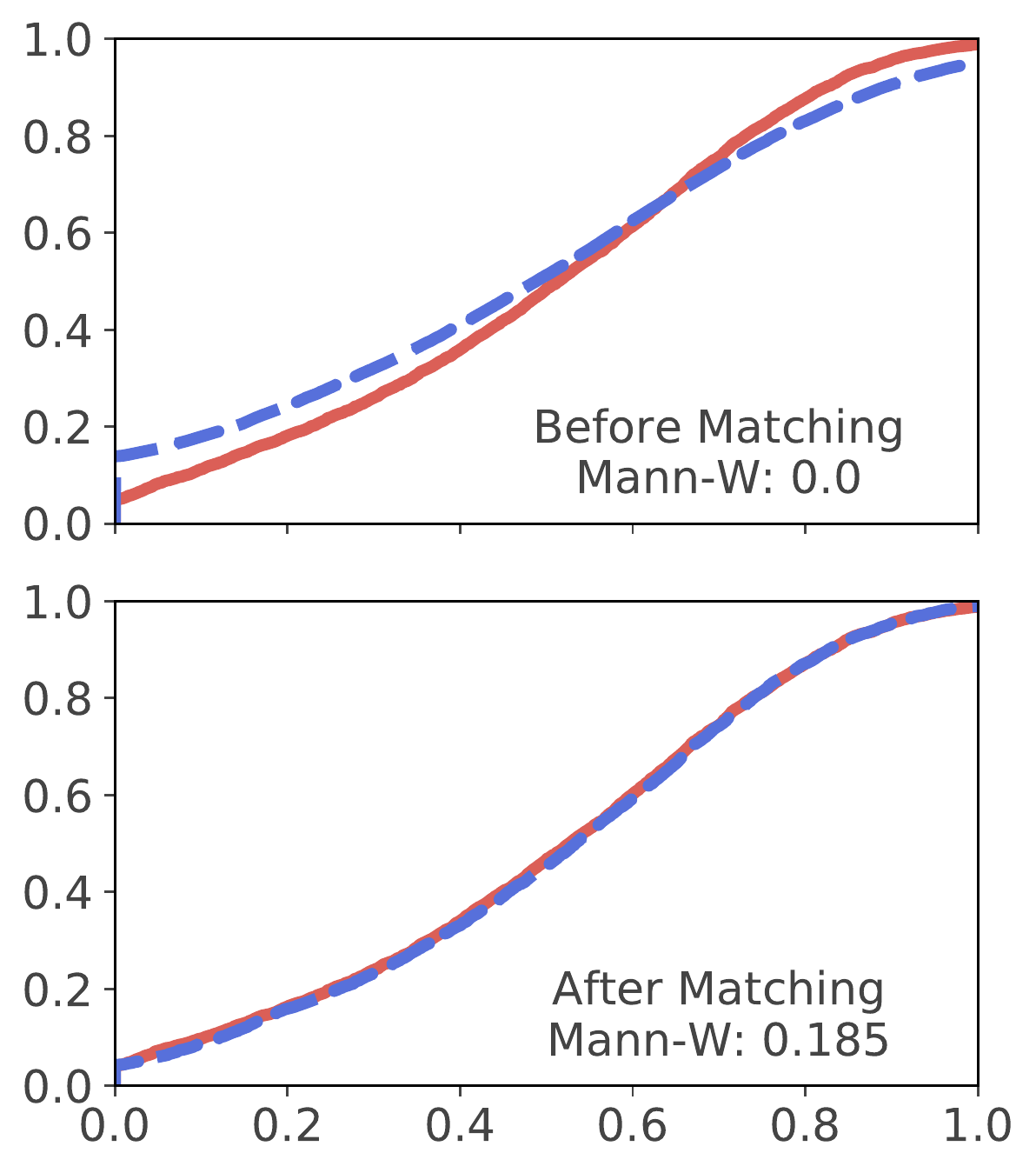}
        \includegraphics[width=0.19\textwidth]
        {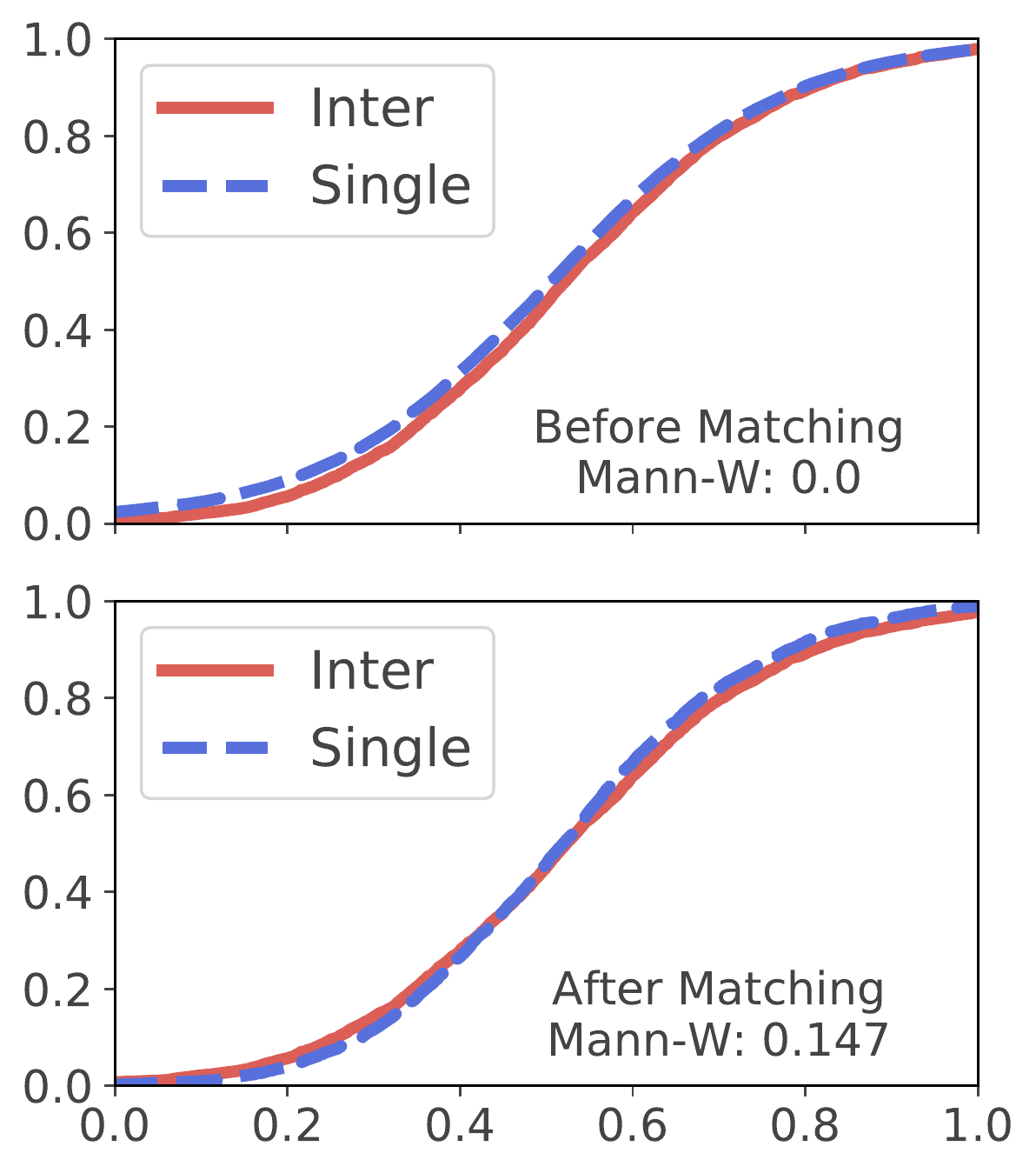}
        \includegraphics[width=0.19\textwidth]
        {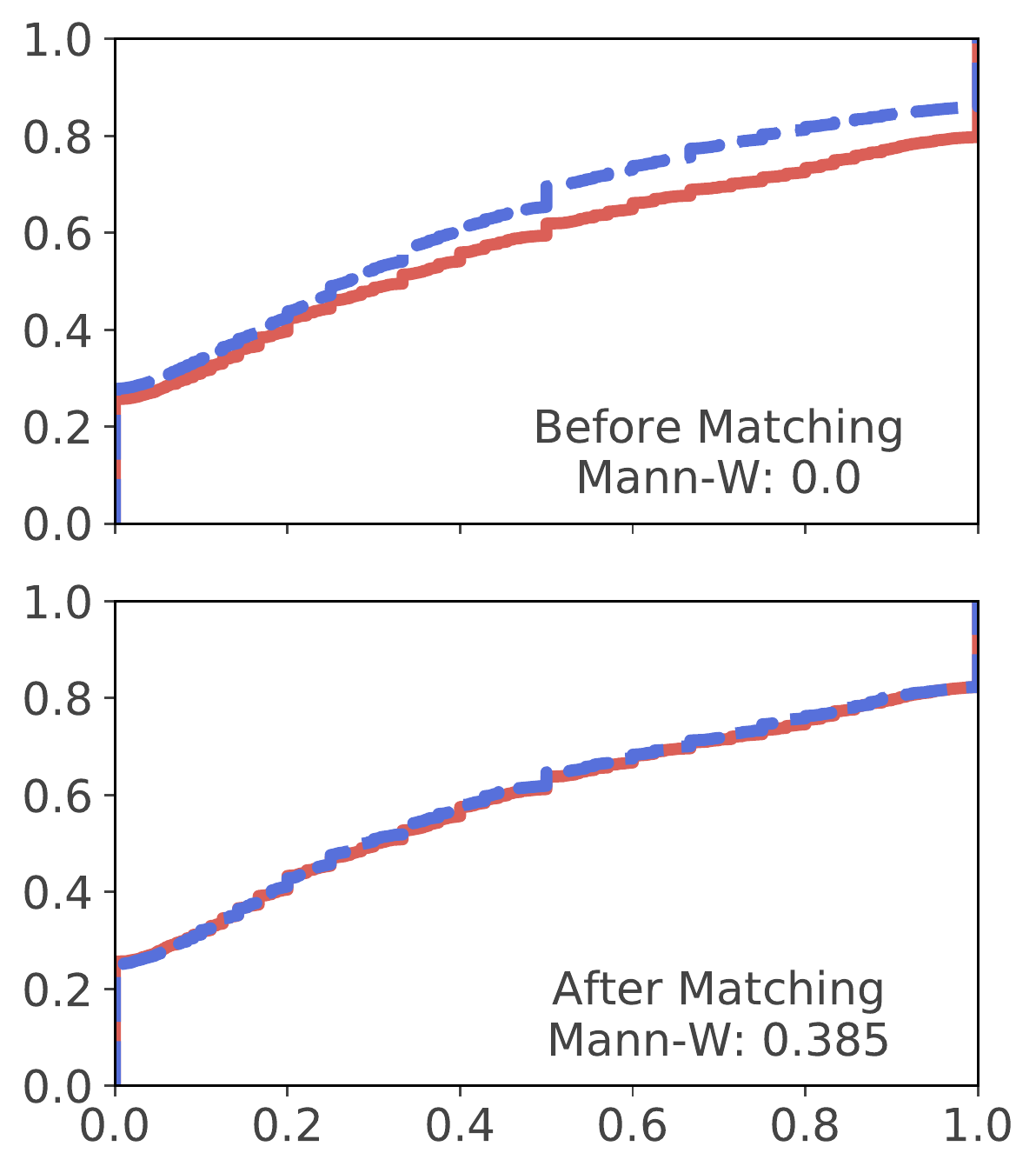}
        \includegraphics[width=0.19\textwidth]
        {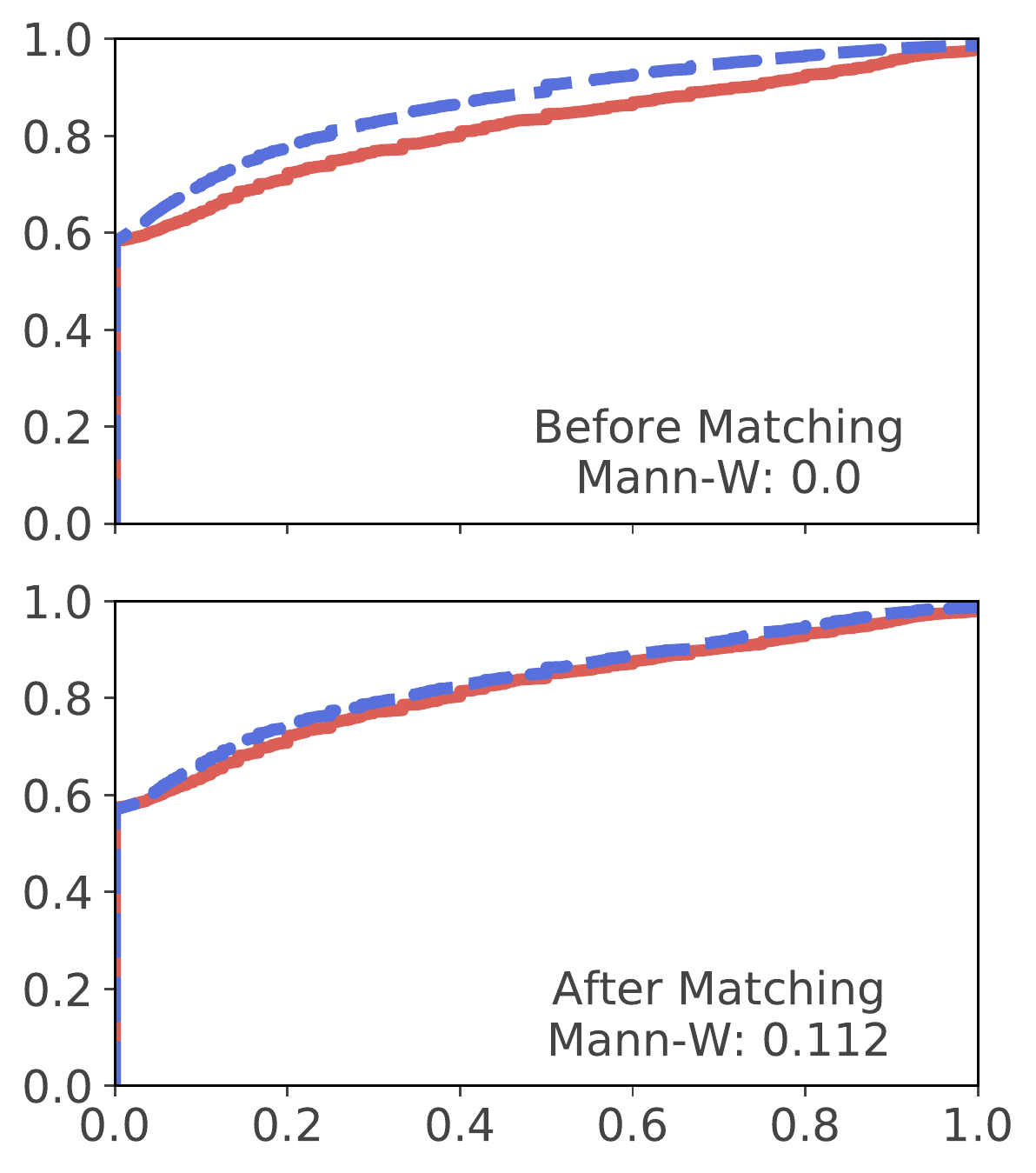}
        \caption{Matching results at level 2.}
    \end{subfigure}
    \hfill
    \begin{subfigure}[t]{\textwidth}
        \includegraphics[width=0.19\textwidth]
        {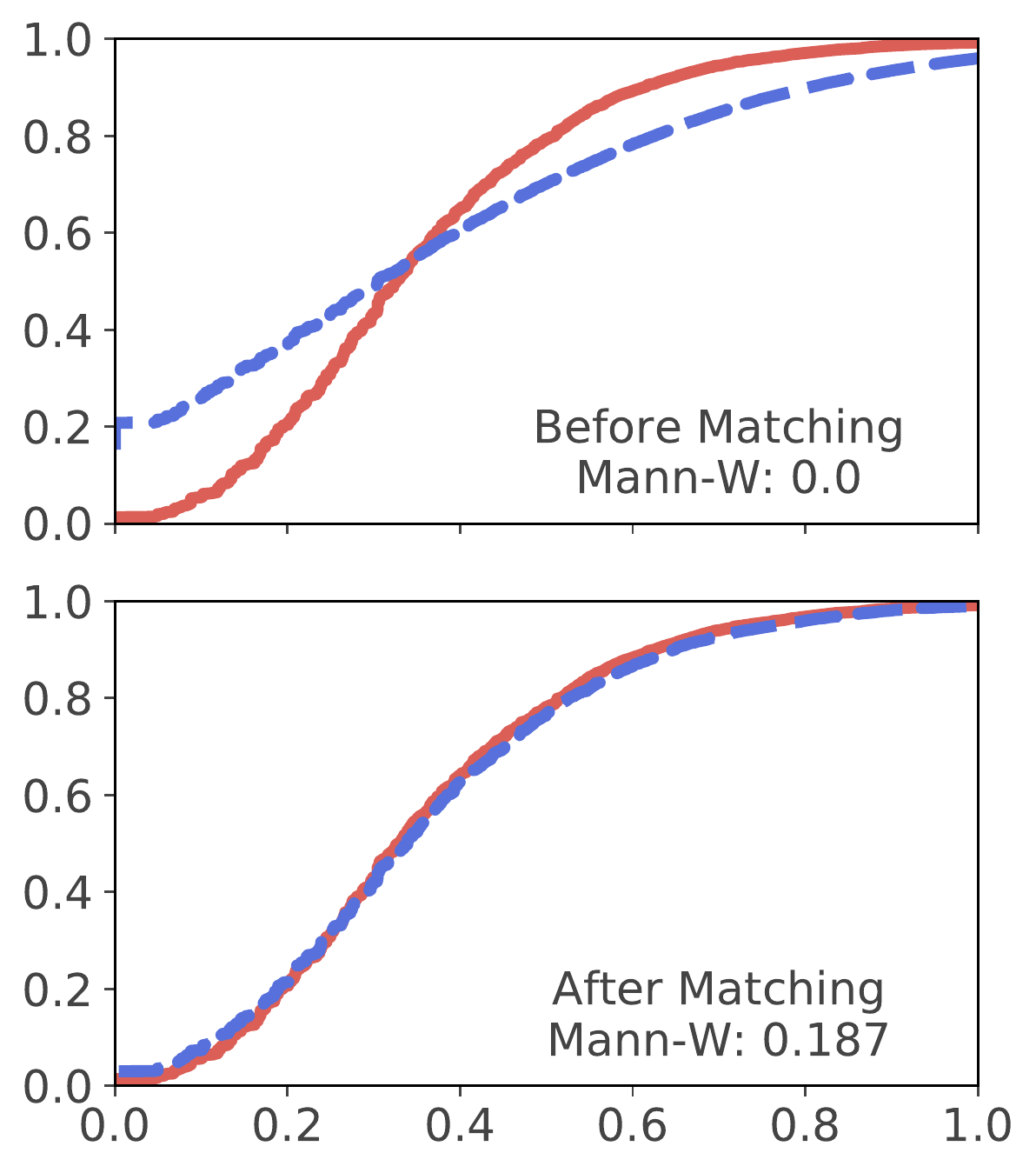}
        \includegraphics[width=0.19\textwidth]
        {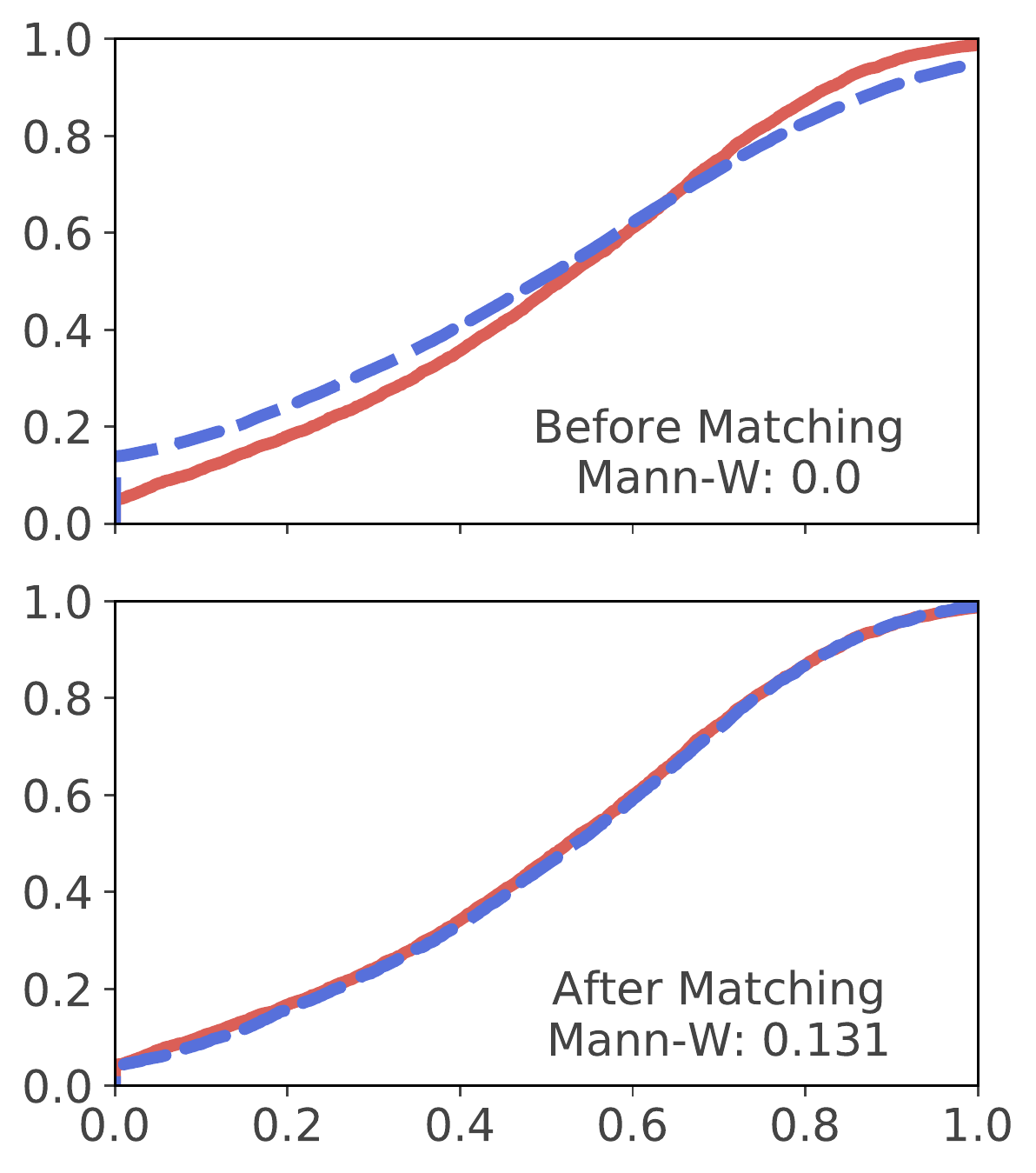}
        \includegraphics[width=0.19\textwidth]
        {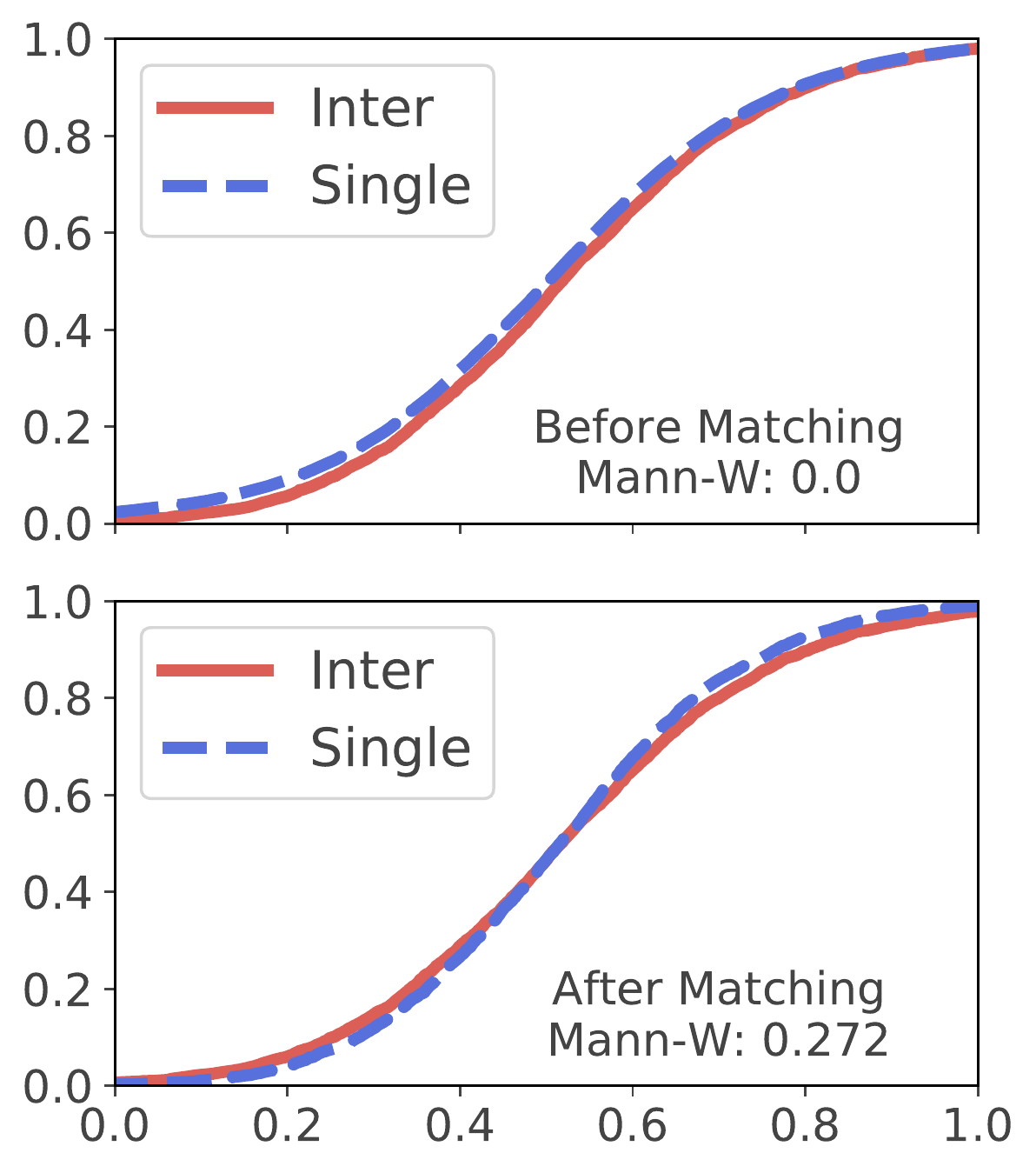}
        \includegraphics[width=0.19\textwidth]
        {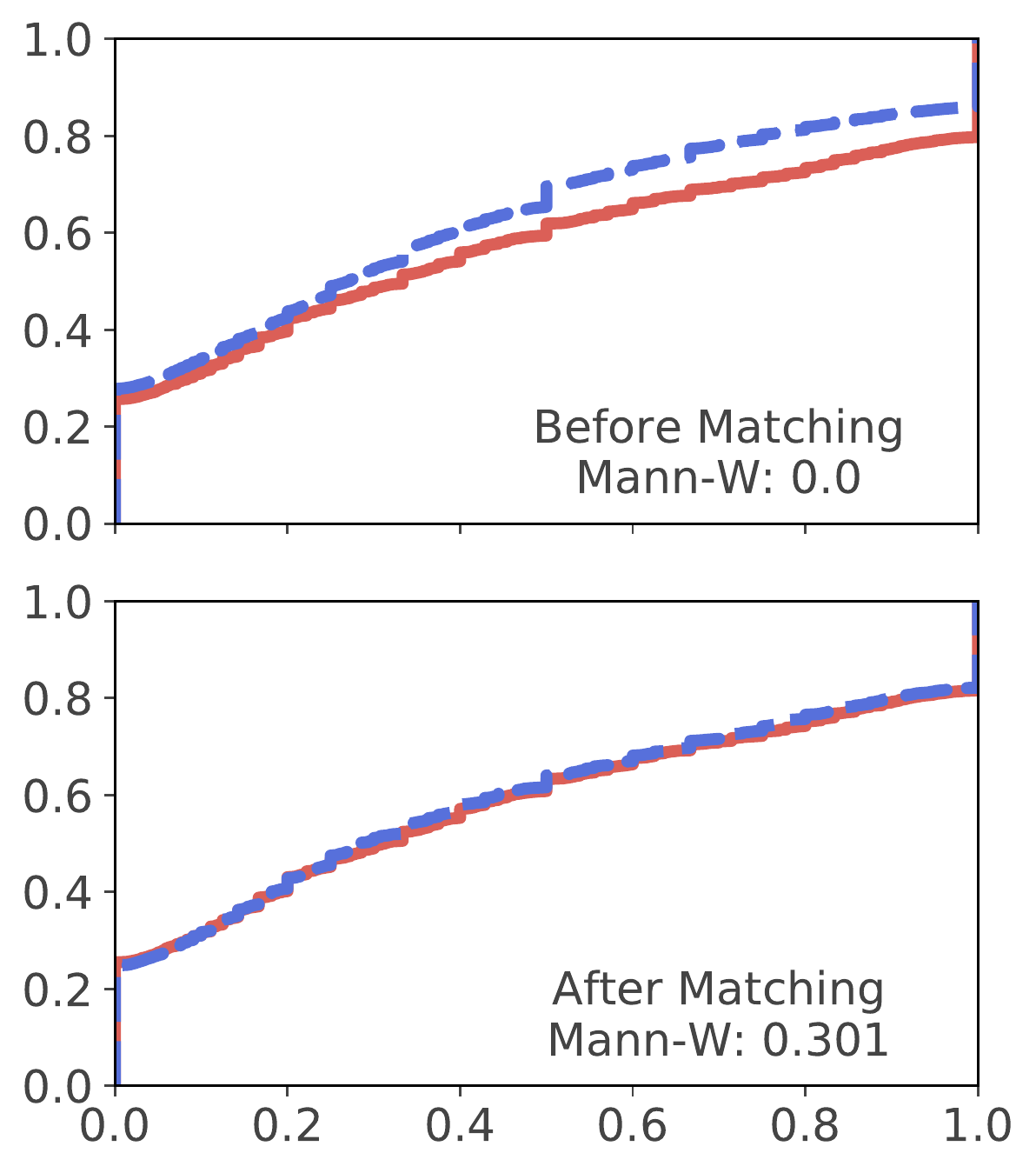}
        \includegraphics[width=0.19\textwidth]
        {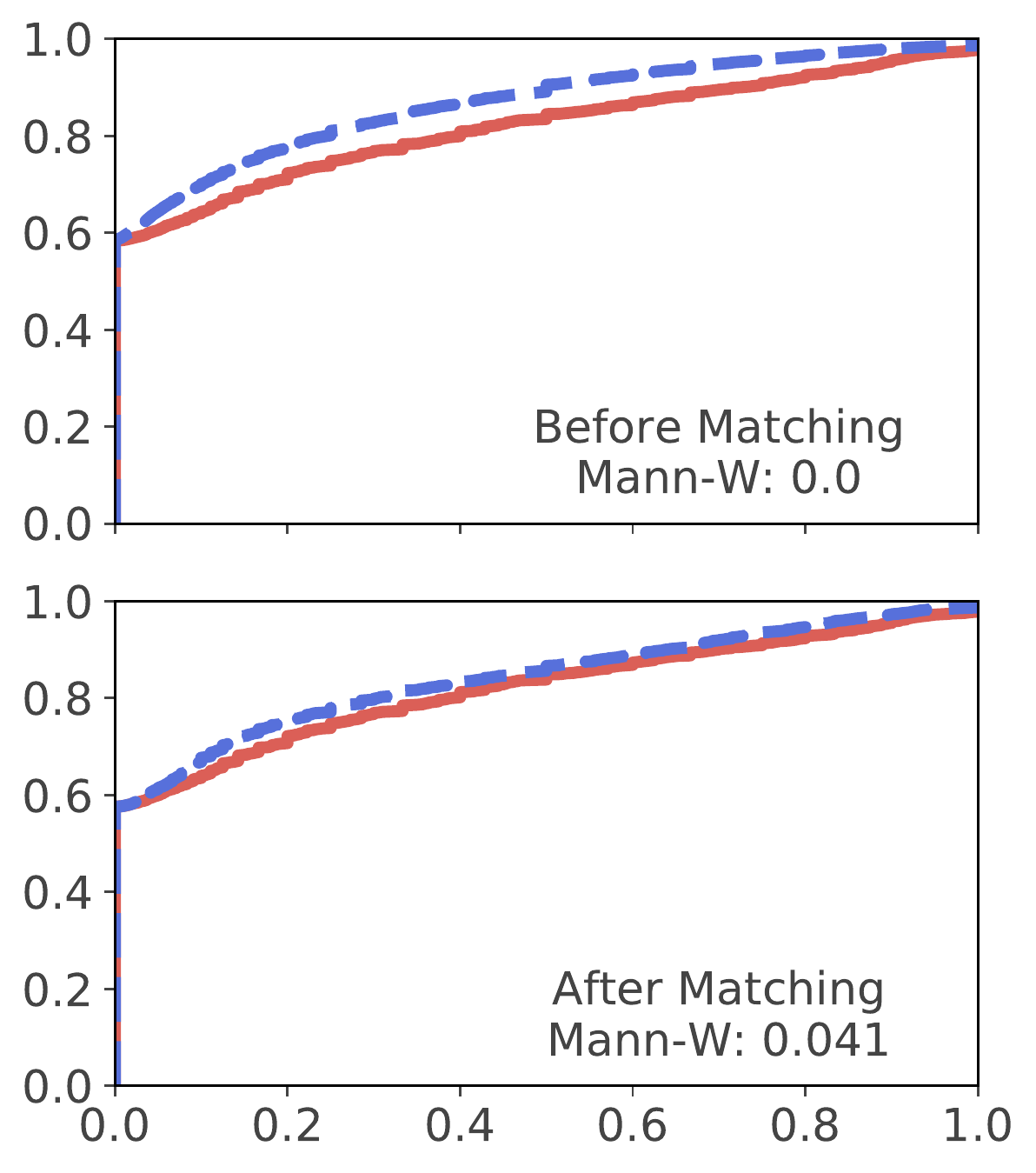}
        \caption{Matching results at level 3.}
    \end{subfigure}
    \hfill
    \begin{subfigure}[t]{\textwidth}
        \includegraphics[width=0.19\textwidth]
        {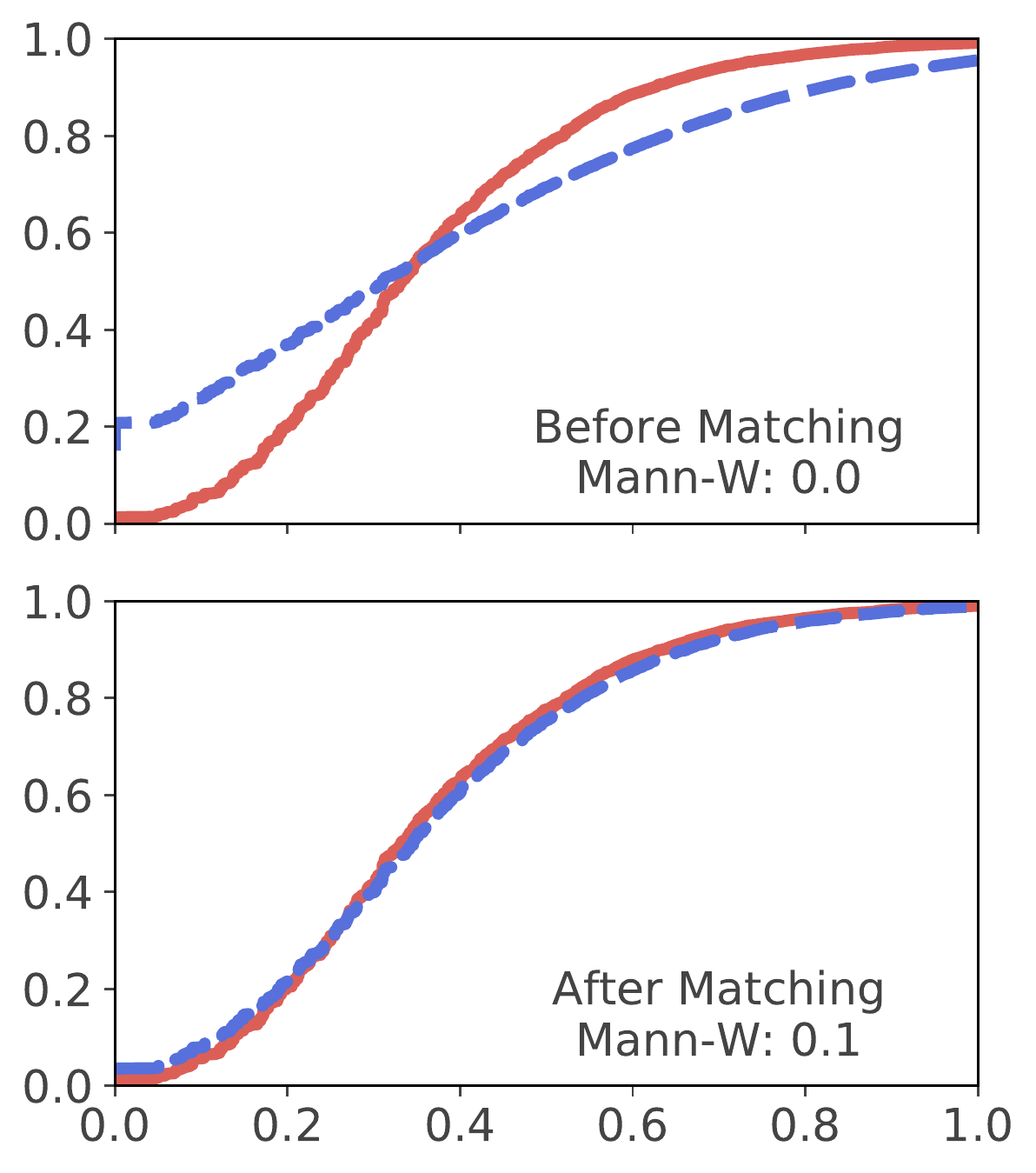}
        \includegraphics[width=0.19\textwidth]
        {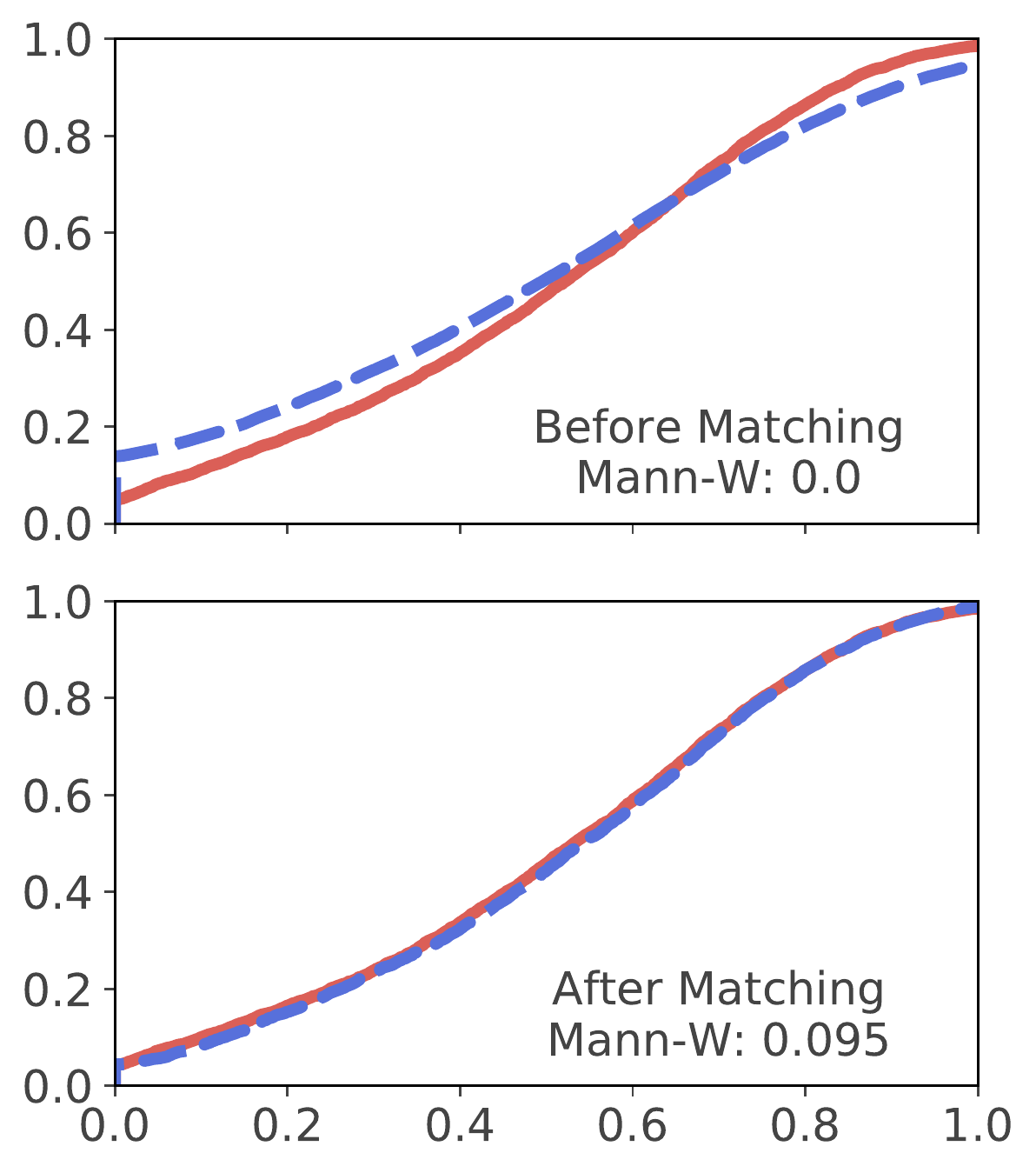}
        \includegraphics[width=0.19\textwidth]
        {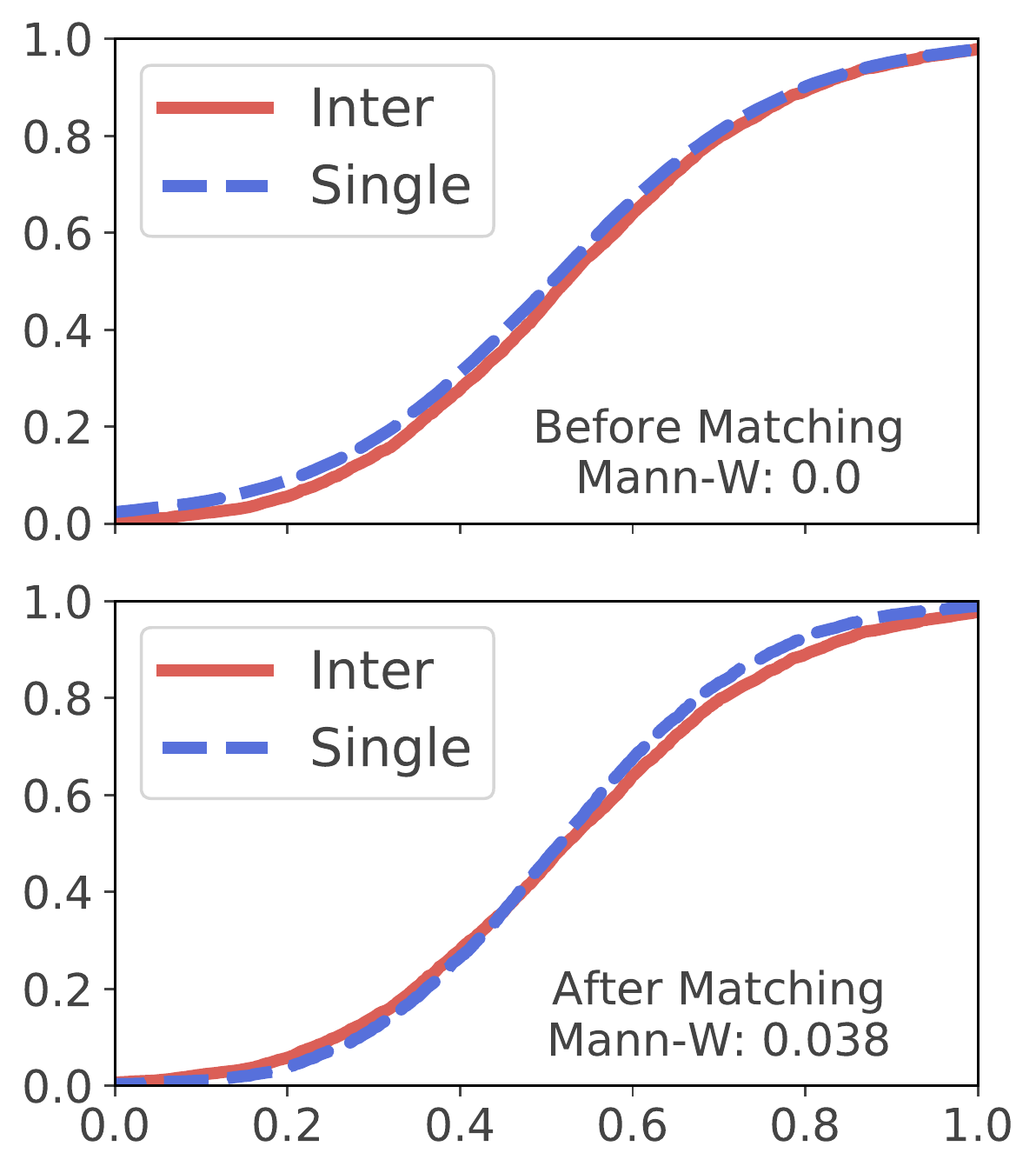}
        \includegraphics[width=0.19\textwidth]
        {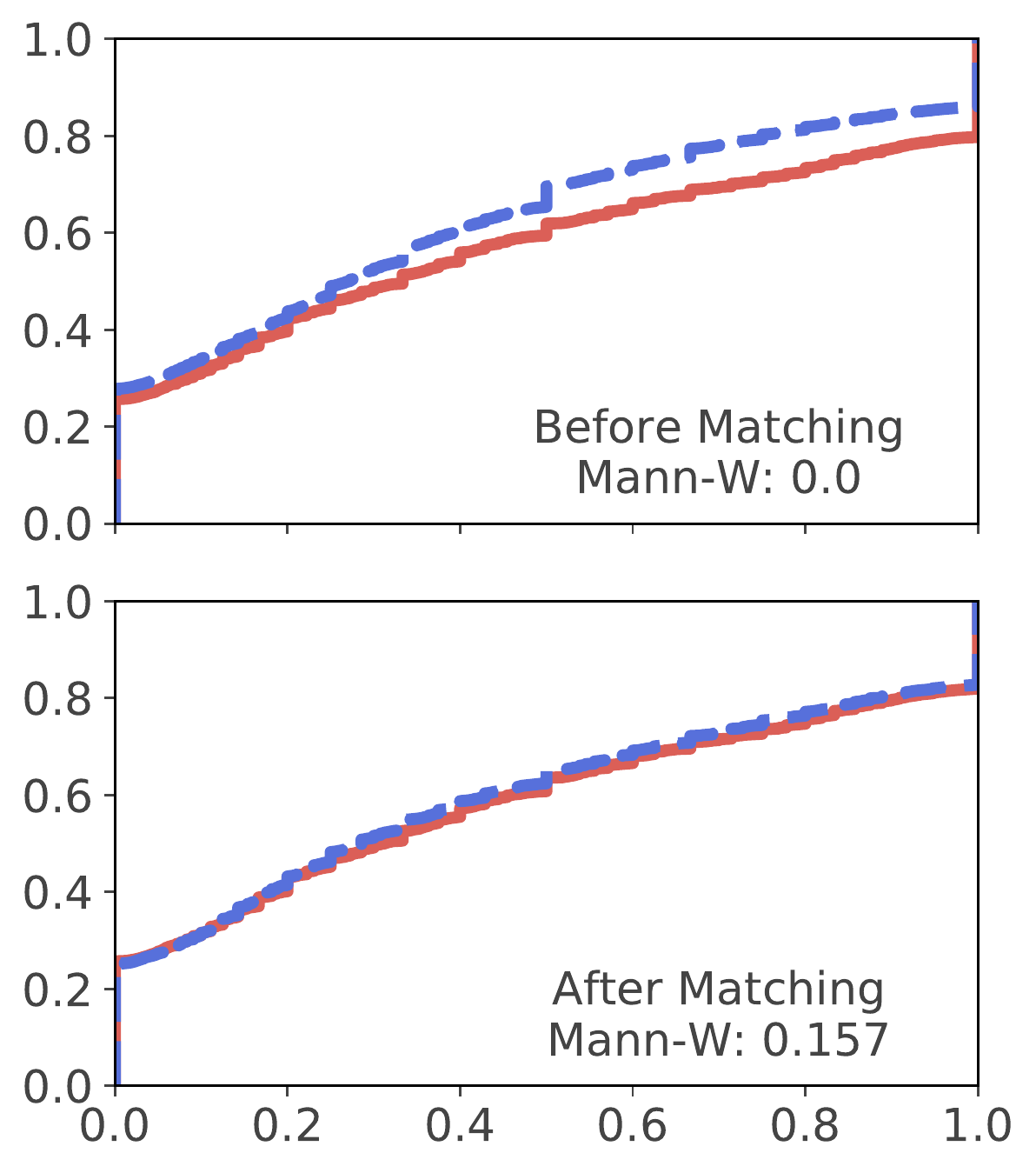}
        \includegraphics[width=0.19\textwidth]
        {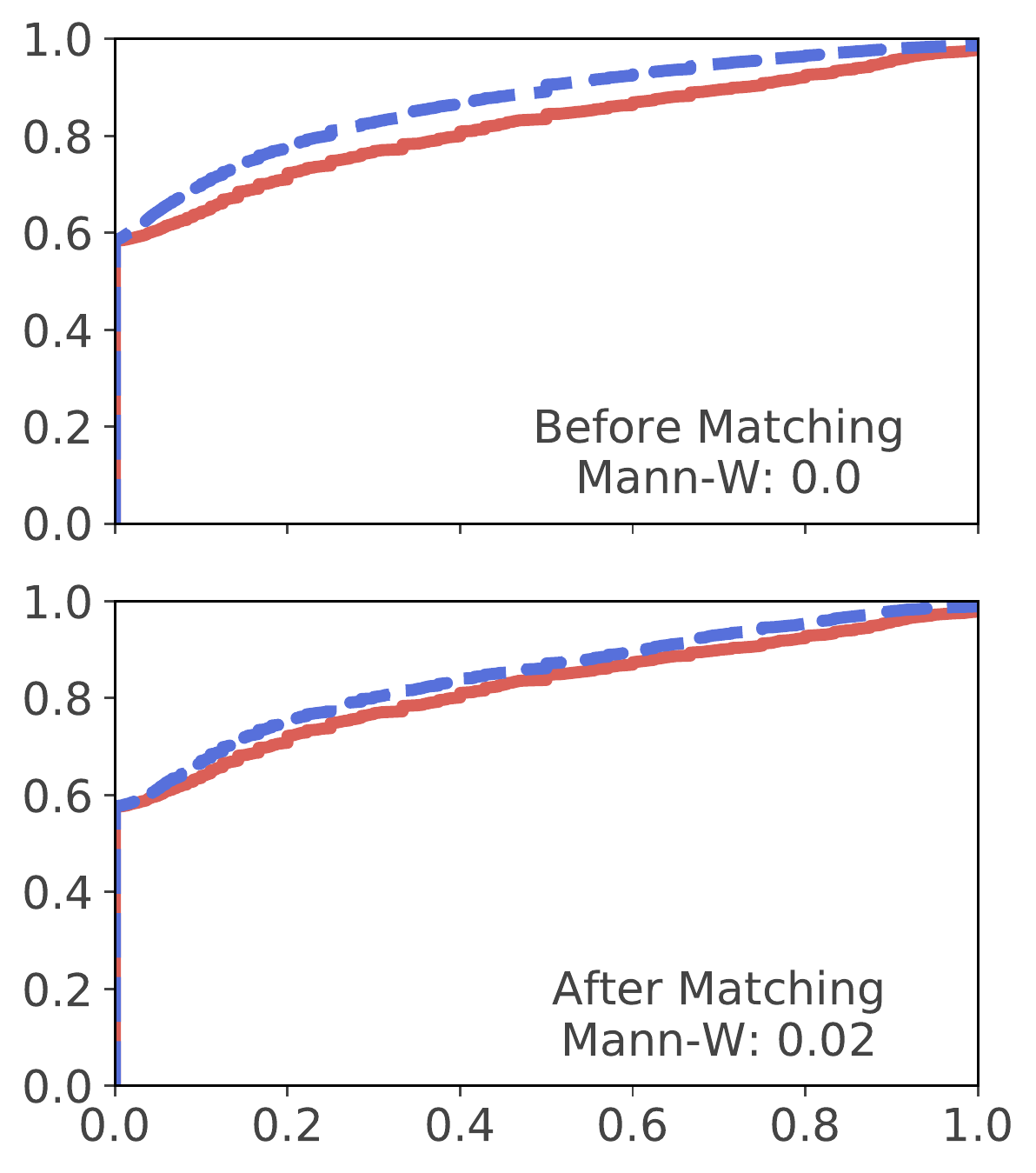}
        \caption{Matching results at level 4.}
    \end{subfigure}
    \hfill
    \begin{subfigure}[t]{\textwidth}
        \includegraphics[width=0.19\textwidth]
        {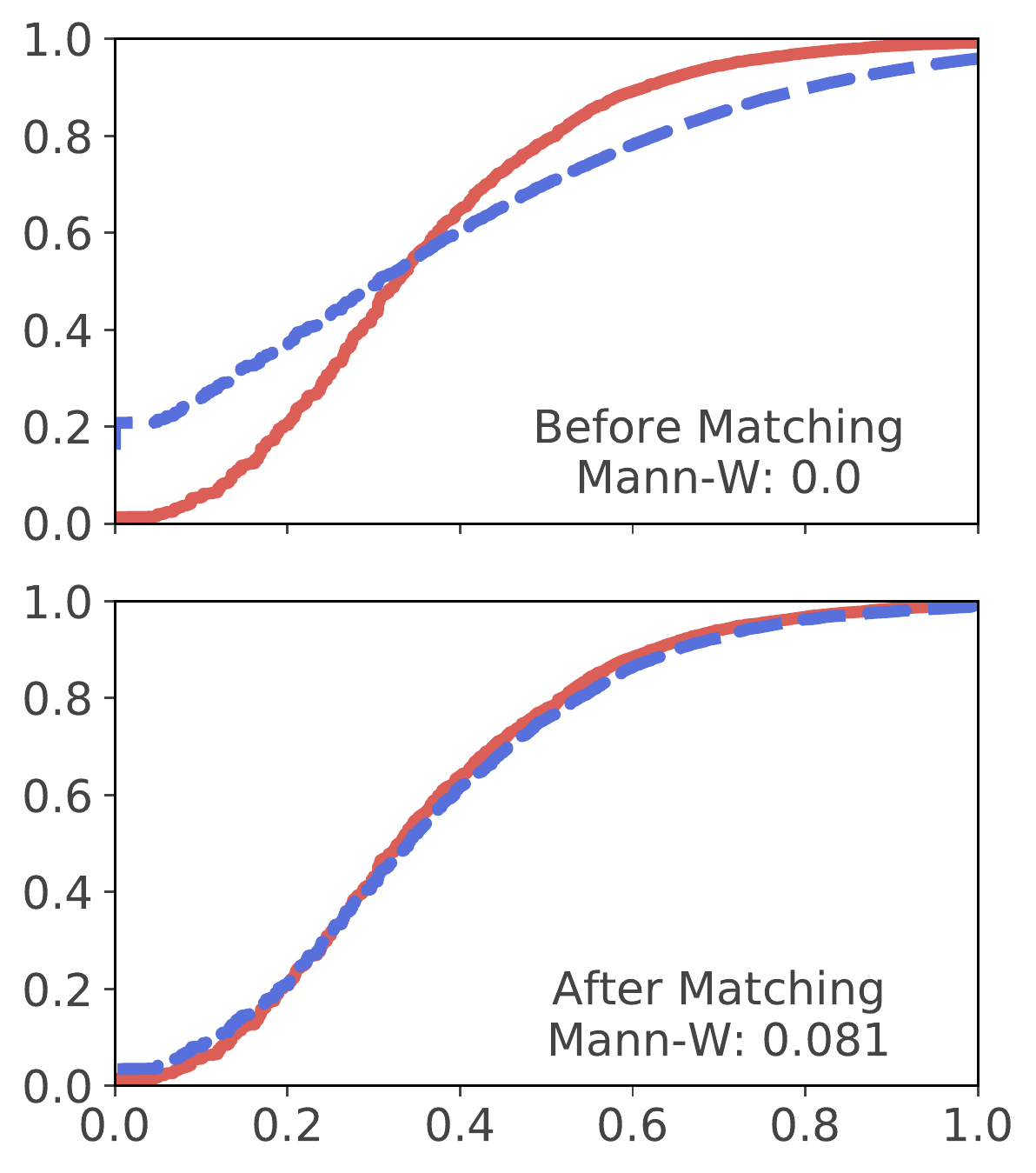}
        \includegraphics[width=0.19\textwidth]
        {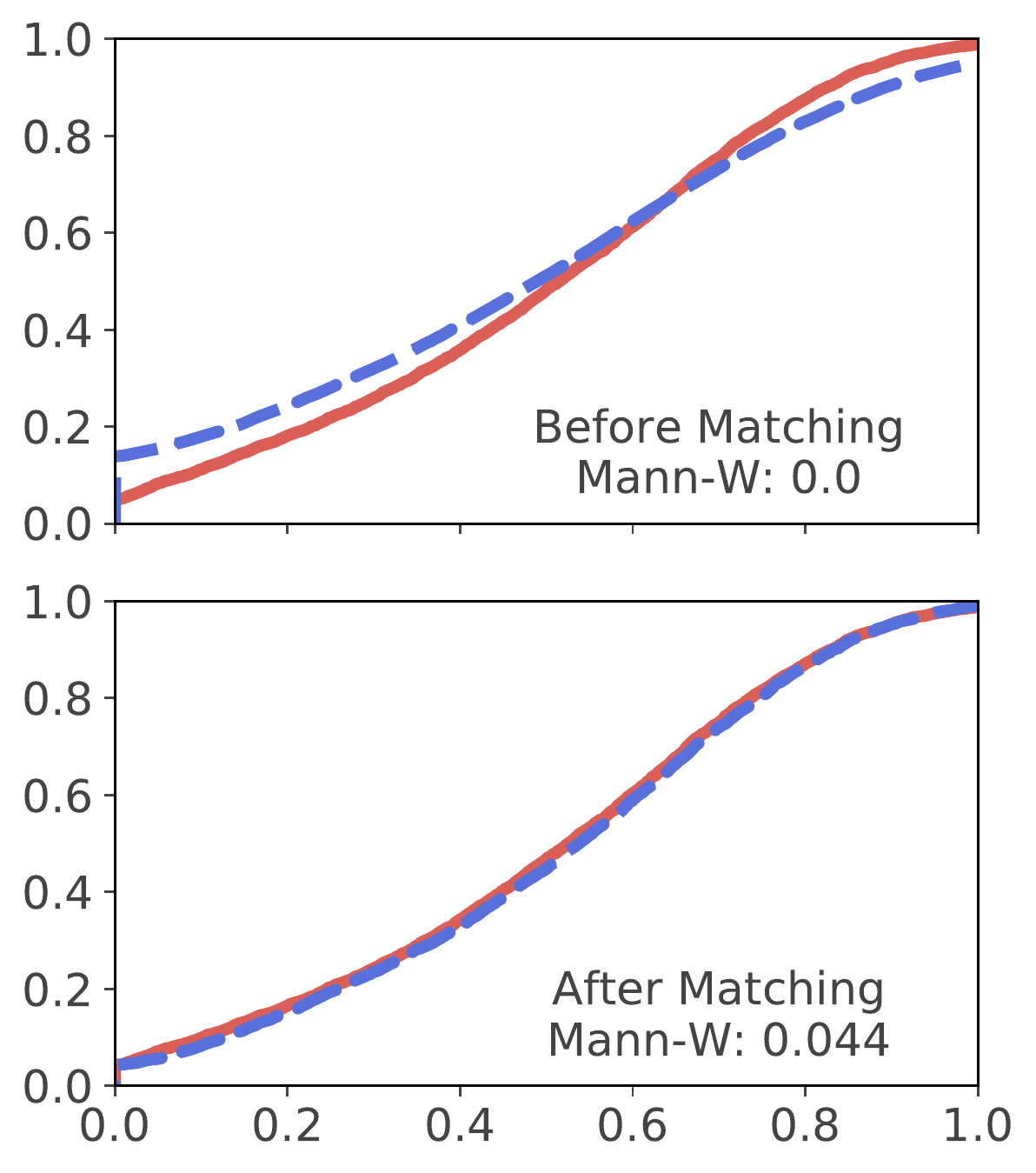}
        \includegraphics[width=0.19\textwidth]
        {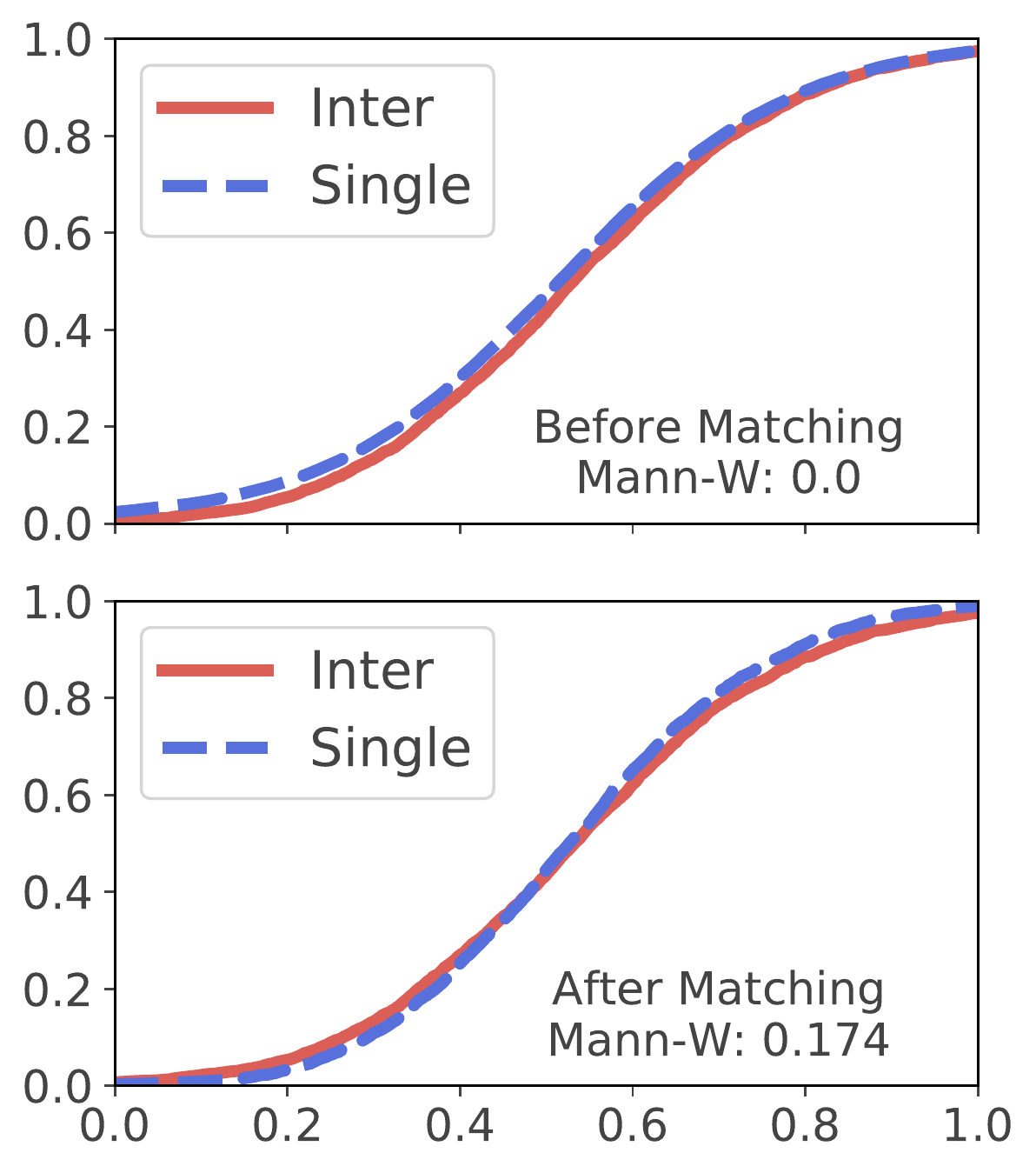}
        \includegraphics[width=0.19\textwidth]
        {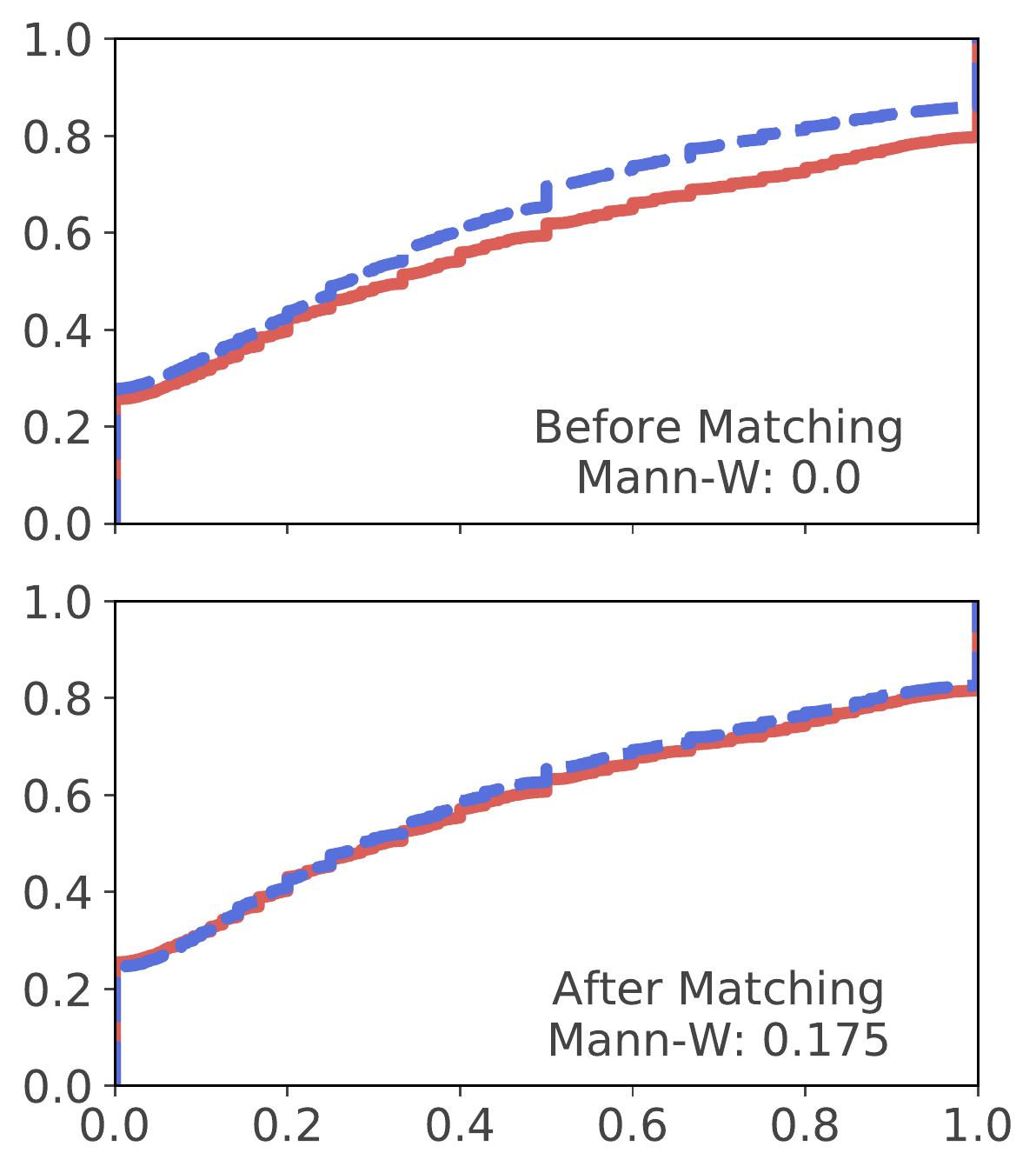}
        \includegraphics[width=0.19\textwidth]
        {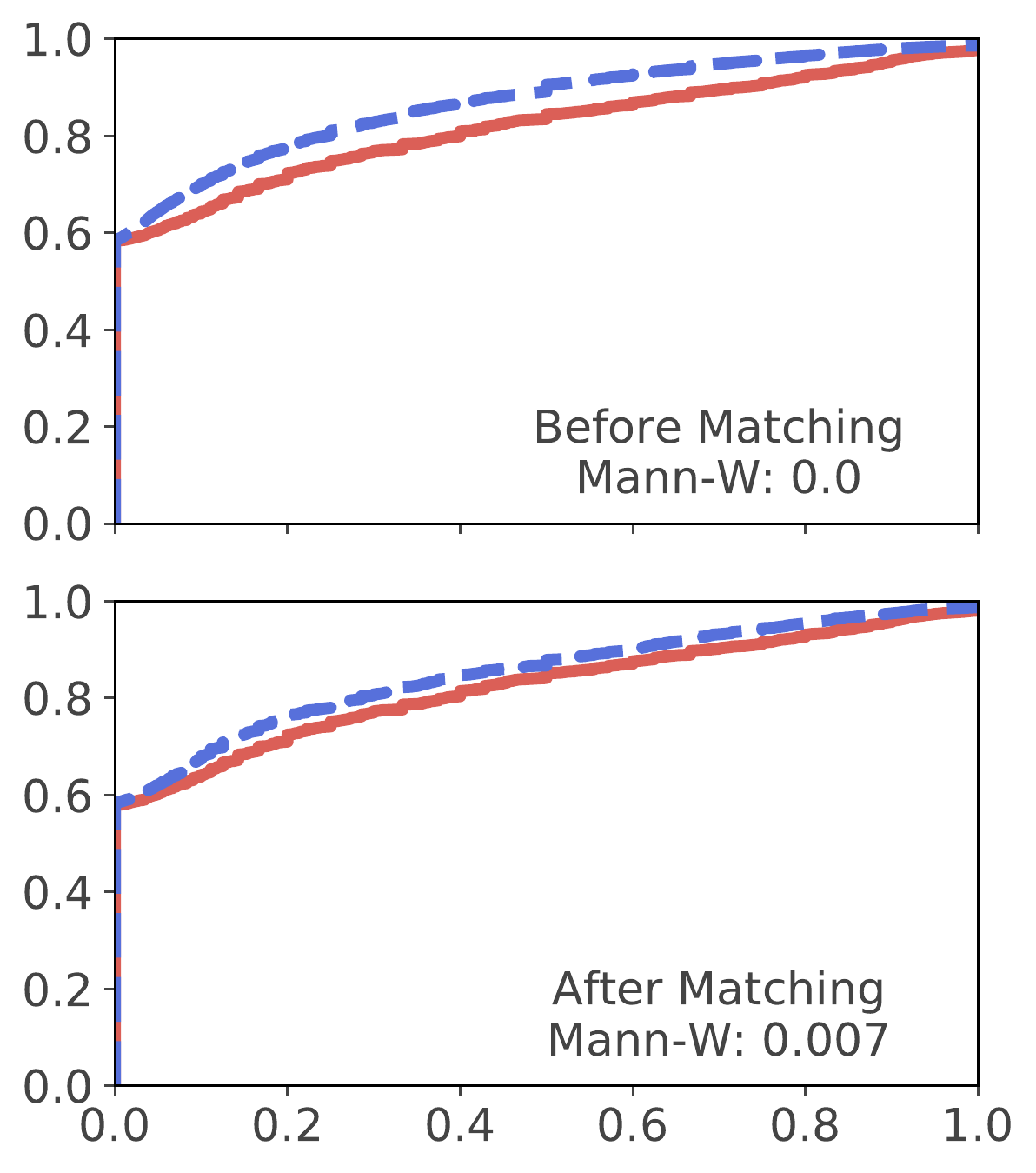}
        \caption{Matching results at level 5.}
    \end{subfigure}
    \caption{Empirical cumulative distribution of each activity feature before and after the matching technique from level 1 to level 5 in the 2016 season.
    The activity features from left to right are the number of comments,
    the average hour gap between comments, the average comment length, 
    the proportion of playoff comments, and the proportion of game thread comments.}
    \label{fig:matchinglevel2016}
\end{figure}

\begin{figure}
    \centering
    \begin{subfigure}[t]{0.32\textwidth}
        \includegraphics[width=\textwidth]
        {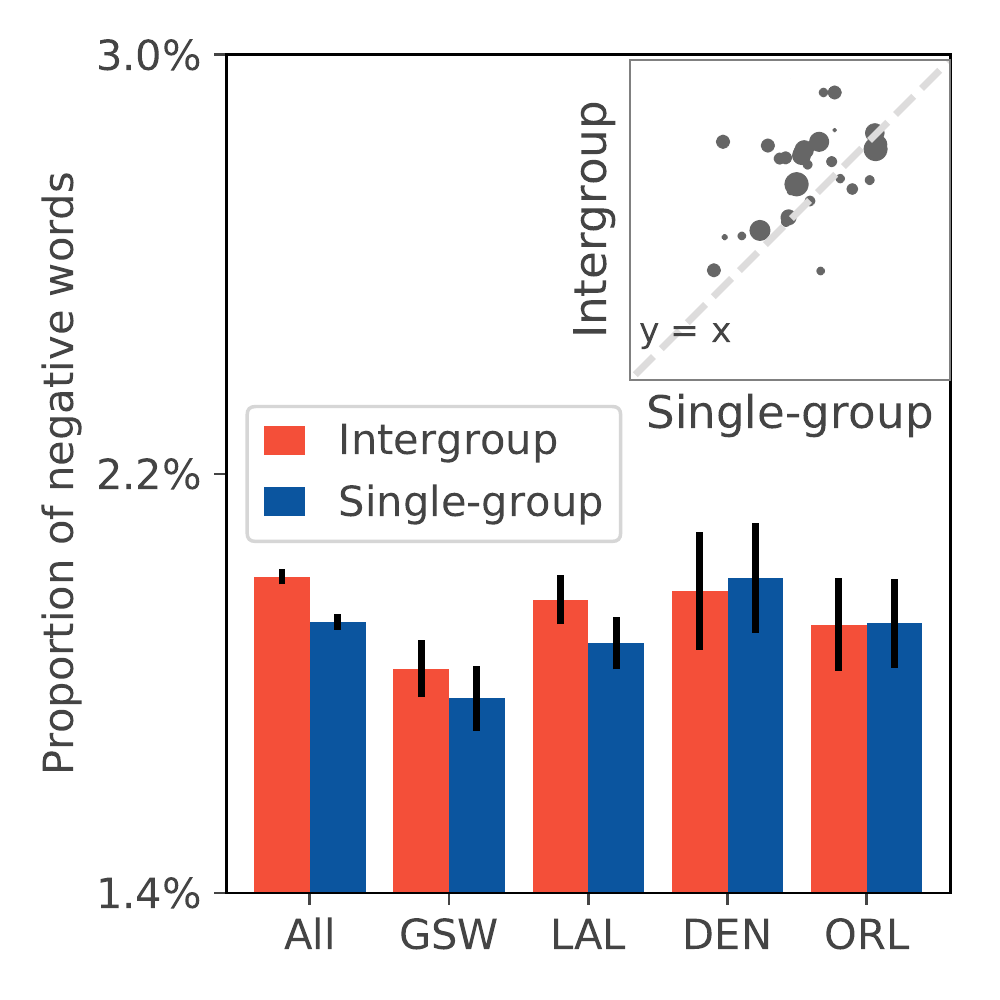}
        \caption{Proportion of negative words.}
        \label{fig:effectneg2017}
    \end{subfigure}
    \hfill
    \begin{subfigure}[t]{0.32\textwidth}
        \includegraphics[width=\textwidth]
        {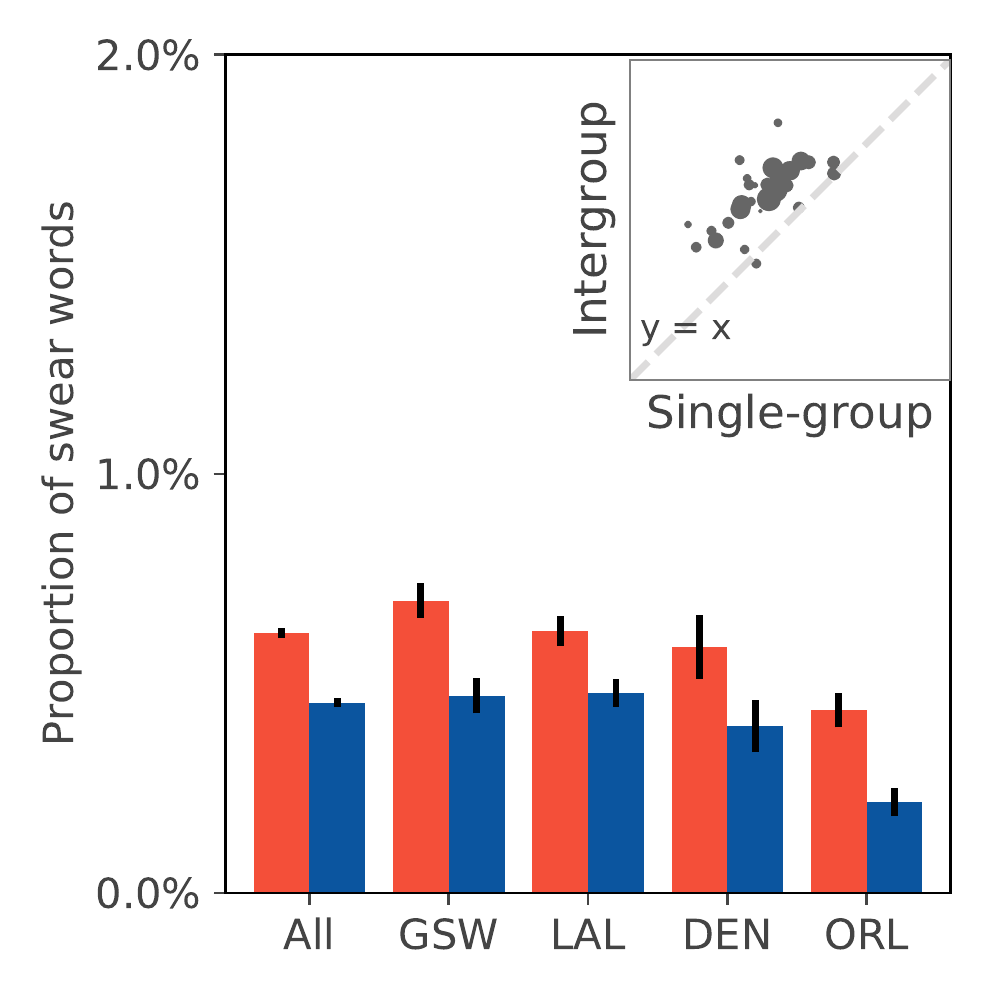}
        \caption{Proportion of swear words.}
        \label{fig:effectswear2017}
    \end{subfigure}
    \hfill
    \begin{subfigure}[t]{0.32\textwidth}
        \includegraphics[width=\textwidth]
        {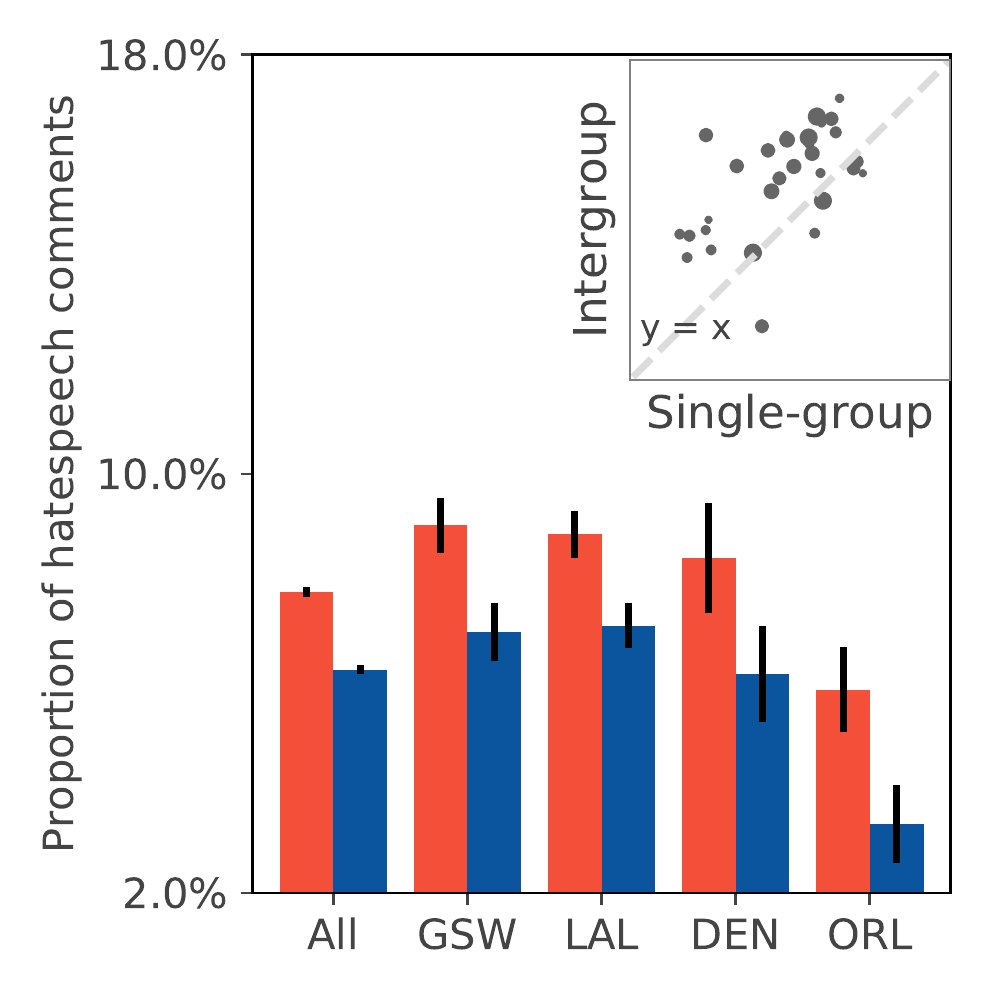}
        \caption{Proportion of hate speech comments.}
        \label{fig:effecthate2017}
    \end{subfigure}
    \caption{Comparison of language usage between \intergroup and \teamonly members in the 2017 season.
    \Intergroupusers use more negative words 
    (two-tailed t-test, $t=4.04$, $p<0.001$,  
    95\% CI=0.04\% to 0.13\%; 24 out of 30 teams, two-tailed binomial test $p=0.001$) 
    and swear words 
    (two-tailed t-test, $t=4.17$, $p<0.001$, 
    95\% CI=0.03\% to 0.10\%; 29 out of 30 teams, two-tailed binomial test $p<0.001$) 
    and generate more hate speech comments 
    (two-tailed t-test, $t=11.01$, $p<0.001$, 95\% CI=1.21\% to 1.74\%; 
    24 out of 30 teams, two-tailed binomial test $p=0.001$).
    Error bars represent standard errors.}
    \label{fig:intrasentiment2017}
\end{figure}

\begin{figure}
    \centering
    \begin{subfigure}[t]{0.32\textwidth}
        \includegraphics[width=\textwidth]
        {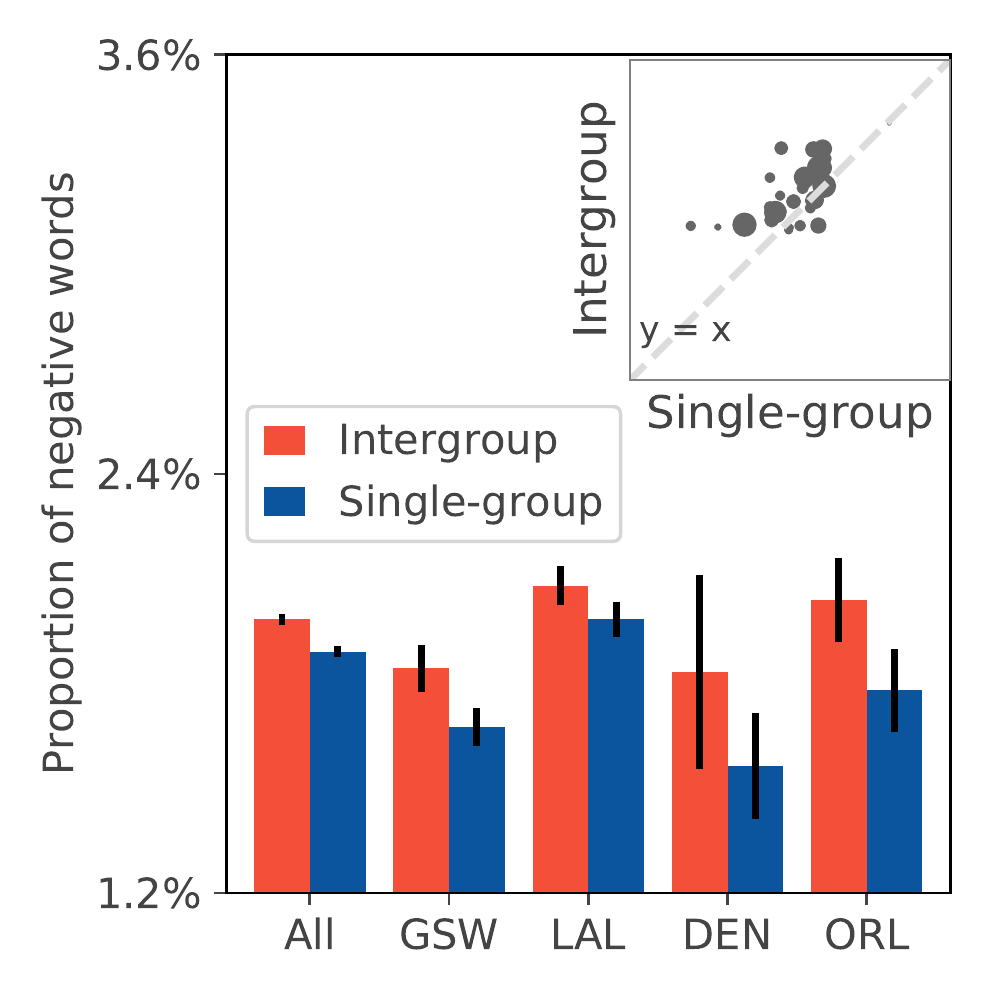}
        \caption{Proportion of negative words.}
        \label{fig:effectneg2016}
    \end{subfigure}
    \hfill
    \begin{subfigure}[t]{0.32\textwidth}
        \includegraphics[width=\textwidth]
        {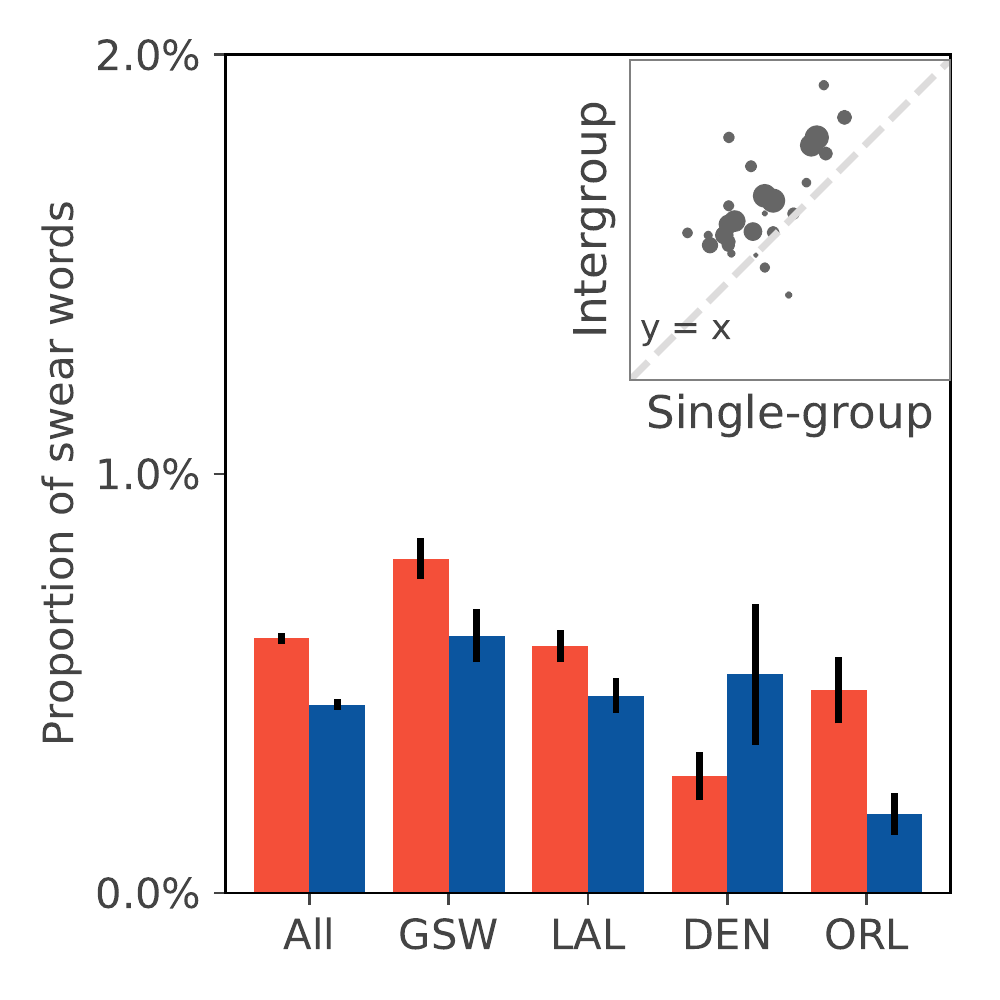}
        \caption{Proportion of swear words.}
        \label{fig:effectswear2016}
    \end{subfigure}
    \hfill
    \begin{subfigure}[t]{0.32\textwidth}
        \includegraphics[width=\textwidth]
        {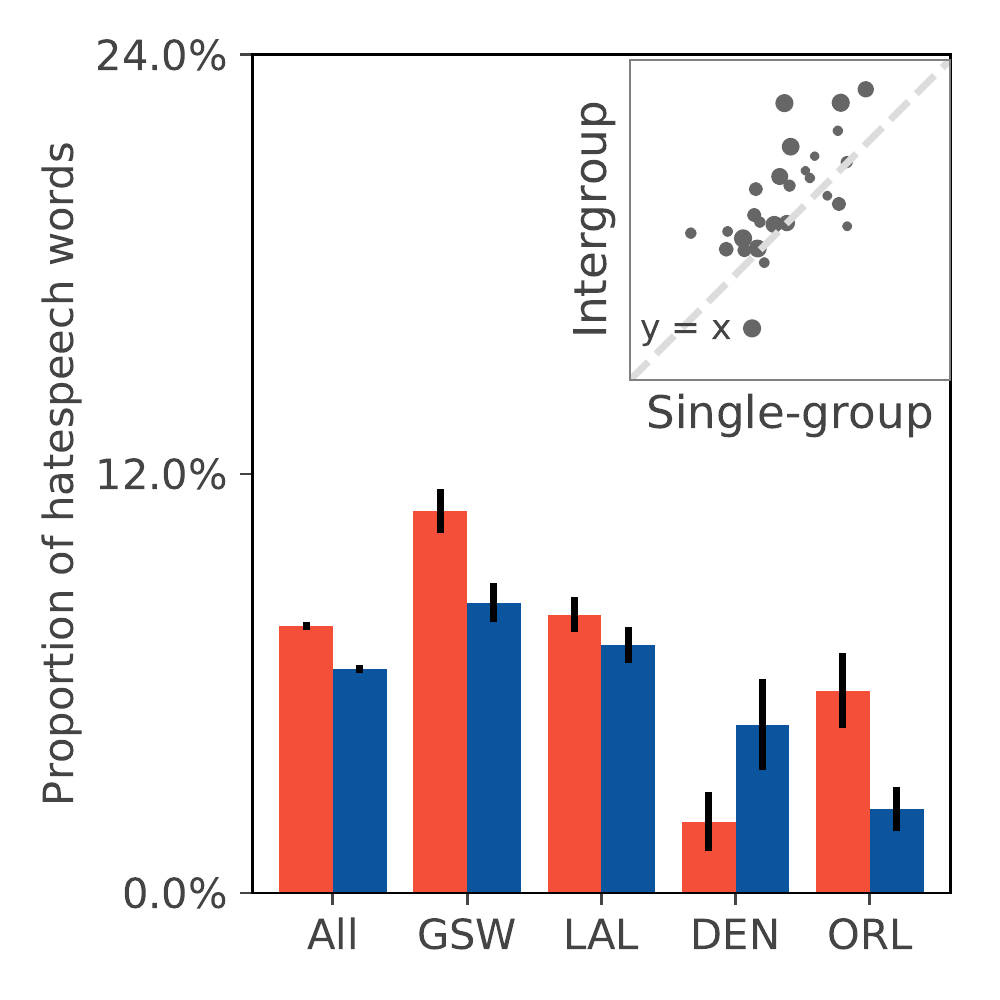}
        \caption{Proportion of hate speech comments.}
        \label{fig:effecthate2016}
    \end{subfigure}
    \caption{Comparison of language usage between \intergroup 
    and \teamonly members in the 2016 season.
    \Intergroupusers use more negative words 
    (two-tailed t-test, $t=3.93$, $p<0.001$,  
    95\% CI=0.05\% to 0.14\%; 23 out of 30 teams, two-tailed binomial test $p=0.005$) 
    and swear words (two-tailed t-test, $t=3.10$, $p=0.002$, 95\% CI=0.02\% to 0.09\%; 
    28 out of 30 teams, two-tailed binomial test $p<0.001$) 
    and generate more hate speech comments (two-tailed t-test, $t=7.95$, $p<0.001$, 
    95\% CI=0.92\% to 1.53\%; 25 out of 30 teams, two-tailed binomial test $p<0.001$).
    Error bars represent standard errors.}
    \label{fig:intrasentiment2016}
\end{figure}

\begin{figure}
    \centering
    \begin{subfigure}[t]{0.32\textwidth}
        \includegraphics[width=\textwidth]
        {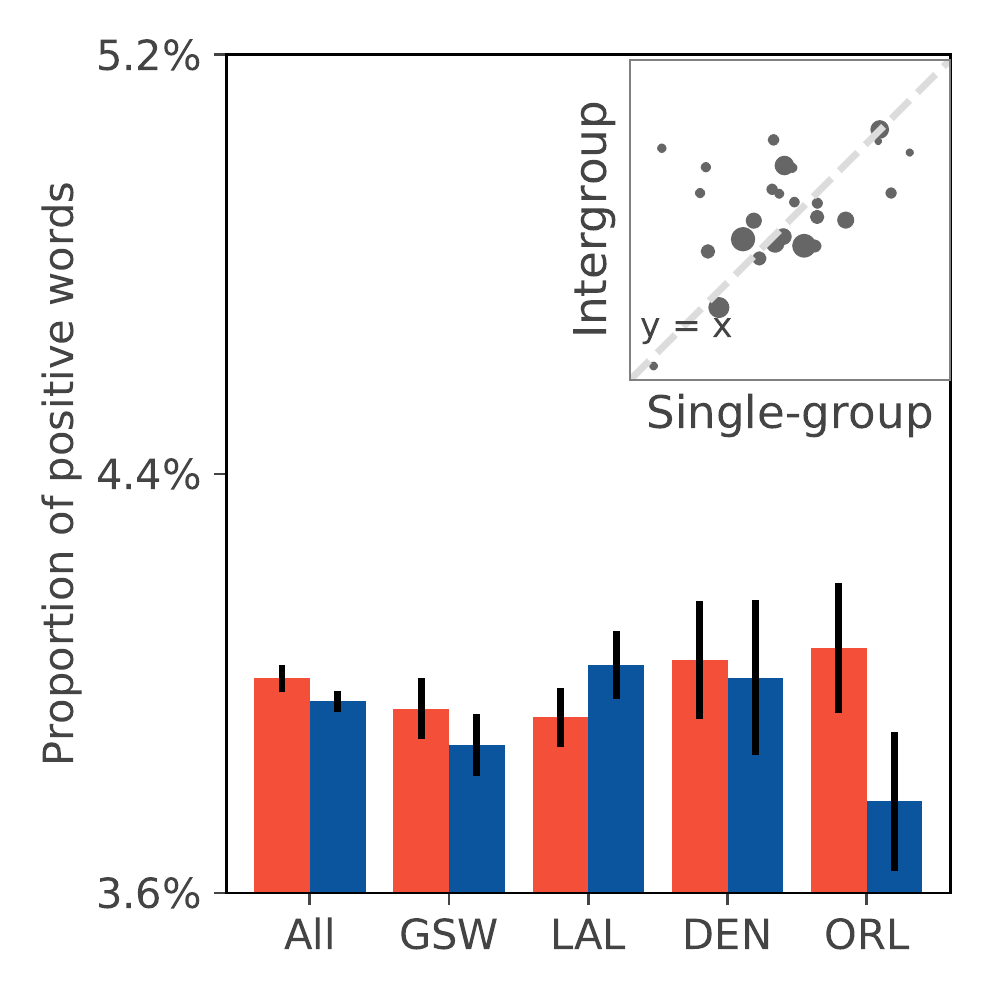}
        \caption{The 2018 season.}
        \label{fig:pos2018}
    \end{subfigure}
    \hfill
    \begin{subfigure}[t]{0.32\textwidth}
        \includegraphics[width=\textwidth]
        {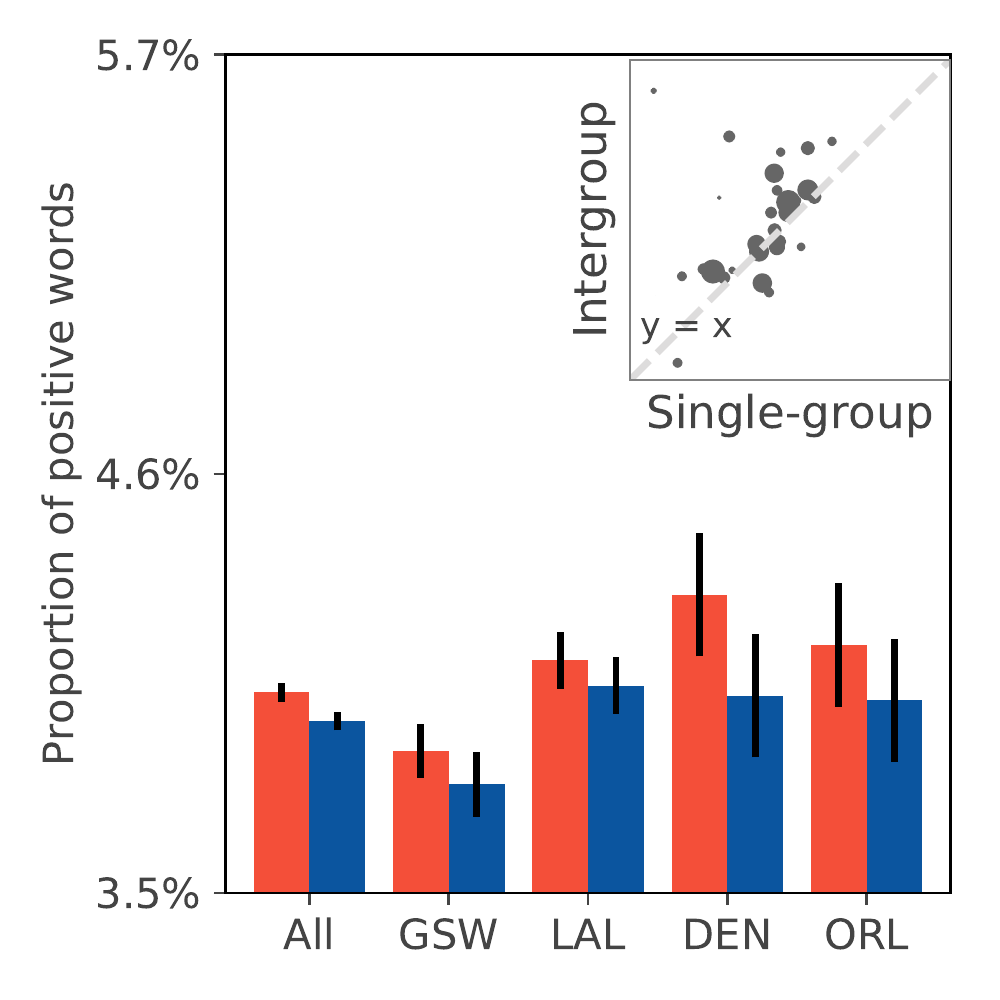}
        \caption{The 2017 season.}
        \label{fig:pos2017}
    \end{subfigure}
    \hfill
    \begin{subfigure}[t]{0.32\textwidth}
        \includegraphics[width=\textwidth]
        {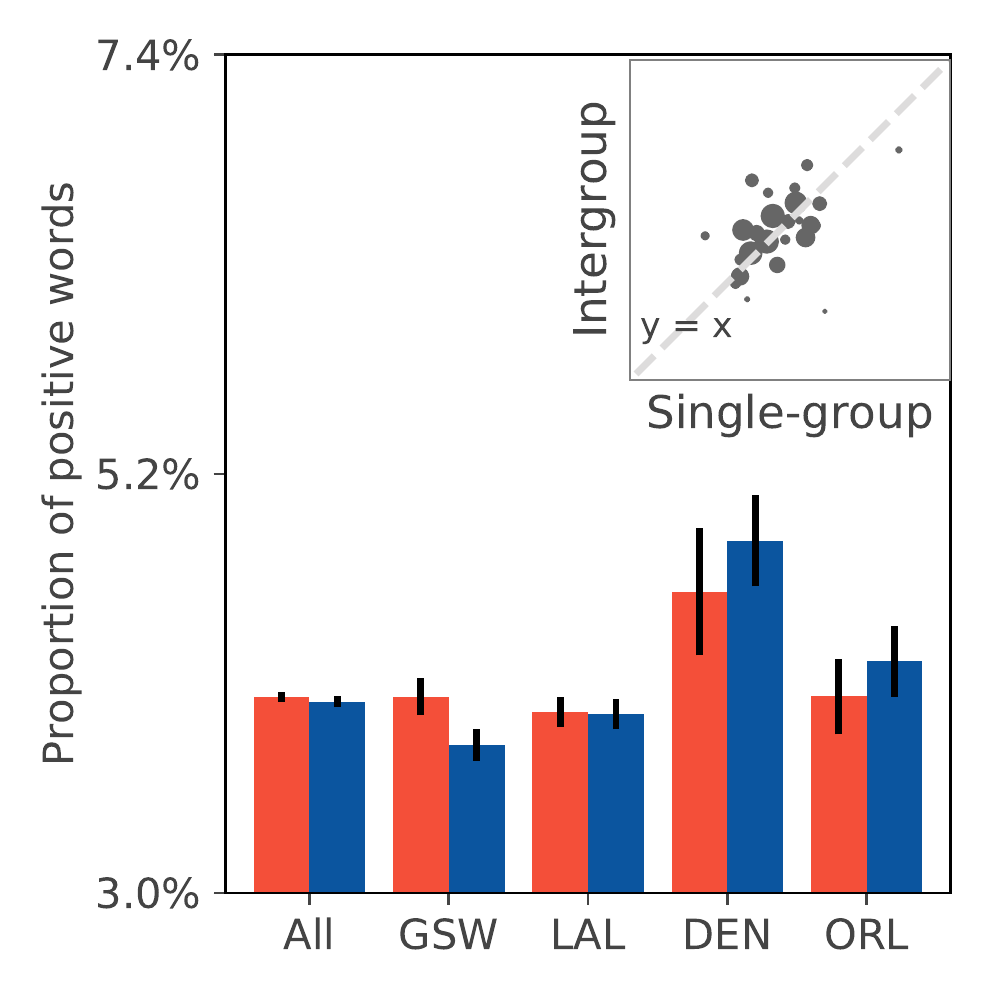}
        \caption{The 2016 season.}
        \label{fig:pos2016}
    \end{subfigure}
    \caption{The comparison of positive language usage 
    between \intergroup and \teamonlyusers in the 2018, 2017, and 2016 seasons.
    We find no consistent trend at the 5\% significance level ($\alpha$ = 0.05)
    (two-tailed t-test, $t=1.36$, $p=0.174$,
    95\% CI=0.019\% to 0.107\%, 16 out of 30 teams, 
    two-tailed binomial test $p=0.856$ for the 2018 season;
    two-tailed t-test, $t=2.15$, $p=0.03$,
    95\% CI=0.000\% to 0.144\%, 16 out of 30 teams, 
    two-tailed binomial test $p=0.856$ for the 2017 season;
    two-tailed t-test, $t=0.66$, $p=0.508$,
    95\% CI=-0.050\% to 0.102\%, 14 out of 30 teams, 
    two-tailed binomial test $p=0.856$ for the 2016 season).
    Error bars represent standard errors.}
    \label{fig:positive}
\end{figure}

\begin{figure}
    \center
    \begin{subfigure}[t]{0.32\textwidth}
        \includegraphics[width=\textwidth]
        {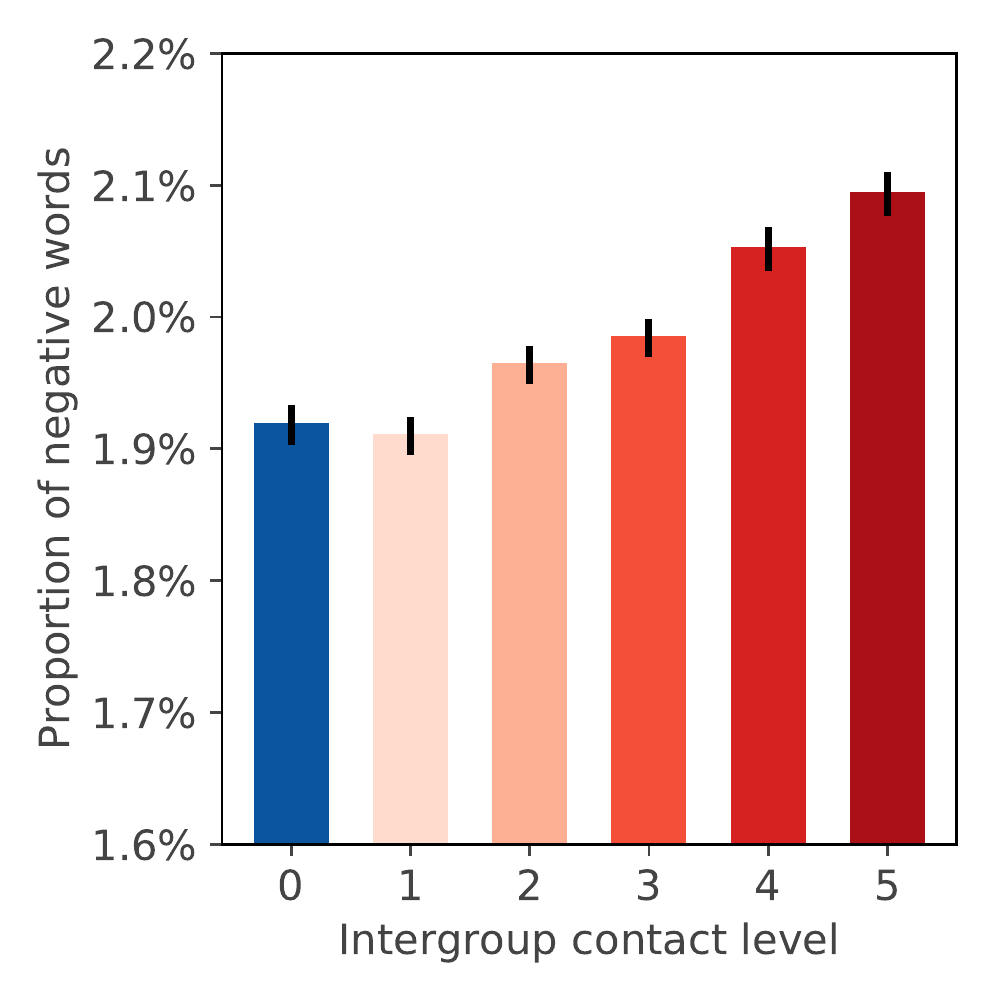}
        \caption{Proportion of negative words.}
        \label{fig:levelneg2017}
    \end{subfigure}
    \hfill
    \begin{subfigure}[t]{0.32\textwidth}
        \includegraphics[width=\textwidth]
        {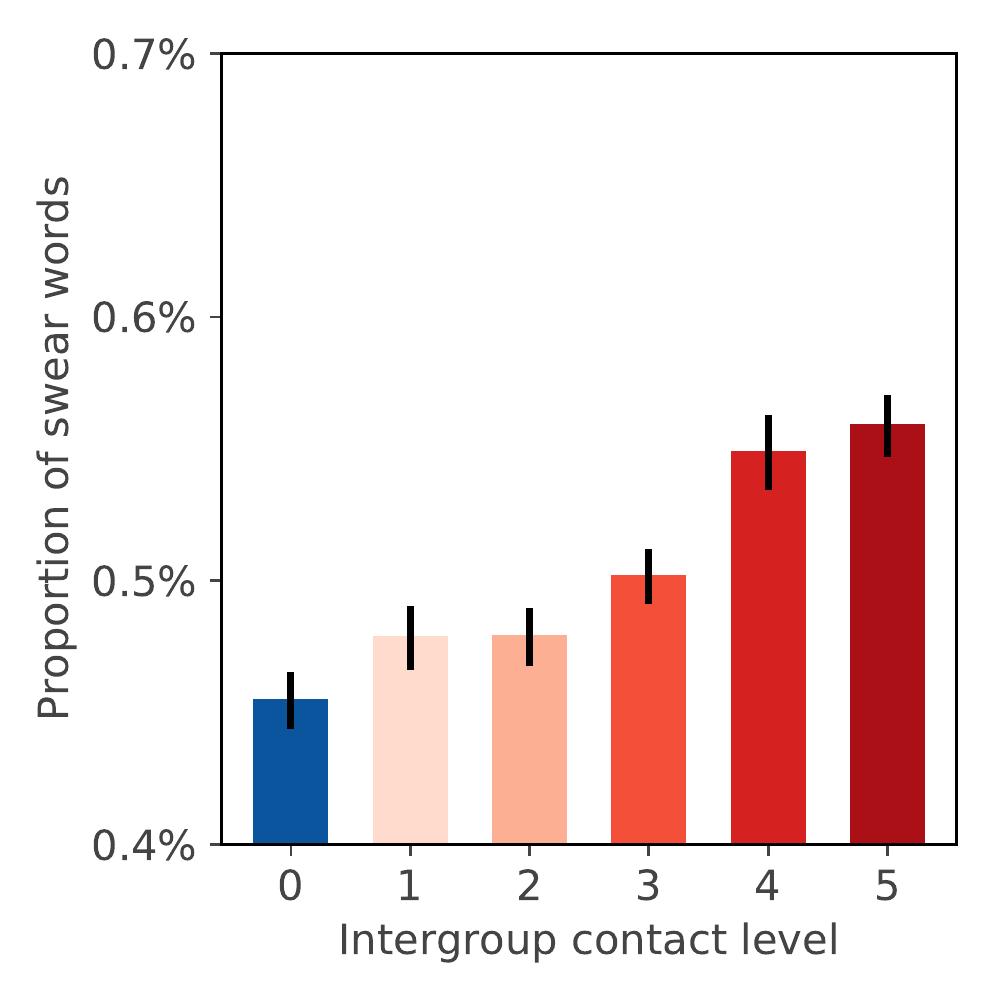}
        \caption{Proportion of swear words.}
        \label{fig:levelswear2017}
    \end{subfigure}
    \hfill
    \begin{subfigure}[t]{0.32\textwidth}
        \includegraphics[width=\textwidth]
        {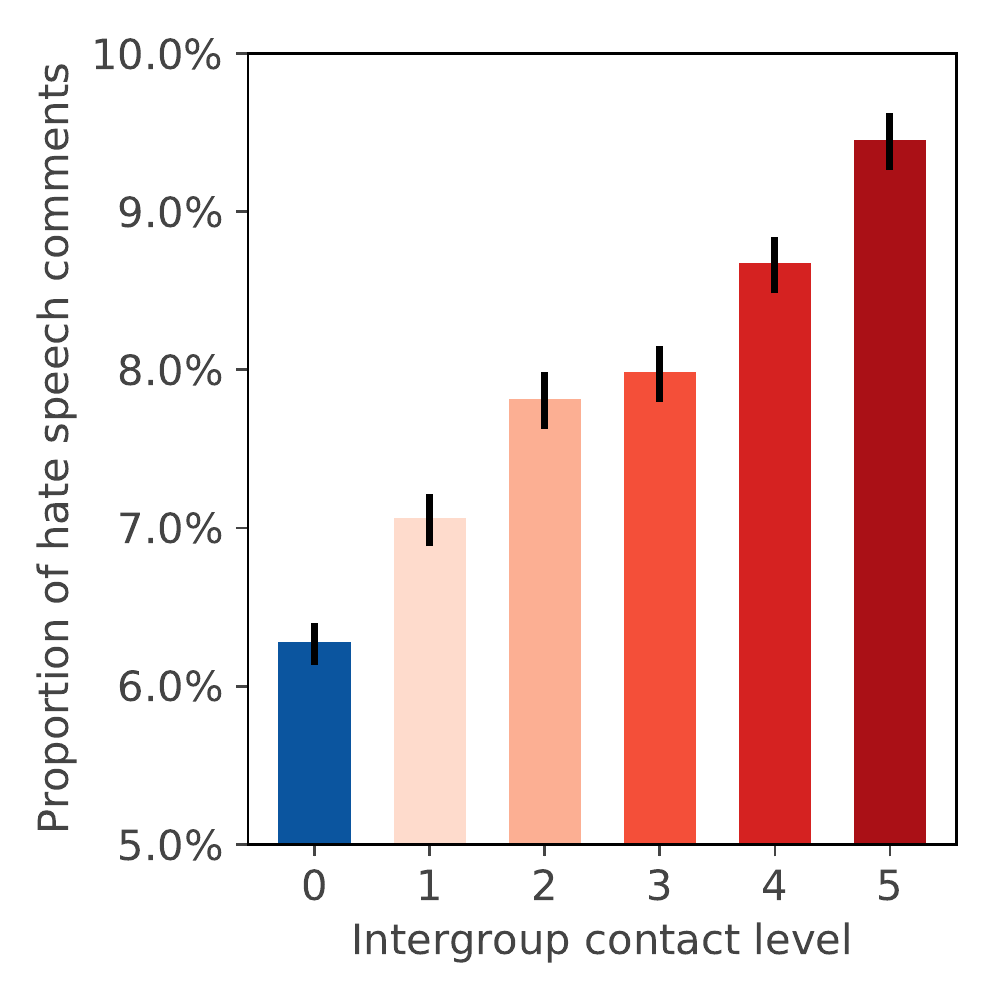}
        \caption{Proportion of hate speech comments.}
        \label{fig:levelhate2017}
    \end{subfigure}
    \caption{Intragroup language usage differences of members with different
    intergroup contact levels in the 2017 season.
    x-axis represents intergroup levels determined by the fraction of 
    comments in \communityname{/r/NBA}. 
    We observe a consistent monotonic pattern in the proportion of 
    negative words (mean = 1.92\%, 1.91\%, 1.96\%,
    1.98\%, 2.05\%, and 2.09\%, respectively for labels from 0 to 6), 
    swear words (mean = 0.45\%, 0.48\%, 0.48\%, 0.50\%, 0.55\%, and 
    0.56\%, respectively for labels from 0 to 6), 
    and hate speech comments
    (mean = 6.27\%, 7.05\%, 7.81\%, 8.97\%, 8.67\%, and 9.44\%,  
    respectively for labels from 0 to 6).      
    }
    \label{fig:levellanguage2017}
\end{figure}

\begin{figure}
    \center
    \begin{subfigure}[t]{0.32\textwidth}
        \includegraphics[width=\textwidth]
        {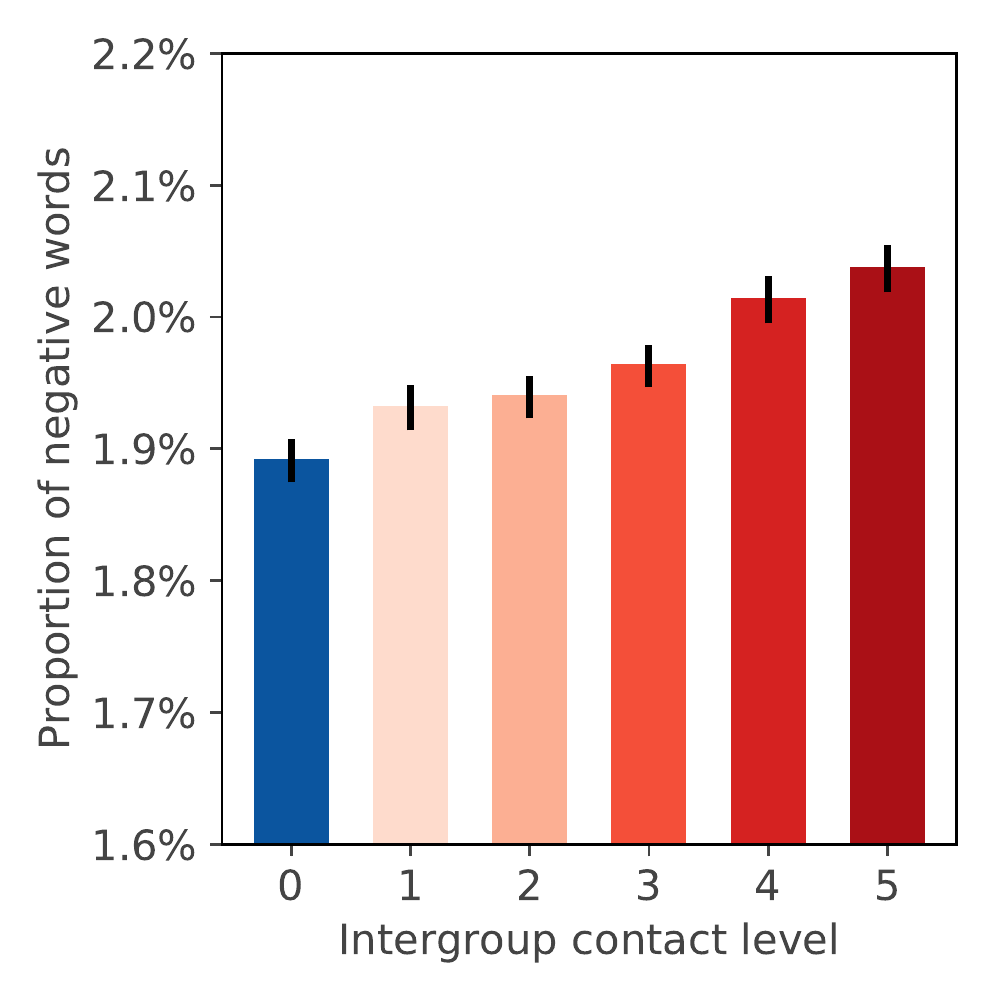}
        \caption{Proportion of negative words.}
        \label{fig:levelneg2016}
    \end{subfigure}
    \hfill
    \begin{subfigure}[t]{0.32\textwidth}
        \includegraphics[width=\textwidth]
        {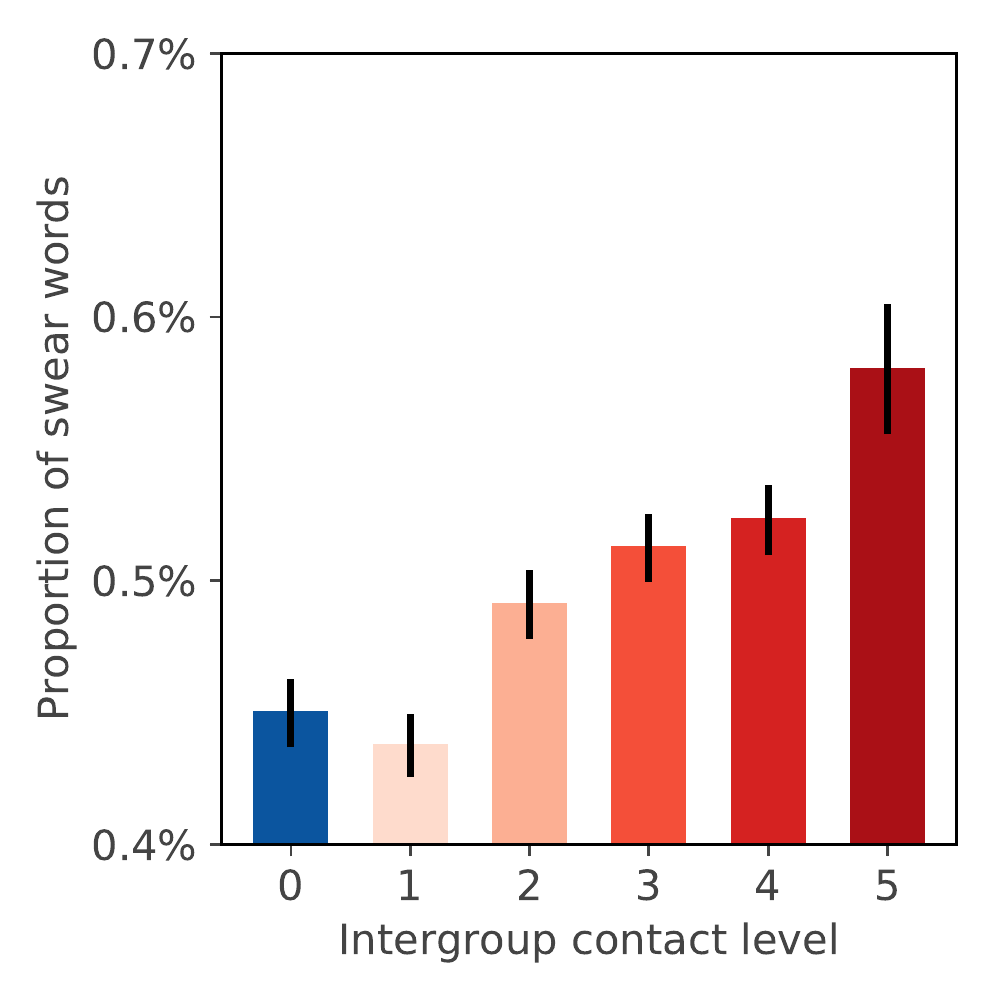}
        \caption{Proportion of swear words.}
        \label{fig:levelswear2016}
    \end{subfigure}
    \hfill
    \begin{subfigure}[t]{0.32\textwidth}
        \includegraphics[width=\textwidth]
        {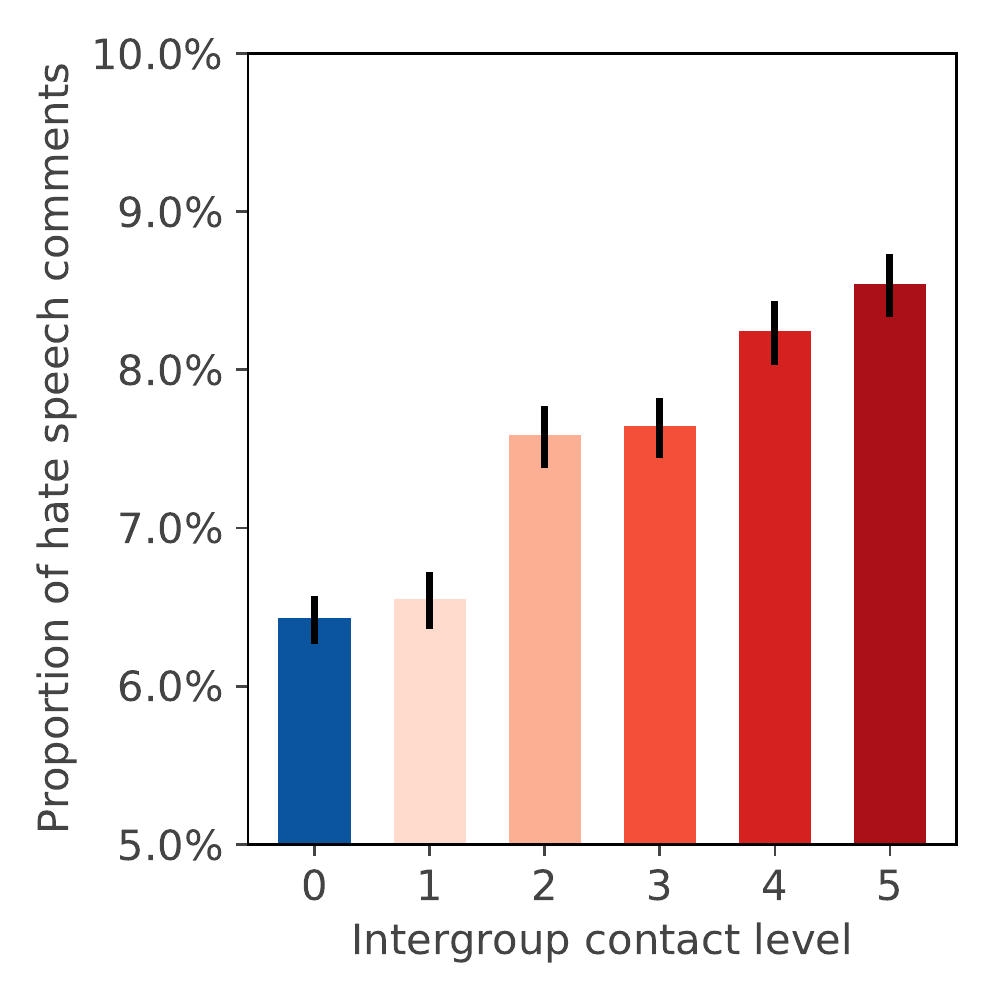}
        \caption{Proportion of hate speech comments.}
        \label{fig:levelhate2016}
    \end{subfigure}
    \caption{Intragroup language usage differences of members with different
    intergroup contact levels in the 2016 season.
    x-axis represents intergroup levels determined by the number of 
    comments in \communityname{/r/NBA}. 
    We observe a consistent monotonic pattern in the proportion of 
    negative words (mean = 1.89\%, 1.93\%, 1.94\%,
    1.96\%, 2.01\%, and 2.04\%, respectively for labels from 0 to 6), 
    swear words (mean = 0.45\%, 0.44\%, 0.49\%, 0.51\%, 0.52\%, and 
    0.58\%, respectively for labels from 0 to 6), 
    and hate speech comments
    (mean = 6.42\%, 6.54\%, 7.57\%, 7.63\%, 8.23\%, and 8.53\%,  
    respectively for labels from 0 to 6).        
    }
    \label{fig:levellanguage2016}
\end{figure}

\begin{figure}
    \center
    \begin{subfigure}[t]{0.32\textwidth}
        \includegraphics[width=\textwidth]
        {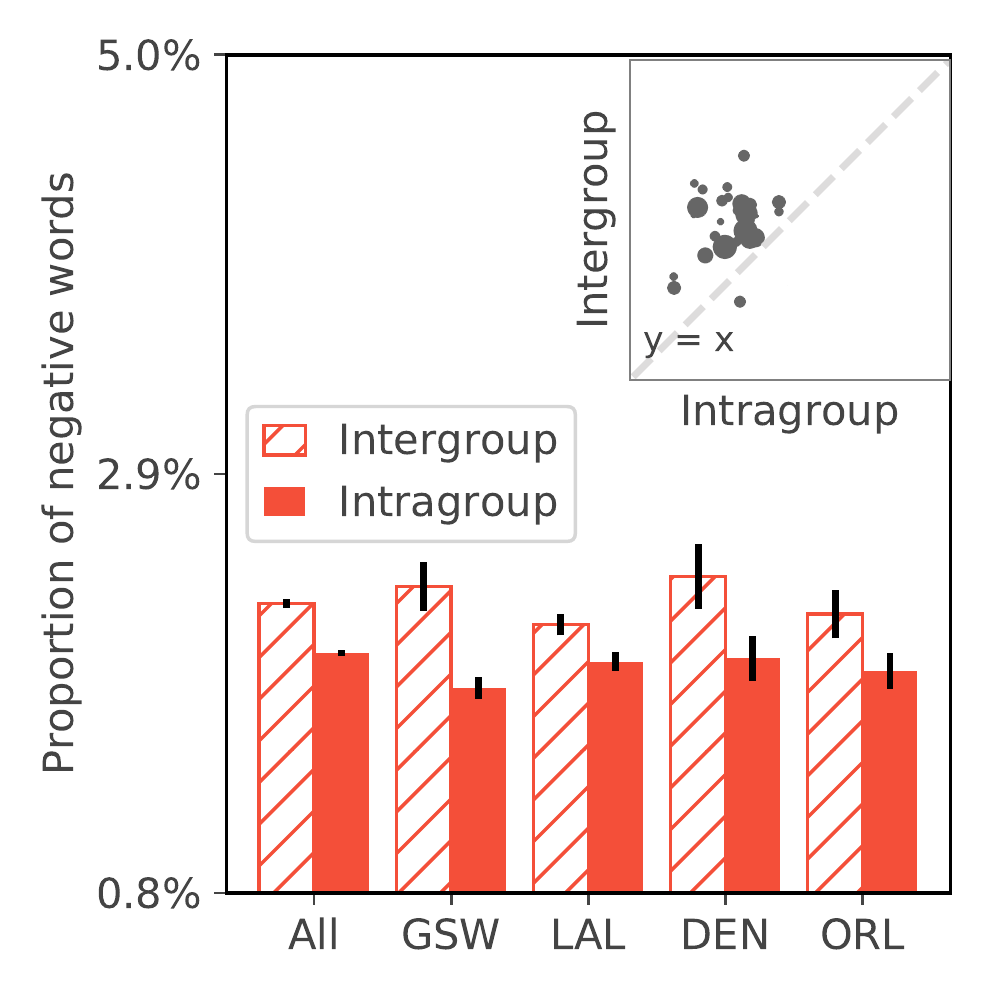}
        \caption{Proportion of negative words.}
        \label{fig:interneg2017}
    \end{subfigure}
    \hfill
    \begin{subfigure}[t]{0.32\textwidth}
        \includegraphics[width=\textwidth]
        {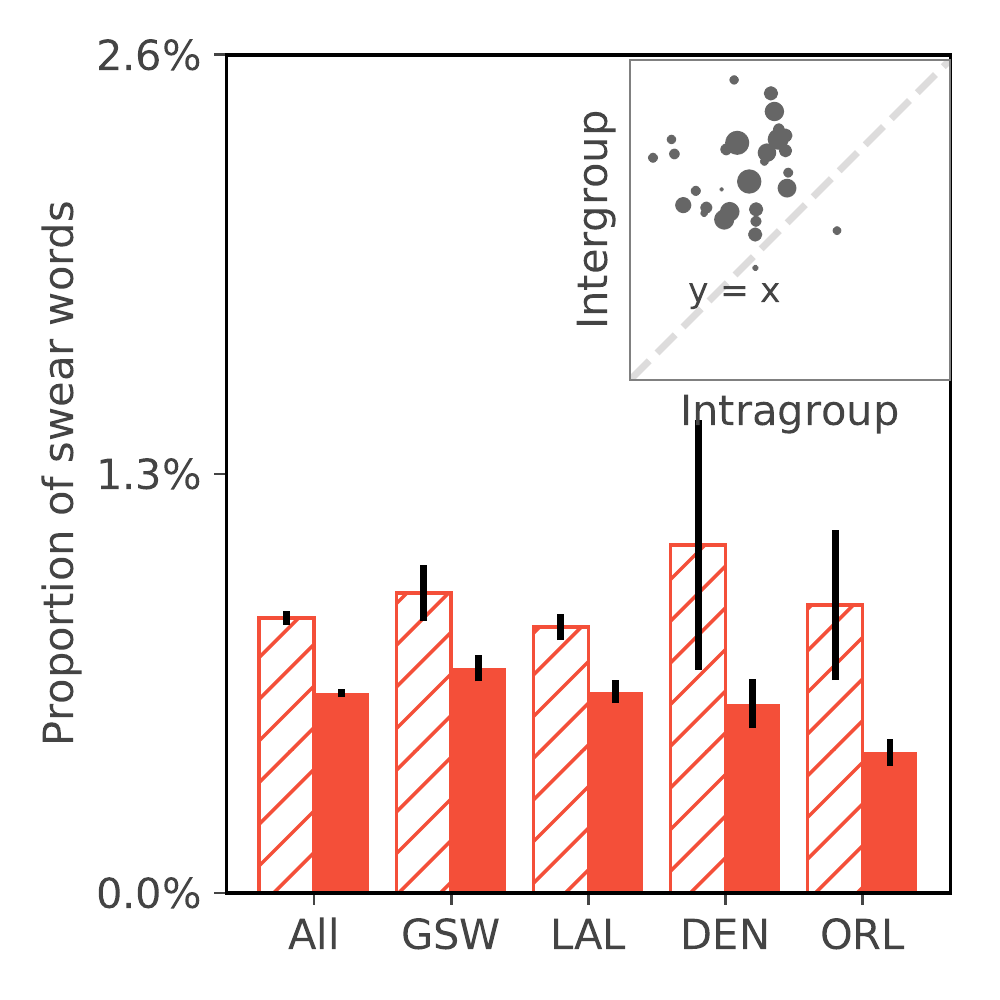}
        \caption{Proportion of swear words.}
        \label{fig:interswear2017}
    \end{subfigure}
    \hfill
    \begin{subfigure}[t]{0.32\textwidth}
        \includegraphics[width=\textwidth]
        {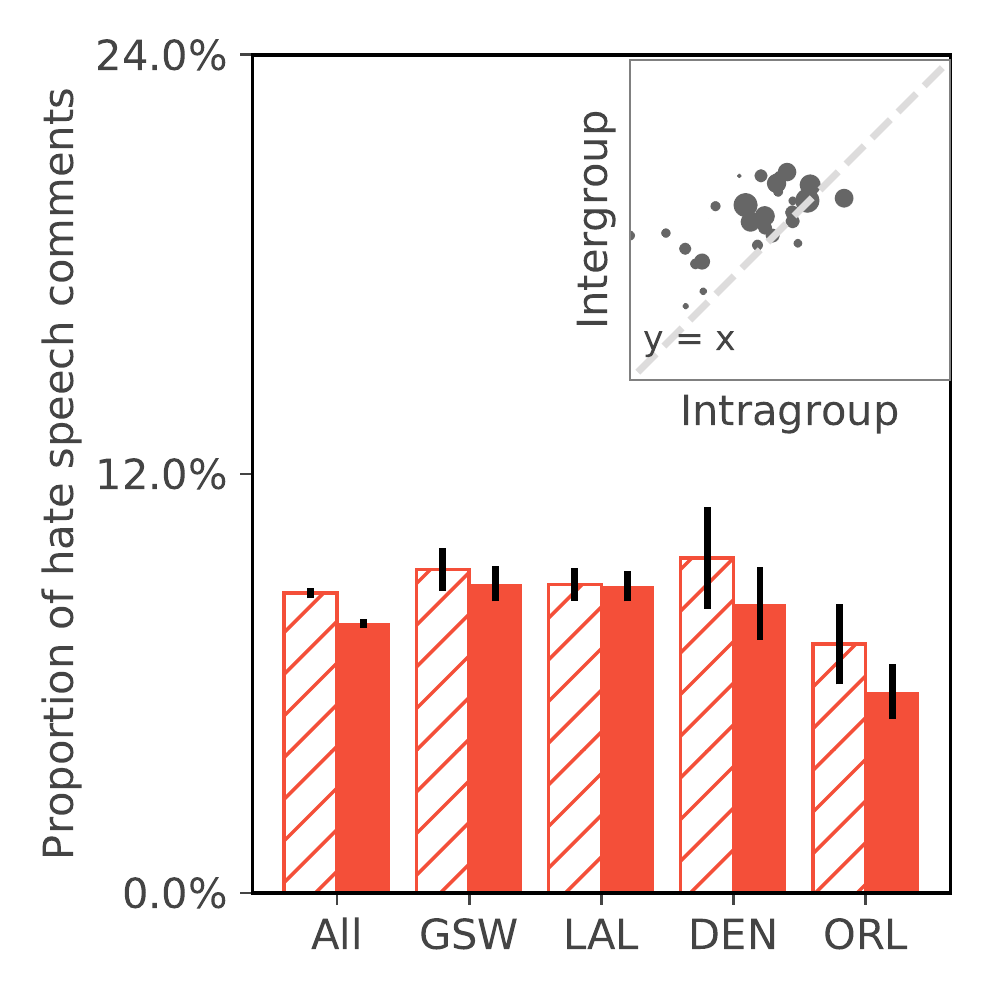}
        \caption{Proportion of hate speech comments.}
        \label{fig:interhate2017}
    \end{subfigure}
    \caption{\Intergroupusers use more negative language in the intergroup setting
    than in the intragroup setting in the 2017 season. 
    They use more negative words
    (two-tailed t-test, $t=9,36$, $p<0.001$, 95\% CI=0.19\% to 0.30\%; 
    29 out of 30 teams, two-tailed binomial test $p<0.001$)
    and swear words 
    (two-tailed t-test, $t=13.80$, $p<0.001$, 95\% CI=0.29\% to 0.38\%; 
    28 out of 30 teams, two-tailed binomial test $p<0.001$)
    and generate more hate speech comments
    (two-tailed t-test, $4.43$, $p<0.001$, 
    95\% CI=0.48\% to 1.25\%; 
    26 out of 30 teams, two-tailed binomial test $p=0.005$).
    Error bars represent standard errors.
    }
    \label{fig:intersentiment2017}
\end{figure}

\begin{figure}
    \center
    \begin{subfigure}[t]{0.32\textwidth}
        \includegraphics[width=\textwidth]
        {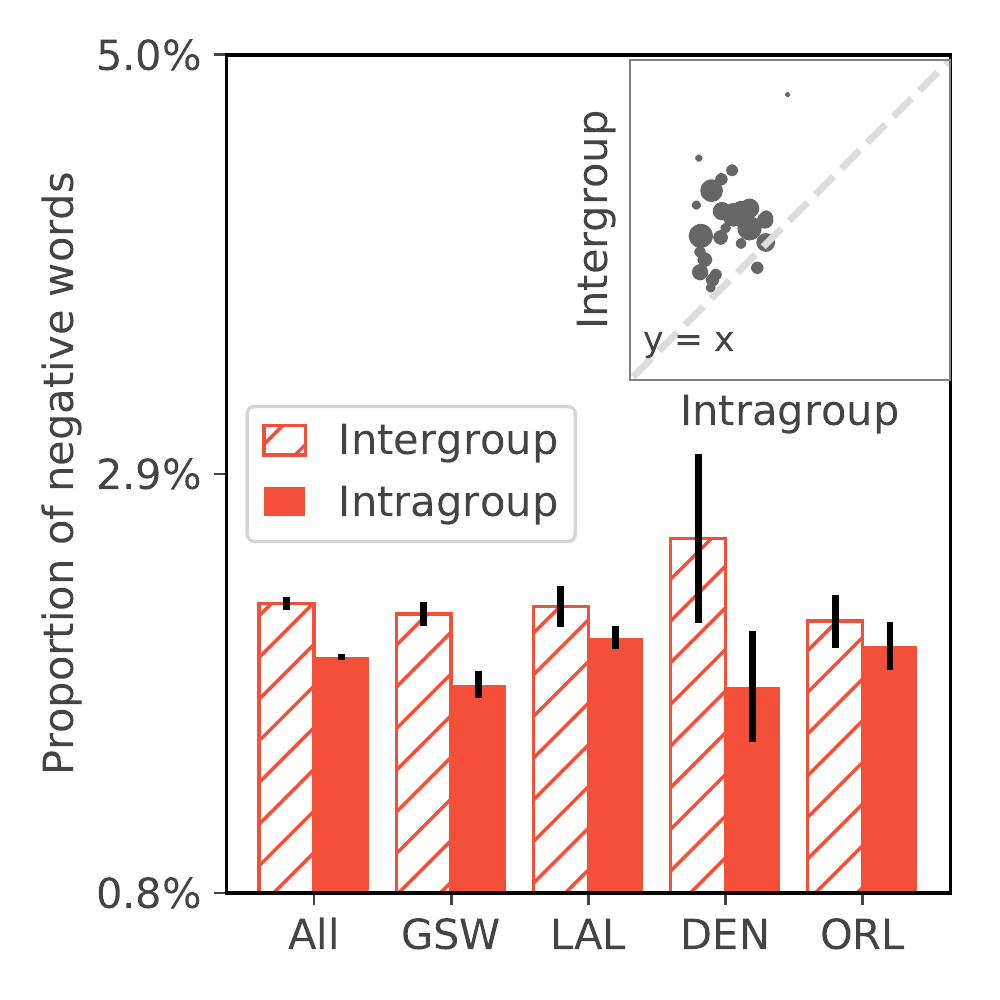}
        \caption{Proportion of negative words.}
        \label{fig:interneg2016}
    \end{subfigure}
    \hfill
    \begin{subfigure}[t]{0.32\textwidth}
        \includegraphics[width=\textwidth]
        {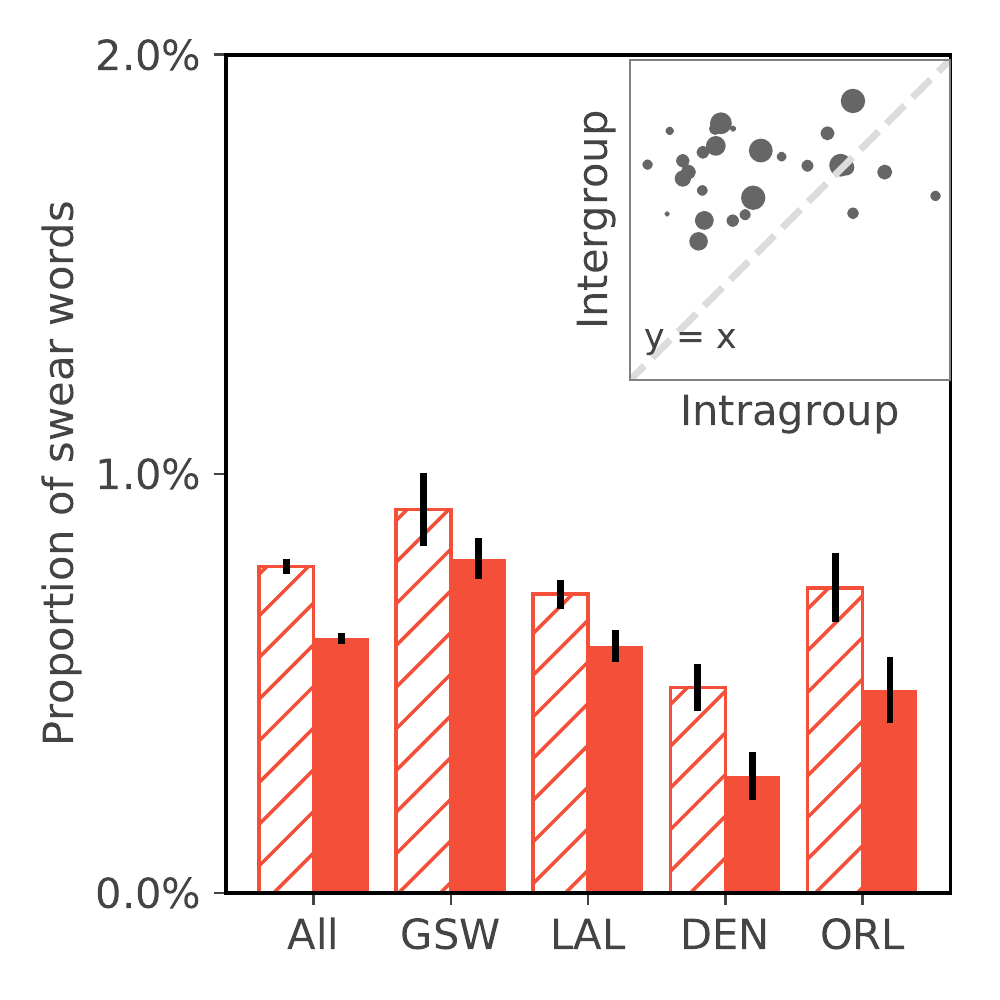}
        \caption{Proportion of swear words.}
        \label{fig:interswear2016}
    \end{subfigure}
    \hfill
    \begin{subfigure}[t]{0.32\textwidth}
        \includegraphics[width=\textwidth]
        {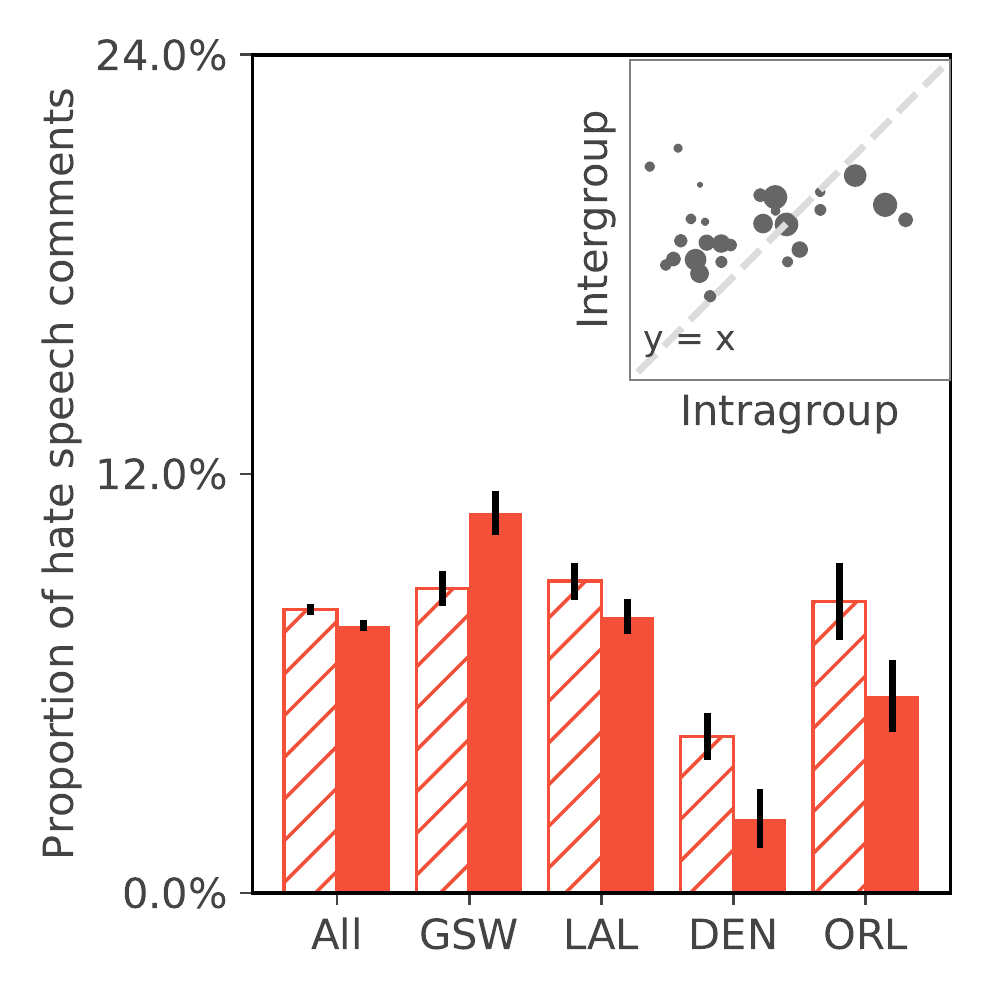}
        \caption{Proportion of hate speech comments.}
        \label{fig:interhate2016}
    \end{subfigure}
    \caption{\Intergroupusers use more negative language in the intergroup setting
    than in the intragroup setting in the 2016 season. 
    They use more negative words
    (two-tailed t-test, $t=7.21$, $p<0.001$, 95\% CI=0.20\% to 0.34\%; 
    28 out of 30 teams, two-tailed binomial test $p<0.001$)
    and swear words 
    (two-tailed t-test, $t=12.22$, $p<0.001$, 95\% CI=0.23\% to 0.32\%; 
    27 out of 30 teams, two-tailed binomial test $p<0.001$)
    and generate more hate speech comments
    (two-tailed t-test, $t=2.17$, $p=0.030$, 
    95\% CI=0.04\% to 0.89\%; 
    22 out of 30 teams, two-tailed binomial test $p=0.016$).
    Error bars represent standard errors.
    }
    \label{fig:intersentiment2016}
\end{figure}

\end{document}